\documentclass[aps,prx,twocolumn,superscriptaddress,floatfix]{revtex4-2}
\usepackage[table,svgnames]{xcolor}
\definecolor{darkpink}{rgb}{1, 0.3, 0.4}
\definecolor{darkgreen}{rgb}{0, 0.5, 0}
\definecolor{darkblue}{rgb}{0, 0, 0.5}
\usepackage[colorlinks = true, urlcolor = darkpink, linkcolor = NavyBlue, citecolor = darkgreen, bookmarks = false]{hyperref}
\usepackage{amsfonts,amsmath,amssymb,mathrsfs}
\usepackage{color, colortbl}
\usepackage{graphicx}
\usepackage{dsfont}
\usepackage[normalem]{ulem}
\usepackage{adjustbox}
\usepackage[skins,theorems]{tcolorbox}
\usepackage{framed}
\usepackage{stmaryrd}

\tcbset{highlight math style={enhanced,
  colframe=teal,colback=white,arc=0pt,boxrule=1pt}}

\makeatletter
\def\@ssect@ltx#1#2#3#4#5#6[#7]#8{%
  \def\H@svsec{\phantomsection}%
  \@tempskipa #5\relax
  \@ifdim{\@tempskipa>\z@}{%
    \begingroup
      \interlinepenalty \@M
      #6{%
       \@ifundefined{@hangfroms@#1}{\@hang@froms}{\csname @hangfroms@#1\endcsname}%
       {\hskip#3\relax\H@svsec}{#8}%
      }%
      \@@par
    \endgroup
    \@ifundefined{#1smark}{\@gobble}{\csname #1smark\endcsname}{#7}%
  }{%
    \def\@svsechd{%
      #6{%
       \@ifundefined{@runin@tos@#1}{\@runin@tos}{\csname @runin@tos@#1\endcsname}%
       {\hskip#3\relax\H@svsec}{#8}%
      }%
      \@ifundefined{#1smark}{\@gobble}{\csname #1smark\endcsname}{#7}%
      \addcontentsline{toc}{#1}{\protect\numberline{}#8}%
    }%
  }%
  \@xsect{#5}%
}%
\makeatother

\usepackage{braket}
\usepackage{slashed}
\usepackage{mathtools}
\usepackage{hyperref}
\usepackage{multibib}
\usepackage{enumerate}
\usepackage{booktabs}

\usepackage{pdfpages} 
\makeatletter
\AtBeginDocument{\let\LS@rot\@undefined}
\makeatother

\usepackage{tikz}
\usetikzlibrary{arrows,topaths,shapes.geometric,calc,decorations.markings,decorations.pathmorphing,shadows}
\usepackage{pgfplots}

\def\expect#1{\langle#1\rangle}

\def\ol#1{\bar{#1}}
\newcommand{\ii}{{\rm i}}
\newcommand{\dd}{{\rm d}}

\newcommand{\xx}{{\rm x}}
\newcommand{\yy}{{\rm y}}
\newcommand{\zz}{{\rm z}}
\newcommand{\uu}{{\rm u}}
\newcommand{\vv}{{\rm v}}

\newcommand{\llb}{{\llbracket}}
\newcommand{\rrb}{{\rrbracket}}

\pgfplotsset{compat=1.18}

\begin{document}

\title{Universal anomalous fluctuations in charged single-file systems}

\author{\v{Z}iga Krajnik}
\affiliation{Faculty for Mathematics and Physics,
University of Ljubljana, Jadranska ulica 19, 1000 Ljubljana, Slovenia}
\affiliation{Department of Physics, New York University, 726 Broadway, New York, NY 10003, United States}

\author{Johannes Schmidt}
\affiliation{Technische Universit\"at Berlin, Institute for Theoretical Physics, Hardenbergstr. 36, D-10623 Berlin, Germany}
\affiliation{Bonacci GmbH, Robert-Koch-Str. 8, 50937 Cologne, Germany}

\author{Vincent Pasquier}
\affiliation{Institut de Physique Th\'{e}orique, Universit\'{e} Paris Saclay, CEA, CNRS UMR 3681, 91191
Gif-sur-Yvette, France}

\author{Toma\v{z} Prosen}
\affiliation{Faculty for Mathematics and Physics,
University of Ljubljana, Jadranska ulica 19, 1000 Ljubljana, Slovenia}

\author{Enej Ilievski}
\affiliation{Faculty for Mathematics and Physics,
University of Ljubljana, Jadranska ulica 19, 1000 Ljubljana, Slovenia}

\begin{abstract}

Introducing a general class of one-dimensional single-file systems (meaning that particle crossings are prohibited) of interacting hardcore particles {with internal degrees of freedom (called charge)}, we exhibit a novel type of dynamical universality reflected in anomalous statistical properties of macroscopic fluctuating observables such as charge transfer. We find that stringent dynamical constraints lead to universal anomalous statistics of cumulative charge currents manifested both on the timescale characteristic of typical fluctuations and also in the rate function describing rare events.
By computing the full counting statistics of net transferred charge between two extended subsystems, we establish a number of unorthodox dynamical properties in an analytic fashion. Most prominently, typical fluctuations in equilibrium are governed by a universal distribution that markedly deviates from the expected Gaussian statistics, whereas large fluctuations are described by an exotic large-deviation rate function featuring an exceptional triple critical point. Far from equilibrium, competition between dynamical phases leads to dynamical phase transitions of first and second order and spontaneous breaking of fluctuation symmetry of the univariate charge large-deviation function.
The rich phenomenology of the outlined dynamical universality is exemplified on an exactly solvable classical cellular automaton of charged hardcore particles. We determine the dynamical phase diagram in the framework of Lee--Yang's theory of phase transitions and
exhibit a hyper-dimensional diagram of distinct dynamical regimes. Our findings lead us to conclude that the conventional classification of dynamical universality classes based on the algebraic dynamical exponents and asymptotic scaling functions that characterize hydrodynamic relaxation of the dynamical structure factor is incomplete and calls for refinement.
\end{abstract}

\maketitle

\tableofcontents

\section{Introduction}
\label{sec:introduction}

Explaining the microscopic foundation and providing a complete classification of macroscopic phenomenological laws that
emerge from highly complex time-reversible evolution laws still remains one of the central unrealized goals in statistical physics.
At the heart of this endeavor one encounters \emph{universality}, a notion which in the broadest sense signifies
loss of microscopic information at an emergent macroscopic scale. The concept of dynamical universality usually refers to
an effective equation of motion governing the late-time relaxation of conservation laws on a hydrodynamic scale
in which microscopic model-dependent details are almost entirely washed out, only entering implicitly through coupling coefficients.
The most prominent and widespread example is Fick's law of normal diffusion which is most
commonly understood as a coarse-grained description of randomly walking Brownian particles.
The law of diffusion is nevertheless omnipresent, arising in a wide array of different chaotic dynamical systems with one or few conservation laws (e.g. energy, particle or charge conservation). Diffusive dynamics of a globally conserved quantity $Q$
is conventionally characterized on the basis of the dynamical structure factor,
$S(x,t) \equiv \expect{q(x,t)q(0,0)}^{c}$ of charge density $q(x,t)$,
where $\expect{\bullet}^{c}$ denotes the connected correlation function in a stationary equilibrium state.
More precisely, a conserved charge $Q$ is said to undergo normal diffusion in $d$ spatial dimensions whenever the corresponding $S(x,t)$
(normalized by static charge susceptibility $\chi$) exhibits asymptotic decay on a characteristic timescale $t \sim x^{z}$
with a Gaussian scaling profile and $z=2$.

Ubiquity of normal diffusion in non-relativistic systems is not particularly striking as it essentially only rests on the assumption that 
the flux density is proportional to the gradient of charge density. We nevertheless know of numerous exceptions to this rule,
particularly in one spatial dimension. Presently we know of several universal laws characterized by algebraic dynamical exponent different from $z=2$. For large times, the dynamical structure factor (DSF) admits an asymptotic decay of the form
$S(x,t) \simeq (\lambda\,t)^{-1/z} f_{\rm sc}[(x-vt)/(\lambda\,t)^{1/z}]$, for some (typically non-Gaussian) stationary scaling function $f_{\rm sc}$, algebraic dynamical exponent $z$,
coupling parameter $\lambda$ and drift velocity $v$.
One of the most celebrated examples of a non-diffusive dynamics is the universality class of the Kardar--Parisi--Zhang equation \cite{KPZ}
often found in models of growing interfaces in one spatial dimension. The KPZ universality class is associated with a superdiffusive exponent
$z = 3/2$ and universal function $f_{\rm PS}$ obtained by Pr\"{a}hofer and Spohn \cite{Prahofer}. On the other hand,
much richer behavior can be found in systems with multiple local conservation laws.
The mode-coupling theory of nonlinear fluctuating hydrodynamics \cite{MendlSpohn15,Schutz18} indeed predicts an infinite family of universal superdiffusive exponents in the range $3/2 \leq z < 2$, coinciding with `Kepler ratios'
formed by consecutive Fibonacci numbers \cite{Popkov15,Schutz18}. The associated scaling functions $f_{\rm sc}$ can be either
Pr\"{a}hofer--Spohn or one of the stable Levy distributions.
Hence, algebraic dynamical exponents by themselves do not uniquely determine a dynamical universality class.
In homogeneous models, subdiffusive dynamical exponents $z>2$ are less common; they can appear, for example,
in certain types of models with non-trivial kinetic constraints \cite{Singh21}, yielding a fractional exponent
$z = 8/3$, or in systems with conserved multipole charge moments \cite{Sala20,Morningstar20,Zhang20,Iaconis21} and `fracton matter'
\cite{Gromov20,Grosvenor21}.

Integrable systems display a somewhat exceptional transport behavior.
Owing to ballistically propagating quasiparticles, the conserved charges generically spread with a ballistic exponent $z=1$.
The variance of the DSF $S(x,t)$ thus grows as $\sim t^{2}$, while the variance's magnitude defines the
Drude weight \cite{transport_review}. The asymptotic scaling function on the Euler scale depends in general on the spectrum of effective velocities through the model's dispersion relation and therefore does not assume any universal form. Nonetheless, it admits a universal mode decomposition \cite{DS17,IN17} expressed in terms of the state functions of generalized hydrodynamics \cite{CDY16,BCDF16}.
Charges associated with  global continuous (N\" other) symmetries behave exceptionally: in states invariant under charge conjugation
(or particle-hole symmetry), the Drude peak exactly vanishes \cite{IN_Drude} and in interacting systems one typically finds a Gaussian asymptotic form of $S(x,t)$ with variance growing linearly with time, apart from exceptional situations that are inherently linked with unbroken non-abelian symmetries that give rise to the `KPZ physics' cf. \cite{Ilievski18,Ljubotina19,PhysRevLett.123.186601,superuniversality} and
a review \cite{superdiffusion_review}. Hence, normal charge diffusion is not a priori incompatible with strong ergodicity breaking and may occur even in integrable interacting systems.

{\em Revisiting dynamical universality.}---{The notion of diffusive dynamics is most commonly identified with the
Gaussian scaling form of $S(x,t)$ associated to the local density of a conserved quantity computed in an appropriate time-invariant measure (state).
Here we argue, however, that hydrodynamic decay of density fluctuations encoded in the dynamical structure factor is not always
adequate for diagnosing dynamical universality: the dynamical exponent and asymptotic scaling form of $S(x,t)$ alone are not (in general) 
enough for an unambiguous classification of nonequilibrium universality classes. To resolve this shortcoming, we propose to study statistical properties of macroscopic fluctuating quantities. We provide concrete examples of dynamical systems which despite featuring the dynamical exponent $z=2$ and a Gaussian DSF, regarded as the hallmark properties of diffusive dynamics, display distinctly non-diffusive behavior that goes beyond dynamical two-point functions. As we demonstrate, in certain types of systems one finds that universal properties of charge dynamics only become manifest at the level of fluctuating macroscopic (i.e. extensive) observables such as cumulative currents that encode the full counting statistics (FCS) of charge transfer. The current operational definition of normal diffusion is therefore incomplete. That is, whenever the late-time (i.e. hydrodynamic) behavior of dynamical correlations comprising charge or current densities
on the characteristic diffusive timescale deviate from those found in the stochastic diffusion equation it is not legitimate to speak of normal (Gaussian) diffusion. Another necessary condition for normal diffusion is, for instance, the validity of the central limit property,
signifying that (at late times on a typical timescale $t^{1/2z}$) the transmitted charge between two semi-infinite halves of the system
yields a Gaussian distribution. We may in principle further demand that statistical properties of macroscopic fluctuating observables associated with exponentially rare events also behave universally at late times. Indeed, the macroscopic fluctuation theory (MFT) stipulates that
the large-deviation rate function of chaotic dynamics with a single conservation law corresponds to the variational minimum of a \emph{universal} `MFT action' \cite{Bertini01,Bertini02} that only takes as an input the diffusivity and mobility as functions of the equilibrium charge density.}

Finally, it is important to stress that many of the nuances that pertain to dynamical universality are in fact rather general, i.e. not limited
to the phenomenon of diffusion. For example, the centered time-integrated current in stationary states of the KPZ universality class is distributed according to the Baik--Rains distribution \cite{BaikRains} with a finite skewness \cite{Takeuchi18} such as e.g. in the Nagel-Schreckenberg model \cite{deGier05}. By contrast, detailed balance guarantees that superdiffusive (N\"{o}ther) charges (which are universal in integrable systems with non-abelian symmetries \cite{superuniversality,MatrixModels}) instead have symmetrically distributed fluctuations  \cite{KIP22}.

{\em Anomalous fluctuations in single-file systems.}---Despite a widespread belief that in generic, strongly chaotic systems
statistical properties of conserved quantities obey the \emph{central limit property}, that is exhibit Gaussian fluctuations on the typical (diffusive) timescale, we currently lack an analytic proof (or even empirical evidence) to confirm this expectation. At present, we
only know of certain formal sufficient requirements \cite{Bryc93,Jaksic12} which are, however, difficult to explicitly verify in practice. It appears plausible that \emph{ergodicity} provides a sufficient condition for the onset of normal diffusion (apart from $z=2$ and Gaussian DSF which are necessary). Whether ergodicity is necessary is less obvious, however.

In this work, we explore a different route and examine the role of ergodicity breaking. Specifically, our aim is to study
statistical properties of charge fluctuations that arise due to strict \emph{kinetic constraints},
for both diffusive and ballistic (i.e. integrable) particle dynamics. Most strikingly, we demonstrate that lack of ergodicity can have profound consequences for the central limit property. To advance our standpoint, we introduce and examine a general class of one-dimensional models if hardcore classical particles equipped with positive or negative charge. By employing rigorous analytic techniques,
we explicitly compute the full counting statistics of joint particle-current fluctuations, thereby unveiling
a surprisingly rich phenomenology of charge transport at the level of fluctuations. Specifically, the models under consideration
are distinguished by two defining properties: (i) a \emph{`single-file constraint'} and (ii) \emph{`charge inertness'}. The former implies that trajectories of particles cannot cross, while the latter signifies that (internal) charge degrees of freedom carried by the particles are uncorrelated with their trajectories. 

The study of ergodicity-breaking phenomena has been a fruitful area of theoretical research
in recent years, including the effects of kinetic constraints \cite{Sala20,Gromov20,Grosvenor21,Feldmeier22}.
The models we consider in this work indeed belong to a wider class of so-called `pattern-conserving systems' that feature
classical phase-space fragmentation, i.e. foliation of the configuration space into exponentially many decoupled dynamical sectors.
The other defining property, namely inertness of charge, refers to absence of dynamics in the internal (charge) space and its decoupling from particle dynamics. While such charged single-file systems are generically non-integrable they include, as a subclass,
exactly solvable models with free or interacting ballistic particles. In the latter case, the single-file property
implies that charge experiences a slowdown and spreads on a diffusive timescale. A particularly simple example of such dynamics is realized by a classical reversible deterministic cellular automaton describing interacting charged particles. In our previous work \cite{Krajnik22}, we have obtained the exact FCS of the transferred charge in equilibrium. Quite remarkably, it turns out that the probability distribution of the cumulative charge current is \emph{non-Gaussian}, indicating violation of the central limit property. In this paper, we provide a more comprehensive understanding of this unorthodox behavior and its intimate relation to absence of `regularity conditions'. Moreover, we establish its universality for the entire family of charge single-file systems and proceed to demonstrate that consequences of the lack of regularity are even more pronounced away from equilibrium.

Dynamical properties of kinetically constrained systems have been recently investigated in
ref.~\cite{Feldmeier22} for a broad class of systems. Focusing on the two-point function, it is shown that pattern conservation in generic (i.e. non-integrable) systems results in a Gaussian distribution of a tracer particle, albeit with variance growing subdiffusively as $\sigma^{2}(t) \asymp 2\sqrt{D\,t}$. Notably, the latter has to be distinguished from the `canonical' subdiffusion with dynamical exponent $z=4$ which arises
in dipole conserving systems without pattern conservation. In contrast, charge correlators in integrable systems instead
decay with exponent $z=2$ and Gaussian scaling profile of variance  $\sigma^{2}(t) \asymp 2 D\,t$
indicative of normal diffusion and `diffusion constant' $D$. A more detailed information about charge transport, e.g. the
statistics of charge transfer captured by the FCS, is however still lacking at the moment.
To fill this gap, we here carry out a comprehesive examination of charge single-file systems.

{\em Universal properties of charged single-file systems.}---The main technical contribution of this work is the exact form of finite-time moment generating function for the full counting statistics of joint particle-charge transfer away from equilibrium. With aid of asymptotic analysis, we analyze its late-time behavior and deduce the stationary probability density of charge fluctuations in equilibrium. We discover a number
of anomalous dynamical properties at the fluctuating level.
{Depending on the algebraic exponent of particle dynamics, the models belong to one of the two subclasses:
integrable (free or solitonic) systems with a diffusive charge DSF or generic diffusive particle systems (governed by either deterministic
or stochastic dynamical laws) with a Gaussian subdiffusive charge DSF.} Both of the subclasses possess divergent scaled cumulants, signifying inapplicability of the MFT \cite{MFT} and its ballistic counterpart \cite{DoyonMyers20,MBHD20,PerfettoDoyon21,BMFT} to describe fluctuations on the typical scale.
Most prominently, we find that charge fluctuations behave anomalously on both typical and large scales, which
we establish on very general grounds and independently of a model's microscopic details.
On typical timescale, we find that equilibrium charge transport in ballistic charge single-file systems  differs from normal (Fickian) diffusion in a fundamental way: in absence of charge bias charge undergoes diffusive relaxation (described by a Gaussian DSF) while statistics of charge fluctuations reveal several dynamical features that go beyond normal diffusion.

{\em Possible generalizations.}---The recent studies of anomalous fluctuations in certain integrable quantum spin chains
featuring kink excitations \cite{KIP22,Sarang_FCS,KSIP23} indicate that inertness of charge may not be essential for some
of the universal properties encountered in charged single-file systems. The inertness condition is quite restrictive and dropping it would allow us to encompass a much broader class of single-file models. The core purpose of our study is however
to provide a rigorous analysis, and the main advantage of imposing the inertness of charge is that it permits us derive explicit closed-form results without resorting to any approximation. Relaxing the inertness condition is nonetheless an important direction of study for future research.

{\em Note added.}---Shortly after the completion of this work, a related work \cite{Kormos22} appeared  that
reports the exact FCS for a class of integrable models by employing an effective low-energy theory based on a semiclassical approximation,
obtaining the same non-Gaussian typical distribution of the transmitted soliton charge (property \ref{prop1}) transported by kinks.
The same distribution has indeed been retrieved earlier in ref. \cite{Altshuler06} using form-factor techniques.
Kink scattering in such an effective semiclassical theory is purely reflecting and hence such systems belong to the class of charged single-file systems.
All other listed properties, however, have not (to our knowledge) been noticed or discussed so far in the literature
(aside from property \ref{prop4}, which we inferred in a particular model in our previous work \cite{Krajnik22}).

\medskip

\paragraph*{Outline.}
The paper is structured as follows. In Section \ref{sec:setting} we first introduce a novel class of classical one-dimensional systems comprising hard-core particles carrying internal degrees of freedom that are subjected to a non-crossing (i.e.~single-file) constraint. To set the stage we proceed by outlining the general setting which includes a quick introduction to the basic concepts of
the large-deviation principle and the Gallavotti--Cohen relation.  For concreteness, we also briefly discuss two simple representative examples. Section \ref{sec:results} contains an exposition of our main results. We open the section by a non-technical overview of the most salient universal features and subsequently zero in on various aspects. We first outline the `dressing formalism' and provide a succinct summary
of key results inferred from asymptotic analysis, independently at the level of the moment generating and large-deviation rate functions. We close the section by discussing several remarkable interrelated features, including universal non-Gaussian fluctuations in equilibrium and several competing coexisting dynamical phases.
Section \ref{sec:hardcore} is devoted to exhibiting our formalism on a concrete model. We choose an exactly solvable classical cellular automaton, which has a remarkably simple (analytic) structure. We conclude in Section \ref{sec:conclusion} by summarizing the key results and briefly discuss why our results can likely be generalized further by relaxing the underlying assumptions.
We also include appendices \ref{sec:exact_FCS}, \ref{sec:dressing_technical}, \ref{sec:LeeYang}, \ref{sec:equal_densities},
containing additional technical details and derivations.

\section{Setting and background}
\label{sec:setting}

We begin by outlining the general setting and introducing the basic concepts of fluctuation theory.
We then proceed with a concise summary of the main results and discuss the key physical features.

We specialize this study exclusively to classical dynamical systems of interacting distinguishable particles in one spatial dimension
that conserve the number of particles. Accordingly let $x_{\ell}=x_{\ell}(t)$ label the trajectory of the $\ell$-th particle
and ${\bf x}(t)=(x_{1}(t),x_{2}(t),\ldots,x_{Q_p}(t))$ be the position vector of $Q_p$ particles at time $t$. Space and time can be
continuous or discrete (our main working example, introduced below in Sec.~\ref{sec:models}, is a fully discrete model).
In addition, particles carry \emph{discrete} internal degrees of freedom ${\rm c}_{\ell}$, e.g. charge or color, which can take $N$ discrete
values. For definiteness, in this paper we adopt binary charges, ${\rm c}_{\ell}\in \mathbb{Z}_{2}$.
Charges are assigned to particles in an unconstrained manner.
{While there exist generalizations to continuous internal degrees of freedom, in the following we consider
exclusively the discrete binary case.}

The dynamics can be either deterministic (not necessarily Hamiltonian) or stochastic.
In deterministic models, dynamics of particles is governed by the time-local bijective phase-space mapping (propagator) $\mathcal{U}_{t}$, formally expressed as ${\bf x}(t)=\mathcal{U}_{t}\circ {\bf x}(0)$, that constitutes a group,
$\mathcal{U}_{t+t'}=\mathcal{U}_{t} \, \circ \, \mathcal{U}_{t'}$,
for both continuous ($t\in\mathbb R$) or discrete ($t\in\mathbb Z$) times; in the discrete time case the dynamics corresponds to iteration of the one-step propagator $\mathcal{U}_{1}$. For definiteness, we shall confine ourselves to dynamical systems with short range interaction.

More importantly, we now impose two additional dynamical constraints:
\begin{framed}
\begin{enumerate}[(I)]
\item {\tt single-file condition}: particles are not allowed to jump across (or pass by) one another, meaning that their
trajectories are subject to the ordering condition
\begin{equation}
x_{\ell-1}(t) < x_{\ell}(t)
\label{eqn:single_file}
\end{equation}
at all times $t$ and for all $\ell \in \{2,\ldots,Q_p\}$.
\item {\tt charge inertness}: charge degrees of freedom remain attached to particles at all times and have no dynamics of their own.
Therefore, the ${\rm c}_{\ell}$ carried by $\ell$-th particle does not depend on time and the total charge is conserved.
Equipping particles with charges does not affect the underlying dynamics of particles governed by the propagator $\mathcal{U}_{t}$.
\end{enumerate}
\end{framed}
In conjunction properties (I) and (II) imply that \emph{any} configuration of charges ${\bf c}=({\rm c}_{1},{\rm c}_{2},\ldots,{\rm c}_{Q_p})$ in an initial configuration remains preserved throughout the time evolution. In this view, interaction among the charged particles can be regarded as `purely reflective'. Inertness of charge can be also perceived as an extreme version of `charge separation' with trivial (charge) dynamics.

Neglecting the internal degrees of freedom, condition (I) is the defining property of the so-called \emph{single-file dynamics}
(see e.g. Refs.~\cite{Krapivsky14,Krapivsky15,Krapivsky_tagged,Imamura17} discussing the exact FCS), referring in general to quasi-1D systems of particles clogged in a narrow channel (so that they are unable to overtake each other). Conservation of the initial charge pattern implies that the entire phase space foliates into $2^{Q_p}$ dynamical subsectors, closely related to the notion of Hilbert space fragmentation studied in the context of certain non-ergodic quantum dynamics \cite{Khemani20,Sala20}.

Separation of charge from matter permits us to integrate out the charge degrees of freedom in an exact analytic fashion and will thus be of pivotal importance for computing the moment generating function (MGF) encoding the statistics of charge transfer.

\medskip

\paragraph*{Time-reversal symmetry.}
We additionally demand the time evolution to be \emph{time-reversible}: in deterministic dynamics, this requires an involutive phase-space mapping $\Theta$ such that
\begin{equation}
\Theta \circ \mathcal{U}_{t}=\mathcal{U}_{-t}\circ \Theta.
\end{equation}
In stochastic systems, described by a master equation $\dd P_{t}/\dd t = \mathscr{M} P_{t}$ with a Markov operator $\mathscr{M}$,
Time-reversal invariance follows directly from \emph{detailed balance} (see e.g. \cite{Mallick15})
\begin{equation}
\mathscr{M}(C,C^{\prime})P_{\rm eq}(C^{\prime}) = \mathscr{M}(C^{\prime},C)P_{\rm eq}(C),
\end{equation}
where $C$ describes the complete configuration of the system (phase space point), e.g. $C=(\bf{x},\bf{c})$, and $M(C^{\prime},C)\dd t$ is the probability of transition $C\to C^{\prime}$ during $\dd t$
and $P_{\rm eq}$ is the stationary probability measure, $(\dd/\dd t) P_{\rm eq} = 0$, representing e.g. the Gibbs equilibrium
state, $P_{\rm eq}\simeq \exp{(-\beta\,E(C))}$. By virtue of detailed balance, the probability of
finding a `forward' trajectory $C_{\rightarrow}(t)$ is equal to that of the time-reversed trajectory $C_{\leftarrow}(t)$, that is
$\mathbb{P}(C_{\rightarrow}(t))=\mathbb{P}(C_{\leftarrow}(t))$.

\medskip

\paragraph*{Grand-canonical equilibrium.}

The total particle number and total charge, denoted by $Q_{p}=\sum_\ell 1$ and $Q_{c}=\sum_\ell c_\ell$ respectively, are two general conserved quantities of our models
(expressible as spatial sums of local densities). There could in principle be additional local conservation laws, which nevertheless have no impact on our findings.
By coupling the two conserved charges to chemical potentials $\beta_{p}$ and $\beta_{c}$, we consider grand-canonical ensembles corresponding to the stationary measure $P_{\rm gc} = \mathcal{Z}^{-1} \exp{(\beta_{p}Q_{p}+\beta_{c}Q_{c})}$, normalized by the partition sum $\mathcal{Z}$. Since both conserved quantities $Q_{p,c}$ represent homogeneous sums of local (one-site) densities, the partition sum factorizes into one-body terms $\mathcal{Z}=\mathcal{Z}^{N}_{1}$, $\mathcal{Z}_{1}=\exp(f_{\rm eq})$, where $f_{\rm eq}$ is free energy density per unit length,
\begin{equation}
f_{\rm eq}(\beta_{p},\beta_{c}) = \log{[1+e^{\beta_{p}}(e^{\beta_{c}}+e^{-\beta_{c}})]}.
\end{equation}
For our convenience, we shall parametrize equilibrium states by specifying averages,
$\rho \equiv \expect{{\rm q}_{p}}_{\rm gc} = \partial_{\beta_{p}} f_{\rm eq}$, and
$\rho\,b \equiv \expect{{\rm q}_{c}}_{\rm gc} = \partial_{\beta_{c}}f_{\rm eq}$, related to chemical potentials via
\begin{equation}
\beta_{p} = \log{\left[\frac{\rho\sqrt{1-b^{2}}}{2(1-\rho)}\right]},\qquad
\beta_{c} = \frac{1}{2}\log{\left[\frac{1+b}{1-b}\right]}.
\label{eqn:chemical_potentials}
\end{equation}
By inverting the second relation, we have $b=\tanh{(\beta_{c})}$.
The covariance matrix $\chi$ of (static) charge susceptibilities is a $2\times 2$ matrix with matrix elements
$\chi_{ij}=\partial_{\beta_{i}}\partial_{\beta_{j}}f_{\rm eq}(\beta_{p},\beta_{c})$ (for $i,j \in \{p,c\}$), reading
\begin{equation}
\chi =
\begin{pmatrix}
\chi_{pp} & \chi_{pc} \\
\chi_{cp} & \chi_{cc}
\end{pmatrix}
=
\begin{pmatrix}
\rho(1-\rho) & \rho(1-\rho)b \\
\rho(1-\rho)b & \rho - (\rho b)^{2}
\end{pmatrix}.
\end{equation}

\paragraph*{Full counting statistics: bipartioning protocol.}

There exist two widely popular settings to compute the full counting statistics of cumulative currents far away from equilibrium.
In the mesoscopic approach, a system of finite length is attached between two effective baths which drive it away from equilibrium.
At late times, the system relaxes into a stationary current-carrying state, and one can monitor the net charge transfer during a finite interval of time. To eliminate the role of stochastic baths, we shall use an alternative technique.
In this work, we consider infinitely extended systems prepared initially in a nonstationary initial state consisting of two thermalized semi-infinite partitions joined at the origin that are released to evolve under a time-reversal invariant equation of motion. Each partition is initialized in a stationary ensemble characterized by a finite density of particles and charge. The density of particles in the left (right) partition is $\rho_{-}, (\rho_+) \in [0,1]$, while $\ol{\rho}_{\pm}\equiv 1-\rho_{\pm} \in [0,1]$ are the corresponding densities of vacancies. Similarly, we introduce the biases $b_{\pm} \in [-1,1]$, such that $\rho_{\pm}(1+b_{\pm})/2$ are the densities of positively charged particles in the left or right partition.

In the general case of unequal densities and biases, $\rho_{-}\neq \rho_{+}$ and $b_{-}\neq b_{+}$, we find a finite net current of particles and charge flowing across the origin, leaving behind an ever-expanding dynamical interface.
The time-integrated particle and charge currents will accordingly grow as $\mathcal{O}(t^{\alpha})$ at late times, for some (model-dependent) exponent $\alpha\le 1$.

\subsection{Representative models}
\label{sec:models}

The entire class of models that respect the defining conditions (I) and (II) (as specified in Sec.~\ref{sec:introduction})
display \emph{universal} anomalous fluctuations of macroscopic transferred charge.
Before summarizing our key findings in Sec.~\ref{sec:results}, we wish to emphasize that
all of our main conclusion hold irrespective of the underlying particle dynamics (apart from minor technical assumptions to ensure that fluctuations of particle trajectories behave sufficiently regularly). Before delving into technical aspects,
we find it instructive to introduce two simple models that belong to this particular class of single-file systems.

\begin{figure}[htb]
\includegraphics[width=0.9\columnwidth]{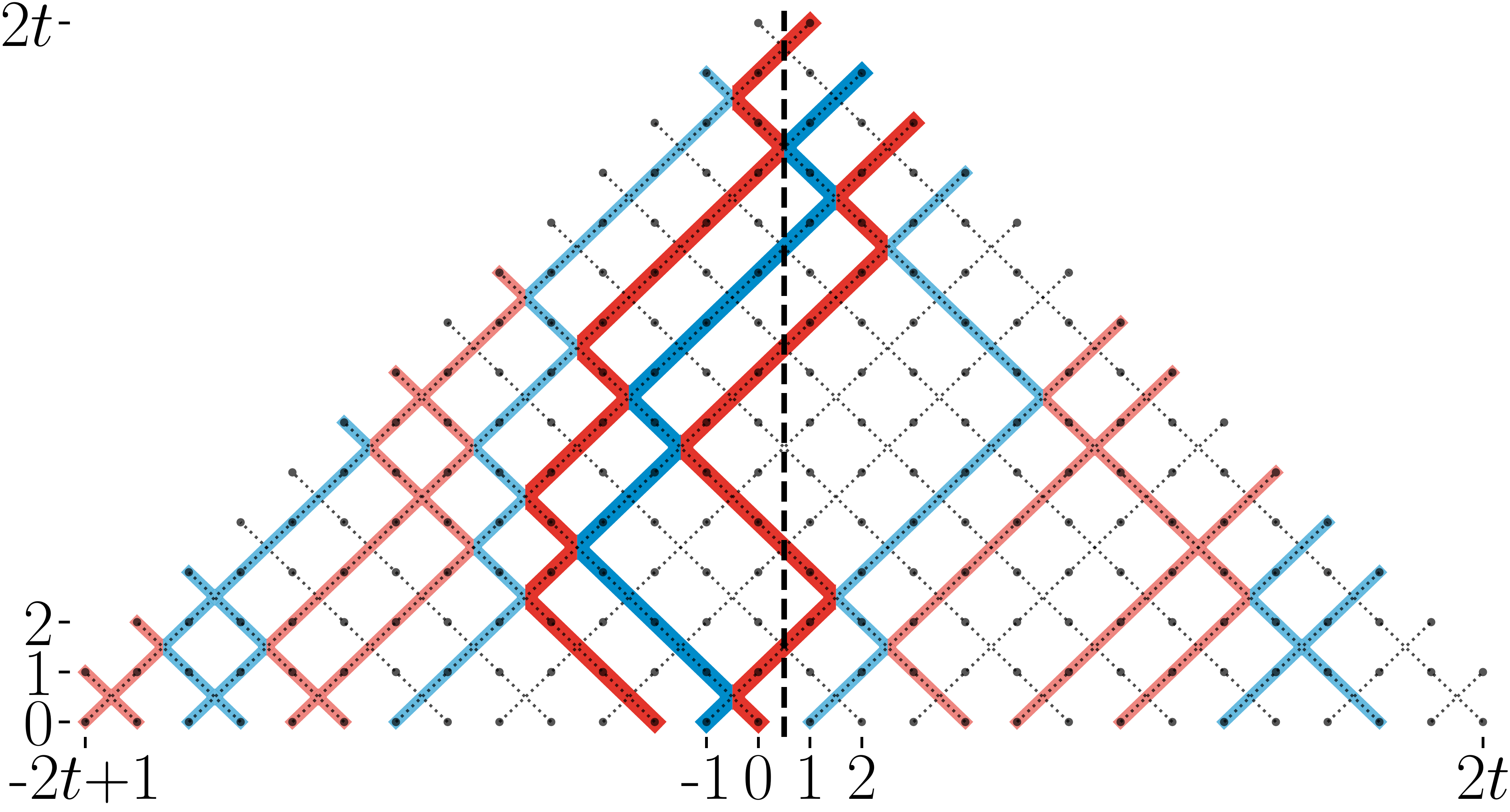}
\caption{Hardcore charged gas in discrete space-time, representing a classical cellular automaton of interacting charged particles
that propagating ballistically between their collisions. Scattering between positively (red) and negatively (blue) charged particles
is purely reflective, yielding the single-file property of non-intersecting particles' worldlines.}
\label{fig:hardcore}
\end{figure}

\medskip

\paragraph*{Exactly solvable hardcore charged gas.}
Arguably the simplest dynamical system complying with all of the above requirements are free, ballistically propagating `matter' particles that carry an internal binary ($\mathbb{Z}_{2}$) degree of freedom (which one may think of as charge or color).
A discrete space-time version of such a model is the \emph{hardcore gas cellular automaton}, illustrated in Fig.~\ref{fig:hardcore}, introduced and studied initially in Ref.~\cite{Medenjak17}
(see also the follow-up works \cite{Klobas2018,Medenjak19,Medenjak_OTOC}).

The $\mathbb{Z}\times \mathbb{Z}$ space-time lattice is occupied with one of the three local states:
a positive  charge ($+$), a negative  charge ($-$), or a charge-neutral vacant site $\emptyset$. The time evolution is realized by
sequential `brickwork' application of local two-body maps given by the following rules,
the `interaction vertex' $({\rm q}_{c},{\rm q}^{\prime}_{c})\leftrightarrow ({\rm q}_{c},{\rm q}^{\prime}_{c})$ for
${\rm q}_{c},{\rm q}^{\prime}_{c}\in \{+,-\}$, and a `free vertex' $(\emptyset,{\rm q}_{c})\leftrightarrow ({\rm q}_{c},\emptyset)$
for ${\rm q}_{c}\in \{+,-,\emptyset\}$ (see Ref.~\cite{Medenjak17} for more precise definitions).
The propagator $\mathcal{U}_{2t}$, consisting of alternating odd $\mathcal{U}^{\rm o}$ and even layers $\mathcal{U}^{\rm e}$
(satisfying $\mathcal{U}^{\rm o} \, \circ \, \mathcal{U}^{\rm o} = \mathcal{U}^{\rm e} \, \circ \, \mathcal{U}^{\rm e}={\rm Id}$),
$\mathcal{U}_{2t}=(\mathcal{U}^{\rm e} \, \circ \, \mathcal{U}^{\rm o})^{t}$, is invariant under  time-reversal, with
$\Theta = \mathcal{U}^{\rm o}$.

The model, which belongs to a family of (super)integrable classical cellular automata \cite{GP22}, exhibits ballistic charge transport with diffusive corrections. The basic properties of charge transport can be inferred from the asymptotic scaling form of
the dynamical structure factor of charge density, $S_{c}(x,t)\equiv\expect{{\rm q}_{c}(x,t){\rm q}_{c}(0,0)}_{\rm gc}$
(normalized by $\sum_{x}\,S_{c}(x,t)=\chi_{cc}$), computed in Ref.~\cite{Klobas2018}
\begin{equation}
S_{c}(x,t) = \rho(1-b^{2})\,\mathscr{G}_{\rm dif}(x,t) + \mathcal{D}_{c}\,\mathscr{G}_{\rm bal}(x,t),
\end{equation}
where $\mathscr{G}_{\rm dif}(x,t)\equiv (2\pi t\,D_{c})^{-1/2} \exp{[-x^{2}/(2D_{c}t)]}$ denotes a central Gaussian peak that
broadens diffusively with diffusion constant $D_{c}=(1-\rho)/\rho$, whereas
$\mathscr{G}_{\rm bal}(x,t)\equiv \tfrac{1}{2}(\delta(x-t)+\delta(x+t))$ are two side ballistic `sound' peaks
whose magnitude corresponds to the charge Drude weight $\mathcal{D}_{c} = \rho(1-\rho)b^{2}$.

The charge diffusion constant $D_{c}$ (determined from the variance of the asymptotic DSF) and the direct current charge conductivity
(related to $D_{c}$ via Einstein's relation) can likewise be retrieved from a quenched inhomogeneous profile or in the
mesoscopic setup with boundary driving (see Refs.~\cite{Medenjak17,Klobas2018,Medenjak19}). The fluctuation-dissipation relation
therefore remains intact.

In order to detect signatures of the fragmented phase space we have to look beyond the DSF $S_{c}(x,t)$. To this end,
we compute the full counting statistics of the cumulative charge current. In our previous study \cite{Krajnik22}, we have established that the stationary PDF of the time-integrated charge current in equilibrium at finite density and no charge bias takes a \emph{non-Gaussian} form.
This result is somewhat at odds with the general prediction of uniformly scaling cumulants in ballistic fluctuation theory for integrable models developed in Refs.~\cite{MBHD20,DoyonMyers20}.

Stimulated by the peculiarities observed in Ref.~\cite{Krajnik22}, in Sec.~\ref{sec:hardcore} we revisit the problem of
anomalous charge-current fluctuations in the hardcore gas and use this opportunity to clarify the theoretical underpinnings behind the observed anomalous behavior in full capacity. By generalizing the computation of the FCS to nonequilibrium states, we map out
an unexpectedly rich phenomenology of the charge-current LDF associated with rare fluctuations.

\begin{figure}[htb]
\includegraphics[width=\columnwidth]{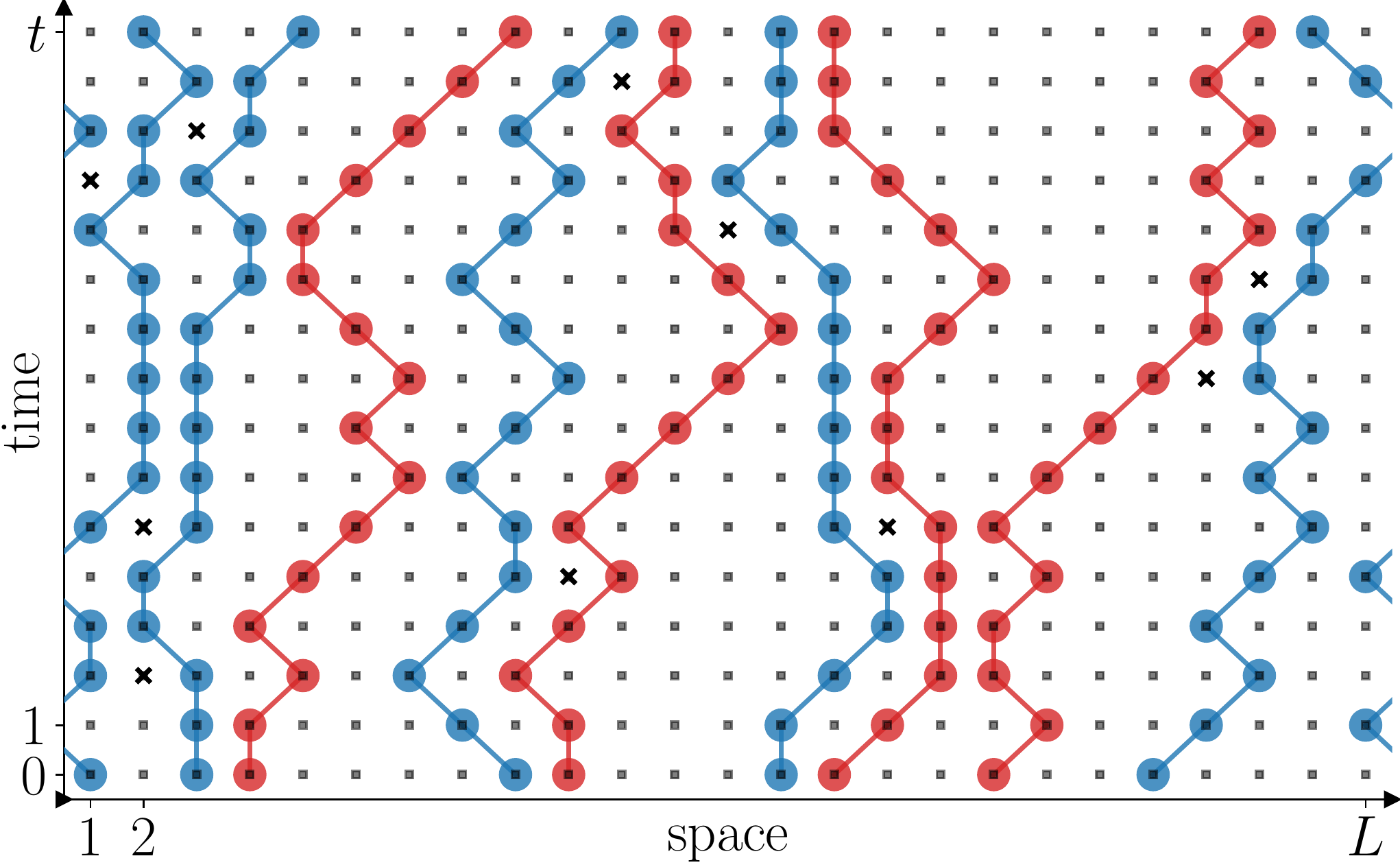}
\caption{Stochastic two-species simple symmetric exclusion process in discrete time with a parallel update scheme: at each timestep,
a particle is activated with a finite probability (here $p_{\rm act}=1$) and jumps in a random unbiased direction to the target neighboring
site provided the target site is unoccupied. Potential conflicts (marked by crosses) are resolved democratically.}
\label{fig:SSEP_dynamics}
\end{figure}

\medskip

\begin{figure}[htb]
\includegraphics[width=\columnwidth]{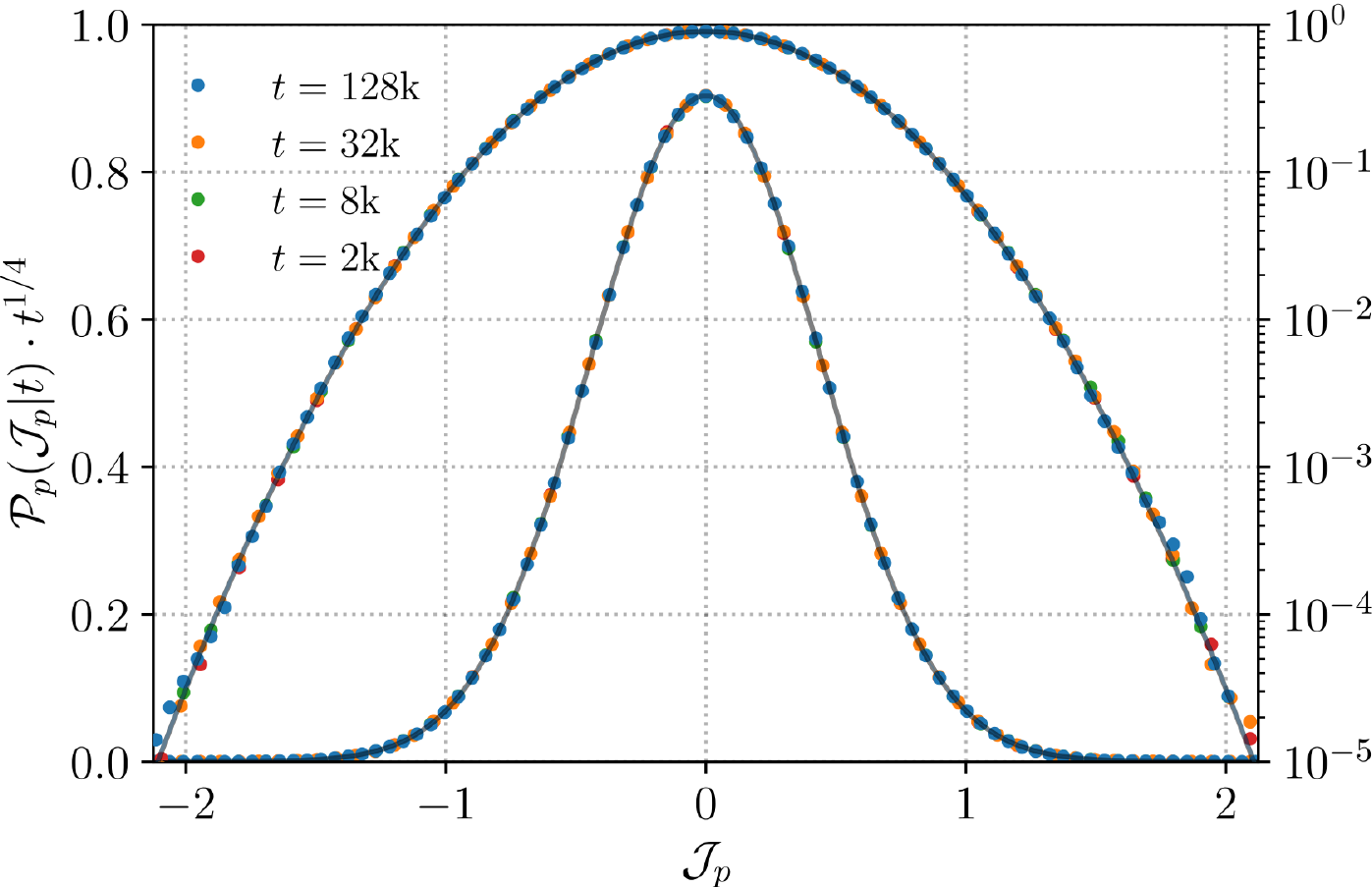}
\caption{Typical fluctuations of the particle transfer in the two-species symmetric simple exclusion process in discrete time with
a parallel update rule, showing the time-dependent PDF $\mathcal{P}_{p}(\mathcal{J}_{p}|t)$ of the rescaled cumulative particle current
$\mathcal{J}_{p}(t)=t^{-1/2z_{p}}{\rm J}_{p}(t)$ for different times (in normal and logarithmic scale),
with system size $L=10^{6}$ and particle density $\rho=0.5$.
Particles undergo diffusive motion with dynamical exponent $z_{p}=2$. Black curves correspond to
the asymptotic stationary PDF $\mathcal{P}^{\rm typ}_{p}(j_{p})=\lim_{t\to \infty}t^{1/4}\mathcal{P}_{p}(j_{p}=\mathcal{J}_{p}|t)$.
A least-squares fit to a Gaussian profile gives an estimated standard deviation $\sigma_{p}\approx 0.441 \pm 0.002$.}
\label{fig:ssep_particle_current}
\end{figure}

\paragraph*{Two-species symmetric exclusion process with parallel updates.}
The other representative example is a stochastic model. We consider the simplest two-species variant of the \emph{symmetric simple exclusion process} (SSEP), realized with a parallel update rule (see Fig.~\ref{fig:SSEP_dynamics} for an illustration): by randomly distributing charged particles on a discrete lattice, at every time-step each particle attempts to jump with probability $p$ to one of its adjacent sites (drawn uniformly at random), subjected to the exclusion rule that prevents particles from jumping to non-vacant sites. When two different particles attempt to jump to the same site, the `winner' is chosen randomly with equal probability (ensuring detailed balance)
We have explicitly verified (see Fig.~\ref{fig:ssep_particle_current}) that typical fluctuations on a timescale $t^{1/4}$
are normally distributed.

We stress that the proposed model differs crucially from the two-component AHR model \cite{AHR98} where charge-carrying particles are allowed to swap their positions with a finite probability. Despite the extra selection rule, dynamics of particles remain diffusive and, similarly to the ordinary simple (symmetric) exclusion process \cite{Arratia83}, fluctuations of the transmitted particles are normally distributed.

\subsection{Full counting statistics}
\label{sec:FCS}

We now make a slight digression to introduce the key objects for computing the full counting statistics of  charge transfer.
In the following we assume space and time to be continuous. We make this choice solely for compactness of notation,
and adapting our construction to lattice models in continuous time or to the fully discrete setting is straightforward.

We consider an extended system with a finite number $N$ of globally conserved charges ${\rm Q}_{i}=\int \dd x\,{\rm q}_{i}(x,t)$ enumerated by label $i$, with ${\rm q}_{i}(x,t)$ denoting the local densities at position $x$ and time $t$. The associated current densities, denoted by ${\rm j}_{i}(x,t)$, are determined from the local continuity relations,
$\partial_{t}{\rm q}_{i}(x,t)+\partial_{x}{\rm j}_{i}(x,t)=0$. 

The aim is to quantify fluctuations of the total time-integrated current density flowing through the origin in 
a time interval $t$. To this end, we introduce cumulative currents (also known as \emph{Helfand moments} \cite{Helfand60})
\begin{equation}
{\rm J}_{i}(t) = \int^{t}_{0}\dd \tau \,{\rm j}_{i}(0,\tau).
\end{equation}
The full set of temporally extensive dynamical variables $\mathbf{J}(t)=({\rm J}_{1}(t),\ldots,{\rm J}_{N}(t))$
quantifies the net transferred charges through the origin over a time period $t$. By prescribing an initial state (an ensemble of phase space points i.e. configurations), the task is to derive the joint PDF $\mathcal{P}(\mathbf{J}|t)$.

We focus our attention to two physically distinguished timescales.
Helfand moments ${\rm J}_{i}(t)$ associated to \emph{typical} trajectories are of the order $\mathcal{O}(t^{1/2z_{i}})$, corresponding to
the standard deviation, i.e. the square root of the second cumulant (variance)
\begin{equation}
c^{(i)}_{2}(t) = \Big\langle {\rm J}^{2}_{i}(t) \Big\rangle^{\rm c}_{\rm init} \sim t^{1/z_{i}}.
\end{equation}
In diffusive systems, the characteristic dynamical exponent equals $z=2$.
By contrast, due to ballistically propagating quasiparticles in integrable systems one generically finds ballistic scaling $z=1$, see e.g. Refs.~\cite{MBHD20,DoyonMyers20,PerfettoDoyon21}.
There are nevertheless important exceptions to this generic behavior, for example,
charge transport in the presence of charge-conjugation symmetry \cite{KIP22,Krajnik22}.

Coupling ${\rm J}_{i}(t)$ to `counting fields' $\lambda_{i}$, we introduce the \emph{multivariate moment generating function} (MMGF),
\begin{equation}
G(\boldsymbol{\lambda}|t) = \Big\langle e^{\boldsymbol{\lambda}\cdot \mathbf{J}(t)}\Big\rangle_{\rm init},
\end{equation}
where $\boldsymbol{\lambda}=(\lambda_{1},\ldots,\lambda_{N})$ and
the ensemble averaging is done with respect to a prescribed \emph{non-stationary} initial state.
For discrete variable systems, ${\rm J}_{i}(t) \in \mathbb{Z}$, hence periodicity in the imaginary direction permits us to restrict $\lambda_{i}$
to the infinite cylinder, $\lambda_{i} \in \mathbb{R}\times S^{1} \subset \mathbb C$ (we shall largely restrict our considerations to real $\lambda_{i}$,
and only relax this condition in Appendix \ref{sec:LeeYang}).
The multivariate MGF is given by the multivariate Laplace transform of the time-dependent joint PDF $\mathcal{P}(\mathbf{J}|t)$,
\begin{equation}
G(\boldsymbol{\lambda}|t)=\int \dd \mathbf{J}\,e^{\boldsymbol{\lambda}\cdot \mathbf{J}(t)}\mathcal{P}(\mathbf{J}|t),
\end{equation}
where $\dd \mathbf{J}\equiv \prod_{i=1}^{N}\dd {\rm J}_{i}$.
All the higher (finite-time) cumulants $c^{(i)}_{n>2}(t)$ of the integrated current density ${\rm J}_{i}(t)$ can be computed
with aid of the \emph{multivariate cumulant generating function} (MCGF),
\begin{equation}
F(\boldsymbol{\lambda}|t)=\log G(\boldsymbol{\lambda}|t).
\end{equation}
via
\begin{equation}
c^{(i)}_{n}(t) = \frac{\partial^{n} F(\boldsymbol{\lambda}|t)}{\partial \lambda^{n}_{i}}\Big|_{\boldsymbol{\lambda}=0}.
\end{equation}

Furthermore, let $\mathcal{P}_{i}({\rm J}_{i}|t)$ denote the univariate (time-dependent) PDFs obtained by marginalization of
the full joint PDF $\mathcal{P}(\mathbf{J}|t)$,
\begin{equation}
\mathcal{P}_{i}({\rm J}_{i}|t) = \int \prod_{l\neq i} \dd {\rm J}_{l}\mathcal{P}(\mathbf{J}|t). 
\end{equation}
The associated univariate MGFs $G_{i}(\lambda_{i}|t)$ are obtained by setting all $\lambda_{l\neq i}$ to zero,
$G_{i}(\lambda_{i}|t)\equiv G(\boldsymbol{\lambda}|t)|_{\lambda_{l\neq i}\to 0}$; they
correspond to the Laplace transform of the PDFs $\mathcal{P}_{i}({\rm J}_{i}|t)$
\begin{equation}
G_{i}(\lambda_{i}|t) = \int \dd {\rm J}_{i}e^{\lambda_{i}\,{\rm J}_{i}}\mathcal{P}_{i}({\rm J}_{i}|t).
\end{equation}

\medskip

\paragraph*{Typical fluctuations.}

To infer the \emph{stationary} PDFs $\mathcal{P}(\mathcal{J}_{i}|t)$ associated with typical fluctuations,
we first rescale the cumulative current ${\rm J}_{i}(t)$ as $\mathcal{J}_{i}(t)\equiv t^{-1/2z_{i}}{\rm J}_{i}(t)$,
and subsequently take the large-time limit,
\begin{equation}
\mathcal{P}_{i}(j_{i})=\lim_{t\to \infty}t^{1/2z_{i}}\mathcal{P}_{i}(\mathcal{J}_{i}=j_{i}|t).
\end{equation}

\medskip

\paragraph*{Large deviation principle.}
The theory of large deviations \cite{Ellis_book,Touchette_LDT,Esposito_review} deals with probabilities of exponential form.
According to the large deviation principle, atypical fluctuations of the cumulative currents ${\rm J}_{i}(t)$
decay exponentially with time,
\begin{equation}
\mathbb{P}({\rm J}_{i}(t)) \asymp e^{-t^{\alpha_{i}}\,I_{i}(j)},
\end{equation}
where exponents $\alpha_{i}$ are referred to as `speeds'.
Large fluctuations are in general reserved for fluctuations of the largest magnitude $\mathcal{O}(t^{\alpha_{i}})$
(whereas fluctuations larger than typical are sometimes referred to as `moderate fluctuations').

Under certain technical conditions, the LDFs $I_{i}(j_{i})$ can be extracted from the associated MGFs.
The asymptotic scaling of the univariate MGFs,
\begin{equation}
G_{i}(\lambda_{i}|t) \asymp e^{t^{\alpha_{i}}F_{i}(\lambda_{i})},
\end{equation}
is governed by \emph{scaled} cumulant generating functions (SCGFs)
\begin{equation}
F_{i}(\lambda_{i})=\lim_{t\to \infty}t^{-\alpha_{i}}\log G_{i}(\lambda_{i}|t).
\label{eqn:scaled_CGF}
\end{equation} 
Note that dynamical exponents $\alpha_{i}$ are (by definition) the largest exponents such that $F_{i}(\lambda_{i})$ exists and are non-trivial.
Moreover, $F_{i}(\lambda_{i})$ are convex functions of the counting field $\lambda_{i} \in \mathbb{R}$.
Extended diffusive systems are characterized by $\alpha=1/2$, see e.g. \cite{Derrida09,Imamura17}.
By contrast, systems that support long-lived (free or interacting) quasiparticle excitations generically exhibit growth
with ballistic exponent $\alpha=1$.

By taking into account that all the currents are mutually coupled we subsequently put $\alpha_{i}=\alpha$ for all $i$.
To properly exhibit the symmetry properties of the counting process, it is crucial to treat all the (cumulative) currents on equal footing.
The joint LDF $I(\boldsymbol{j})\equiv I(j_{1},\ldots,j_{N})$ associated to $\mathcal{P}(\mathbf{J}|t)$ is accordingly given by
\begin{equation}
I(\boldsymbol{j} ) = -\lim_{t\to \infty}t^{-\alpha}\log \mathcal{P}(\boldsymbol{j}=\boldsymbol{\mathcal{J}}|t).
\end{equation}
Provided that the multivariate SCGF $F(\boldsymbol{\lambda})$ is everywhere differentiable on its domain $\mathbb{R}^{N}$,
the G\"{a}rtner--Ellis theorem ensures that the LDF $I(\boldsymbol{j})$ is given by the Legendre--Fenchel transform
\begin{equation}
I(\boldsymbol{j}) = F^{\star}(\boldsymbol{j})
= {\rm sup}_{\boldsymbol{\lambda} \in \mathbb{R}^{N}}\big\{\boldsymbol{\lambda}\cdot \boldsymbol{j}-F(\boldsymbol{\lambda})\big\},
\end{equation}
representing a convex, lower-semicontinuous and non-negative multivariate function obeying
${\rm inf}_{\boldsymbol{j}\in \mathbb{R}^{N}}I(\boldsymbol{j})=0$.
At late times, the expectation values of currents are encoded in the first moment,
\begin{equation}
\ol{\boldsymbol{j}} = \lim_{t\to \infty}t^{-\alpha}\expect{\mathbf{J}(t)}_{\rm init}
= \frac{\dd F(\boldsymbol{\lambda})}{\dd \boldsymbol{\lambda}}\Big|_{\boldsymbol{\lambda}=\mathbf{0}},
\end{equation}
such that $I(\ol{\boldsymbol{j}})=0$.

\medskip

\paragraph*{Regularity.}
It is instructive to briefly discuss the formal properties of the univariate SCGFs $F(\lambda)$ (suppressing the subscript label).
In physics literature, $F(\lambda)$ is introduced as the generating function of \emph{scaled} cumulants $s_{n}$ through the series expansion
\begin{equation}
F(\lambda)=\sum_{n=0}^{\infty}\frac{\lambda^{n}}{n!}s_{n}.
\end{equation}
It is important to keep in mind, however, that this is \emph{not} unconditionally true. Instead, only when all cumulants $c_{n}(t)$ grow asymptotically with a common algebraic exponent, that is $c_{n}(t)\sim t^{\alpha}$, we have that $s_{n}=\lim_{t\to \infty}t^{-\alpha}c_{n}(t)$. As emphasized in \cite{Krajnik22}, only provided that the large-time limit $F(\lambda)=\lim_{t\to \infty}t^{-\alpha} F(\lambda|t)$ can be exchanged with an infinite sum, it is guaranteed that $s_{n}$ exist and correspond to the series coefficients of the SCGF $F(\lambda)$.
Establishing that $F(\lambda)$ is a \emph{real} analytic function around the origin $\lambda=0$ (with a finite radius of convergence) is,
perhaps unintuitively, not enough to ensure interchangeability of limits. Rather, a stronger sufficient `regularity condition' is required, as first pointed out by Bryc \cite{Bryc93}: if $F(\lambda|t)$ is \emph{holomorphic} at all times in some finite fixed neighborhood around the origin in the complex $\lambda$-plane, then $F(\lambda)$ represents a \emph{faithful} generating function of scaled cumulants (see also \cite{Jaksic12}).
To the best of our knowledge, Bryc regularity does not follow from a more general principle and thus it remains an open question whether faithfulness of $F(\lambda)$ can be formally established without invoking any model-specific information. Lastly, we note that even an unfaithful SCGF is still physically meaningful. Assuming it is everywhere differentiable on its domain, it provides the LDF
via the Legendre transform, $I(j)=F^{\star}(j)\equiv{\rm sup}_{\lambda \in \mathbb{R}}\{\lambda\,j-F(\lambda)\}$.

Even though establishing faithfulness of $F(\lambda)$ might at this point appear an unnecessary hindrance,
lack of regularity can profoundly influence the structure of fluctuations.
It appears however that this subtle aspect has been entirely disregarded in physics applications thus far, until our recent
work \cite{Krajnik22}; by computing exact charge-current fluctuations in the hard-core automaton, we have
shown that the MGF in equilibrium (at finite particles density and general bias) fails to satisfy the aforementioned regularity condition, which can be traced back to emergent dynamical criticality (attributed to Lee--Yang zeros colliding with the real axis at the origin of the $\lambda$-plane, see Appendix \ref{sec:LeeYang}).

\subsection{Multivariate fluctuation relation}
\label{sec:MFR}

In the context of (quasi)stationary current-carrying steady states, there is an emergent symmetry principle
that concerns the structure of temporal fluctuations of macroscopic charge transfer at late times, originally empirically discovered
in shear fluids in Ref.~\cite{ECM93}. Afterwards, Gallavotti and Cohen \cite{GC95} provided a rigorous derivation for a certain class of strongly chaotic dynamical systems known as Anosov systems (on compact manifolds). Building on the results of \cite{Kurchan98}, the Gallavotti--Cohen relation (GCR) has soon afterwards been established in Ref.~\cite{LS99} as a general property of finite-state irreducible and aperiodic (ergodic and mixing) Markov chains. The GCR is a symmetry that emerges in nonequilibrium states at late times and, unlike the transient fluctuation theorem, is generally not valid at finite times. As an introduction to the subject, we can recommend Ref.~\cite{Hurtado11}, while more comprehensive
and technical expositions can be found in Refs.~\cite{Maes99,Andrieux06,Andrieux08,Jaksic12,Gaspard13}.

By virtue of \emph{detailed balance} there is no average current flow in equilibrium states. Consequently,
fluctuations (both typical and large) of magnitudes $\pm j_{i}$ away from the mean value $\ol{j}_{i}=0$ are equiprobable.
This is no longer the case away from equilibrium as $\ol{j}_{i}\neq 0$ implies a preferred direction for particle and current fluxes.
Accordingly, observing a large deviation from the mean current in the direction of the flow is exponentially
more likely than observing a current of the same magnitude flowing in the opposite direction.
Remarkably, the probabilities of the two events can be related by exploiting time-reversibility of the microscopic evolution law,
yielding a \emph{universal} ratio of the form for asymptotically large times
\begin{equation}
\frac{\mathbb{P}(\boldsymbol{j})}{\mathbb{P}(-\boldsymbol{j})} \asymp e^{t^{\alpha}\boldsymbol{\sigma}_{\boldsymbol{\varepsilon}}(\boldsymbol{j})},
\label{eqn:MFR0}
\end{equation}
with a linear form $\boldsymbol{\sigma}_{\boldsymbol{\varepsilon}}(\boldsymbol{j}) = \boldsymbol{\varepsilon} \cdot \boldsymbol{j}$ and the vector of `thermodynamic forces'
$\boldsymbol{\varepsilon} \equiv (\varepsilon_{1},\ldots,\varepsilon_{N}) \in \mathbb{R}^{N}$, customarily called \emph{affinities}.
In the rest of the paper, we shall refer to Eq.~\eqref{eqn:MFR0} as the \emph{multivariate fluctuation relation} (MFR), see e.g. Ref.~\cite{Gaspard13}. The average of the exponent in Eq.~\eqref{eqn:MFR0} can be interpreted as
the rate of thermodynamic entropy production $\sigma_{S}\equiv \dd S(t)/\dd t \geq 0$,
\begin{equation}
\sigma_{S} = \boldsymbol{\varepsilon}\cdot \ol{\boldsymbol{j}}
= \lim_{t\to \infty} t^{-\alpha}D_{\rm KL}\big(\mathcal{P}(\mathbf{J}|t)||\mathcal{P}(-\mathbf{J}|t)\big),
\label{eqn:entropy_production}
\end{equation}
coinciding with the rescaled (in units of $t^{\alpha}$) \emph{relative entropy} (also known as the Kullback--Leibler divergence) of
$\mathcal{P}(\mathbf{J}|t)$ and its time-reversed counterpart $\mathcal{P}(-\mathbf{J}|t)$,
$D_{\rm KL}\big(\mathcal{P}(\mathbf{J}|t)||\mathcal{P}(-\mathbf{J}|t)\big)=\int \dd \mathbf{J}\mathcal{P}(\mathbf{J}|t)\log{\left[\frac{\mathcal{P}(\mathbf{J}|t)}{\mathcal{P}(-\mathbf{J}|t)}\right]}$. The LDF $I(\boldsymbol{j})$ therefore satisfies the relation
\begin{equation}
\tcbhighmath[drop fuzzy shadow]{I(-\boldsymbol{j}) - I(\boldsymbol{j}) = \boldsymbol{\varepsilon}\cdot \boldsymbol{j}.}
\label{eqn:GCR}
\end{equation}
Expressed in terms of the multivariate SCGF, the MFR is manifested as an \emph{inversion symmetry}
around $\boldsymbol{\lambda=}-\boldsymbol{\varepsilon}/2$,
\begin{equation}
F(\boldsymbol{\lambda}) = F(-\boldsymbol{\lambda}-\boldsymbol{\varepsilon}).
\end{equation}
This is a mutivariate generalization of the celebrated Gallavotti--Cohen fluctuation relation \cite{LS99,Gaspard13}.

In stochastic systems, the Gallavotti--Cohen relation is a corollary of the additivity principle \cite{BodineauDerrida04}.
The GCR is however obeyed even in the absence of the additivity principle (e.g. in systems supporting dynamical phase transitions),
see. Ref.~\cite{MFT}. In higher-dimensional time-reversal invariant systems there exists more general, so-called isometric, fluctuation relations \cite{Hurtado11}.
Within the scope of MFT, the fluctuation symmetry for particle-conserving time-reversal invariant diffusive systems takes a universal form (see Ref.~\cite{Derrida07}),
reading $\varepsilon(\rho_{-},\rho_{+}) = \int^{\rho_{-}}_{\rho_{+}}\dd \rho\,[2{\rm D}(\rho)/\sigma(\rho)]
=\int^{\rho_{-}}_{\rho_{+}}\dd \rho \, f^{\prime \prime}_{\rm eq}(\rho)=f^{\prime}_{\rm eq}(\rho_{-})-f^{\prime}_{\rm eq}(\rho_{+})$.

\medskip

\paragraph*{Univariate fluctuation relations.}

Unlike the multivariate SCGF $F(\boldsymbol{\lambda})$, univariate SCGFs $F_{i}(\lambda_{i})$ will not in general
display any particular symmetry property in spite of time-reversal invariance \cite{LS99}. This asymmetry is simply due to the fact that all currents flip sign under the time-reversal, that is $\rm{j}_i \circ \Theta = -\rm{j}_i$.
One can nonetheless identify situations when even the marginalized SCGFs $F_{i}(\lambda_{i})$ possess
a Gallavotti--Cohen symmetry of the form
\begin{equation}
F_{i}(\lambda_{i}) = F_{i}(-\lambda_{i}-\tilde \varepsilon_{i}),
\label{eqn:UFR}
\end{equation}
for some `effective' univariate affinity $\tilde \varepsilon_{i}$ (in general differing from the affinity component $\varepsilon_{i}$).
This happens, for example, when (i) the nonequilibrium state is induced by a single thermodynamic force
(corresponding to bias a $\delta \mu_{i}\neq 0$), (ii) whenever the $i$th cumulative current ${\rm J}_{i}(t)$ is parametrically slower or faster than other cumulative currents (see e.g. Ref.~\cite{BulnesCuetara13}) or (iii) under
the `tight-coupling condition' ${\rm J}_{i}\simeq {\rm J}_{l}$ for all $\ell \neq i$ (see e.g. Ref.~\cite{GG07}).

In Section \ref{sec:results}, we describe another, different, dynamical mechanism involving two coupled currents
that obey both the univariate and joint fluctuation relations. However, while the univariate fluctuation relation (UFR) associated to the particle current is always obeyed, the UFR of the charge current can be spontaneously broken. 

\subsection{Central Limit Theorem}
\label{sec:CLT}

The Central Limit Theorem is one of the most celebrated results in probability theory. The theorem states that empirical means of independent random variables with finite variances become normally distributed when the number of samples grows large. More remarkably, even in strongly interacting particle systems subjected to highly non-trivial (temporal) correlations (as commonly found in physics applications), one empirically finds that temporal clustering of correlations is typically strong enough to preserve central limit behavior. In other words, if the memory of all initially correlated local observables (the current density, for example) decays sufficiently fast, fluctuations of the associated macroscopic dynamical quantity (the time-integrated current density)
cannot be distinguished from those of a random process.

Typical values of the cumulative current ${\rm J}(t)$ at large $t$ are proportional to the standard deviation $\sqrt{c_{2}(t)} \sim t^{1/2z}$.
To infer the associated \emph{stationary} PDF, the time-integrated current density has to be rescaled as
$\mathcal{J}(t)=t^{-1/2z}{\rm J}(t)$, yielding
 \begin{equation}
\mathcal{P}^{\rm typ}(j)\equiv \lim_{t\to \infty} t^{1/2z}\mathcal{P}(j=\mathcal{J}|t).
\end{equation}
Cumulants of $\mathcal{P}^{\rm typ}(j)$, denoted by $\kappa_{n}$, are (assuming the limits exist, i.e. $\kappa_{n}<\infty$ for all $n$) accordingly given by
\begin{equation}
\kappa_{n}(t)\equiv t^{-n/2z}c_n(t),\qquad \kappa_{n}=\lim_{t\to \infty}\kappa_{n}(t).
\end{equation}
Then, the central limit behavior (CLT property) holds if and only if $\kappa_{2}$ is finite and non-zero ($0<\kappa_{2}<\infty$)
and $\kappa_{n>2}=0$, implying $\mathcal{P}^{\rm typ}(j)$ is a Gaussian distribution of zero mean and finite variance $\kappa_{2}$.
Fluctuations whose PDFs deviate from Gaussianity can be regarded as \emph{anomalous}.

Bryc regularity provides a \emph{sufficient} condition of the CLT property. More specifically, complex analyticity of $G(\lambda|t)$ within a disc centered around $\lambda=0$ guarantees existence of scaled cumulants, i.e. $s_{n}<\infty$ for all $n\in \mathbb{N}$, which in turn implies that $\lim_{t\to \infty}\kappa_{n>2}=0$,
while $\kappa_{2}=s_{2}$ stays finite. Lack of regularity on the other hand opens the door for anomalous statistics of typical events. Beware however, that absence of regularity is not necessary detrimental to the CLT property (see Ref.~\cite{Krajnik22} for an example). 
As pointed out previously in Refs.~\cite{KIP22,Krajnik22}, singular scaled cumulants arise if $z>1/\alpha$ (assuming a generic SCGF with $\dd^2F /\dd \lambda^2|_{\lambda = 0} > 0$),
signifying that $c_{2}(t)\sim t^{1/z}$ grow asymptotically \emph{slower} than $t^{\alpha}$, namely the exponent governing the asymptotic growth of $G(\lambda|t)$.
 Absence of regularity is manifested, for example, in certain widely studied integrable systems that support subballistic
(either diffusive \cite{DeNardis18,GHD_review,transport_review} or superdiffusive \cite{Ljubotina19,MatrixModels,superdiffusion_review}) charge transport. It is nonetheless not inherently linked to integrability.
For instance, the proposed parallel-update SSEP is one of the simplest stochastic models violating Bryc regularity: while particles diffuse through the system ($\alpha=1/2$), charge is slowed down by the exclusion rule and instead spreads subdiffusively with dynamical exponent $z=4$.

\section{Results}
\label{sec:results}

{In this section we expound the main findings of our study. We being by familiarizing the reader with the main concepts and
spelling out all the universal features of charged single-file systems. In Sec.~\ref{sec:dressing_approach} we provide
a brief summary of the dressing approach, which we further detail out in Appendix \ref{sec:dressing_technical}, and proceed
to describe the mechanisms that lead to dynamical phase transitions of first and second order (Secs.~\ref{sec:first-order} and \ref{sec:second-order}, respectively). We conclude with a classification of all the dynamical regimes in Sec.~\ref{sec:regimes}.}

\subsection{Summary}
\label{sec:summary}

{
Charged single-file systems display anomalous dynamical behavior that arises a consequence of the two defining dynamical constraints.
The most stringent constraint is the non-intersecting rule for particle trajectories, which already leads to profound restrictions on the
charge dynamics: in a given time interval $t$, ${\rm J}_{p}(t)$ can only increase due to right-moving particles arriving from the left of the origin ($x<0$) and, vice-versa, any decrease of the charge transfer must come from particles that have arrived from the right half of the system ($x>0$). Futhermore, owing to the inertness property, the same logic indeed applies to charge degrees of freedom, where we have to additionally account for the charge-dependent sign. As a useful analogy, one can picture both partitions as two separate sources of fluctuations, each attempting to enforce its own fluctuations on the whole system. Despite conceptual simplicity there is nevertheless no easy way to determine
in advance which of the partitions eventually prevails after performing an ensemble average over initial configurations.
The large charge-current fluctuations encoded in the rate function can be thus viewed as a minimization problem involving two branches
(one per each initial partition) whose solution determines the so-called physical branch, i.e. the branch that dominantes at late times.
It turns out, somewhat unexpectedly, that the physical branch depends quite intricately on the initial condition (i.e. the density of particles and average charge) \emph{and} the magnitude of fluctuations (equivalently, on the counting field).

Competition between dynamical phases ascribed to different branches gives rise to dynamical phase transitions.
In Secs.~\ref{sec:first-order} and \ref{sec:second-order}, we discuss this phenomenon
at the level of the moment generating function $G(\lambda|t)$. The positive (depicted in red color) branch and negative (blue color) branch are associated with the two \emph{local} extrema of the time-asymptotic $G(\lambda|t)$ located in the bulk of its domain.
It may happen however that in certain cases the global extremum is attained along the diagonal of the domain, i.e. when the difference
of the charge transfer from the left and right partition scales subextensively with the time interval $t$.
In this situation one finds another, third branch that is \emph{flat} (i.e. constant in $\lambda$) branch.
Depending on the initial condition and the value of counting field $\lambda$, dynamical phase transition can occur between the two regular branches (first order) or between a regular and the flat branch (second order). 

We find various exotic universal features which we subquently describe in a systematic fashion:
\begin{enumerate}
\item\label{prop1} a universal non-Gaussian probability distribution (see Eq.~\eqref{eqn:universal_typical_distribution}) of net charge transfer on the typical timescale, signifying a violation of the central limit property,
\item\label{prop2} a rich and intricate structure of the scaled cumulant generating function,
governed by coexistent (meta)stable dynamical phases leading to several emergent dynamical regimes (see Sec.~\ref{sec:regimes}),
\item\label{prop3} dynamical phase transitions of first and second order in the large-deviation rate function (see Sec.~\ref{sec:first-order} and
Sec.~\ref{sec:second-order}) that cannot be captured by conventional approaches such as MFT,
\item \label{prop4} the onset of dynamical criticality in equilibrium at vanishing charge bias (see Appendix \ref{sec:triple}),
\item\label{prop5} spontaneous breaking of the Gallavotti--Cohen symmetry (cf. Sec.~\ref{sec:fluctuation_symmetry})
in the univariate rate function of charge transfer.
\end{enumerate}
}

\medskip 

\subsubsection{Dressing approach}
\label{sec:dressing_approach}

Our main objective is to compute the full time-dependent joint PDF $\mathcal{P}_{c, p}({\rm J}_{c},{\rm J}_{p}|t)$ and to subsequently infer from it the joint LDF $I_{c,p}(j_{c},j_{p})$. Here variables $j_{i}$ pertain to dynamically
rescaled cumulative (i.e. time-integrated) particle and charge currents, $\mathcal{J}_{i}(t)=t^{-\alpha}{\rm J}_{i}(t)$, where
$\alpha$ quantifies the timescale of large (exponentially rare) events.
 
Computing exact finite-time rate function in genuinely \emph{interacting} systems seems a rather hopeless task.
Even in `exactly solvable' models, computing the MGF at finite times presents a daunting challenge.
Fortunately however, the principal characteristic features of the considered dynamically constrained models can be described in a fully analytic and rigorous fashion, provided that the underlying statistical properties
of particle dynamics are supplied as a phenomenological input (similarly as in MFT, where one provides the diffusion constant and conductivity).

Using that charges have no effect on the underlying particle dynamics, the counting statistics for the charge degrees of freedom
can be resolved in a purely combinatorial fashion. For this reason, we suggestively call this technique the ``dressing'' approach.
Here we only briefly describe the basic idea and leave a detailed analysis to Section \ref{sec:dressing_technical}.
The main object is the time-independent `dressing factor' $I_{c|p}(j_{c},j_{p})$, representing
the conditional probability for observing $j_{c}$ for a given value of $j_{p}$.
By adjoining the particle-current rate function $I_{p}(j_{p})$, we obtain a joint bivariate LDF of the form
$I_{c,p}(j_{c},j_{p}) = I_{c|p}(j_{c},j_{p})+I_{p}(j_{p})$. In Sec.~\ref{sec:fluctuation_symmetry}, we establish that a fluctuation relation of the form $I_{c, p}(-j_{c},-j_{p})-I_{c, p}(j_{c},j_{p})=\varepsilon_{c}j_{c}+\varepsilon_{p}j_{p}$ is satisfied provided that $I_{p}(j_{p})$ obeys the univariate Gallavotti--Cohen relation $I_{p}(-j_{p})-I_{p}(j_{p})=\tilde \varepsilon_{p}j_{p}$.

In the following, we confine our analysis mostly to the univarite LDF $I_{c}(j_{c})$. The reason is two-fold.
Firstly, since $I_{c}(j_{c})$ inherits the most salient qualitative properties of $I(j_{c},j_{p})$, we find it better suited to exhibit
the underlying dynamical criticality. Secondly, $I_{c}(j_{c})$ can undergo spontaneous breaking of fluctuation symmetry.
To compute $I_{c}(j_{c})$, the joint LDF $I_{c, p}(j_c, j_{p})$ has to be minimized over the range of $j_{p}$.
The biphasic structure of the dressing factor allows us to perform a `chiral decomposition'
into two separate optimizations $I^{(\pm)}_{c,p}(j_{c},j_{p})$ associated with two branches of the rate function,
$I^{(\pm)}_{c}(j_{c})={\rm sup}_{\lambda}\{\lambda\,j_{c}-F_{\pm}(\lambda)\}$.
In practice, one first carries out `inner optimizations' on $I^{(\pm)}_{c,p}(j_{c},j_{p})$ yielding $I^{(\pm)}_{c}(j_{c})$, and finally selects the optimal global value (for fixed $j_c$), namely $I_{c}(j_{c})={\rm min}\{I^{(\pm)}_{c}(j_{c})\}$.

\subsubsection{Coexisting dynamical phases and first-order dynamical phase transition}
\label{sec:first-order}

We now describe the main universal characteristics of the charge LDF $I_{c}(j_{c})$.
We find it convenient to discuss it in terms of its Legendre-dual function $F_{c}(\lambda_{c})$, which we formally view
as the dynamical free energy density governing the asymptotic growth of the dynamical partition sum $G_{c}(\lambda_{c}|t)$.
To lighten our notation, we subsequently drop the subscript label `$c$' from the univariate charge MGF and LDF
(while making the identifications $\lambda_{c}\to \lambda$, $j_{c}\to j$).

During a finite window of time $t$, the associated `dynamical free energy' $F(\lambda|t)$ receives contributions $F_{\pm}(\lambda)$ from two distinct \emph{dynamical phases}. We can picture them as distinct branches of the dynamical free energy (measured in
units of $t^{\alpha}$), see Eq.~\ref{eqn:scaled_CGF}. However, only the larger (in magnitude) of the two branches is physically relevant at late times.
The other phase is subleading and merely visible as a transient finite-time correction that fades away exponentially with time. The exception to this are equilibrium states, where detailed balance ensures that both branches contribute equally.

Characterizing the nature of charge-current fluctuations boils down to determining which of the two competing branches $F_{\pm}$ dominates for any specified value of the counting field $\lambda \in \mathbb{R}$. Based on this, we can thus anticipate two intervals,
denoted by $\mathscr{I}_{\pm}\subset \mathbb{R}$, along which the respective branches $F_{\pm}$ dominate the asymptotic growth of (charge) MGF $G(\lambda|t)$. On purely formal grounds, we can regard these two (meta)stable branches $F_{\pm}(\lambda)$ as distinct dynamical phases.
We thus deal with a scenario that closely resembles the physics of first-order thermodynamic phase transitions \cite{Biskup00}.

Suppose that $F_{\pm}(\lambda)$ exchange dominance at $\lambda=\lambda_{\bowtie}$. If $F_{\pm}(\lambda)$ are
strictly convex, then $\lambda_{\bowtie}$ represents a non-differentiable (corner) point in $F(\lambda)$. 
Non-differentiable point are typically a precursor of a \emph{first-order} (dynamical) phase transition.

How emergence of non-differentiable points affects the large-deviation rate function is less obvious and requires a careful analysis.
To begin with, presence of a corner no longer guarantees that the rate function $I(j)$ coincides with
the Legendre dual $F^{\star}(j)$ of the SCGF $F(\lambda)$. In general, $F^{\star}(j)$ is only a convex hull that bounds the physical rate function $I(j)$ from below \cite{Touchette_LDT}. Conversely, the Legendre--Fenchel transform of $I(j)$ \emph{always}, regardless of convexity, yields the physical $F(\lambda)$.  A non-differentiable corner point in $F(\lambda)$ translates to an \emph{affine} (i.e. linear) segment
in $F^{\star}(j)={\rm sup}_{\lambda \in \mathbb{R}}\{\lambda\,j-F(\lambda)\}$,
spanning a contiguous range of values in $j$ between the left and right derivatives of $F(\lambda)$ at the corner $\lambda_{\bowtie}$.
Any non-differentiable point in $F(\lambda)$ therefore erases some information about the rate function, meaning that
computing the rate function $I(j)$ necessitates additional information beyond that provided by $F(\lambda)$ alone.
In general, $F(\lambda)$ with a corner point implies that the charge-current rate function $I(j)$ cannot be strictly convex everywhere;
it is either non-convex or it contains an affine part, in formal analogy to non-concave microcanonical entropies that imply inequivalent (thermodynamic) ensembles \cite{Costeniuc05,Costeniuc06}.

Based on the general formal analysis of the solutions to the optimization problem (see Section \ref{sec:dressing_formalism} for details) we conclude that $I(j)$ never develops an affine parts. Instead, it simply consists of two `patches' of \emph{locally} convex branches,
meaning that convexity of $I(j)$ will not be preserved globally (i.e. for the entire admissible range of large integrated
currents $j$). Until the critical value $j_{\bowtie}$, large charge-current fluctuations are realized
by one of the partitions (branches) $I^{(\pm)}(j<j_{\bowtie})$, beyond which the events from the other branch
$I^{(\mp)}(j>j_{\bowtie})$ become more probable and take over. We have thus eliminated the possibility of stable phase coexistence.
Coexistence of dynamical phases, emerging in certain systems supporting second-order DPTs associated with particle-hole symmetry breaking, is associated with \emph{convex} rate functions possessing affine 
parts, corresponding to the Legendre--Fenchel transform $F^{\star}(j)$. In our case, the single-file constraint on particle trajectories prohibits phase coexistence. We find that the rate function $I(j)$ is strictly convex everywhere, except at the critical point $j=j_{\bowtie}$ where both branches intersect, $I^{(+)}(j_{\bowtie})=I^{(-)}(j_{\bowtie})$.

\begin{figure}[htb]
\includegraphics[width=0.9\columnwidth]{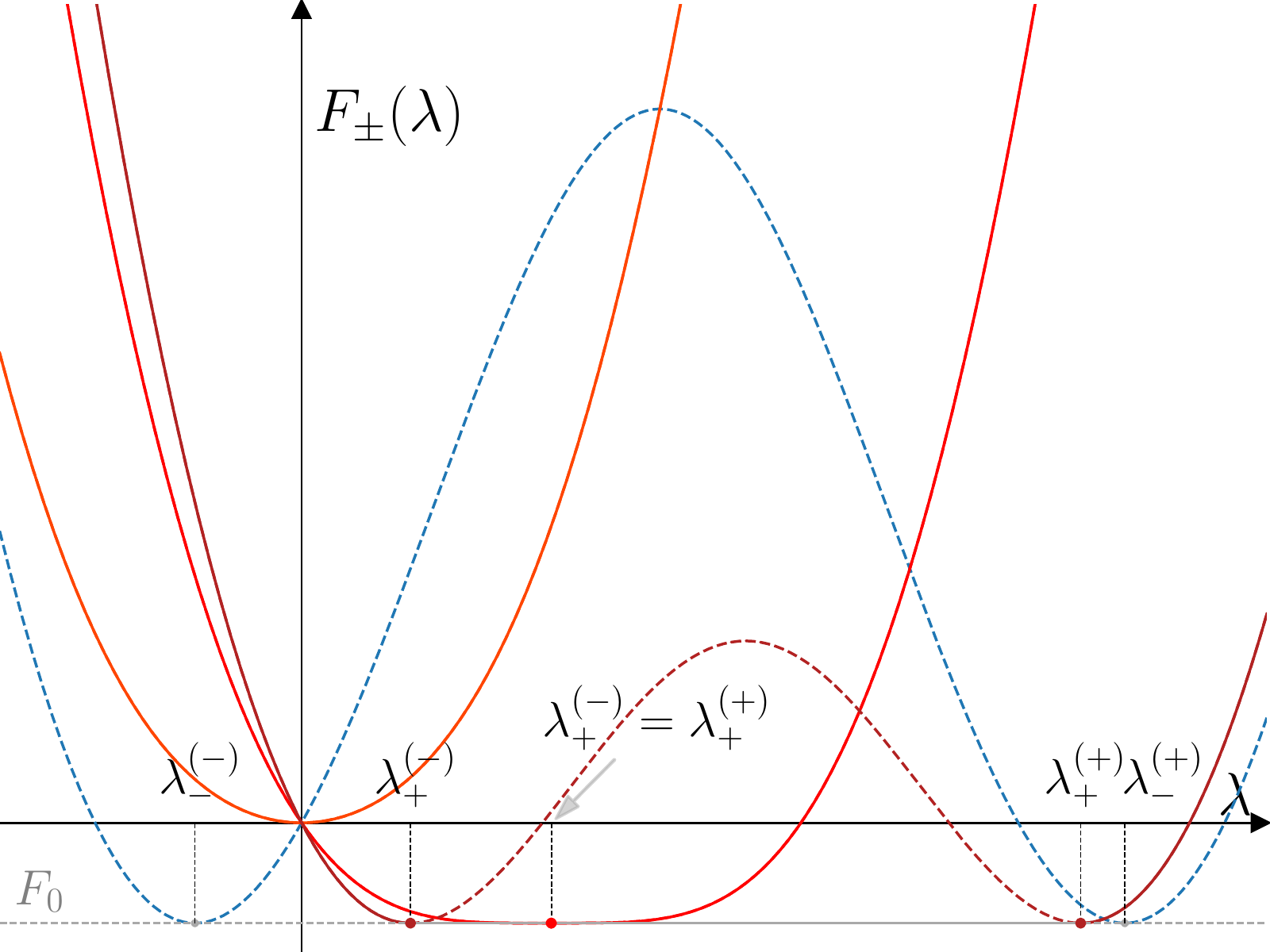}
\caption{Continuous symmetry breaking of convexity of a regular bulk branch: upon smoothly varying $b_{+}$,
a (strictly) convex bulk branch $F_{+}(\lambda)$ (light red) with a single minimum undergoes a continuous transition
to a symmetry-broken non-convex shape (dark red) with a doubly degenerate minimum at value $F_{0}$. The middle red curve shows the critical
shape when the two minima $\lambda^{(\pm)}_{+}$ of $F_{+}(\lambda)$ (either real or purely imaginary) coalesce. 
The other bulk branch $F_{-}(\lambda)$ (blue) always remains in a symmetry-broken form and is unaffected by the change of $b_{+}$.
}
\label{fig:branch_breaking}
\end{figure}

The central question now is whether non-differentiable points in $F(\lambda)$ have any adverse consequence for the UFR.
A crucial observation in this respect is that the first-order DPTs emerge due to appearance of a single critical point.
This is to be contrasted with other symmetry-breaking scenarios discussed previously in the literature (see Refs.~\cite{Baek17,Baek18}))
where critical points are produced in pairs, that is symmetrically with respect to inversion point of an unbroken phase.
In absence of second-order critical points, the UFR will still hold locally for all $j$ that are smaller in magnitude
than the distance of the non-differentiable point in $I(j)$ from the origin. In contrast, for large charge-current fluctuations in the direction of the flow that in magnitude exceed the critical value $j_{\bowtie}$, fluctuations in the opposite direction of the same size are realized by a different bulk branch and hence the GCR \eqref{eqn:GCR} will no longer be satisfied \emph{globally} for all values of $j$.

\subsubsection{Dynamical phase transition of second order}
\label{sec:second-order}

The class of models we consider supports another type of dynamical phase transitions. There is a subtle symmetry-breaking mechanism that induces a DPT of \emph{second order}, signaled by the emergence of (strictly) flat segments in $F(\lambda)$. In fact, individual (meta)stable branches $F_{\pm}(\lambda)$ may develop such a flat piece, arising as a consequence of a \emph{continuous} (i.e. second-order) phase transition of Landau--Ginzburg type from a strictly convex, called regular, form to a \emph{symmetry-broken} non-convex shape with a doubly-degenerate minimum. Such a transition from a regular to a symmetry-broken branch
is illustrated in Fig.~\ref{fig:branch_breaking}.

Under certain (mild) `regularity assumptions' on the particle rate function $I_{p}(j_{p})$ (see Sec.~\ref{sec:dressing_formalism}),
\emph{at least} one of the branches $F_{\pm}(\lambda)$ has broken symmetry.
This property can again be deduced by investigating the formal structure of solutions to the outlined optimization problem.
It can be shown that convexity at $j_{c}=0$ enforces that at least one of the branches attains the minimum at the boundary of the optimization domain at $j_{p}=0$. Lack of convexity precludes a symmetry-broken branch to be physical for the entire range of
real counting fields $\lambda \in \mathbb{R}$. We must then distinguish between the following cases: (i) one of the bulk branches $F_{\pm}$ is regular and thus directly corresponds to $F(\lambda)$, or (ii) only the strictly convex parts of $F_{\pm}(\lambda)$ that has experienced a symmetry-breaking transition are physical. In the latter case, the missing range of $\lambda$ (lying along the unphysical part of the broken branch) must then be identified with the flat branch $F_{0}<0$,
being the dominant contribution to the MGF $G(\lambda|t)$ at late times.

We next discuss the distinguished role of the constant branch. Unlike $F_{\pm}(\lambda)$, it appears (for physical values $\lambda \in \mathbb{R}$) only away from equilibrium. The corresponding `interval of dominance' $\mathscr{I}_{0}\subset \mathbb{R}$ is a \emph{single} compact interval on the real $\lambda$-axis (always excluding the origin) located in between two doubly-degenerate minima of the physical branch
$F_{\pm}(\lambda \in \mathbb{R}\setminus \mathscr{I}_{0})$.
Another important general property (see Sec.~\ref{sec:phases} for details) is that only a regular dominant (i.e. physical) branch
can undergo a symmetry-breaking transition, whereas the subdominant branch remains broken throughout.
This scenario is visualized in Fig~\ref{fig:branch_breaking}.
By continuously varying the counting field along the real $\lambda$-axis,
the physical bulk branch $F_{\pm}$ reaches its minimum $\lambda=\lambda^{(-)}_{\pm}$,
`jumps' over to $F_{0}$ and, upon reaching another minimum at $\lambda^{(+)}_{\pm}>\lambda^{(-)}_{\pm}$,
returns back to the same branch $F_{\pm}$. There is however another possible scenario
when both branches $F_{\pm}$ have broken symmetry; it may occur that the closest
(degenerate) minimum to $\lambda^{(-)}_{\pm}$ belongs to the other branch at $\lambda^{(+)}_{\mp}$, in which case
for $\lambda>\lambda^{(+)}_{\mp}$ (assuming absence of first-order phase transitions upon further increasing $\lambda$)
the dominant branch is $F_{\mp}$.

We have thus far established the following picture. Either of the bulk branches $F_{\pm}(\lambda)$ may, upon varying the densities or biases in the initial state, experience a second-order transition to a symmetry-broken form with two-fold degenerate minima. The transition occurs when the unique minimum of an unbroken physical branch $F_{\pm}(\lambda)$ decreases to $F_{0}$ (see Figure \ref{fig:branch_breaking}), giving rise a constant branch in $F(\lambda)$.
The \emph{physical} $F(\lambda)$ thus becomes continuously degenerate along a compact interval (extending between two adjacent minima of $F_{+}(\lambda)$ or $F_{-}(\lambda)$) where $F_{0}$ dominates the growth of $G(\lambda|t)$. The boundaries of $\mathscr{I}_{0}$, marking transitions between $F_{\pm}$ and $F_{0}$, are critical points associated with a second-order DPT.

The main distinction with the first-order transitions is that $F(\lambda)$ now remains differentiable everywhere, including at the two second-order critical points. Since $F^{\prime}_{0}=0$ and $F^{\prime \prime}(\lambda)>0$, the second derivatives however experience a discontinuity at the minima. This time (unlike in the case of first-order DPTs) differentiability of $F(\lambda)$ ensures that its Legendre dual
$F^{\star}(j)={\rm sup}_{\lambda \in \mathbb{R}}\{\lambda\,j-F(\lambda)\}$ coincides with the LD rate function, $F^{\star}(j) = I(j)$.
Recall that upon performing a Legendre transform of $F(\lambda)$, any flat or affine part with slope $j_{\rm aff}$ is mapped to a \emph{single} point $I(j_{\rm aff})$.
The Legendre counterpart of $F(\lambda)$ with a flat segment will thus feature a \emph{corner} at the origin $j=0$.
This further means that the pair of dynamical critical points associated with the second-order DPT manifests itself
as an isolated non-differentiable point in the rate function.
Following Ehrenfest's classification scheme, such points would correspond to critical point of first order. In this work, we follow the `canonical' terminology of criticality and classify phase transitions{in terms of differentiability (or lack thereof)} of the dynamical free energy (sometimes referred to as `$\lambda$-ensembles', as e.g. in Ref.~\cite{Baek18}).

We can offer another, perhaps more physically suggestive, perspective on the emergence of a flat part. The constant branch $F_{0}$ dominates along an interval $\mathscr{I}_{0}$, signifying that the main contributions to $G(\lambda\in \mathscr{I}_{0}|t)$ at late times are due to phase-space trajectories that differ in a \emph{subextensive} (i.e. for $\rm{J}_i$ on scales asymptotically smaller than $\sim t^{\alpha}$) amount of transported charge.
For any finite large current $j\neq 0$, the rate function $I_{c}(j)$ is instead differentiable and consequently the relevant rare trajectories concentrating around the maximum of MGF $G(\lambda|t)$ carry integrated currents of the order $\mathcal{O}(j\,t^{\alpha})$. By contrast, the subextensive rare events associated to the second-order criticality are not associated with bulk-extremum contributions but rather stem from the global maxima at the boundary of the integration domain (see Sec.~\ref{sec:joint_MGF}). Despite being exponentially unlikely, with a probability decaying with a rate of $I_{0}(j)=-F_{0}>0$, one would need to look at subleading orders in time to gain further insight into the finer structure of such trajectories.

Lastly, we examine the validity of the univariate fluctuation relation. We need to explicitly distinguish between the following two cases (i) the flat part connect both the degenerate minima $\lambda^{(\pm)}_{\pm}$ on the same bulk branch $F_{\pm}$ or (ii) $F_{0}$ interpolates between two degenerate minima of different bulk branches. In case (i), the UFR for the charge LDF $I(j)$
remains intact (provided that $F_{\pm}$ individually obey the symmetry), as evidently both degenerate minima appear symmetrically with respect to the inversion point $\lambda^{(0)}_{\pm} = -\tilde{\varepsilon}_{c}/2$, irrespective of the extent of the flat branch $F_{0}$. Analogously, the appearance of a flat part preserves the UFR of the LDF in spite of a corner at $j=0$.  The situation is different if the flat part connects two degenerate minima on different bulk branches. Even when $F_{\pm}$ both have inversion points $\lambda = -\tilde{\varepsilon}^{(\pm)}_{c}/2$,
the UFR ceases to hold simply because the two reflection points in general do not coincide,
$\tilde{\varepsilon}^{(+)}_{c}\neq \tilde{\varepsilon}^{(-)}_{c}$.
In this case, the presence of $F_{0}$ spoils the inversion symmetry of $F(\lambda)$. What is less obvious is that there are no direct transitions from regime (i) to (ii), or vice-versa, but only via the regime that features a first-order criticality. Finally, we also mention that out of equilibrium with uniform particle density, $\rho_{-}=\rho_{+}$, and arbitrary charge biases $b_{\pm}$, the UFR is always violated.

\medskip
\onecolumngrid
\subsubsection{Dynamical regimes}
\label{sec:regimes}

Upon continuously changing the counting field along the real $\lambda$-axis,
dynamical phases $F_{+}(\lambda)$, $F_{-}(\lambda)$ and $F_{0}$ shown, respectively,  as red, blue and gray curves in the left column  figures and the corresponding rate functions $I_ +(j), I_-(j)$, shown as red and blue curves in the right column figures, 
display different interweaving patterns. We shall explicitly distinguish between \emph{four} different scenarios which we hereafter refer 
to as `dynamical regimes':
\medskip

\begin{itemize}

\item {\tt regular regime:} one of the convex bulk branches $F_{\pm}(\lambda)$ dominates over the entire physical range of counting fields
$\lambda \in \mathbb{R}$. Correspondingly, the LD rate function is Legendre dual to $F(\lambda)$, $I(j)=F^{\star}(j)$, and involves a single physical (strictly convex) branch $I(j)=I^{(\pm)}(j)$.

\begin{figure}[h!]
\includegraphics[width=0.3\textwidth]{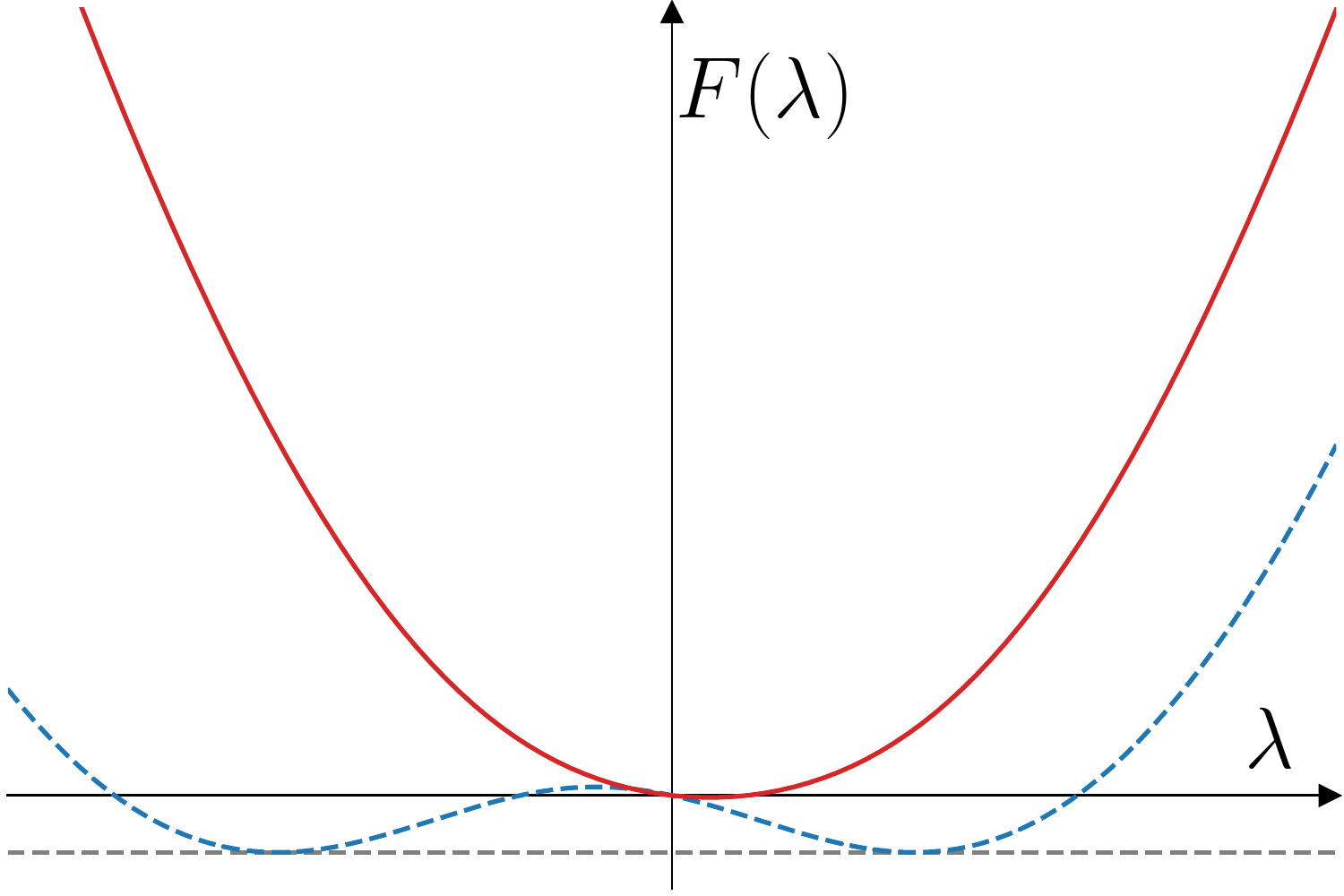}
\hspace*{2cm}
\includegraphics[width=0.3\textwidth]{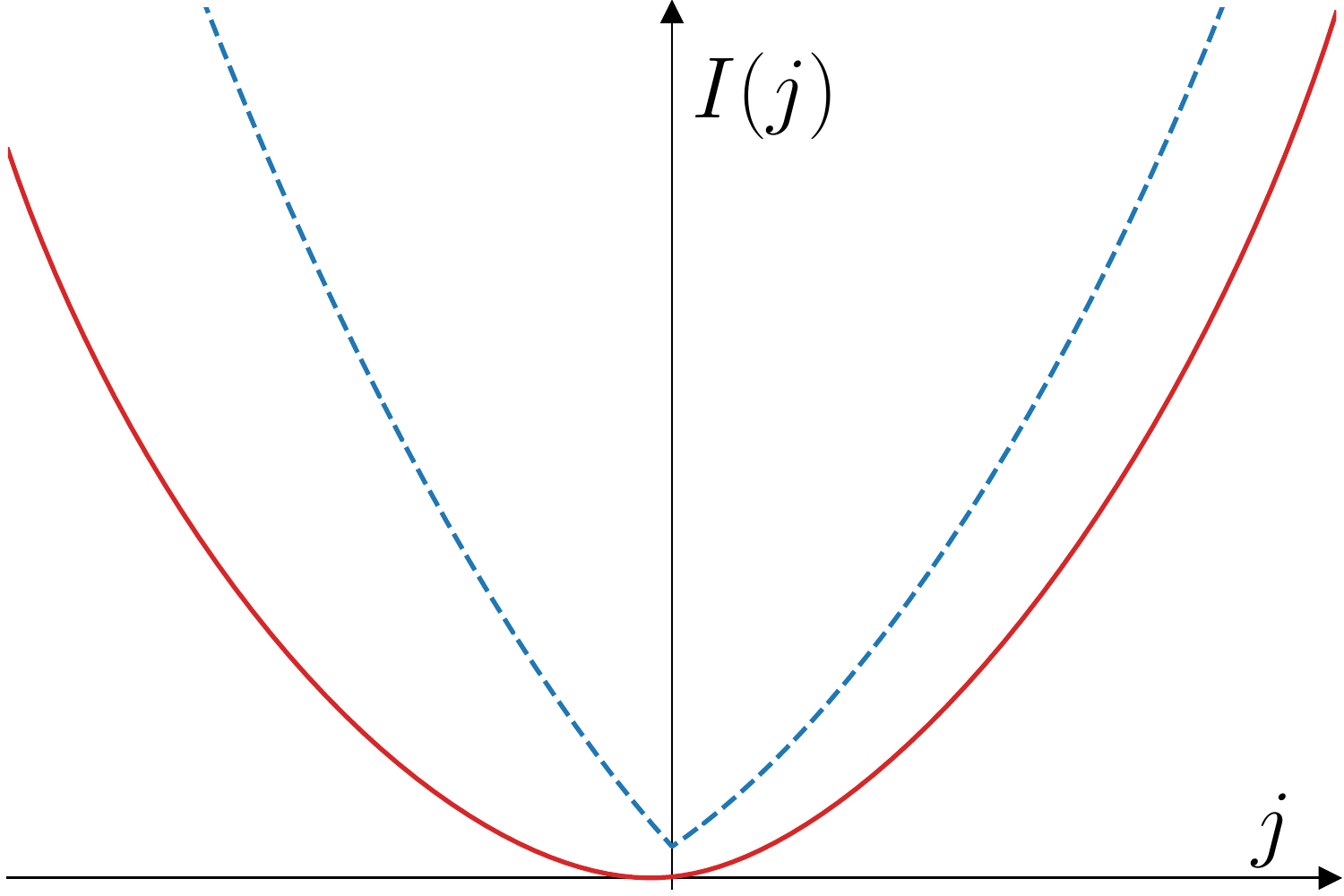}
\end{figure}

\item {\tt corner regime:} there is an exchange of dominance between the bulk branches $F_{\pm}(\lambda)$. Their respective regions of dominance $\mathscr{I}_{\pm}$ meet at the critical point $\lambda_{\bowtie} \in \mathbb{R}$ where the physical SCGF $F(\lambda)$ develops a non-differentiable (corner) point. The latter is the critical point of a \emph{first-order} dynamical phase transition. The rate function $I(j)$ is no longer the Legendre transform of $F(\lambda)$ but instead a \emph{non-convex} function with a non-differentiable (corner) points at the critical large current $j_{\bowtie}$, corresponding to the minimum of the two branches $I(j)={\rm min}_{j}\{I^{(\pm)}(j)\}$.

\begin{figure}[h!]
\includegraphics[width=0.3\textwidth]{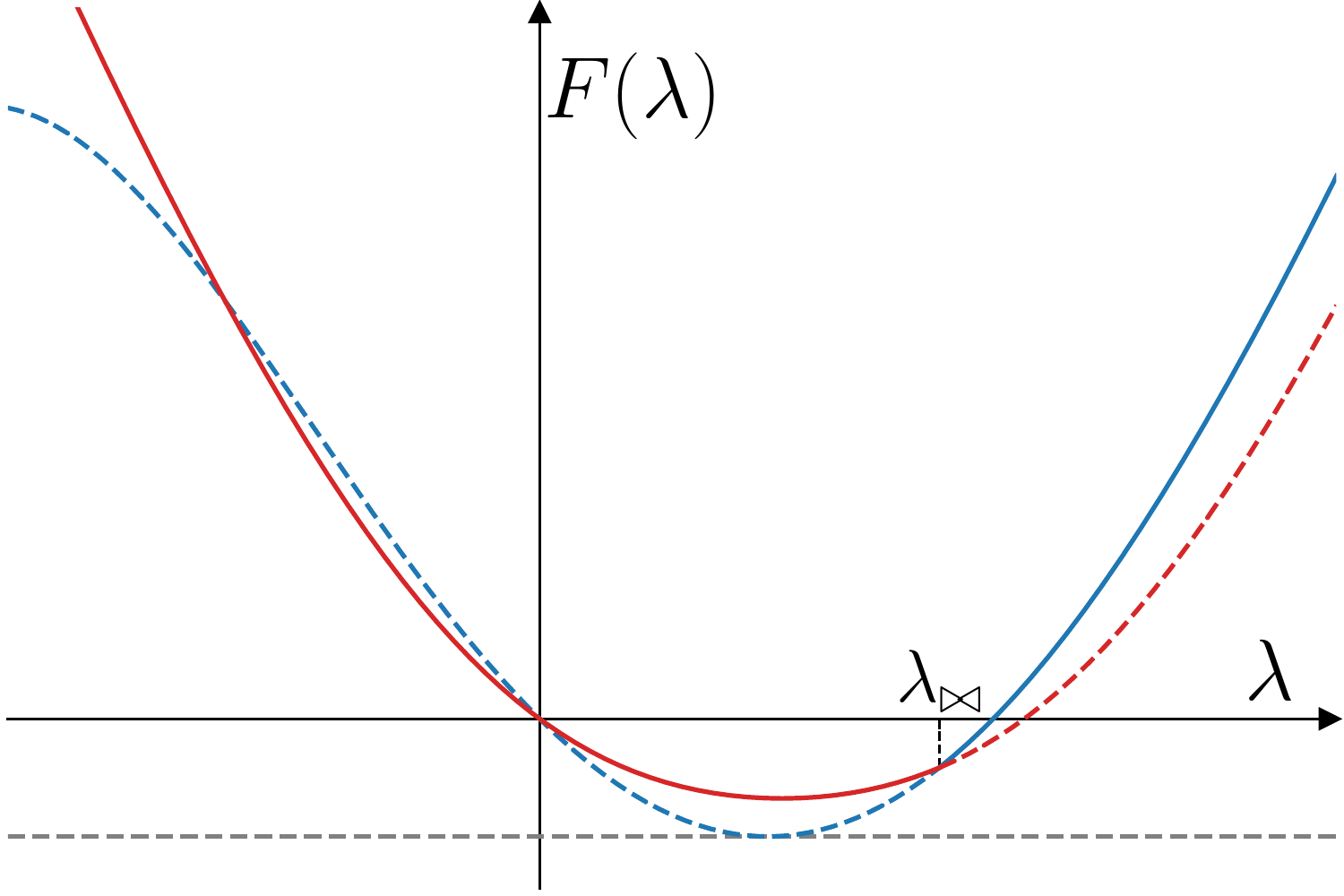}
\hspace*{2cm}
\includegraphics[width=0.3\textwidth]{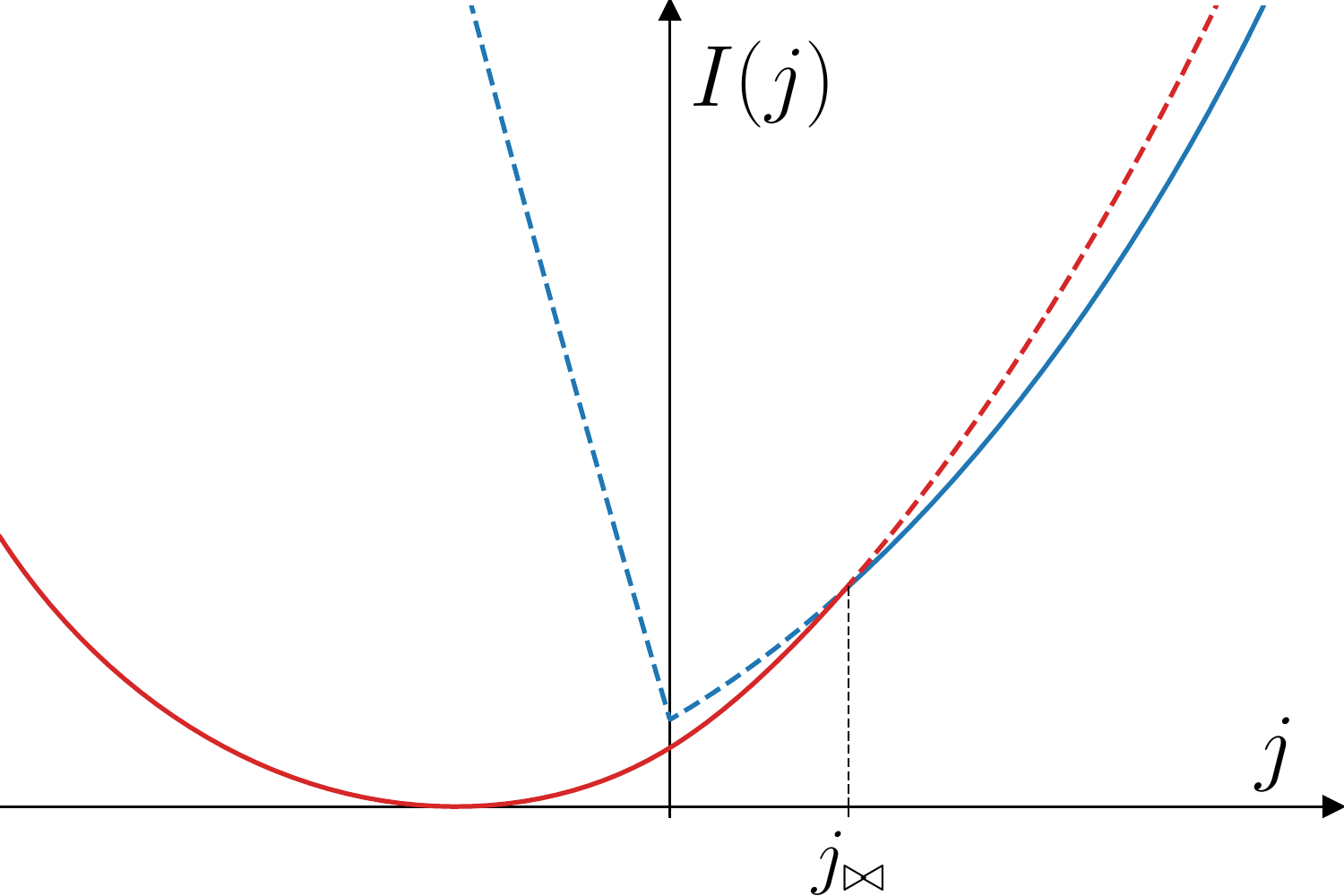}
\end{figure}

\item {\tt tunneling regime:} the flat branch $F_{0}$ sets in, arising from a continuous symmetry-breaking transition of a convex regular branch into a non-convex form with a doubly degenerate minimum. The flat part of $F(\lambda)$ connects between two degenerate minima of symmetry-broken branches $F_{\pm}$. We call this phenomenon `tunneling' (and symbolize it
by $\asymp$). There are two subregimes: the tunneling transition via $F_{0}$ connecting two degenerate minima of the same branch $F_{\pm}$, labeled by $\llb \pm \asymp \pm \rrb$, and the transition connecting the left minimum of $F_{\pm}$ to the right minimum of another branch $F_{\mp}$, labeled by $\llb \pm \asymp \mp \rrb$. Every transition between the minima of $F_{\pm}$ and $F_{0}$ is a dynamical  phase transitions of second order. In both subregimes, the LD rate function $I(j)$ is the minimum of both convex branches, $I(j)={\rm min}_{j}\{I_{\pm}(j)\}$, each of which has a non-differentiable point (corner) at $j=0$.

\begin{figure}[h!]
\includegraphics[width=0.3\textwidth]{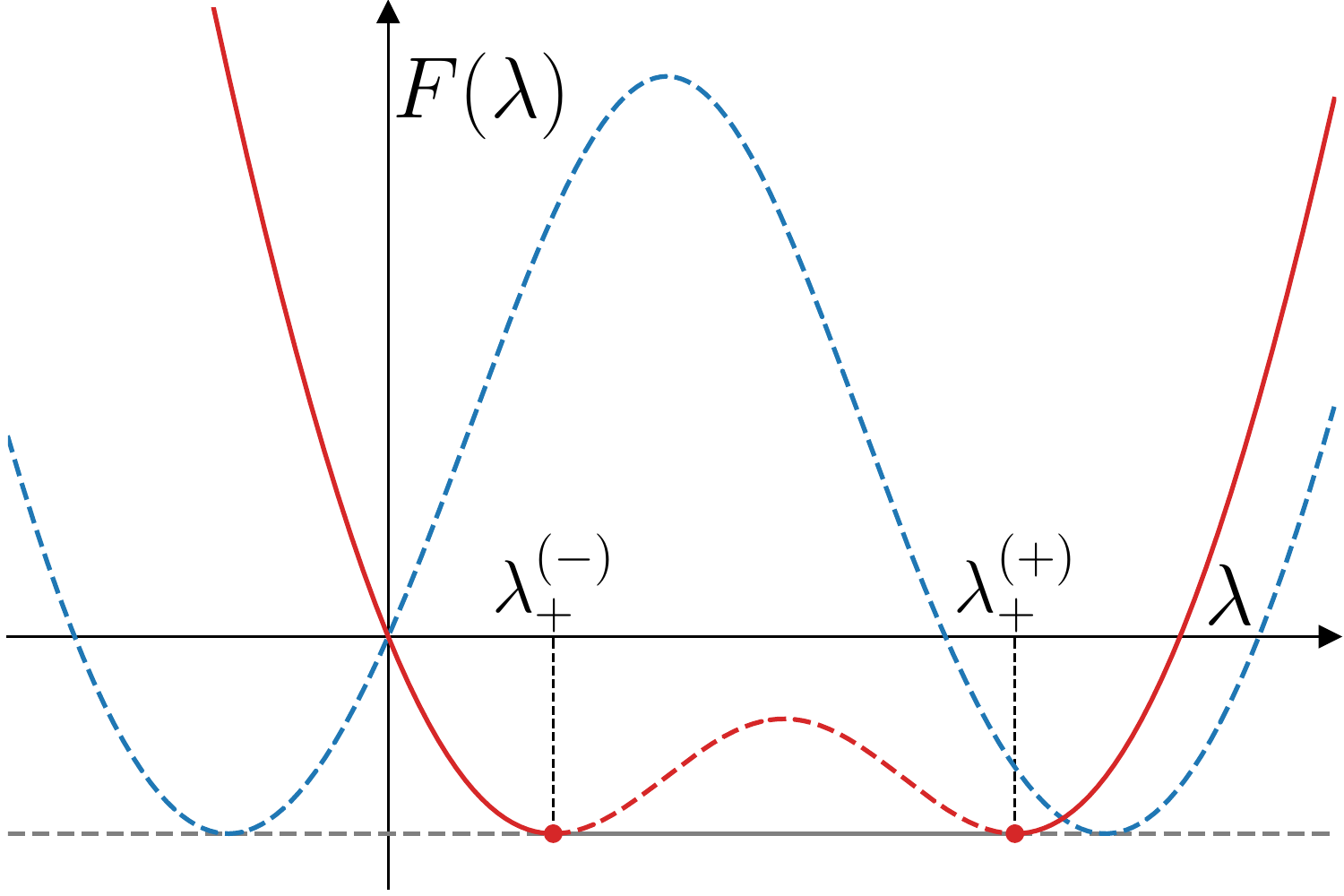}
\hspace*{2cm}
\includegraphics[width=0.3\textwidth]{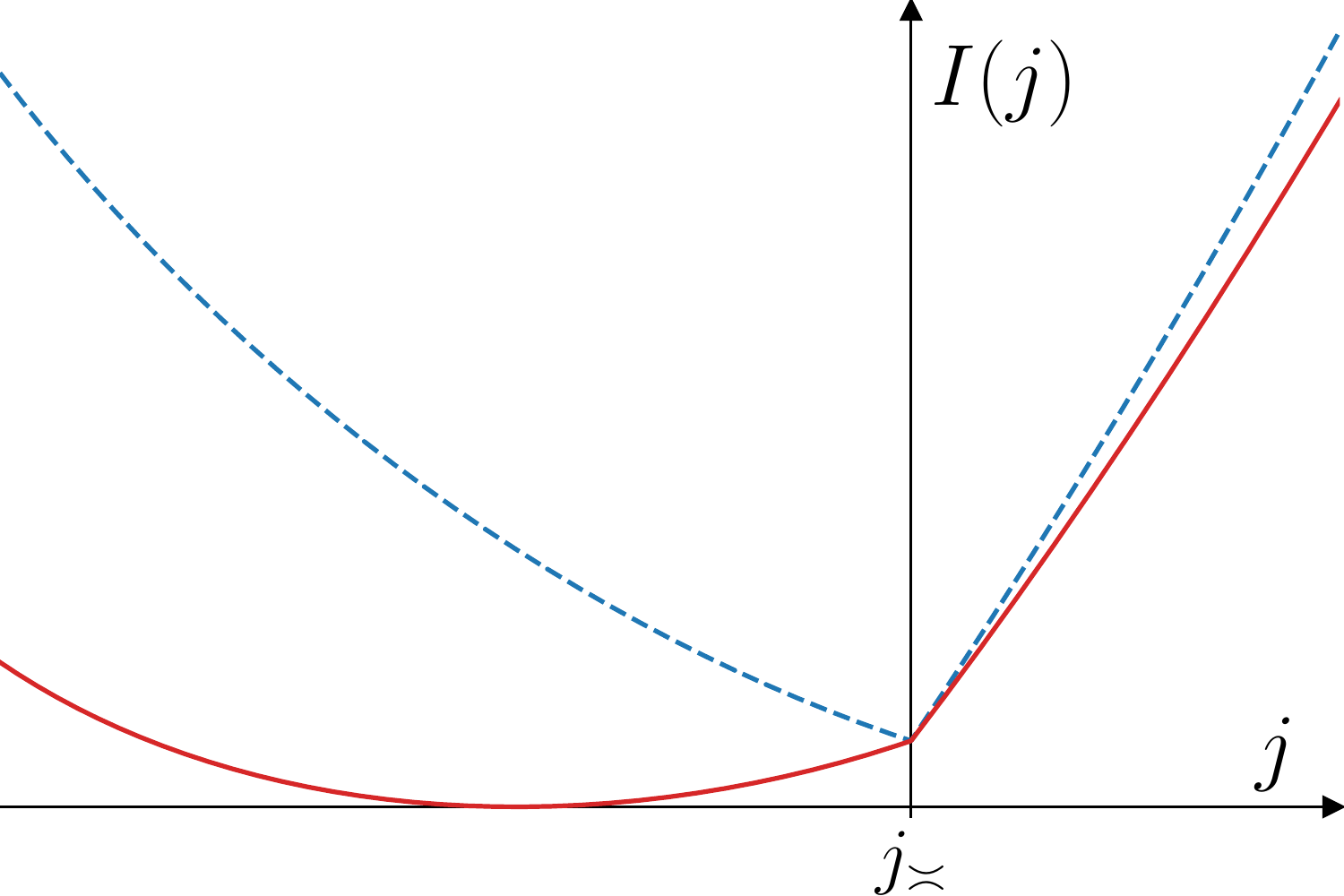}\\
\vspace*{0.5cm}
\includegraphics[width=0.3\textwidth]{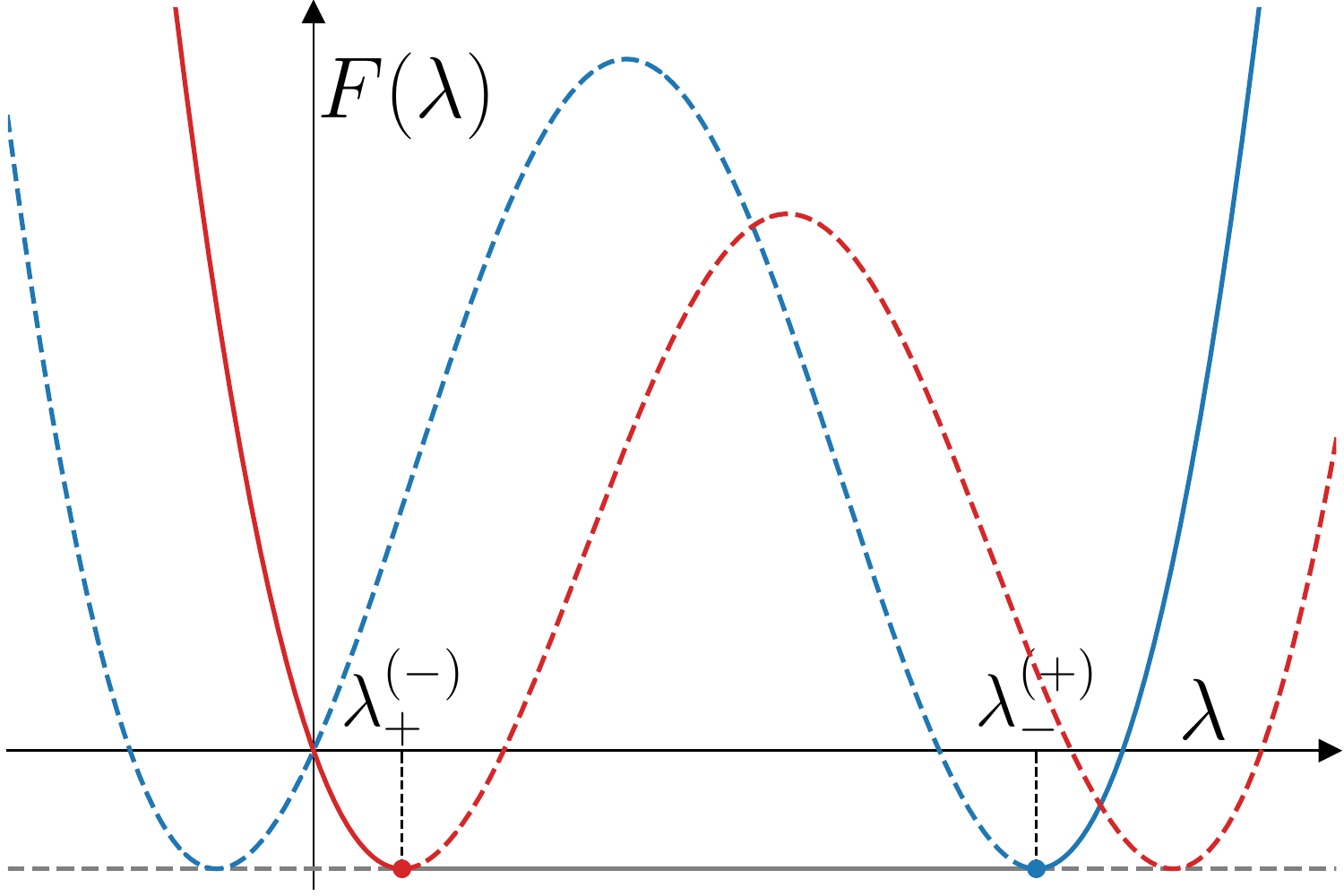}
\hspace*{2cm}
\includegraphics[width=0.3\textwidth]{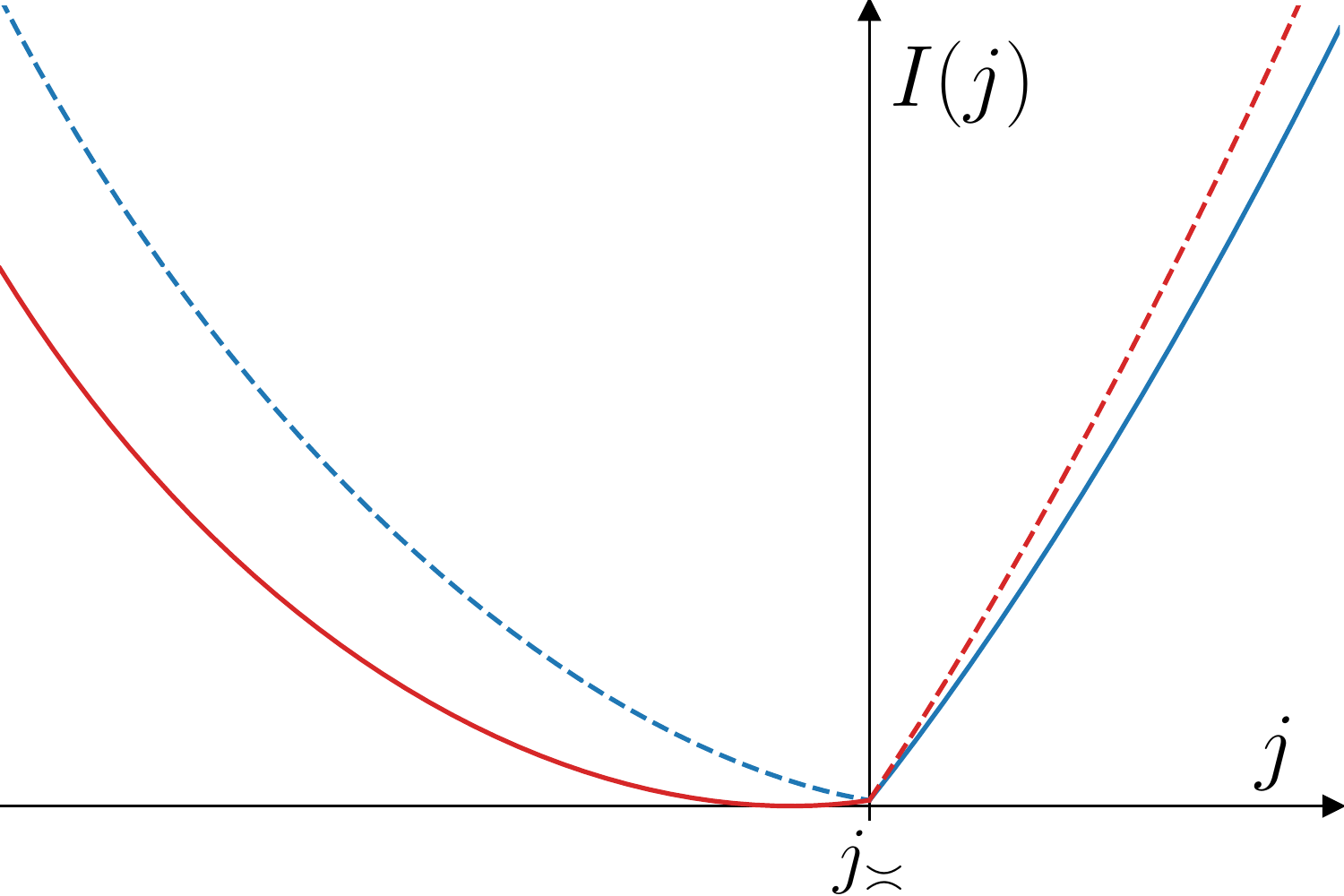}
\end{figure}

\item {\tt mixed regime:} apart from tunneling between two degenerate minima of the same bulk branch $F_{\pm}$, there is a transfer to another branch via a non-differentiable corner point, or in the opposite order. Intersection of bulk branches $F_{\pm}$ can never coexist with tunneling to another branch.
\end{itemize}

\twocolumngrid

\medskip

\paragraph*{Recap.}
Before detailing out various formal aspects of the outlined dynamical criticality, we take an opportunity to succinctly summarize the
key findings and state the general conclusions:
\begin{itemize}
\item[$\spadesuit$] The single-file property (I) combined with inertness of charge (II) implies fragmentation of the classical phase space, foliating into exponentially many sectors characterized by conserved charge patterns. Dynamical systems of this type display strongly non-ergodic behavior manifested through the competition of dynamical phases. For a given value of the counting field $\lambda \in \mathbb{R}$, the finite-time CGF $F(\lambda|t)$ receives contributions from
both branches $F_{\pm}(\lambda)$ attributed to each of the two partitions in the initial non-stationary state. While away from equilibrium only one of them dominates the growth of MGF at late times, this delicately depends on the average value of
particle and charge densities characterizing initial non-stationary states.
An exchange of dominance between $F_{+}(\lambda)$ and $F_{-}(\lambda)$ introduces a non-differentiable (corner) point in the SCFG $F(\lambda)$, signaling a DPT of the first order. In this event, the Gallavotti--Cohen relation breaks for large currents $|j|>|j_{\bowtie}|$, but survives for subcritical values $|j|<|j_{\bowtie}|$. The large-deviation rate function $I(j)$ corresponds to taking the minimum of
two convex branches $I_{\pm}(j)$.

\item[$\clubsuit$] There exists a region in the parameter space of initial bipartitioned states where a new constant branch $F_{0}$ emerges as a part of the physical SCGF $F(\lambda)$. The flat segments arise when the physical branch
experiences a symmetry-breaking phase transition into a non-convex form.
The flat branch interpolates between two degenerate minima -- either of the same symmetry-broken branch $F_{\pm}(\lambda)$
or two adjacent minima of the opposite branches -- along a compact interval $\mathscr{I}_{0}$ where it dominates over $F_{\pm}(\lambda)$.
While transitions from the bulk branches $F_{\pm}(\lambda)$ to the flat branch $F_{0}$ (or vice-versa) do not spoil differentiability of $F(\lambda)$, the second derivative $F^{\prime \prime}(\lambda)$ features a discontinuity at the minima
of $F_{\pm}(\lambda)$. The boundaries of $\mathscr{I}_{0}$ are accordingly interpreted as critical points of a second order DPT, transcribing into a single non-differentiable (corner) point in the associated LDF $I(j)$ at $j=0$.


\item[$\blacklozenge$] Depending on the presence and type of dynamical criticality,
there are four qualitatively distinct dynamical regimes, dubbed as \texttt{regular} (of type $\llb + \rrb$ or $\llb - \rrb$), \texttt{tunneling}
(of types $\llb \pm \asymp \pm \rrb$ or $\llb \pm \asymp \mp \rrb$), \texttt{corner} $\llb \pm \bowtie \mp \rrb$ and
finally \texttt{mixed} $\llb \pm \bowtie \pm \asymp \mp \rrb$. These four regimes provide the full partitioning of the parameter space.
The UFR of the SCGF $F(\lambda)$ is globally preserved in $\llb \pm \rrb$ and $\llb \pm \asymp \pm \rrb$ regimes, locally preserved for subcritical large currents in $\llb \pm \bowtie \mp \rrb$ regime and fully violated in $\llb \pm \asymp \mp \rrb$ regime.
\end{itemize}

\subsection{Large fluctuations: the dressing formalism}
\label{sec:dressing_formalism}

We now describe the \emph{dressing procedure} that permits one to compute the charge-current rate function from that of the particle-current. This can be achieved, remarkably, without resorting to any model-specific input. For technical reasons, we shall only assume certain minimal `regularity properties' on the particle rate function. We first outline the procedure at the level of the rate function by expressing the charge-current rate function as the solution to a convex optimization problem \eqref{eqn:Ic_infimum} and systematically examine its structure. Here we only provide a succinct summary. Futher details can be found in Appendix \ref{sec:dressing_technical}.

\subsubsection{Dressing the particle rate function}
\label{sec:dressing_rate}

By prescribing $\mathcal{P}_{p}({\rm J}_{p}|t)$, the joint PDF $\mathcal{P}_{c,p}({\rm J}_{p},{\rm J}_{c}|t)$ can be computed with aid of
the conditional PDF $\mathcal{P}_{c|p}({\rm J}_{c}|{\rm J}_{p})$ according to the main axiom of probability,
\begin{equation}
\mathcal{P}_{c,p}({\rm J}_{c},{\rm J}_{p}|t) = \mathcal{P}_{c|p}({\rm J}_{c}|{\rm J}_{p})\mathcal{P}_{p}({\rm J}_{p}|t).
\label{eqn:joint_probability}
\end{equation}
This can be formally viewed as an operator (with the kernel $\mathcal{P}_{c|p}({\rm J}_{c}|{\rm J}_{p})$) which we
regard as the dressing operator $\mathfrak{D}_\mathcal{P}$, namely
\begin{equation}
\mathcal{P}_{c, p}({\rm J}_{c}, {\rm J}_p|t)=\mathfrak{D}_{\mathcal{P}}[\mathcal{P}_{p}({\rm J}_{p}|t)]({\rm J}_{c}).
\label{eqn:dressing_P}
\end{equation}
The univariate PDF $\mathcal{P}({\rm J}_{c}|t)$ can be obtained by integrating out the particle current,
\begin{equation}
\mathcal{P}_c({\rm J}_{c}|t) = \int \dd {\rm J}_{p}\mathcal{P}_{c,p}({\rm J}_{c}, {\rm J}_{p}).
\label{eqn:conditional_probability}
\end{equation}
Since the assignment of internal charge degrees of freedom is, by virtue of the inertness property, uncorrelated with particles' positions,
$\mathcal{P}_{c|p}({\rm J}_{c}|{\rm J}_{p})$ is indeed merely a combinatorial factor.
We shall suggestively refer to it as the ``dressing factor''. Below we compute its general form for the class of bipartitioned grand-canonical ensembles.

We begin by introducing a few auxiliary objects. For simplicity, we assume in the following that space is discrete.
Fixing a window of time $[0,t]$, we denote by $\Lambda_{\pm}$ the sublattices occupied by particles at initial time $t=0$ that have crossed the origin during that time interval, with $n_{-}=|\Lambda_{-}|$ particles initially in the left subsystem and similarly $n_{+}=|\Lambda_{+}|$ particles in the right subsystem. Moreover, we denote by $p_{{\rm L},\pm}\equiv(1\pm b_{-})/2$ and $p_{{\rm R},\pm}\equiv (1\pm b_{+})/2$ the probabilities of finding a charge
$+$ or $-$ in the left (${\rm L}$) and right (${\rm R}$) partitions, respectively. Combinatorial counting yields 
\begin{align}
\label{eqn:uncrestricted_dressing}
\mathcal{P}_{c|p}({\rm J}_{c}|{\rm J}_{p}) &= \sum_{n_{-},n_{+}=0}^{\infty}\delta_{n_{-}-n_{+},{\rm J}_{p}}
\sum_{m_{-}}^{n_{-}}\sum_{m_{+}}^{n_{+}}\binom{n_{-}}{m_{-}}\binom{n_{+}}{m_{+}} \nonumber \\
& \times p^{m_{-}}_{L,+}p^{n_{-}-m_{-}}_{L,-}p^{m_{+}}_{R,-}p^{n_{+}-m_{+}}_{R,+}\delta_{2m-n,{\rm J}_{c}},
\end{align}
where the outer double summation over $n_{-}$ and $n_{+}$ goes over all possible combinations of crossings, while the inner double summation counts over all the colorings of charges. Note that we have also incorporated the Kronecker $\delta$-constraint to ensure the difference $n_{-}-n_{+}={\rm J}_{p}$, with $n\equiv n_{+}+n_{-}$, $m\equiv m_{+}+m_{-}$.

By further demanding the single-file property \eqref{eqn:single_file}, the general form of Eq.~\eqref{eqn:uncrestricted_dressing} simplifies significantly. The central observation is that \emph{at most one} of the sets $\Lambda_\pm$ is non-empty which,
after explicitly resolving the Kronecker constraints, brings us to a far simpler, factorizable expression for the conditional probability,
\begin{equation}
\mathcal{P}_{c|p}({\rm J}_{c}|{\rm J}_{p}) = \mathcal{P}^{[0]}_{c|p}({\rm J}_{c}|{\rm J}_{p})\, \mathcal{B}({\rm J}_{p},{\rm J}_{c}).
\end{equation}   
Largely for convenience, we have separated out the conditional probability in the absence of biases ($b_\pm=0$),
\begin{equation}
\mathcal{P}^{[0]}_{c|p}({\rm J}_{c}|{\rm J}_{p}) \equiv 2^{-|{\rm J}_{p}|} \binom{|{\rm J}_{p}|}{\tfrac{|{\rm J}_{p}|+{\rm J}_{c}}{2}},
\label{eqn:P0_def}
\end{equation}
and introduced the `biasing factor' $\mathcal{B}({\rm J}_{p},{\rm J}_{c})$ of the form
\begin{equation}
\mathcal{B}({\rm J}_{p},{\rm J}_{c}) \equiv 2^{|{\rm J}_{p}|}\left(\frac{1-b^{2}_{\varsigma}}{4}\right)^{|{\rm J}_{p}|/2}
\left(\frac{1-b_{\varsigma}}{1+b_{\varsigma}}\right)^{\varsigma\,{\rm J}_{c}/2},
\end{equation}
depending on bias parameters $b_{\pm}$ implicitly through the direction of the integrated particle current ${\rm J}_{p}$
via signature $\varsigma \equiv -{\rm sgn}({\rm J}_{p})$. Owing to the single-file property, positive (negative) number of transferred particles are associated with the negative (positive) partitions.

\medskip

\paragraph*{Large fluctuations.}

We are now in a position to infer the exact LDF of the transferred charge from the asymptotic behavior
of the dressing kernel $\mathcal{P}_{c|p}({\rm J}_{c}|{\rm J}_{p})$. To this end, we first pick two arbitrary dynamical exponents $\zeta_{i}$
in the range $1/2z_{i}\leq \zeta_{i}\leq \alpha_{i}$ (with $i\in \{p,c\}$)
and introduce the corresponding rescaled cumulative currents, $\mathcal{J}_{i}(t)=t^{-\zeta_{i}}{\rm J}_{i}(t)$.
In terms of the rescaled currents, we have the following asymptotic formula
(abusing notation for PDFs with scaled arguments)
\begin{equation}
t^{-\zeta_{c}}\mathcal{P}_{\zeta_{c}}(j_{c}|t) \asymp
\int \dd j_{p}\mathcal{P}_{c|p}(j_{c}|j_{p})\mathcal{P}_{\zeta_{p}}(j_{p}|t).
\label{eqn:Pjc_asymptotic}
\end{equation}
We are mainly interested in asymptotic behavior associated with the \emph{largest} timescale $\zeta_{p}=\alpha_{p}\equiv \alpha$,
pertaining to rare space-time trajectories in which the transferred particle number scales asymptotically
as $\sim t^{\alpha}$. We remind the reader that the timescale $\alpha$ is fixed by the rate of growth
of $G_{p}(\lambda|t)\equiv \expect{e^{\lambda\,{\rm J}_{p}(t)}}_{\rm init}$ at late times.
Inertness of charge immediately implies that the net charge current carried by those rare events is of the same order,
that is $\alpha_{c}=\alpha_{p}$. When we wish to infer the statistics of large charge-current fluctuations we therefore set $\zeta_{c}=\zeta_{p}={\alpha}$.

\medskip

In the following, we further make the following assumption on the cumulative particle current ${\rm J}_{p}(t)$:
\begin{enumerate}[(a)]
\item ${\rm J}_{p}(t)$ obeys the LD principle on the \emph{large} timescale $\alpha$, with the SCGF given by
\begin{equation}
F_{p}(\lambda_p)=\lim_{t\to \infty}t^{-\alpha}\log G_{p}(\lambda_p|t).
\end{equation}
Let moreover $\mathcal{P}_{\alpha}(\mathcal{J}_{p}(t)|t)$ denote the PDF 
associated to the rescaled time-integrated particle current $\mathcal{J}_{p}(t) = t^{-\alpha}{\rm J}_{p}(t)$.
At late times, the probability for observing a value $j_{p}$ of the rescaled cumulative particle current $\mathcal{J}_{p}(t)$ 
is characterized by the rate function
\begin{equation}
I_{p}(j_{p}) = -\lim_{t\to \infty}t^{-\alpha}\log \mathcal{P}_{\alpha}(\mathcal{J}_p =j_{p}|t).
\end{equation}
\item in equilibrium, \emph{typical} fluctuations of the cumulative particle current ${\rm J}_{p}(t)$, characterized
by scaling exponent $\zeta_{p}=1/2z_{p}$ (where $z_{p}=1/\alpha$ is the algebraic dynamical exponent associated with
the asymptotic temporal growth of the second cumulant of $G_{p}(\lambda_p|t)$),
are \emph{Gaussian} with zero mean and variance of $\sigma^{2}_{p}$,
\begin{equation}
\mathcal{P}_{1/2z_{p}}(j_{p}) = \frac{1}{\sqrt{2\pi \sigma^{2}_{p}}}\exp{\left[-\frac{j^{2}_{p}}{2\sigma^{2}_{p}}\right]}.
\end{equation}
\end{enumerate}

We proceed by approximating the binomial weights in Eq.~\eqref{eqn:P0_def} using the Stirling formula. To facilitate the computation, it is convenient to introduce a new dynamical variable
\begin{equation}
\xi(t)\equiv t^{\zeta_{c}-\zeta_{p}}\frac{j_{c}}{|j_{p}|},
\label{eqn:xi_def}
\end{equation}
in terms of which the exact asymptotic expression
for the conditional probability (suppressing subexponential terms) takes the form
\begin{equation}
\mathcal{P}^{[0]}_{c|p}(j_{c}|j_{p}) \asymp \exp{\left[-t^{\zeta_{p}}|j_{p}|\Xi(\xi)\right]},
\label{eqn:rescaled_conditional}
\end{equation}
with
\begin{equation}
\Xi(\xi) \equiv \frac{1}{2}\sum_{\epsilon \in \{\pm\}}\left[(1+ \epsilon \, \xi)\log(1+ \epsilon \, \xi)\right]-\log{(2)}.
\end{equation}
Similarly, the (rescaled) biasing weight, denoted hereafter by $\mathcal{B}(j_{p},j_{c})$, can be presented in a factorized form,
\begin{equation}
\mathcal{B}(j_{p},j_{c}) \asymp \mathcal{B}_{p}(j_{p})\mathcal{B}_{c}(j_{c})
\label{eqn:rescaled_biasing_weight}
\end{equation}
with
\begin{align}
\mathcal{B}_{p}(j_{p}) &= \exp{\left[t^{\zeta_{p}}\frac{|j_{p}|}{2}\log{\left(\frac{1-b^{2}_{\varsigma}}{4}\right)}\right]},\\
\mathcal{B}_{c}(j_{c}) &= \exp{\left[t^{\zeta_{c}}\frac{\varsigma\,j_{c}}{2}\log{\left(\frac{1-b_{\varsigma}}{1+b_{\varsigma}}\right)}\right]}.
\end{align}

We now observe that for moderate fluctuations associated with timescales $\zeta_{c}<\zeta_{p}$, only the lowest non-trivial order in $\xi$ in Eq.~\eqref{eqn:rescaled_conditional} remains relevant at late times,
yielding a remarkably simple result
\begin{equation}
\mathcal{P}^{[0]}_{c|p}(j_{c}|j_{p}) \asymp \exp{\left[-t^{2\zeta_{c}-\zeta_{p}}\frac{j^{2}_{c}}{|j_{p}|}\right]}.
\label{eq:asymp_P0}
\end{equation} 
By contrast, in the case of large fluctuations $\zeta_{c}=\zeta_{p}$ we have $\xi \in \mathcal{O}(t^{0})$, i.e.
$\xi\equiv \xi(0)=j_{c}/|j_{p}|$ becomes independent of time. The charge-current univariate PDF $\mathcal{P}_{c}(\mathcal{J}_{c}|t)$
of the dynamically rescaled cumulative charge current $\mathcal{J}_{c}(t)=t^{-\alpha}\,J_{c}(t)$ is accordingly given by
the following asymptotic expression (for compactness suppressing irrelevant subexponential terms in the integrand)
\begin{equation}
e^{-t^{\alpha}I_{c}(j_{c})} \asymp t^{-\alpha/2}
\int \dd j_{p} \, e^{-t^{\alpha}I_{c,p}(j_{c},j_{p})}.
\label{eqn:Ic_from_joint_rate_function}
\end{equation}
The above expression can be viewed as marginalization of the \emph{joint} rate function $I_{c,p}(j_{c},j_{p})$, see e.g.~\cite{Chaganty97}.
The latter can be naturally decomposed as
\begin{equation}
\tcbhighmath[drop fuzzy shadow]
{
I_{c,p}(j_{c},j_{p}) = I_{c|p}(j_{c},j_{p}) + I_{p}(j_{p}),
}\label{eqn:I_dressing}
\end{equation}
where $I_{p}(j_{p})$ is interpreted as the \emph{marginal} rate function, while $I_{c|p}(j_{c}|j_{p})$ is the \emph{conditional rate function}
\begin{equation}
I_{c|p}(j_{c}|j_{p}) = \sum_{\epsilon \in \{\pm\}}\frac{|j_{p}|}{2}
(1+\epsilon \, \xi)\log{\left[\frac{1+\epsilon \, \xi}{1-\epsilon \varsigma b_{\varsigma}}\right]},
\end{equation}
with signature $\varsigma = -{\rm sgn}(j_{p})$.

In summary, provided the particle SCGF $F_{p}(\lambda_{p})$ as an input,
one can retrieve the joint LDF $I_{c,p}(j_{c},j_{p})$ and SCGF $F_{c,p}(\lambda_{c},\lambda_{p})$ via the following sequence of
explicit transformations
\begin{equation}
F_{p}(\lambda_{p}) \xrightarrow{\!\star\!} I_{p}(j_{p}) \xrightarrow{\mathfrak{D}_\mathcal{P}} I_{c, p}(j_{c}, j_p)
\xrightarrow{\!\star\!} F_{c, p}(\lambda_{c}, \lambda_p)
\end{equation} 
where the action of $\mathfrak{D}_\mathcal{P}$ on a rate function is given by Eq.~(\ref{eqn:I_dressing}).
Moreover, if $F_{p}(\lambda_{p})$ fulfills the assumptions of the G\"{a}rtner--Ellis theorem, the corresponding rate
function $I_{p}(j_{p})$ is simply given by the Legendre transform of the particles SCGF $F_{p}(\lambda)$,
i.e. $I_{p}(j_{p}) = F_{p}^{\star}(j_{p})$.

The univariate charge-current LDF $I_{c}(j_{c})$ can be straightforwardly retrieved by marginalization.
By invoking the Laplace principle, in the large-time limit $t\to \infty$ the integral \eqref{eqn:Ic_from_joint_rate_function} localizes around the extremum, implying
\begin{equation}
\tcbhighmath[drop fuzzy shadow]
{I_{c}(j_{c}) = {\rm inf}_{j_{p}} I_{c,p}(j_{c},j_{p}) .}
\label{eqn:Ic_infimum}
\end{equation}
To finally obtain the SCGF $F_{c}(\lambda)$ one can make use of the Legendre--Fenchel transform,
\begin{equation}
F_{c}(\lambda) = I_c^\star(\lambda) \equiv {\rm sup}_{j_{c}}\{\lambda j_{c}-I_{c}(j_{c})\}.
\end{equation}

\subsubsection{Dressing the moment generating function}

The dressing procedure described in Sec.~\ref{sec:dressing_rate} can be alternatively formulated at the level of the moment generating functions.
Here we derive a simple correspondence between the finite-time particle-current MGF $G_{p}(\lambda_{p}|t)$ and the joint
particle-charge MGF $G_{c,p}(\lambda_{c},\lambda_{p}|t)$. In the following computations, we employ
the multiplicative counting fields $\zz_{p}\equiv e^{\lambda_{p}}$ and $\zz_{c}\equiv e^{\lambda_{c}}$ and
assume that the integrated currents ${\rm J}_{p}$, ${\rm J}_{c}$ take only integer values (as is the case {for point particle or discrete variable systems}).

Computing the joint \emph{finite-time} MGF $G_{c,p}(\lambda_{c},\lambda_{p}|t)$ amounts to acting with the dressing operator $\mathfrak{D}_{G}$,
\begin{equation}
G_{c,p}(\zz_{c},\zz_{p}|t) = \mathfrak{D}_{G}[G_{p}(\zz_{p}|t)],
\label{eqn:Dg}
\end{equation}
on the particle MFG
\begin{equation}
G_{p}(\zz_{p}|t)=\expect{\zz^{{\rm J}_{p}(t)}_{p}} = \mathfrak{L}[\mathcal{P}({\rm J}_{p}|t)](\zz^{-1}_{p}).
\end{equation}
The dressing operator can be most conveniently expressed as a composition
$\mathfrak{D}_{G}=\mathfrak{L} \, \circ \, \mathfrak{D}_{\mathcal{P}} \, \circ \, \mathfrak{L}^{-1}$,
representing conjugation of $\mathcal{P}({\rm J}_{c}|t)$ by the \emph{bilateral} Laplace transform $\mathfrak{L}$,
\begin{equation}
\mathfrak{L}[\mathcal{P}({\rm J}_{p}|t)](\zz^{-1}_{p})\equiv \int \dd {\rm J}_{p} \zz^{{\rm J}_{p}} \mathcal{P}({\rm J}_{p}|t),
\end{equation}
whose inverse satisfies $\mathcal{P}({\rm J}_{p}|t)=\mathfrak{L}^{-1}[G_{p}(\zz_{p}|t)]({\rm J}_{p})$.

Evaluating the action of the dressing operator \eqref{eqn:Dg} requires a few technical steps which are spelled out
in Appendix \ref{sec:MGF_dressing}. There we demonstrate that acting with $\mathfrak{D}_{G}$ corresponds to applying
the following simple \emph{substitution rule}:
\begin{equation}
\tcbhighmath[drop fuzzy shadow]{
G_{c,p}(\zz_{c},\zz_{p}|t) = G_{p}(\zz_{p}|t) \Big|_{\zz^{\pm n}_{p}\to \zz^{\pm n}_{p}[\mu_{\mp}(\zz_{c})]^{n}},}
\label{eqn:replacement_rule}
\end{equation}
with the `dressed counting fields'
\begin{equation}
\mu_\pm(\zz_{c}) = \tfrac{1}{2}(\zz_{c}  + \zz^{-1}_{c}) \mp b_{\pm}\tfrac{1}{2}(\zz_{c} - \zz^{-1}_{c}).
\end{equation}

In summary, we have thus established that
\begin{quote}
\emph{the finite-time joint MGF $G_{c, p}(\zz_c, \zz_p|t)$ is given by the Laurent series expansion of the particle MGF $G_{p}(\zz_{p}|t)$
upon multiplying all positive and negative integral powers of (exponential) counting fields $\zz^{\pm n}_{p}$ by
the corresponding dressed counting fields $[\mu_{\pm}(\zz_{c})]^{n}$.}
\end{quote}

\subsection{Universal anomalous fluctuations in equilibrium}

In this section, we consider the univariate PDF of the cumulative charge current rescaled
to the timescale of typical fluctuations $\zeta_{c}=1/2z_{c}$,
\begin{equation}
\mathcal{P}^{\rm typ}_{c}(j_{c})=\lim_{t\to \infty}t^{-\zeta_{c}}\mathcal{P}_{\zeta_{c}=1/2z_c}(\mathcal{J}_{c}|t).
\end{equation}
We shall now establish the following remarkable property: in \emph{equilibrium} ensembles with finite particle density and without
charge bias ($b=0$), the PDF $\mathcal{P}^{\rm typ}_{c}(j_{c})$ takes a universal \emph{non-Gaussian} form in spite of detailed balance.
This property has been previously observed and explained in our recent paper \cite{Krajnik22},
where we computed the FCS for an exactly solvable classical automaton of hardcore charged particles.
We revisit the model in Sec.~\ref{sec:hardcore} and compute the FCS of charge transfer
with respect to non-stationary bipartitioned initial states.

We wish to stress that the observed anomalous fluctuations found in unbiased equilibrium ensembles are a general feature of
dynamical systems that are subjected to the constraints specified in Sec.~\ref{sec:setting}.
In other words, absence of the so-called CLT property is a corollary of the imposed constraints, namely (I) the single-file property and (II) inertness of charge.

We can infer directly from Eqs.~\eqref{eqn:Pjc_asymptotic} and Eq.~\eqref{eq:asymp_P0} that convergence of the rescaled charge-current
PDF $t^{-\zeta_{c}}\mathcal{P}_{\zeta_{c}}(\mathcal{J}_{c}|t)$ towards a non-trivial \emph{stationary} PDF can be achieved only provided that
the particle and charge dynamical exponents obey
\begin{equation}
\tcbhighmath[drop fuzzy shadow]{z_{c} = 2z_{p}.}
\end{equation}
We have thus inferred that inert charges are slowed down and spread through the system on a timescale given by
the square root of that associated with particle transport.

The PDF $\mathcal{P}^{\rm typ}_{c}(j_{c})$ takes the \emph{universal} form with the following integral representation
\begin{equation}
\mathcal{P}^{\rm typ}_{c}(j_{c}) = \frac{1}{2\pi \sigma_p}\int^{\infty}_{-\infty}\frac{\dd j_{p}}{\sqrt{|j_{p}}|}
\exp{\left[-\frac{j^{2}_{p}}{2\sigma^{2}_{p}}-\frac{j^{2}_{c}}{2|j_{p}|}\right]}.
\end{equation}
The corresponding MGF $G^{\rm typ}_{c}(\eta)$ is given by the bilateral Laplace transform of $\mathcal{P}^{\rm typ}_{c}(j_{c})$, namely
$G^{\rm typ}_{c}(\eta)=\mathfrak{L}[\mathcal{P}^{\rm typ}_{c}(j_{c})](\eta)
=\int^{\infty}_{-\infty}\dd j_{c}e^{-\eta\,j_{c}}\mathcal{P}^{\rm typ}_{c}(j_{c})$,
yielding the following integral representation
\begin{equation}
G^{\rm typ}_{c}(\eta) = \frac{1}{\sqrt{2\pi} \sigma_p} \int^{\infty}_{-\infty}\dd j_p
\exp{\left[-\frac{j_p^2}{2\sigma^{2}_{p}}+\frac{|j_p|\eta^{2}}{2}\right]}.
\end{equation}
Splitting the integral into two separate integrals over the real semi-axes $\mathbb{R}_{\pm}$,
and using the identity $\int^{\infty}_{0}\dd u\,\pi^{-1/2}\exp{[-v\,u-(u/2)^{2}]}=\exp{(v^{2})}{\rm erfc}(v)$,
we arrive at the compact explicit expression
\begin{equation}
\tcbhighmath[drop fuzzy shadow]{G^{\rm typ}_{c}(\eta) = E_{1/2}\Big(\frac{\eta^{2} \, \sigma_{\star}}{4}\Big),}
\end{equation}
where $E_{1/2}(y)$ belong to a one-parameter family of Mittag-Leffler functions $E_{a}(y)$ (see e.g. \cite{Haubold11})
\begin{equation}
E_{a}(y) = \sum_{n\geq 0}\frac{y^{n}}{\Gamma(a\,n+1)},
\end{equation}
widespread in applications of fractional calculus \cite{Mainardi_book,Gorenflo_book} ($\Gamma$ denotes Euler's Gamma function), while
$\sigma_{\star}\equiv \sqrt{2}\sigma_{p}$ sets the characteristic width. In particular, the second cumulant $\kappa_{2}$
of $\mathcal{P}^{\rm typ}_{c}(j_{c})$ equals $\kappa_{2}=\sqrt{2/\pi}\sigma_{p}$.
For example, in the hardcore automaton $\sigma_{p}(\rho)=\sqrt{\rho(1-\rho)}$.

\begin{figure}[htb]
\includegraphics[width=1.02\columnwidth]{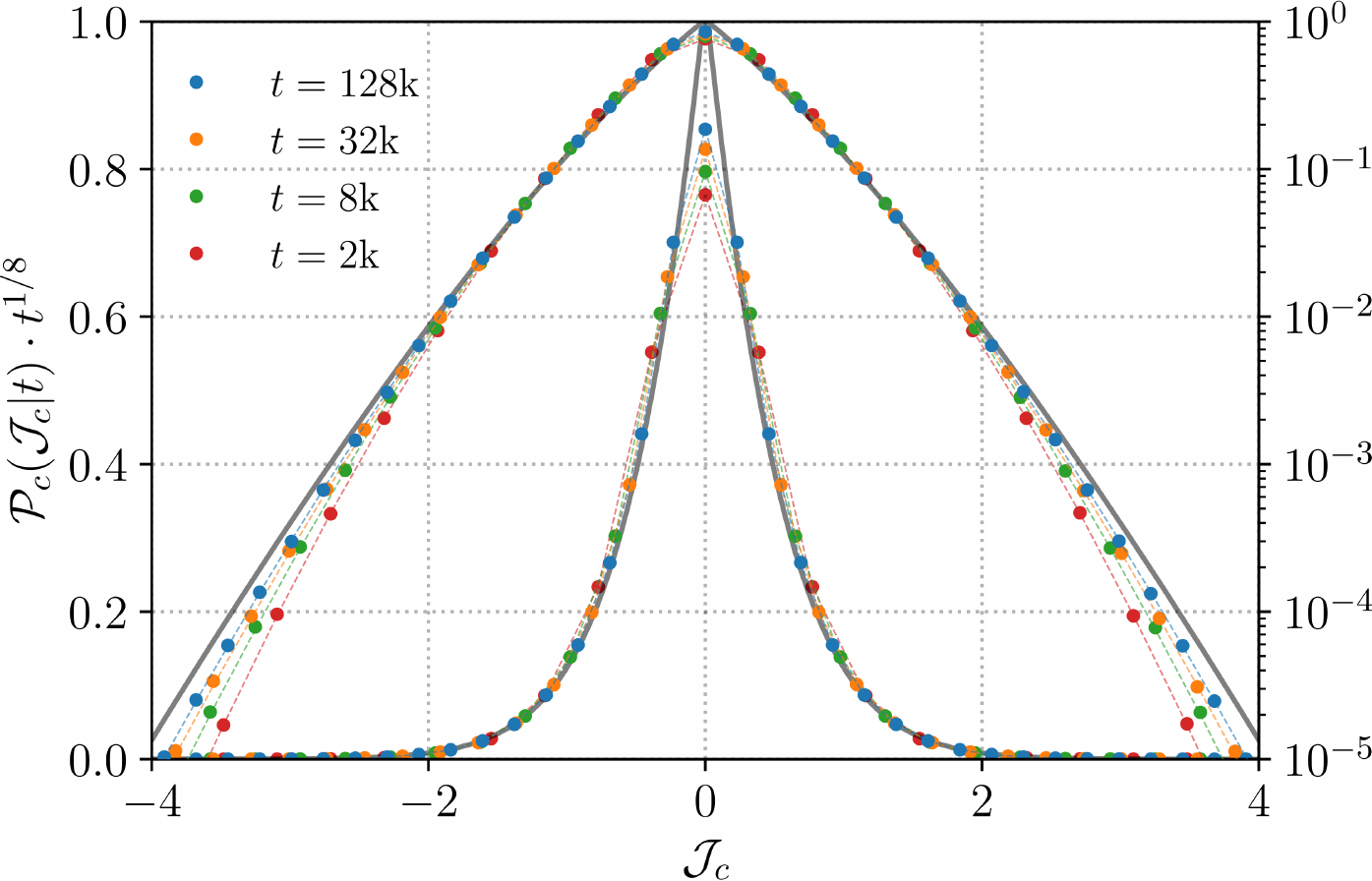}
\caption{Time-dependent probability density functions $t^{1/2z_{c}}\mathcal{P}_{c}(\mathcal{J}_{c}|t)$ (colored dashed curves, computed using Eqs.~\eqref{eqn:joint_probability}, \eqref{eqn:uncrestricted_dressing} and a Gaussian distribution of ${\rm J}_p$)
of the rescaled cumulative charge current $\mathcal{J}_{c}(t)=t^{-1/2z_{c}}{\rm J}_{c}(t)$ associated with
typical fluctuations with $z_{c}=4$, shown for the two-species simple symmetric exclusion process with parallel update rule at different times (displayed on normal and logarithmic scale). Solid black curves are the theoretically predicted M-Wright distribution corresponding to the stationary PDF $\mathcal{P}^{\rm typ}_{c}(j_{c})=\lim_{t\to \infty}t^{1/8}\mathcal{P}(\mathcal{J}_{c}|t)$.}
\label{fig:coloredSSEP}
\end{figure}

Since $E_{a}(y)$ are entire functions, their inverse (bilateral) Laplace transform is essentially the Fourier transform,
yielding $\mathfrak{L}^{-1}[E_{2a}(y^{2})]=\tfrac{1}{2}M_{a}(|y|)$.
Here $M_{a}(y)$ denote a one-parameter family of PDFs known by
the name of \emph{symmetrized M-Wright function} \cite{Mainardi20},
\begin{equation}
M_{a}(y)=\sum_{n\geq 0}\frac{(-y)^{n}}{n!\,\Gamma((1-a)-n\,a)},
\end{equation}
belonging to a subfamily of special functions called Wright functions. The final result is a closed-form universal expression for the PDF,
\begin{equation}
\tcbhighmath[drop fuzzy shadow]
{\mathcal{P}^{\rm typ}_{c}(j_{c}) = \frac{1}{\sqrt{\sigma_{\star}}} M_{1/4}\Big(\frac{2|j_{c}|}{\sqrt{\sigma_{\star}}}\Big),}
\label{eqn:universal_typical_distribution}
\end{equation}
shown in Fig.~\ref{fig:coloredSSEP}.
The M-Wright function of the scaling variable $2J_{c}/\sqrt{\sigma_{\star}}t^{1/2z_{c}}$ indeed plays the role of
the Green's function of the Cauchy problem associated with the time-fractional (in Caputo sense) diffusion equation
$\partial^{\beta}\psi/\partial t^{\beta}=D_{\beta}\partial^{2}\psi/\partial x^{2}$ of fractional order $\beta = 2/z_{c}$ (see also \cite{PhysRevE.61.132,Barkai2002} for a connection between fractional diffusion and continuous-time random walks).
The analogy is not exact, however; in the above PDF, the index of the function is always (i.e. irrespective of exponent $z_{p}$) equal to $z_{p}/(2z_{c})=1/4$. This ratio is presently uniquely fixed by demanding time-stationarity of the appropriately rescaled
dynamical PDF $\mathcal{P}_{c}({\rm J}_{c}|t)$.

\subsection{Dynamical phases}
\label{sec:phases}

In this section, we take a closer look at the univariate SCGF $F_{c}(\lambda_{c})\equiv F(\lambda)$.
Given that the function $F(\lambda)$ governs the asymptotic growth of the charge-current MGF $G_{c}(\lambda|t)\equiv G(\lambda|t)$, we may regard it -- \emph{mutatis mutandis} -- as a dynamical analogue of (thermodynamic) free energy. Supplying the particle univariate
MGF $G_{p}(\lambda|t)$, the exact $G(\lambda|t)$ is obtained by substituting the counting fields $\exp{(\pm n\lambda)}$ with their biased counterparts $(\mu_{\pm}(\lambda))^{n}$, see Eq.~\eqref{eqn:replacement_rule}. The task boils down to inferring the asymptotic behavior of
$G(\lambda|t)$, by e.g. applying the methods of localization or steepest descent.
To this end, one first replaces the discrete summation and combinatorial weights with appropriate continuous counterparts, yielding a two-dimensional integral over a (possibly non-compact) integration domain $\mathscr{D}$.
Then, we take into account that \emph{for asymptotically large times, the integral concentrates about the extremal points (local maxima) in the integration domain $\mathscr{D}$.} In general, we have to distinguish between two types of extrema:
(i) local maxima in the interior of the domain $\mathscr{D}$, and (ii) maxima located at the boundary $\partial \mathscr{D}$.

Whenever the MFG $G(\lambda|t)$ receives contributions from multiple extrema, we may asymptotically approximate it by
\begin{equation}
G(\lambda|t) \asymp \sum_{k=1}^{K} W_{k}(\lambda)\,e^{t^{\alpha} F_{k}(\lambda)}.
\label{eqn:G_branches}
\end{equation}
Here $k=1,\ldots,K$ enumerates all (possibly degenerate) critical points (including the boundary extrema), while expansion coefficients $W_{k}(\lambda)$ will be referred to as weighting functions. Functions $F_{k}(\lambda)$ in the exponents can be suggestively interpreted as coexisting (stable or metastable) \emph{branches} of the dynamical free energy $F(\lambda)$.

By fixing the value of the counting field $\lambda$, a single branch $F_{k}(\lambda)$ eventually dominates at late times (save for degenerate cases which we neglect for the time being). All other subdominant branches are only visible in the transient dynamics
(in the form of corrections that are exponentially suppressed as $t \to \infty$).
Nothing however prevents a subdominant branch from taking over the dominant one upon varying the counting field. It is thus conceivable that by virtue of multiple coexisting competing dynamical phases, different branches dominate the late-time growth of $G(\lambda|t)$, depending of the value of the counting field $\lambda$.
As a matter of fact, such behavior may depend on the parameters of the initial state.

Allowing for the most general scenario, we can envisage a partitioning of the \emph{complex} $\lambda$-plane
$\cup_{k=1}^{K}\mathscr{R}_{k} \cong \mathbb{C}$, where $\mathscr{R}_{k} \in \mathbb{C}$ denote the corresponding `regions of dominance'.
If we are only interested in reconstructing the LD rate function $I(j)$, it suffices to compute
$F(\lambda)$ along the real $\lambda$-axis (with an extra technical assumption that that $F(\lambda)$ is everywhere differentiable).
Let accordingly $\mathscr{I}_{k}$ denote the regions of dominance pertaining to $F_{k}(\lambda)$. Intervals $\mathscr{I}_{k}$ are given by intersections of $\mathscr{R}_{k}$ with the real $\lambda$-axis, $\cup_{k=1}^{K}\mathscr{I}_{k} \cong \mathbb{R}$.
Certain $\mathscr{I}_{k}$ can be empty, attributed to those $F_{k}(\lambda)$ that remain subdominant for all values of $\lambda\in \mathbb{R}$.
The structure of the associated LD rate function can be, as it turns out, quite non-trivial; although
there are $K$ dynamical phases involved, the total number of distinct sequences of interweaving branches $F_{k}(\lambda)$
in the direction of the (fictitious) `$\lambda$-flow' (see Fig.~\ref{fig:lambda_flow} for an illustration), can be much larger.

Transitions between dynamical phases occur at the boundaries of intervals $\mathscr{I}_{k}$, identified with dynamical critical points.
While the SCGF $F(\lambda)$ is a continuous and everywhere convex function of $\lambda \in \mathbb{R}$,
it will in general feature non-analytic behavior at the critical points.
Judging from the asymptotic form \eqref{eqn:G_branches} of the multibranched structure of the MGF, we anticipate DPTs of the \emph{first order} with one of several critical points $\lambda_{\bowtie}$.
Such point show up as \emph{non-differentiable} (corner) point in the physical $F(\lambda)$.
We remind the reader that Legendre duality is no longer guaranteed to hold for any SCGF $F(\lambda)$ that develops a non-differentiable point \cite{Touchette_LDT}; it is only when $F(\lambda)$ is differentiable everywhere on its domain that there is a guarantee (in the form of the G\"{a}rtner--Ellis theorem) that the Legendre transform of $F(\lambda)$ yields the large-deviation rate function, i.e. that $I(j) = F^{\star}(j)$. Otherwise, $F^{\star}(j)$ is merely the convex hull of $I(j)$.

As we explain next, exploiting additional analytic input stored in the weighting functions $W_{k}(z)$ provides an elegant route that
bypasses the need for the G\"{a}rtner--Ellis theorem or a fully-fledged steepest descent analysis. By adopting the asymptotic representation \eqref{eqn:G_branches} as a starting point, the inverse Laplace transform can be computed for each branch individually as follows \cite{THT10}: by decomposing the PDF $\mathcal{P}(j|t)=\sum_{k}\mathcal{P}_{k}(j|t)$ and introducing potentials $\varPhi_{k}(\lambda;j)\equiv \lambda\,j-F_{k}(\lambda)$, each term is asymptotically of the form
\begin{equation}
\mathcal{P}_{k}(j|t)\asymp \frac{1}{2\pi \ii}\int_{\mathcal{D}_{k}}\dd \lambda W_{k}(\lambda)e^{t^{\alpha}\,\varPhi_{k}(\lambda;j)},
\label{eqn:P_branch}
\end{equation}
where $\mathcal{D}_{k}$ denote \emph{deformed} integration (Bromwich) contours passing through saddle points $\lambda^{\star}_{k}$ of $\varPhi_{k}(\lambda;j)$, given by the unique solutions to $F^{\prime}_{k}(\lambda)=j$. We have assumed that functions $W_{k}(\lambda)$ are either free of poles (in the complex $\lambda$-plane) or that in the process of deforming the original Bromwich integration contours to $\mathcal{D}_{k}$ we have not crossed any poles. In the opposite case, the asymptotics of $\mathcal{P}_{k}(j|t)$ would pick up additional contributions stemming from the residues of the integrand in Eq.~\eqref{eqn:P_branch} (see Ref.~\cite{THT10} for more details). At late times, we thus have
\begin{equation}
\mathcal{P}_{k}(j|t) \asymp e^{t^{\alpha}\,\varPhi_{k}(\lambda^{\star}_{k};j)},
\end{equation}
and the rate function $I(j)=-\lim_{t\to \infty}t^{-\alpha}\log \mathcal{P}(j|t)$ is simply the infimum over the branches
\begin{equation}
I(j) = {\rm inf}_{k}\{\varPhi_{k}(\lambda^{\star}_{k};j)\}.
\end{equation}
We note that a similar type of large-deviation rate functions describe mixtures of Bernoulli trials \cite{Dinwoodie92}.

\section{Hardcore cellular automaton}
\label{sec:hardcore}

Having finally put the formal framework in place, we now turn to practical applications and
exemplify how the outlined techniques can be applied to a specific model.
For demonstrative purposes, we present the full solution of the classical deterministic time-reversible classical cellular automaton
introduced in Sec.~\ref{sec:setting}.

We devote the remainder of the paper to a comprehensive analysis of charge fluctuations \emph{away} from equilibrium,
thereby expanding on our previous work \cite{Krajnik22}. We first briefly present, as a warm-up, how to infer the full counting statistics associated with particle number. By applying the dressing method, we then compute the exact statistics of charge and exhibit its most salient features.
Finally, we carry out a systematic analysis of various dynamical phases and their interweaving patterns by studying the corresponding
phase diagrams in the complex $\lambda$-plane within the scope of Lee--Yang theory.

\medskip

\subsection{Particle fluctuations}

We consider an inhomogeneous initial state consisting of two semi-infinite partitions, each initialized in a grand-canonical equilibrium state with respective particles densities $\rho_{-}$ and $\rho_{+}$.
For later convenience, we introduce the ratio of hole to particle densities,
$\nu=\ol{\rho}/\rho$. The latter obeys $1/\nu = \rho^{-1} \partial f_{\rm eq}(\rho)/\partial \rho$,
where $f_{\rm eq}=\log{(1+e^{\beta_{p}})}$ is the free-energy density parametrized by chemical potential $\beta_{p}=\log{(1/\nu)}$.
The static susceptibility accordingly reads $\chi_{p}(\rho)=\partial^{2}f_{\rm eq}(\rho)/\partial \beta_{p}^{2}=\rho(1-\rho)$.

To compute fluctuations of net particle transfer, we may simply ignore the charge degrees of freedom. In the hardcore automaton,
this effectively eliminates interaction among the particles, and hence computing the corresponding FCS becomes a simple exercise; in the absence of relaxation, fluctuations can be read off directly from the initial condition, yielding the univariate SCGF of the form
\begin{equation}
F_{p}(\lambda_{p}) = \log{\left[(\rho_{+}+e^{\lambda_{p}}\ol{\rho}_{+})(\rho_{-}+e^{-\lambda_{p}}\ol{\rho}_{-})\right]},
\label{eqn:Fp}
\end{equation}
associated with a sum of independent Bernoulli processes characterized by state densities $\rho_{\pm}$.
The result is in agreement with the celebrated Levitov--Lesovik formula \cite{LL94,LLL96} (specialized for the case of perfect transmission).
Since particles propagate freely, the MGF $G_{p}(\lambda_{p}|t)=\exp{(t\,F_{p}(\lambda_{p}))}$ is indeed exact at all times and not just asymptotically. Moreover, $F_{p}(\lambda_{p})$ depends only on a single parameter
\begin{equation}
\varpi \equiv \rho_{+}\ol{\rho}_{-}(e^{-\lambda_{p}}-1)+\rho_{-}\ol{\rho}_{+}(e^{\lambda_{p}}-1),
\end{equation}
in terms of which it reads {$F_p(\lambda_p)=\log{(1+\varpi(\lambda_p))}$}.
It is instructive to compare this result with the SCGF of the simple symmetric exclusion process, reading \cite{Derrida09}
${F_{\rm SSEP}(\lambda_p)}=\tfrac{1}{\pi}\int^{\infty}_{-\infty}\dd u\,\log{[1+\varpi(\lambda_p) \exp{(-u^{2})}]}$.

The first two scaled cumulants as functions of particle densities read
\begin{align}
s^{(p)}_{1} &= \lim_{t\to \infty}\frac{c^{(p)}_{1}(t)}{t} = \ol{j}_{p} = \rho_{-} - \rho_{+},\\
s^{(p)}_{2} &= \lim_{t\to \infty}\frac{c^{(p)}_{2}(t)}{t} = \rho_{-}\ol{\rho}_{-} + \rho_{+}\ol{\rho}_{+}.
\end{align}
Setting $\rho_{-}-\rho_{+}\equiv \delta \rho$ and expanding $F_{p}(\lambda)$ around equilibrium, $\delta \rho \to 0$, 
we can extract `diffusivity' ${\rm D}_{*}(\rho)=1$ and `mobility' $\sigma_{*}(\rho)=2\rho(1-\rho)$, obeying the local
Einstein relation $2\chi_p(\rho) {\rm D}_{*}(\rho)=\sigma_{*}(\rho)$.
Curiously, we have retrieved the exact same dependence as found in the SSEP (on an infinite line, see Ref.~\cite{Derrida09}),
with the proviso that SSEP is a diffusive system ($\alpha=1/2$) while free particles are ballistic ($\alpha=1$). Beware
not to confuse $\sigma_{*}(\rho)$ with conductivity $\sigma(\rho)$; in ballistic systems, the second cumulant instead gives
the first absolute moment of the DSF called the Drude self-weight
$s^{(p)}_{2}=\lim_{t\to \infty}\int^{t}_{-t}\dd \tau \expect{{\rm j}_{p}(0,\tau){\rm j}_{p}(0,0)}^{c}$, see Ref.~\cite{DS17}.

Fluctuations of the particle transfer evidently satisfy the GCR.
In analogy with the SSEP, the fluctuation symmetry is inherited from reflection symmetry of the reduced variable $\varpi$.
Moreover, the univariate affinity $\tilde \varepsilon_{p}$ is simply given by the difference of chemical potentials,
\begin{equation}
\tilde{\varepsilon}_{p} = \beta_{p,-} - \beta_{p,+} = \log \nu_{+} - \log \nu_{-} = \log \kappa^{2},
\label{eqn:epsilon_tilde_p}
\end{equation}
implying the UFR of the form
\begin{equation}
F_{p}(\lambda) = F_{p}(-\lambda - \tilde{\varepsilon}_{p}).
\label{eqn:UFR_p}
\end{equation}

Using further that freely propagating particles obey pure transmission, the SCGF $F_{p}(\lambda)$ satisfies the
so-called \emph{extended fluctuation relation} \cite{BD13}. The latter states that $F_{p}(\lambda)$ is fully determined
already by the first scaled cumulant $s^{(p)}_{1}=\ol{j}_{p}$, namely
\begin{equation}
F_{p}(\lambda) = \int^{\lambda}_{0}\dd \gamma \, \ol{j}_{p}(\log \nu_{-} - \gamma,\log \nu_{+} + \gamma).
\end{equation} 
Plugging in $\ol{j}_{p}=(1+\nu_{-})^{-1}-(1+\nu_{+})^{-1}$ correctly reproduces Eq.~\eqref{eqn:Fp}.

\medskip

\begin{figure}[htb]
	\includegraphics[width=0.87\columnwidth]{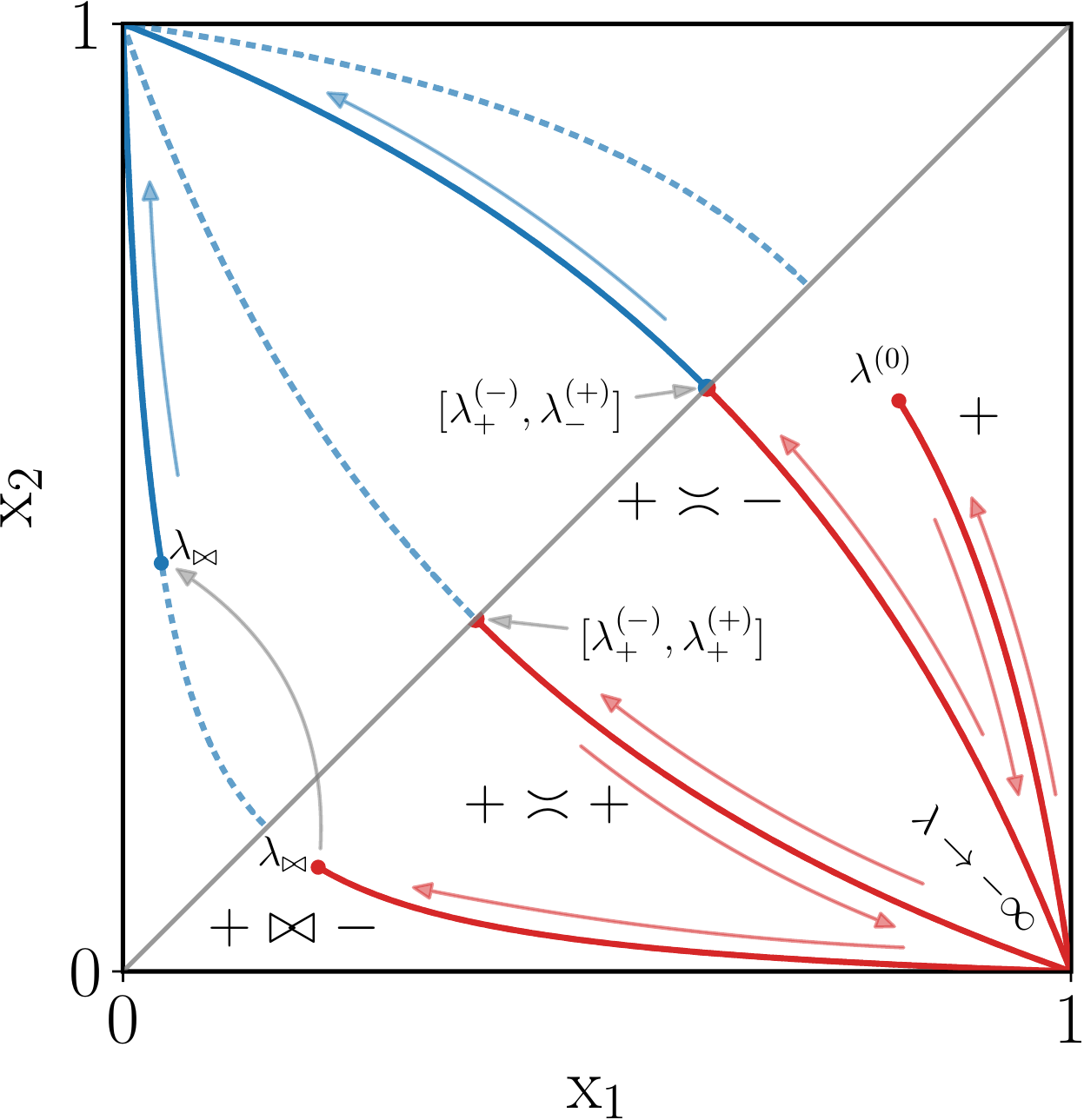}
	\caption{Visualization of the `$\lambda$-flow' over the integration domain $\mathscr{D}_{\square}$ of MGF $G(\lambda|t)$,
		parametrized in terms of continuous scaling coordinates ${\rm x}_{1}=l/t$ and ${\rm x}_{2}=r/t$ (with $l$ and $r$ representing the number of transferred ballistic holes from the left and right partitions, respectively, in units of $t$).
		Trajectories along the direction of arrows represent the motion of bulk maxima from the interior of the integration domain, associated with the bulk branches, namely $F_{+}(\lambda)$ (red) and $F_{-}(\lambda)$ (blue).
		When the bulk maxima reach the diagonal, representing the boundary $\mathscr{D}_{\square}$ for each of the bulk subdomains
		$\mathscr{D}_{\pm}$, the constant (flat) branch $F_{0}$ sets in.
		Solid (dashed) curves designate the physical (subleading, metastable) contributions to
		the SCGF $F(\lambda)$. Four different dynamical regimes are shown: regular regime $\llb +\rrb$, where a single physical branch $F_{+}(\lambda \in \mathbb{R})$ dominates for all $\lambda \in \mathbb{R}$; two types of tunneling (sub)regimes, to the same branch $\llb + \asymp + \rrb$ and to a different branch $\llb + \asymp - \rrb$
		via $F_{0}$ supported on $[\lambda^{(-)}_{\pm},\lambda^{(+)}_{-}]$, and a corner regime $\llb + \bowtie - \rrb$ with $F_{+}$ jumping over to $F_{-}$ at corner point $\lambda_{\bowtie}$.
	}
	\label{fig:lambda_flow}
\end{figure}

\subsection{Joint particle-charge fluctuations}
\label{sec:joint_MGF}

We now consider the FCS of joint particle-charge transfer encoded in $G_{c,p}(\lambda_{c},\lambda_{p}|t)$.
The exact finite-time joint MFG $G_{c,p}(\lambda_{c},\lambda_{p}|t)$ can be calculated from first principles (see Appendix \ref{sec:exact_FCS}), by
following the lines of Ref.~\cite{Krajnik22} adapted to the bipartitioned initial state (see Sec.~\ref{sec:results}).
Alternatively, $G_{c,p}(\lambda_{c},\lambda_{p}|t)$ can be obtained from the univariate particle-current MGF $G_{p}(\lambda_{p}|t)$
by applying the general replacement rule \eqref{eqn:replacement_rule}.
The associated joint SCGF $F_{c,p}(\lambda_{c},\lambda_{c})=\lim_{t\to \infty}t^{-1}\log G_{c,p}(\lambda_{c},\lambda_{p}|t)$
is subsequently computed by applying the Laplace's method (the derivations are presented in Appendix \ref{sec:exact_FCS}).

By introducing the `dressed counting fields'
\begin{equation}
\mu_{\pm}(\lambda_{c}) \equiv \cosh{(\lambda_{c})} \mp b_{\pm}\sinh{(\lambda_{c})},
\label{eqn:dressed_counting_fields}
\end{equation}
and a pair of auxiliary functions depending on the densities and $\lambda_{p}$,
\begin{equation}
\nu_{\pm}(\lambda_{p})\equiv \frac{\ol{\rho}_{\pm}}{\rho_{\pm}}e^{\pm \lambda_{p}},\qquad \nu_{\pm} \equiv \nu_{\pm}(0),
\label{eqn:nu}
\end{equation}
we find $F^{(\pm)}_{c,p}(\lambda_{c},\lambda_{p}) \equiv \log{(f^{(\pm)}_{c,p}(\lambda_{c},\lambda_{p}))}$ with
\begin{equation}
\tcbhighmath[drop fuzzy shadow]{
f^{(\pm)}_{c,p}(\lambda_{c},\lambda_{p})
= \prod_{\epsilon \in \{\pm\}} \frac{[\mu_{\pm}(\lambda_{c})]^{\pm \epsilon} + \nu_{\epsilon}(\lambda_{p})}{1+\nu_{\epsilon}}.
}
\label{eqn:f_pm}
\end{equation}

The `bulk branches' $F^{(\pm)}_{c,p}(\lambda_{c},\lambda_{p})$ are attributed to two isolated local maxima residing in the interior
of the upper (lower) triangular subdomain $\mathscr{D}_{-}$ ($\mathscr{D}_{+}$) of $\mathscr{D}_{\square}$,
of the integration domain $\mathscr{D}_{\square}$ of $G_{c,p}(\lambda_{c},\lambda_{p}|t)$, see Appendix ~\ref{sec:exact_FCS}.
On the other hand, $F^{(0)}\equiv F_{0}$ represents a constant branch that depends solely on particle densities,
\begin{equation}
\tcbhighmath[drop fuzzy shadow]{
F_{0} = 2\log\left[\sqrt{\rho_{-}\rho_{+}}+\sqrt{\ol{\rho}_{-}\ol{\rho}_{+}}\right].
}
\label{eqn:flat_branch}
\end{equation}
Such a flat branch occurs when the maximum of $G_{c,p}(\lambda_{c},\lambda_{p}|t)$ it attained exactly on the diagonal of the integration domain
$\mathscr{D}_{\square}$ (as represented by the `$\lambda$-flow' of the bulk maxima in Fig.~\ref{fig:lambda_flow}).
This may be interpreted as bulk maxima escaping `out-of-bounds' upon colliding with the diagonal, playing
the role of the domain boundary $\partial \mathscr{D}$.
More importantly, the bulk maxima are not unconditionally present in the integration domain $\mathscr{D}_{\square}$. Instead, they only
appear within certain regions of the parameter space specified below.
Introducing $\boldsymbol{\lambda}\equiv (\lambda_{c},\lambda_{p})$ and another auxilairy function
\begin{equation}
\kappa(\lambda_{p}) \equiv \sqrt{\frac{\nu_{+}(\lambda_{p})}{\nu_{-}(\lambda_{p})}},\qquad \kappa \equiv \kappa(0),
\label{eqn:kappa}
\end{equation}
we find that in the regions
\begin{equation}
\mathscr{E}_{\pm} = \{\mu_{\pm}(\lambda_{c}) \geq \kappa^{\pm 1}(\lambda_{p}) \land \mu_{\mp}(\lambda_{c}) < \kappa^{\mp 1}(\lambda_{p}) \},
\label{eqn:E_pm}
\end{equation}
in the two-dimensional $\boldsymbol{\lambda}$-plane, only a single extremum contribution $F_{\pm}(\lambda)$ appears in the bulk of
subdomains $\mathscr{D}_{\pm}$, whereas both extrema coexist within
\begin{equation}
\mathscr{E}_{+-} = \{\mu_{+}(\lambda_{c}) \geq \kappa(\lambda_{p}) \land \mu_{-}(\lambda_{c}) \geq \kappa^{-1}(\lambda_{p})\}.
\label{eqn:E_coexist}
\end{equation}

Competition between the two bulk extrema and the flat branch can be summarized in terms of the following \emph{selection rules}
\begin{align}
\mathscr{I}_{\pm} &= \mathscr{E}_{\pm} \lor \Big(\mathscr{E}_{+-} \land
f^{(\pm)}_{c,p}(\boldsymbol{\lambda}) \geq f^{(\mp)}_{c,p}(\boldsymbol{\lambda})\Big), \\
\mathscr{I}_{0} &= \left\{\mu_{+}(\lambda_{c}) < \kappa(\lambda_{p}) \,\, \land \,\, \mu_{-}(\lambda_{c}) < \kappa^{-1}(\lambda_{p})\right\},
\label{eqn:selection_rules}
\end{align}
which permit us to write compactly
\begin{equation}
F_{c,p}(\boldsymbol{\lambda}) = \left\{F^{(k)}_{c,p}(\boldsymbol{\lambda});\,\boldsymbol{\lambda} \in \mathscr{I}_{k}\right\}_{k\in \{+,0,-\}}.
\end{equation}
We postpone a more detailed study of $F_{c,p}(\boldsymbol{\lambda})$ to Sec.~\ref{sec:MFR_specific}, where we establish the fluctuation relation.

\subsection{Anomalous charge fluctuations}
\label{sec:dynamical_regimes}

In the following, we focus our discussion to the univariate charge-current SCGF $F_{c}(\lambda_{c})$.
Although $F_{c}(\lambda_{c})$ is merely a specialization of the full joint SCGF $F_{c,p}(\lambda_{c},\lambda_{p})$ obtained by
putting $\lambda_{p}=0$ (corresponding to integrating out the cumulative particle current), the univariate function
$F_{c}(\lambda_{c})$ reveals (as already summarized earlier in Sec.~\ref{sec:results}) an intricate structure.
Unlike $F_{c,p}(\lambda_{c},\lambda_{p})$, $F_{c}(\lambda_{c})$ does not necessarily
involve all three dynamical phases when restricted to $\lambda_{c}\in \mathbb{R}$.

To lighten our notations, we shall subsequently drop the subscript by identifying $F_{c}(\lambda_{c})\equiv F(\lambda)$, where
\begin{equation}
	F(\lambda) = \left\{F_{k}(\lambda);\,\lambda \in \mathscr{I}^{\rm c}_{k}\right\}_{k\in \{+,0,-\}},
\end{equation}
for $\mathscr{I}^{\rm c}_{k}\equiv \mathscr{I}_{k}(\lambda_{p}=0)$.
Moreover, in the limit $|\lambda|\to \infty$, the bulk branches $F_{\pm}(\lambda)$ grow linearly as 
\begin{equation}
	F_{\pm}(\lambda) \asymp |\lambda| + \log{\left[\frac{(1\mp b_{\pm})\nu_{\pm}}{2(1+\nu_{+})(1+\nu_{-})}\right]} + \mathcal{O}(e^{-\lambda}). 
\end{equation}
which is a manifestation of strict causality (with unit maximal velocity) of the time evolution.

For any fixed choice of the state parameters $\rho_{\pm},b_{\pm}$, the physical value of the SCGF $F(\lambda)$ for a given counting field
$\lambda$ is given by the dominant branch. Which of the branches $F_{k}(\lambda)$ dominates the growth of the MGF $G(\lambda|t)$
at large times is encoded in the selection rules derived in Sec.~\ref{sec:joint_MGF}.

To obtain an explicit parametrization of the separatrices between different dynamical regimes in the hardcore automaton,
we now explicitly work out the selection rules for univariate charge-current SCGF $F(\lambda)$.
The bulk branches $F_{\pm}(\lambda)$ attain their extrema at $\lambda^{(\epsilon)}_{\pm}$, $\epsilon\in \{\pm \}$,
\begin{equation}
	\lambda^{(\epsilon)}_{\pm} = \log{\left[\frac{\kappa^{\pm 1} + \epsilon \sqrt{\kappa^{\pm 2}+b^{2}_{\pm}-1}}{1\mp b_{\pm}}\right]},
	\label{eqn:degenerate_minima}
\end{equation}
and a central extremum at
\begin{equation}
	\lambda^{(0)}_{\pm} = \frac{1}{2}\log{\left[\frac{1\pm b_{\pm}}{1\mp b_{\pm}}\right]} = \beta_{c}(\mp b_{\pm}).
	\label{eqn:unique_minimum}
\end{equation}
Importantly, while $\lambda^{(0)}_{\pm}$ is always real for any $\lambda \in \mathbb{R}$, the other pair of extrema
$\lambda^{(\pm)}_{\pm}$ can be either real (with $\pm$ referring to the left ($-$) and right ($+$) real minima) or imaginary depending on parameters $\kappa,b_{\pm}$ of the initial state.

For compactness of presentation, we now  assume (with no loss of generality) that $\rho_{+}>\rho_{-}$. Introducing new auxiliary parameters 
\begin{equation}
	\gamma_{\pm} \equiv \kappa^{2}(b_{-} \pm 1),
\end{equation}
the phase boundary for the transition $F_{+} \leftrightarrow F_{-}$ (for $\lambda \in \mathbb{R}$)
is inferred by equating $F_{+}(\lambda)=F_{-}(\lambda)$, yielding $\gamma_{\pm}=\pm 1-b_{+}$ for $\pm \lambda>0$.

\begin{widetext}
	
\begin{figure}[t]
\includegraphics[width=\textwidth]{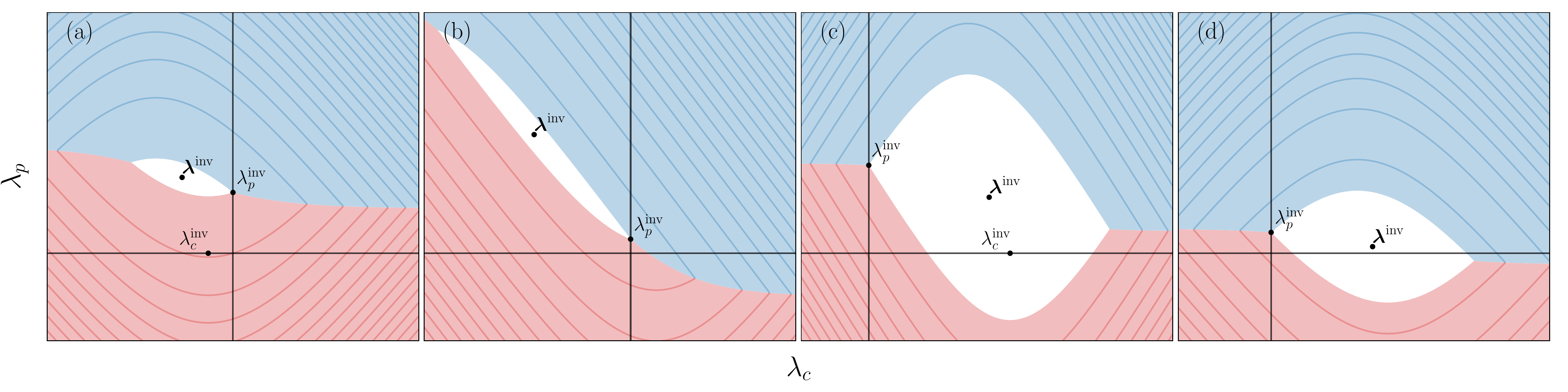}
\caption{Dynamical phase diagram of the joint particle-charge SCGF $F_{c,p}(\lambda_{c},\lambda_{p})$
in the real $\boldsymbol{\lambda}$-plane, exemplified for different dynamical regimes of $F_{c}(\lambda_{c})$:
(a) {\tt regular regime} $\llb + \rrb$,
(b) {\tt corner regime} $\llb + \bowtie - \rrb$,
(c) {\tt tunneling subregime} $\llb + \asymp + \rrb$ and
(d) {\tt tunneling subregime} $\llb + \asymp - \rrb$. The bulk dynamical phases $F^{(\pm)}_{c,p}(\lambda_{c},\lambda_{p})$ 
are shown in red ($+$) and blue ($-$). Their level sets are shown by solid lines of the corresponding color. The constant phase
$F^{(0)}$ is shown in white. Coordinate axes $\lambda_{p}=0$ ($\lambda_{c}=0$) are represented by solid horizontal (vertical) black lines.
The inversion point $\boldsymbol{\lambda}^{\rm inv}=-\boldsymbol{\varepsilon}/2$ of $F_{c,p}(\lambda_{c},\lambda_{p})$ lies in the geometric center of $F^{(0)}$, while $\lambda^{\rm inv}_{p}$ and $\lambda^{\rm inv}_{c}$ are inversion points associated
to the univariate SCGFs $F_{p}(\lambda_{p})$ and $F_{c}(\lambda_{c})$, respectively. Parameters values in (a), (b), (c), (d) read respectively:
$\rho_- = 0.18, 0.34, 0.28, 0.34$, $\rho_+ = 0.4, 0.4, 0.66, 0.43$, $b_-= 0.68, 0.99, -0.9, -0.73$, $b_+=-0.26, 0.3, 0.97, 0.85$.
}
\label{fig:2d_diagram}
\end{figure}

\end{widetext}

Similarly, the phase boundaries between the bulk branches $F_{\pm}(\lambda)$ and flat branch $F_{0}$ occurs when $F_{\pm}=F_{0}$, i.e. at the threshold value $b_{+} = \pm \sqrt{1-\kappa^{2}}$. The limiting curves for the tunneling transitions $F_{\pm} \leftrightarrow F_{\mp}$ are determined from the conditions
\begin{align}
	b_{+}>0:\qquad F_{-}(\lambda^{(+)}_{-})=F_{+}(\lambda^{(\pm)}_{+}),\\
	b_{+}<0:\qquad F_{-}(\lambda^{(-)}_{-})=F_{+}(\lambda^{(\pm)}_{+}),
\end{align}
yielding
\begin{equation}
b_{+}=\sqrt{1+\gamma_{-}},\qquad b_{+}=-\sqrt{1+\gamma_{-}},
\end{equation}
respectively. This allows us to determine the separatrices between different regimes and their support (in the parameter space of initial states):

\begin{enumerate}[(I)]
\item {\tt regular regime} $\llb + \rrb$
\begin{equation}
|b_{+}|<\sqrt{1-\kappa^{2}} \quad \land \quad b_{+} \lessgtr -\gamma_{\pm} \pm 1,
\end{equation}
\item {\tt tunneling regimes}
\begin{itemize}
\item subregime $\llb \pm \asymp \pm \rrb$, with tunneling from $F_{\pm}$ via $F_{0}$ back to the same bulk branch $F_{\pm}$
\begin{equation}
|b_{+}|<\sqrt{1-\kappa^{2}} \quad \land \quad b_{+} \lessgtr -\gamma_{\pm} \pm 1.
\end{equation}
\item subregime $\llb \pm \asymp \mp \rrb$, with tunneling from $F_{\pm}$ via $F_{0}$ to a different bulk branch $F_{\mp}$
\begin{equation}
b_{+} \lessgtr \pm \sqrt{1+\gamma_{-}} \quad \land \quad |b_{+}| > \sqrt{1+\gamma_{-}}.
\end{equation}
\end{itemize}
\item {\tt corner regime} $\llb \pm \bowtie \mp \rrb$
\begin{itemize}
\item \emph{unbroken} branch $F_{+}$ with a unique minimum
\begin{equation}
|b_{+}| < \sqrt{1-\kappa^{2}} \quad \land \quad b_{+} \lessgtr -\gamma_{\pm} \mp 1.
\end{equation}
\item \emph{regular} branch $F_{+}$ with a doubly degenerate minimum
\begin{equation}
|b_{+}| > \sqrt{1-\kappa^{2}} \quad \!\! \land \!\! \quad b_{-} \lessgtr 0 \quad \!\! \land \!\! \quad b_{+} \gtrless \pm \sqrt{1+\gamma_{-}}. 
\end{equation}
\end{itemize}
\item {\tt mixed regime} $\llb\mp \bowtie \pm \asymp \pm \rrb$
\begin{align}
|b_{+}| < \sqrt{1-\kappa^{2}} \quad \land \quad b_{-} &\gtrless 0 \quad \land \nonumber \\
\big(-\gamma_{\mp} \mp 1 &\gtrless b_{+} \gtrless \pm \sqrt{1+\gamma_{-}}\big).
\end{align}
\end{enumerate}

\subsection{Fluctuation symmetry}
\label{sec:fluctuation_symmetry}

In this section, we examine the fluctuation symmetry of the joint MGF $G(\lambda_{c},\lambda_{p}|t)$
(see Appendix \ref{sec:exact_FCS} for the derivation) and univariate charge-current SCGF $F_{c}(\lambda_{c})$.

\subsubsection{Multivariate fluctuation relation}
\label{sec:MFR_specific}

By explicit computations, we established that the
joint particle-charge SCGF $F(\lambda_{c},\lambda_{p})=\lim_{t\to \infty}t^{-1}\log G(\lambda_{c},\lambda_{p}|t)$ obeys
the multivariate fluctuation relation of the form
\begin{equation}
\tcbhighmath[drop fuzzy shadow]{
F(\lambda_{c},\lambda_{p}) = F(-\lambda_{c} - \varepsilon_{c},-\lambda_{p} - \varepsilon_{p}),
}
\label{eqn:MFR} 
\end{equation}

with affinities $\boldsymbol{\varepsilon}\equiv (\varepsilon_{c},\varepsilon_{p})$ reading
\begin{align}
\label{eqn:epsilon_c}
\varepsilon_{c} &= \frac{1}{2}\log{\left[\frac{1+b_{-}}{1-b_{-}}\frac{1-b_{+}}{1+b_{+}}\right]} = \delta \beta_{c},\\
\varepsilon_{p} &=  \tilde{\varepsilon}_p + \frac{1}{2}\log{\left[\frac{1-b^{2}_{-}}{1-b^{2}_{+}}\right]} = \delta \beta_{p}.
\label{eqn:epislon_p}
\end{align}

Recalling that $\delta \beta_{i}=\beta_{i,-} - \beta_{i,+}$ for $i\in \{c,p\}$ correspond to differences of the particle and charge
chemical potentials in the two initial partitions, the affinities indeed take a canonical form \cite{BD13,BD15,DoyonMyers20}.
In the hardcore automaton (and other charged single-file systems) $\tilde{\varepsilon}_{p} = \log \kappa^{2}$.
We have therefore established that in the considered models with time-reversal invariant dynamics
\begin{quote}
\emph{the joint particle-charge fluctuation relation \eqref{eqn:MFR} is unconditionally satisfied in the entire parameter space despite
dynamical phase transitions.}
\end{quote}

To illustrate the geometric meaning of the MFR , we next examine a diagram of dynamical phases in the two-dimensional $\boldsymbol{\lambda}$-plane, depicted in Fig.~\ref{fig:2d_diagram}. Notably, all three distinct dynamical phases {$F^{(k)}(\boldsymbol{\lambda})$}, with $k\in \{+,0,-\}$,
are always present in the joint SCGF $F_{c,p}(\boldsymbol{\lambda})$. The constant phase {$F^{(0)}(\boldsymbol{\lambda})$}
appears as a single compact `island' somewhere between $F^{(\pm)}_{c,p}(\boldsymbol{\lambda})$.
The one-dimensional boundaries $\mathcal{L}_{k\ell}$ separating the $k$th and $\ell$th can be deduced from
the balancing conditions $F^{(k)}(\boldsymbol{\lambda})=F^{(\ell)}(\boldsymbol{\lambda})$ (provided that both phases coexist, i.e. obey the selection rules given by Eq.~\eqref{eqn:selection_rules}). As shown in Fig.~\ref{fig:2d_diagram}, outside of the constant branch $F^{(0)}$,
the level sets of $F_{c,p}(\lambda_{c},\lambda_{p})$ form closed contours. On such contours, any two points symmetric with respect
to the inversion point $\boldsymbol{\lambda}^{\rm inv}$ always belong to distinct bulk phases $F^{(\pm)}_{c,p}(\boldsymbol{\lambda})$.

The MFR implies the existence of an inversion point $\boldsymbol{\lambda}^{\rm inv} = -\boldsymbol{\varepsilon}/2$. Curiously,
the latter is always situated precisely in the geometric center of the constant phase $F_{0}$ (see Fig.~\ref{fig:2d_diagram}),
exactly halfway between two \emph{triple} critical points -- one of them being the inversion point of $F_{p}(\lambda_{p})$ located on the vertical
($\lambda_{c}=0$) axis at $\lambda^{\rm inv}_{p}=-\log{(\kappa)}$, whereas the other is situated at
\begin{equation}
\lambda_{\rm triple} = \left(-\delta \beta_{c},
\lambda^{\rm inv}_{p} + \frac{1}{2}\log\left[\frac{1-b^{2}_{+}}{1-b^{2}_{-}}\right] \right).
\end{equation}
Upon approaching equilibrium ($\delta \beta_{p}=\delta \beta_{c}=0$), both triple points move towards the origin $\boldsymbol{\lambda}=(0,0)$
causing $F^{(0)}$ to disappear. 
It is also worth noticing that the effective affinity $\tilde{\varepsilon}_{p}$ (located on the $\lambda_{c}=0$ axis in the $\boldsymbol{\lambda}$-plane) must be a triple point at the junction of three phase boundaries $\mathcal{L}_{k\ell}$ where
$F^{(+)}(\boldsymbol{\lambda})=F^{(-)}(\boldsymbol{\lambda})=F^{(0)}$.
This readily follows from the fact that $F_{p}(\lambda_{p})$ is (by assumption) strictly convex and twice differentiable.

\begin{figure}[h!]
\includegraphics[width=1.0\columnwidth]{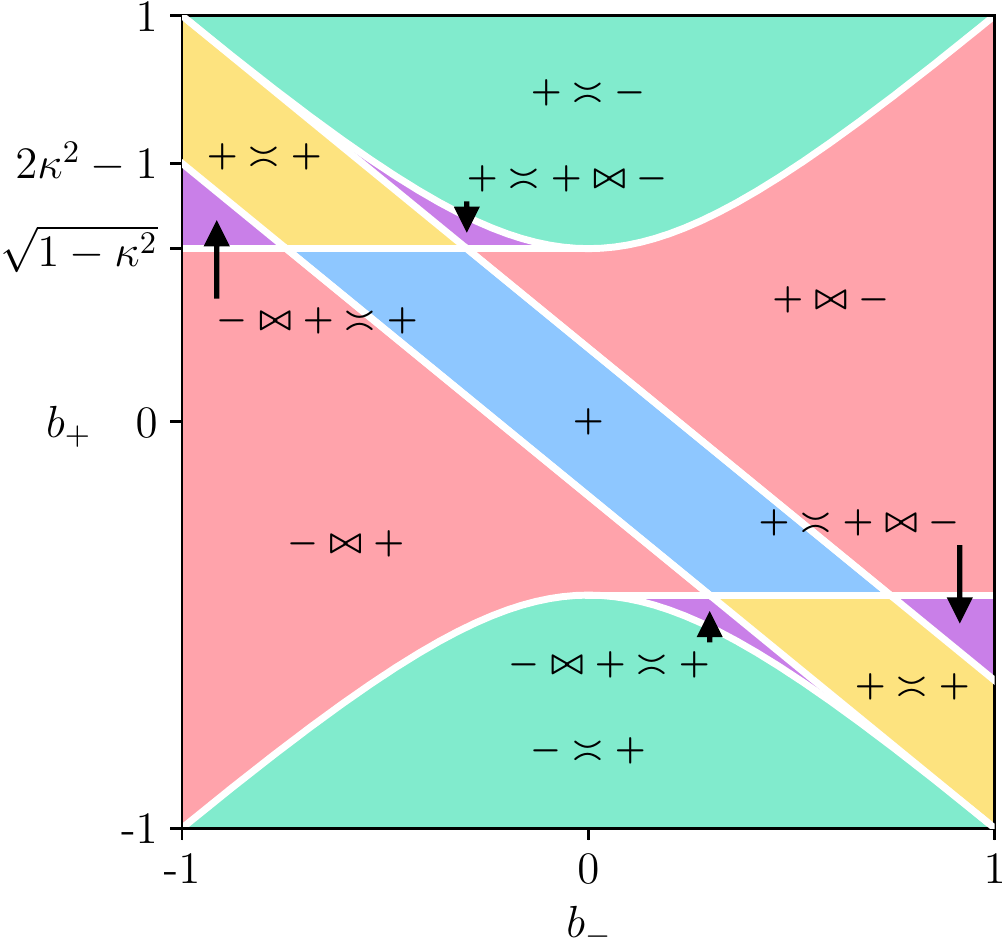}
\caption{Dynamical regimes of the hardcore cellular automaton, represented by a two-dimensional cross section in the $b$-plane cross section
of the parameter space (shown for $\rho_{-}=0.45$ and $\rho_{+}=0.5$) with SCGF $F(\lambda)$ exhibiting different interweaving patterns of dynamical phases: {\tt regular regime} (blue), {\tt corner regime} (red), two distinct types {\tt tunneling regimes} via $F_{0}$  -- to the same branch (yellow) and to another branch (green) -- and {\tt mixed regime} (purple), involving transfer via a corner followed by tunneling
to the same branch or vice-versa.}
\label{fig:regime_plot}
\end{figure}

\subsubsection{Spontaneous breaking of the univariate fluctuation relation}
\label{sec:spontaneous_breaking}

We have already demonstrated that the univariate SCGFs of the particle current $F_{p}(\lambda)$ exhibits the Gallavotti--Cohen
symmetry, cf. Eq.~\eqref{eqn:UFR_p}. We now investigate the symmetry properties of the univariate charge current SCGF
$F_{c}(\lambda_{c})=F_{c,p}(\lambda_{c},\lambda_{p}=0)$ .
In the following, we write shortly $F_{c}(\lambda_{c})\equiv F(\lambda)$, and systematically examine each of the dynamical regimes.

In {\tt regular regimes} $\llb \pm \rrb$, $F(\lambda)$ coincides with one of the bulk branches $F_{\pm}(\lambda)$.
Therefore, $F(\lambda)$ is strictly convex and everywhere differentiable and hence satisfies the following UFR
\begin{equation}
F_{\pm}(\lambda) = F_{\pm}(-\lambda - \tilde{\varepsilon}^{(\pm)}_{c}),\qquad \tilde{\varepsilon}^{(\pm)}_{c} = \mp 2 \beta_{c}(b_{\pm}).
\label{eqn:UFR_c}
\end{equation}
Here the effective affinities $\tilde{\varepsilon}^{(\pm)}_{c}=-2\lambda^{\rm inv}_{c}$, with $\lambda^{\rm inv}_{\pm}=\lambda^{(0)}_{\pm}$
being the unique minima of $F_{\pm}(\lambda)$, are different from the actual charge affinity $\varepsilon_{c}$.
Indeed, $\tilde{\varepsilon}^{(\pm)}_{c}$ only depend on a single chemical potential (that of the dominant dynamical phase) and
not on the canonical `gradient' $\delta \beta_{c}$ given by Eq.~\eqref{eqn:epsilon_c} that enters in the MFR.

In {\tt corner regimes}, denoted by $\llb \pm \bowtie \mp \rrb$, there is a non-differentiable point (a convex corner) in $F(\lambda)$
at $\lambda_{\bowtie} \in \mathbb{R}$ where the bulk branches $F_{\pm}(\lambda)$ interchange their dominance,
signaling a DPT of the first order. In effect, the UFR of $F_{c}(\lambda_{c})$ is invariably violated (seen as the absence of the inversion
point on the $\lambda_{p}=0$ axis in panel (b) in Fig.~\ref{fig:2d_diagram}).

Despite that, the UFR is preserved \emph{locally} within the `subcritical region', i.e. an interval of counting fields $\lambda$ centered at
$\lambda^{(0)}_{\pm} = -\varepsilon^{(\pm)}_{c}/2$, such that

\begin{equation}
\big|\lambda-\lambda^{(0)}_{\pm}\big| < \big|\lambda_{\bowtie}-\lambda^{(0)}_{\pm}\big|.
\label{eqn:subcritical_region}
\end{equation}

Accordingly, the LDF $I(j)$ develops a non-differentiable point (a concave corner) at the critical large current $j_{\bowtie}$,
and hence the UFR is only satisfied locally for subcritical (large) currents $|j|<|j_{\bowtie}|$.
Exceeding the critical current $j_{\bowtie}$, namely for $|j|>|j_{\bowtie}|$, the GCR is no longer satisfied as large fluctuations in the opposite direction belong to a different dynamical phase.

In {\tt tunneling regime}, the appearance of the flat branch $F_{0}$ (assuming $F(\lambda)$ is everywhere differentiable),
require us to differentiate between two scenarios, hereafter referred to as {\tt tunneling  subregimes}:
(i) tunneling from $F_{\pm}$ via $F_{0}$ back to the same phase (denoted by $\llb \pm \asymp \pm \rrb$), or (ii) tunneling to
the other bulk branch (denoted by $\llb \pm \asymp \mp \rrb$). In each to these subregimes, both bulk branches $F_{\pm}$ reside
in a symmetry-broken phase with a two-fold degenerate minima at $\{\lambda^{(\epsilon)}_{\pm}; \epsilon\in\{\pm\}\}$. As we elaborate in turn,
validity of the univariate Gallavotti--Cohen relation depends crucially on the dynamical subregime.

Depending on the state parameters, a second-order DPT takes place at two out of these four degenerate minima. We recall that
at the second-order critical point the second derivative of the physical (dynamical) free energy $F^{\prime \prime}_{\pm}(\lambda)$ experiences a jump discontinuity. The pair of critical points always consists of two adjacent minima. In fact, there are only two possible scenarios:
(i) the pair of second-order critical points coincides with two minima of the same bulk branch or (ii)
there is a tunneling transition from the left minimum $\lambda^{(-)}_{\pm}$ of $F_{\pm}$ to the right minimum $\lambda^{(+)}_{\mp}$
of another branch $F_{\mp}$. Using that the symmetry-broken bulk branches $F_{\pm}(\lambda)$ each enjoys reflection symmetry with respect
to the central maximum $\lambda^{(0)}_{\pm}$, the inversion symmetry $F(\lambda)$ clearly remains intact in case (i) despite the presence
of the constant phase $F_{0}$. In effect, the UFR in the form \eqref{eqn:UFR_c} remain valid.
This is however no longer true in case (ii) where, upon approaching the boundaries of the flat segment $\mathscr{I}_{0}$ from the outside,
we find a mismatch in the second derivatives evaluated in both minima,
\begin{equation}
\lim_{\lambda \to \lambda^{(-)}_{\pm}}F^{\prime \prime}_{\pm}(\lambda)\neq \lim_{\lambda \to \lambda^{(+)}_{\mp}}F^{\prime \prime}_{\mp}(\lambda).
\end{equation}
This means that neither of the effective charge affinities $\tilde{\varepsilon}^{(\pm)}_{c}$ (with the exception of equilibrium, when they are both equal) can be the inversion point of $F(\lambda)$. In effect, the UFR \eqref{eqn:UFR_c} is violated for the entire range of counting fields
$\lambda \in \mathbb{R}$.
Spontaneous breaking of the UFR can be alternatively discussed from the viewpoint of the LDF $I(j)\equiv I_{c}(j_{c})$.
In {\tt tunneling regime}, $I(j)$ is not differentiable at $j=0$ in both subregimes (i) and (ii). In subregime (i), lack of differentiability has no implications on the validity of the UFR since $I(j)$ involves only a single branch $I^{(\pm)}(j)$.
In subregime (ii), positive and negative (large) currents are realized in different dynamical phases and consequently
the UFR breaks down on the entire support of $j$. The same reasoning applies to {\tt mixed regime}, where both types of dynamical criticality are simultaneously present. 

\onecolumngrid

\begin{figure}[htb]
\includegraphics[width=1.0\columnwidth]{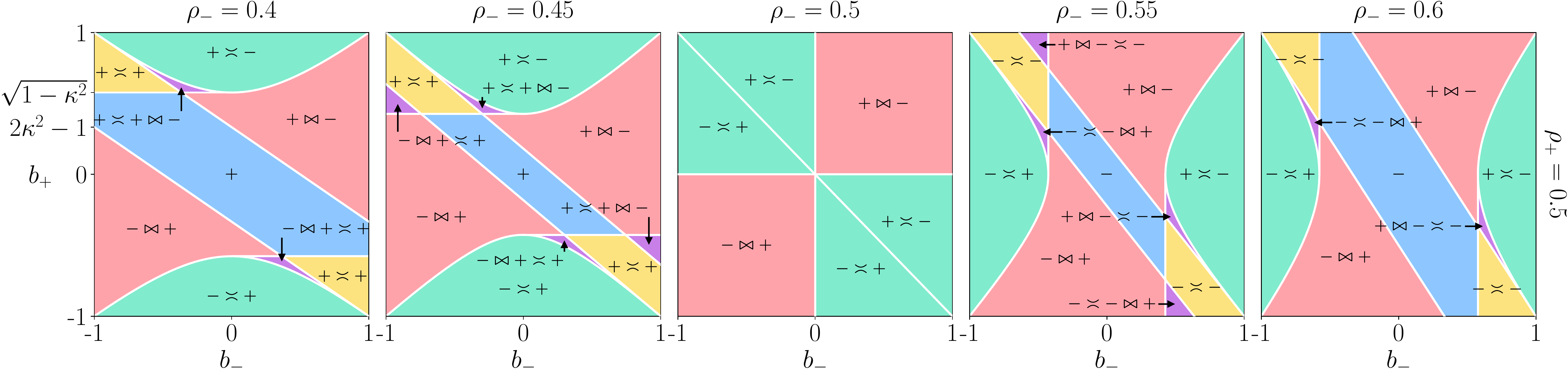}
\caption{The $b$-plane cross section of various dynamical regimes (shown for $\rho_{+}=0.5$ and several values of $\rho_{-}$) in the hardcore automaton: regular $\llb \pm \rrb$ (blue), corner $\llb + \bowtie - \rrb$ (red),
tunnelling regime, with subregimes $\llb \pm \asymp \mp \rrb$ (green) and $\llb \pm \asymp \mp \rrb$ (yellow) and mixed (purple).
Separatrices, marked by white lines, represent phase boundaries associated with dynamical phase transitions of first or second order.
The univariate fluctuation relation of the charge-current large-deviation function holds globally in $\llb \pm \rrb$ and $\llb \pm \asymp \pm \rrb$, and is violated in $\llb \pm \asymp \mp \rrb$ (globally) and $\llb \pm \bowtie \mp \rrb$ (locally).}
\label{fig:dynamical_regimes}
\end{figure}

\twocolumngrid
To conclude the section, we briefly recapitulate the key findings.  Unlike the joint particle-charge SCGF, the univariate charge-current
SCGF $F(\lambda)$ does not \emph{globally} (i.e. for the entire interval of counting fields $\lambda \in \mathbb{R}$) satisfy a fluctuation relation of the Gallavotti--Cohen type. The relation \eqref{eqn:UFR} is found to be obeyed only in a finite fraction of the parameter space, comprising the union of {\tt regular regimes} ($\llb + \rrb$ and $\llb - \rrb$) and {\tt tunneling subregime} $\llb \pm \asymp \pm \rrb$.

As represented in Fig.~\ref{fig:regime_plot}, these regimes are confined within a diagonal belt region in the $b$-plane cross section.
In other words, a continuous (second-order) convexity-breaking transition experienced by regular bulk branches,
inducing dynamical phase transitions between {\tt regular regimes} $\llb \pm \rrb$ and {\tt tunneling subregimes} $\llb \pm \asymp \pm \rrb$,
preserves the UFR.
 By contrast, the UFR is violated globally whenever the system undergoes a dynamical phase transition of first order, namely in {\tt corner regimes} $\llb \pm \bowtie \mp \rrb$ and {\tt mixed regime}. In these two dynamical regime, the UFR still holds \emph{locally} for subcritical large currents.
Finally, the UFR is completely broken in the {\tt tunneling subregime} $\llb \pm \asymp \mp \rrb$. 
We also note that $\llb \pm \asymp \mp \rrb$ cannot be reached from $\llb \pm \rrb$ directly (as seen in Fig.~\ref{fig:regime_plot}) but only via
a first-order DPT to another regime. The dependence of dynamical regimes on state parameters is illustrated in Fig.~\ref{fig:dynamical_regimes}.

\subsection{Phase diagrams}
\label{sec:phase_diagrams}
Thus far we have been almost exclusively concerned with the late-time behavior of the (univariate) charge-current MGF $G(\lambda|t)$
restricted to real-valued counting fields. Using asymptotic analysis, we have inferred the general structure of the SCGF $F(\lambda)$ and the corresponding large-deviation rate function $I(j)$. As already emphasized in Sec.~\ref{sec:setting}, the counting field $\lambda$ is merely a formal complex control variable with no direct physical meaning. As we now demonstrate, the outlined physical phases and interweaving dynamical regime admit analytic continuation to complex $\lambda$. In the conventional theory of thermodynamic phase transitions, such an analytic continuation of couplings or state parameters, such as temperature and chemical potentials, goes under the name of Lee--Yang analysis \cite{LeeYang52}. To elucidate the interplay of the three dynamical phases, the Lee--Yang approach is tremendously helpful. Here we only provide a broader overview of the main results, postponing a detailed analysis to Appendix \ref{sec:LeeYang}.\\

\begin{figure}[h!]
\includegraphics[width=\columnwidth]{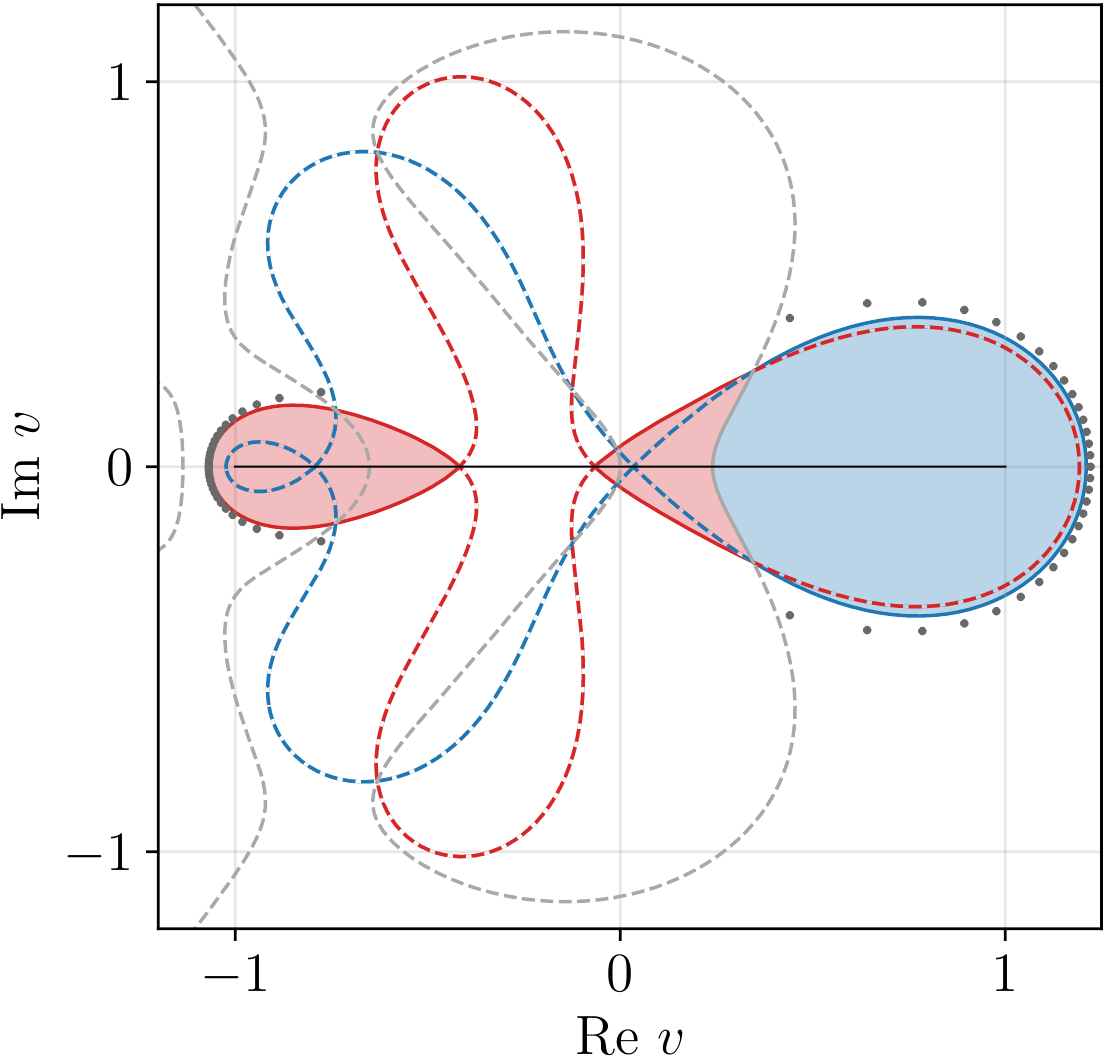}
\caption{A representative example of a phase diagram (selected from the {\tt mixed regime}, with $\rho_{-} = 0.75$, $\rho_{+}=0.77$, $b_{-} =0.78$, $b_{+}=-0.47$.), representing analytic continuation of dynamical phases into the complex $\upsilon$-plane.
Shaded regions indicate the regions of dominance $\mathscr{R}_{\pm}$ belonging to
the bulk dynamical phases ($F_{+}$ in red, $F_{-}$ in blue), while white background is the region of dominance $\mathscr{R}_{0}$
associated with flat branch $F_{0}$. Algebraic curves corresponding to extended anti-Stokes lines that separate distinct dynamical phases
apart from one another: $F_{+} \leftrightarrow F_{0}$ (red), $F_{-}\leftrightarrow F_{0}$ (blue) and $F_{+} \leftrightarrow F_{-}$ (gray).
Physical parts of the curves, marked by solid lines, coincide with the Lee--Yang contours (with dashed lines marking unphysical parts). 
The Lee--Yang zeros (shown for $t=32$) are marked with gray dots.}
\label{fig:LY_mixed}
\end{figure}

The Lee--Yang theory offers a powerful theoretical framework for characterizing thermodynamic phase transitions using the language of complex analysis, based on the singularity structure of the thermodynamic free energy \cite{Suzuki67,JankeKenna02,Bena05}, analytically continued to the complex temperature or fugacity plane. In finite systems, every zero contained in the polynomial part of the partition sum show up as an
isolated logarithmic singularity in the free energy extended to the complex fugacity plane. Although singularities of this type are always away from the physical real axis and thus unphysical, they are nevertheless responsible
for the appearance of critical points at large time. Specifically,  the number of singularities grows extensively (i.e. linearly)
with system size and, in the thermodynamic limit, the singularities tend to form clusters. One commonly finds that the majority of singularities arrange densely along contours or inside two-dimensional subdomains. Such condensates form natural boundaries between distinct thermodynamic phases. 

\begin{figure}[htb]
\includegraphics[width=0.9\columnwidth]{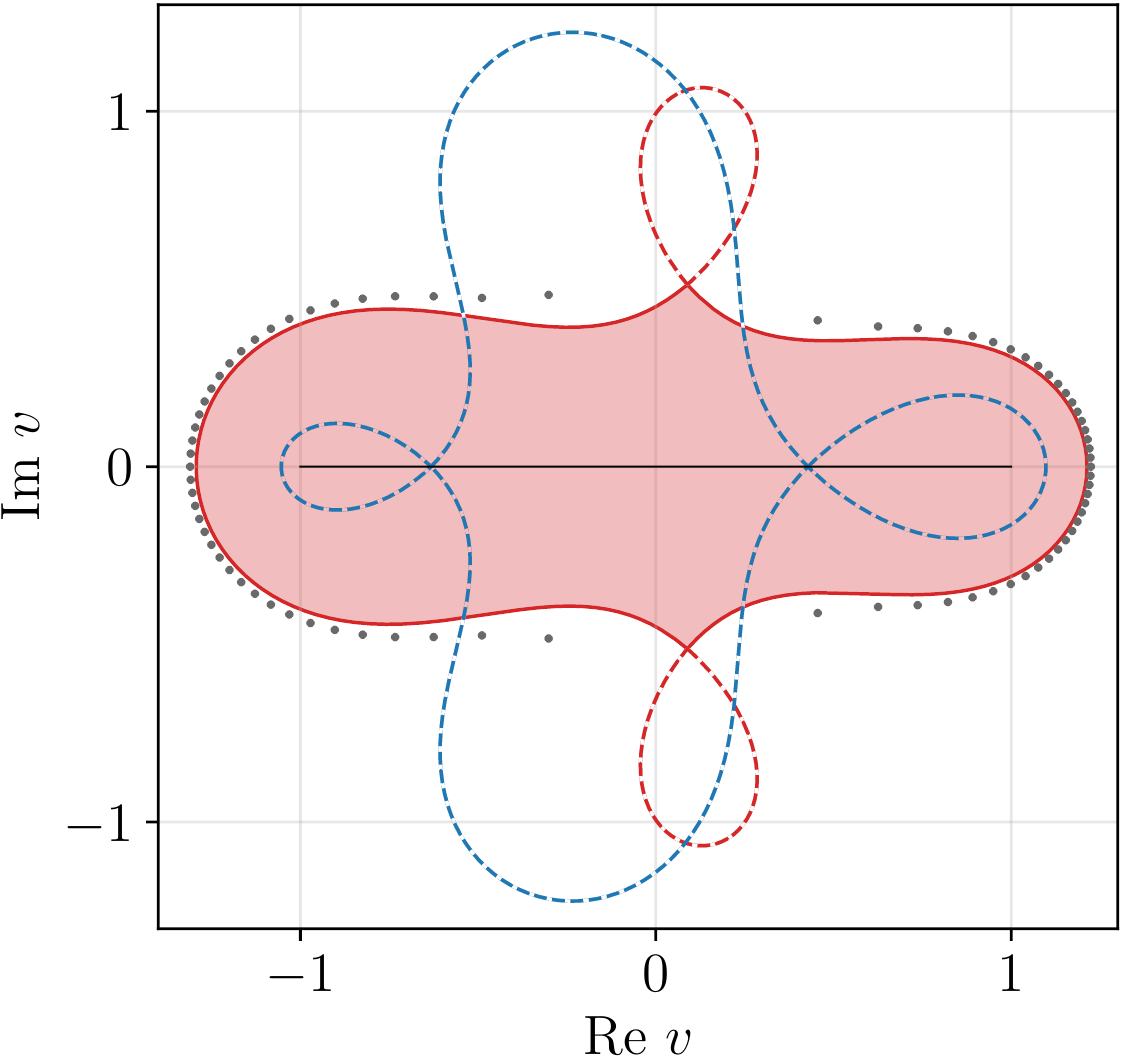}
\caption{
Phase diagram in the {\tt regular regime} $\llb + \rrb$ (with state parameters
$\rho_{-} = 0.18$, $\rho_{+}=0.4$, $b_{-} =0.28$, $b_{+}=-0.14$) in complex $\upsilon$-plane, comprising a single contiguous region of dominance $\mathscr{R}^{(\upsilon)}_{+}$ associated to $F_{+}$ (red),
and the domain $\mathscr{R}^{(\upsilon)}_{0}$ of the constant phase $F_{0}$ (white background).
The extended anti-Stokes line $\mathcal{A}^{(\upsilon)}_{+,0}$ ($\mathcal{A}^{(\upsilon)}_{-,0}$), separating $F_{+}$ ($F_{-}$) from $F_{0}$,
is shown by the red (blue) curve, with solid part corresponding to the Lee--Yang contour $\mathcal{C}^{(\upsilon)}_{+,0}$ (and gray dots marking the Lee--Yang zeros of $G(\upsilon|t)$ at time $t=40$). Dashed blue line is the ghost anti-Stokes line
$\mathcal{A}^{(\upsilon)}_{-,0}$ violating the selection criterion.}
\label{fig:LY_regular}
\end{figure}

Introducing a complex counting field $\zz \equiv e^{\lambda}$, we can apply the Lee--Yang approach to the rescaled CGF $F(\zz|t)$
by maintaining analogy with the thermodynamic free energy.
Taking advantage of the fact that in the hardcore automaton the MGF $G(\zz|t)$ enjoys a particularly simple analytic structure, we managed to obtain a fully general solution. By exploiting the formal analogy with two-dimensional electrostatics \cite{Blythe03,Bena05},
we succeeded in explicitly parametrizing the Lee--Yang contours, permitting us to compute the associated density of zeros from the jump discontinuities of the imaginary potential of the complexified dynamical free energy density $F(\zz)$ upon
transversing the phase boundaries in the $\zz$-plane.
In the hardcore automaton, we find a unit fraction of Lee--Yang zeros condensing along certain closed contours in the complex $\zz$-plane, separating different regions of dominance. This implies that Lee--Yang contours along which the zeros condense naturally play the role of phase boundaries between distinct (complexified) dynamical phases, obstructing analytic continuation of $F(\zz)$ across the phase boundaries. In the language of Morse theory, they are commonly referred to as the anti-Stokes lines (see e.g. Refs.~\cite{Itzykson83,PisaniSmith93}).

\begin{figure}[htb]
\includegraphics[width=\columnwidth]{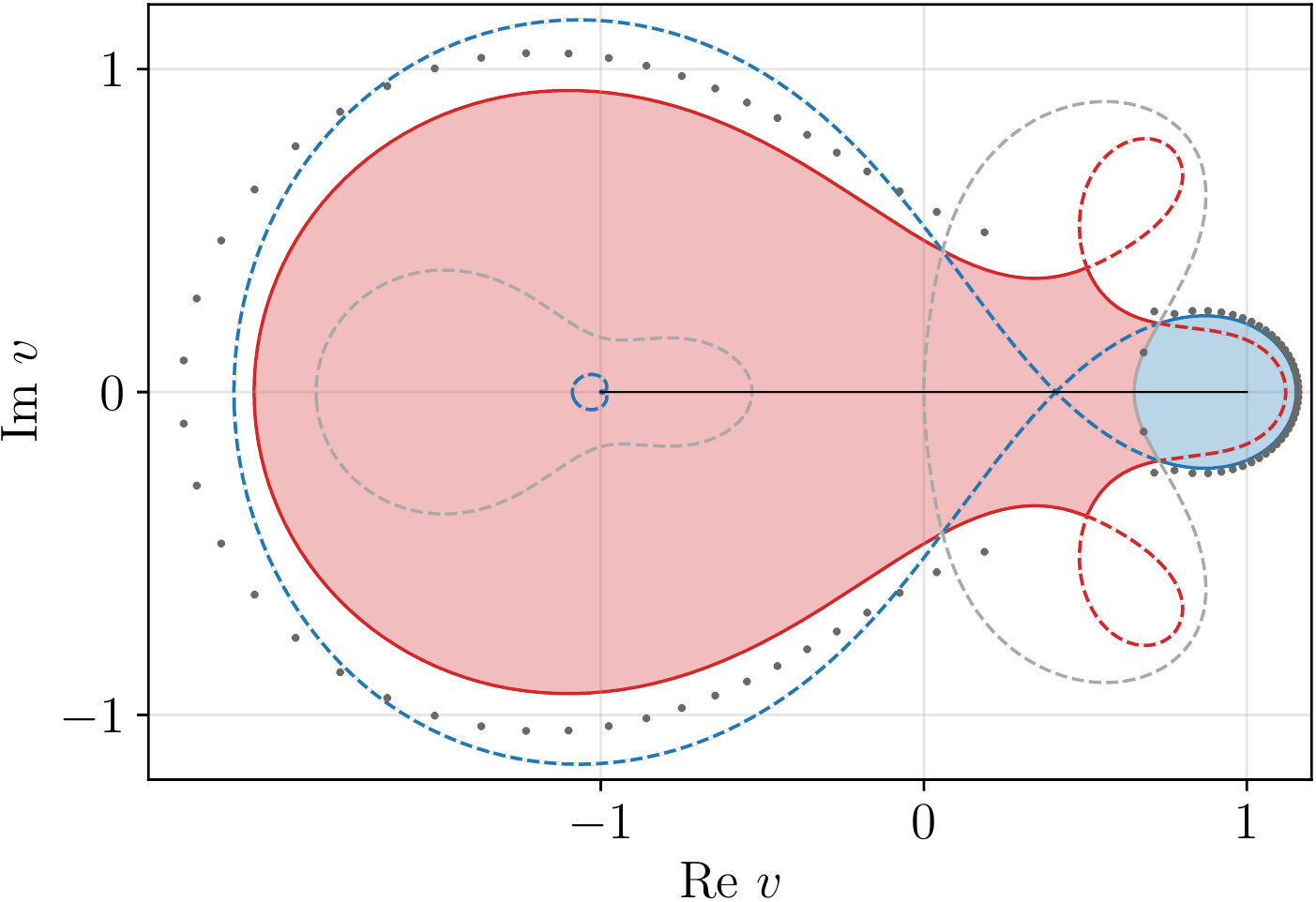}
\caption{
Phase diagram in {\tt corner regime} $\llb + \bowtie - \rrb$ (with state parameters
$\rho_{-} = 0.34$, $\rho_{+}=0.74$, $b_{-} =0.97$, $b_{+}=-0.72$) in complex $\upsilon$-plane,
comprising two disjoint regions of dominance $\mathscr{R}^{(\upsilon)}_{\pm}$ associated to $F_{+}$ (red) and $F_{-}$ (blue),
next to domain $\mathscr{R}^{(\upsilon)}_{0}$ of the constant phase $F_{0}$ (white background).
Curves represent the extended anti-Stokes lines $\mathcal{A}^{(\upsilon)}_{k,\ell}$, separating $F_{k}$ from $F_{\ell}$,
for $k,\ell \in \{+,0,-\}$ with $k\neq \ell$, namely $\mathcal{A}^{(\upsilon)}_{+,0}$ (red), $\mathcal{A}^{(\upsilon)}_{-,0}$ (blue)
and $\mathcal{A}^{(\upsilon)}_{+,-}$ (gray). Solid parts of the curves belong to the Lee--Yang contours.
Gray dots mark the Lee--Yang zeros of $G(\upsilon|t)$ at time $t=46$.}
\label{fig:LY_corner}
\end{figure}
In terms of the multiplicative counting field $\zz$, the MGF $G(\zz|t)$ represents a Laurent polynomial with $2t$ terms. By exploiting the underlying symmetries, there is however a more convenient (bijective) reparametrization in terms of a complex
counting variable $\upsilon$, defined via a conformal (M\"{o}bius) transformation $\upsilon(\zz)=(\zz-1)/(\zz+1)$, which we adopt in the following analysis. The physical part of $F(\upsilon)$ (real $\lambda,\zz$) then maps to the compact interval $\wp \equiv [-1,1]$,
whereas the regions of dominance associated with dynamical phases $F_{k}$ will be denoted by $\mathscr{R}^{(\upsilon)}_{k} \subset \mathbb{C}$.
It is worth noticing that the extended phase diagram in the complex $\upsilon$-plane comprises all three phases, with the sole exception being the regular regimes $\llb \pm \rrb$. In addition, each of the two bulk phases forms a single compact region of dominance, $F_{\pm}(\upsilon \in \mathscr{R}^{(\upsilon)}_{\pm})$,
with the exception of the tunneling regimes $\llb \pm \asymp \pm \rrb$. The flat phase $F_{0}$ may be seen as the background.

In {\tt regular regime}, the physical interval $\wp$ is contained entirely within a single region of dominance $\mathscr{R}^{(\upsilon)}_{k\in \{\pm\}}$, depending on the selection criteria.
Notice that both the dominant and subdominant bulk branches $F_{\pm}$ are always simultaneously present in the $\upsilon$-plane, with each of them becoming physical along $\mathscr{I}^{(\upsilon)}_{\pm} = \wp \cap \mathscr{R}^{(\upsilon)}_{\pm}$
when there is a non-trivial intersection with the physical line $\wp$.

\begin{figure}[htb]
\includegraphics[width=0.9\columnwidth]{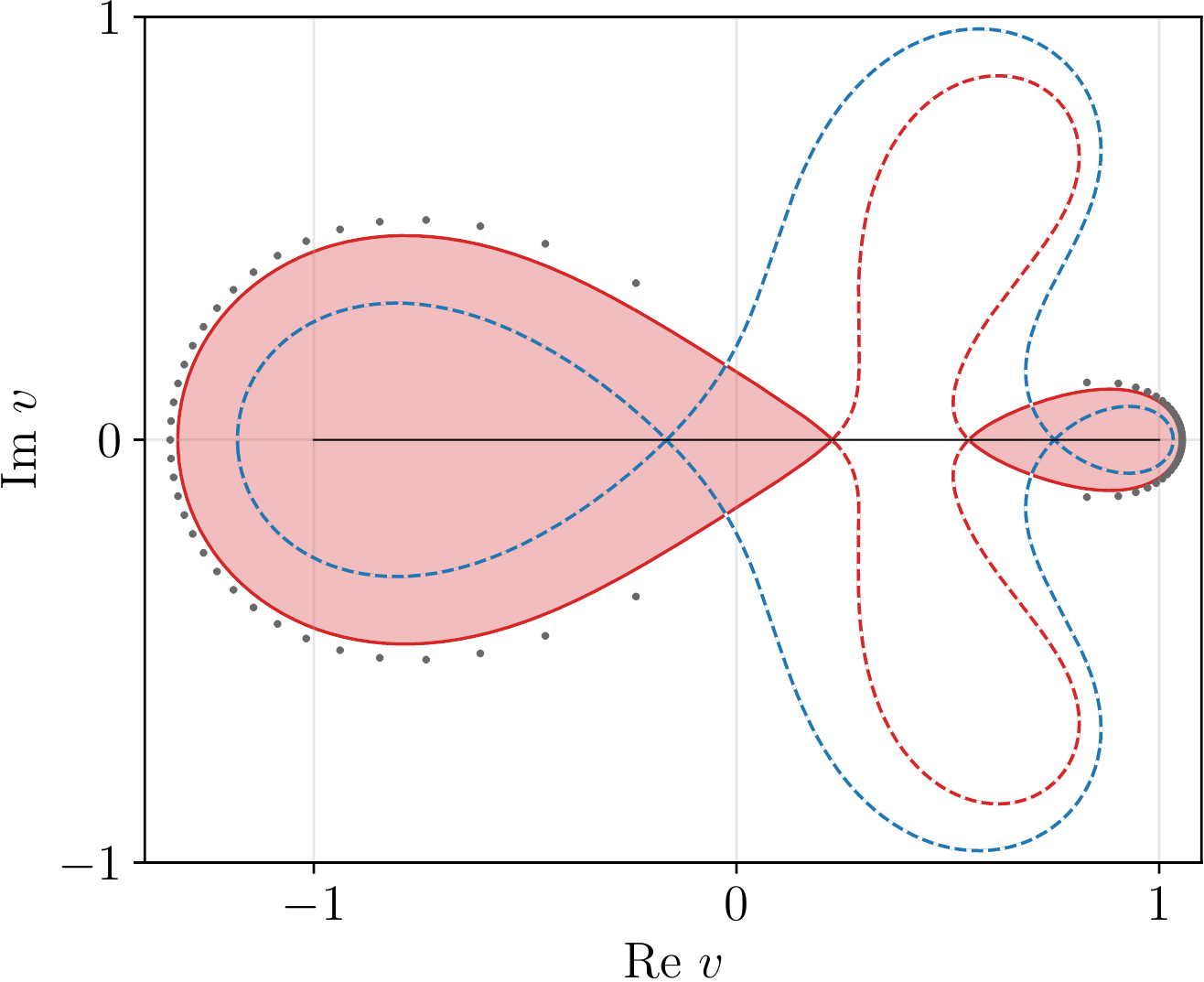}
\caption{
Phase diagram in subregime $\llb + \asymp + \rrb$ of {\tt tunneling regime} (with state parameters $\rho_{-} = 0.23$, $\rho_{+}=0.33$, $b_{-} =-0.67$, $b_{+}=0.69$), comprising two disjoint regions of dominance $\mathscr{R}^{(\upsilon)}_{\pm}$ associated to $F_{+}$ (red), and the domain $\mathscr{R}^{(\upsilon)}_{0}$ of the constant phase $F_{0}$ (white background). The extended anti-Stokes lines $\mathcal{A}^{(\upsilon)}_{+,0}$ ($\mathcal{A}^{(\upsilon)}_{+,0}$), separating $F_{+}$ ($F_{-}$) from $F_{0}$, are shown by red (blue) curve. Solid curves indicates the Lee--Yang contour $\mathcal{C}^{(\upsilon)}_{+,0}$.
Gray dots mark the Lee--Yang zeros of $G(\upsilon|t)$ at time $t=36$.}
\label{fig:LY_tunneling_pp}
\end{figure}

In {\tt corner regime}, the physical interval comprises two regions (intervals) of dominance,
$\wp = \mathscr{I}^{(\upsilon)}_{+}\cup \mathscr{I}^{(\upsilon)}_{-}$,
with $\mathscr{I}^{(\upsilon)}_{\pm}$ meeting at the corner point $\upsilon_{\bowtie}\in\wp$,
being a dynamical critical point of first order where the bulk branches interchange their dominance.
The Lee--Yang contour intersects the physical interval $\wp$ with an impact angle of  $\pi/2$ (see Figure~\ref{fig:LY_corner}),
such that the density of zeros at the intersection is strictly positive, consistent with first-order criticality \cite{JankeKenna02}.
Curiously, the Lee--Yang contour emanating from the physical first-order critical point terminates in the complex $\upsilon$-plane
at a (unphysical) \emph{triple point} where all three distinct phase boundaries meet at one point.

In {\tt tunneling regime}, the flat branch $F_{0}$ acquires a non-empty overlap with the physical interval $\wp$. Disregarding degenerate scenarios, there are only two possibilities of entering tunneling regimes $\llb \pm \asymp \pm \rrb$
by varying parameters of the initial state, either from regular regimes $\llb \pm \rrb$ or from mixed regimes $\llb \mp \bowtie \pm \asymp \pm \rrb$ or $\llb \pm \asymp \pm \bowtie \mp \rrb$. The other tunneling subregimes
$\llb \pm \asymp \mp \rrb$ can be entered from $\llb \pm \asymp \pm \bowtie \mp \rrb$ or $\llb \pm \bowtie \mp \rrb$.

Entering from {\tt regular regime}, the region $\mathscr{R}^{(\upsilon)}_{0}$ approaches close to $\wp$ until it eventually opens up a `corridor' through $\mathscr{R}^{(\upsilon)}_{\pm}$, splitting it into two disjoint compact regions (see Fig.~\ref{fig:LY_tunneling_pp}).
In this process, a physical regular bulk branch experiences a symmetry-breaking transition to a non-convex shape.
Note that the pairs of degenerate minima associated with the bulk branches $F_{\pm}$ show up as singular nodal points of the algebraic curves that parametrize the extended anti-Stokes lines (by disregarding the selection rules). The second-order transition can hence be also be seen as a conjugate pair of complex nodal points of the extended anti-Stokes lines colliding with $\wp$ and thus becoming physical dynamical critical points.

\begin{figure}[htb]
\includegraphics[width=0.9\columnwidth]{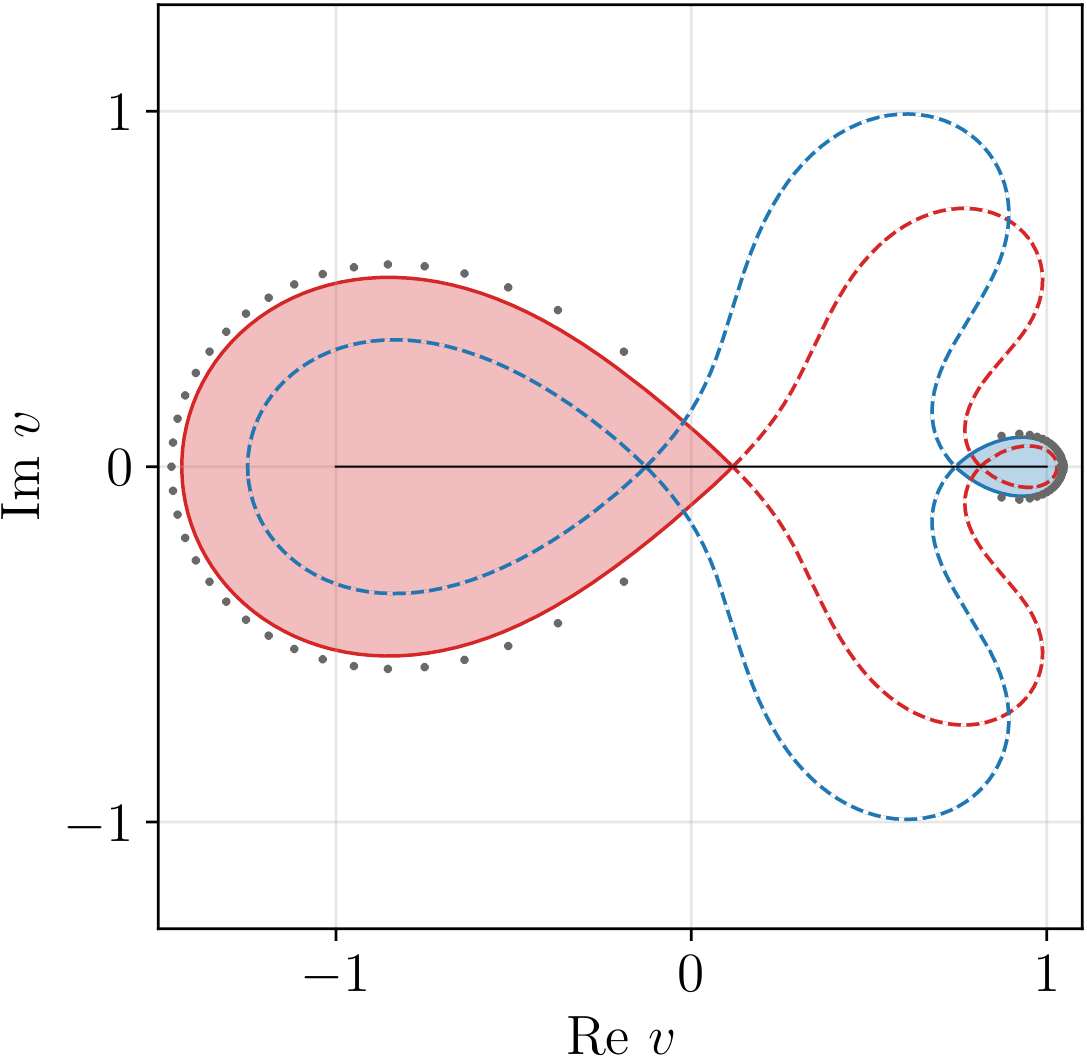}
\caption{
Phase diagram in subregime $\llb + \asymp - \rrb$ of {\tt tunneling subregime} (with state parameters $\rho_{-} = 0.34$, $\rho_{+}=0.43$, $b_{-} =-0.68$, $b_{+}=-0.85$),
comprising two disjoint regions of dominance $\mathscr{R}^{(\upsilon)}_{\pm}$ associated to both bulk dynamical phases $F_{+}$ (red) and $F_{-}$ (blue), and the domain $\mathscr{R}^{(\upsilon)}_{0}$ of the constant phase $F_{0}$ (white background).
The extended anti-Stokes lines $\mathcal{A}^{(\upsilon)}_{\pm,0}$ separating $F_{+}$ ($F_{-}$) from $F_{0}$ are shown by red (blue) algebraic curves. The respective Lee--Yang contour $\mathcal{C}^{(\upsilon)}_{+,0}$ and $\mathcal{C}^{(\upsilon)}_{-,0}$ are marked by solid curves.
Gray dots mark the Lee--Yang zeros of $G(\upsilon|t)$ at time $t=36$.}
\label{fig:LY_tunneling_pm}
\end{figure}
Entering from the mixed regime corresponds to the first and second order critical points swapping places, see Fig.~\ref{fig:LY_mixed}.

Lastly, we note that two tunneling subregimes, depicted in Figures \ref{fig:LY_tunneling_pp} and \ref{fig:LY_tunneling_pm}, are disconnected.
Na\"{i}vely, one would expect that a direct transition between these two regimes takes place when adjacent
degenerate minima of the bulk branches $F_{\pm}$ swap places. This is not what happens, however. Suppose that, for definiteness, we start in
$\llb + \asymp + \rrb$: before $\lambda^{(+)}_{-}$ can pass through $\lambda^{(+)}_{+}$, a transition to $\llb + \asymp + \bowtie - \rrb$ is inevitable simply due to the fact that branch $F_{+}$ is steeper than $F_{-}$ at large $\lambda$.

\section{Conclusion}
\label{sec:conclusion}

In this work, we introduced classical single-file systems of charged particles by imposing two
defining properties: the (I) \emph{single-file property} and (II) \emph{inertness of charges}.
We discovered a new type of universal dynamical behavior that manifests itself through anomalous statistical properties of charge transfer.
In particular, by computing the time-integrated current (charge transfer) in equilibrium and
quasistationary nonequilibrium states, we unveiled several unconventional properties that originate as a consequence of
broken ergodicity caused by classical fragmentation.

We examined the analytic structure of the scaled cumulant generating function and the associated large-deviation rate function
which displays a multifaceted phenomenology. Our results particularly concern two different, physically relevant timescales.
In equilibrium states with vanishing bias, we analytically obtained a universal non-Gaussian probability distribution of typical cumulative charge currents, thereby rigorously establishing violation of the CLT property for the entire class of charged single-file systems.
Anomalous fluctuations however become far more prominent away from equilibrium, where our model undergoes
dynamical phase transitions of two sorts -- either a first-order transition associated with an
exchange of dominance among competing branches {of the moment generating function} attributed to distinct dynamical phases, or a second-order transition induced by a spontaneous breaking of convexity of a stable branch.

The variance of charge transfer is governed by the characteristic dynamical exponent $z_{c}$ which is always twice
the exponent $z_{p}$ of the underlying particle dynamics, namely $z_{c}=2z_{p}$. This means, particularly, that
integrable particle dynamics ($z_{p}=1$) imply diffusive spreading of charge and a Gaussian scaling profile.
Nevertheless, we argue that such spreading of charge does not comply with normal (i.e. Gaussian) diffusion.
More specifically, it yields a non-Gaussian stationary  probability distribution  on the typical diffusive timescale.
With aid of our dressing technique we derived its explicit analytic form and establish its universality for
the considered class of models (irrespectively of the algebraic dynamical exponent $z_{c}$). This shows that
information supplied by the dynamical exponents and the asymptotic scaling function governing the hydrodynamic relaxation of the dynamical structure factor is insufficient for unambiguous classification of dynamical universality classes.
To better substantiate this subtle point, we carried out a comprehensive study of anomalous statistical properties of charge fluctuations associated with atyptical, exponentially rare events, encapsulated by the large-deviation rate function. We classify several dynamical regimes that arise due to competition of distinct competing dynamical phases, resulting from charge-pattern conservation implied by the single-file property. The outlined phenomenology likewise transcends that of stochastic diffusive system whose large fluctuations are captured by the universal action
of Macroscopic Fluctuation Theory.

Fluctuation relations take a rather peculiar form as well. While the multivariate fluctuation relation of
the Gallavotti--Cohen type is unconditionally satisfied as it is inherited from that of the particle current (including in the dynamical regimes featuring dynamical criticality), the structure of the univariate rate function encoding large fluctuations of the cumulative charge current experiences \emph{spontaneous breaking} of the fluctuation relation in spite of time-reversal invariance.
Such breakdown only occurs in a finite region in the global parameter space
and is mediated by dynamical phase transitions of first order.

\medskip

In the second part of the paper we exemplified our methods on a simple representative example of an exactly solvable classical deterministic and reversible cellular automaton. By deriving the exact FCS in bipartitioned nonequilibrium states, we expounded the salient features of the novel non-equilibrium universality and, by complexifying the counting field, computed the phase diagrams in different dynamical regimes.
The equilibrium case, studied already in our previous work \cite{Krajnik22}, is then recovered as a special degenerate case -- the flat branch entirely disappears (and only remains visible for complex counting fields), while first-order critical points flow towards the origin of the complex plane. Restoration of detailed balance ensures that the bulk branches become exactly degenerate. Meanwhile,
the Lee--Yang contours -- separating the bulk branches from the flat one -- meet at the origin,
giving rise to an exceptional triple critical point on the physical axis. The latter is in turn responsible for the lack of regularity, yielding divergent scaled cumulants. In spite of such dynamical criticality, the fluctuation-dissipation relation is preserved.

\medskip

\paragraph*{Outlook.}

A question of pivotal importance that remains to be addressed in future work is whether the exhibited
dynamical universality extends beyond the class of charge single-file systems studied in this paper.
In this respect, it is worth noting that the property of pattern conservation (leading to exponential foliation of the phase space)
is weaker than the combination of constraints (I) and (II). For example, one could consider one-dimensional models where
assignment or dynamics of charge (i.e. internal) degrees of freedom is correlated with particle trajectories.
Our expectation is that such `pattern-conserving' dynamical systems offer promising candidates for displaying
similar anomalous behavior. Various instances of such models have already appeared in the recent literature, mainly in the context of quantum Hamiltonian systems and unitary circuits. Owing to presence of exponentially many dynamical subsectors, the model belongs to a large class of systems exhibiting `Hilbert space fragmentation' \cite{Khemani20,Sala20,Rakovszky20}.

To slightly expand on this point, we briefly discuss the prototypical example of the `folded' XXZ quantum chain -- an exactly solvable model
with a local four-site interaction introduced in Refs.~\cite{Zadnik_foldedI,Zadnik_foldedII}, representing an effective 
model for the large-anisotropy limit of the gapped Heisenberg spin-$1/2$ chain. A classical cellular automaton analogue of the folded XXZ model has been constructed in Ref.~\cite{Pozsgay_folded}. Apart from conserving the total number of particles (magnons), the number of kinks (separating the boundaries of uniformly magnetized domains made out of `frozen' Bethe strings) becomes an additional (emergent) conserved quantity. The model consequently features an exponentially degenerate eigenspectrum owing to immobile spin-conserving patterns, in Ref.~\cite{Zadnik_foldedI} dubbed `topological invariants' (in the classical automaton, these are related to an exponential number of static domain-wall configurations \cite{Pozsgay21}). There exists a dual version of the model with a three-site local update rule (not distinguishing between spin-flip conjugate configurations) obtained by performing the `site-bond transformation' \cite{Pozsgay_folded}, where kinks become immobile particles and (mobile) solitons manifest themselves as bound states of kinks. As shown in Ref.~\cite{Zadnik_foldedI}, the dual folded model can be reinterpreted in terms charged degrees of freedom associated to macrosites obeying the charge-pattern conservation. A related exact mapping, yielding
the so-called Massarani--Mathieu model, was given in Ref.~\cite{Pozsgay21} and employed recently in Ref.~\cite{Feldmeier22} to explain the subdiffusive nature of charge transport. Phenomenological hydrodynamic picture proposed in Ref.~\cite{Sarang_FCS} indeed
predicts that in equilibrium with vanishing magnetization density the statistics of charge transfer obeys the M-Wright distribution,
derived previously in the sine--Gordon theory \cite{Altshuler06,Kormos22} and and in the hardcore automaton \cite{Krajnik22}.
In this work we established the M-Wright distribution as one of the defining universal properties of charged single-file systems.
All this seems to suggest that the folded XXZ automaton, including many other related models such as e.g. \cite{Gamayun23}, belong to the same universality class.
Unfortunately, our `dressing approach' crucially relies on the inertness of charge degrees of freedom and hence
is not directly applicable to such general pattern-conserving systems. Aside from other technical nuances, the main stumbling block is that
particles' worldlines cannot be equipped with charges in an uncorrelated manner.
Finding a direct mapping between the FCS of magnetization in the original folded XXZ automaton charge FCS in
its dual counterpart is further obstructed by length-changing effects \cite{Zadnik_foldedI,Feldmeier22}.

Relatedly, we wish to emphasize that the folded XXZ model, alongside other pattern-conserving systems, violate Bryc regularity.
This can be readily understood from the fact that charge transport (in the unbiased sector) obeys the inequality $z_{c}>1/\alpha$,
where exponent $\alpha$ governs the particle dynamics, see Ref.~\cite{Feldmeier22}. This in turn implies divergent scaled cumulants
caused by an emergent dynamical critical point at the zero counting fields (assuming that the Lee--Yang zeros of the moment generating function approach the origin of the complex $\zz$-plane sufficiently rapidly \cite{Krajnik22}, this would further imply lack of the CLT property).
We accordingly expect that pattern-conserving systems support phase transitions between two or more dynamical phases.
Away from equilibrium, the critical point might depart from the physical (i.e. real) axis in the $\zz$-plane. In the opposite scenario, a system
would sustain a first-order dynamical phase transition leading to a spontaneous breaking of the univariate Gallavotti--Cohen relation. 
The Lee--Yang formalism offers a powerful tool to investigate these aspects.

Even though the considered models are interacting, the imposed conditions allow us to analytically compute the fluctuations of transmitted charge. As such they could serve as a useful minimal example for addressing other questions in interacting dynamics.
For example, the conditional distribution $\mathcal{P}_{p|c}^{[0]}({\rm J}_p|{\rm J}_c)$ is bi-modal as a function of the integrated particle current in unbiased equilibrium.
Interpreting the latter as an order-parameter similarly as in \cite{KPZSchSch2022}, this suggests that the order-parameter
may become bi-modal when conditioning on anomalous currents even in equilibrium.

Before closing, we wish to shortly discuss the experimental relevance of our findings.
Our hope is that the universal anomalous distribution of typical fluctuations is, thanks to
its strongly non-Gaussian character, amenable to direct experimental detection.
By analogy with conventional single-file systems, the aim would be to realize a Brownian motion of two different
species of particles clogged in a narrow channel or particles that spread diffusively on a one-dimensional substrate.
The proposed version of a simple symmetric exclusion process of non-crossing Brownian trajectories of two colors
(or its continuum limit) provides one basic mathematical model of such a process. A direct experimental detection of dynamical phase
transitions in the structure of rare events posits a much greater challenge. Nevertheless, one can hope at least to be able to
detect some traces of dynamical criticality, or emergence of phase boundaries, via probing temporal growth of the higher cumulants, following
the lines of Refs.~\cite{Flindt13,Peng15,Brandner17,Deger18,DF19,PhysRevResearch.3.033206}.

\medskip
{\textbf{Acknowledgments}}.
We thank J. M. Luck and M. Kormos for correspondence. The authors benefited from discussions with K. Mallick, B. Doyon, B. Derrida and H. Spohn at the workshop on ``Integrable and Chaotic dynamics'' in Pokljuka, Slovenia (in July 2022).
The work has been supported by ERC (European Research Council) Advanced grant 694544-OMNES (TP), by ARRS (Slovenian research agency) research program P1-0402 (\v ZK, TP, EI), and by the SFB910 (project number 163436311) of the DFG (German Research Foundation) (JS). \v{Z}K acknowledges support of the Milan Lenar\v{c}i\v{c} foundation and the Simons Foundation via the Simons Junior Fellowship grant 1141511.

\appendix

\section{Exact full counting statistics}
\label{sec:exact_FCS}

Considering a bipartitioned nonequilibrium initial state, the main object of study is the finite-time joint MGF
\begin{equation}
G(\lambda_{c},\lambda_{p}|t) = \Big\langle e^{\lambda_{c}{\rm J}_{c}(t) + \lambda_{p}{\rm J}_{p}(t)} \Big\rangle_{\rm init}.
\end{equation}
By applying methods of localization, we compute large-$t$ limit of $G(\lambda_{c},\lambda_{p}|t)$ and infer the
exact joint SCGF $F(\lambda_{c},\lambda_{p})$. We subsequently restrict our considerations to
the univariate SCGF $F_{c}(\lambda_{c})$ encoding the FCS of charge transfer, while
computing the scaled cumulants and deduce the probability distribution of typical fluctuations.
We proceed afterwards by carefully examining the general solutions of the minimization problem arising from the dressing procedure applied
to the large-deviation rate function \eqref{eqn:Ic_infimum} and outline the main formal properties of such a dressing transformation.
We also include illustrative examples of the emergent dynamical regimes. We provide a short overview of Lee--Yang theory, which we  afterwards apply to the hardcore cellular automaton. The exact finite-time MGF allows us to determine the complex regions of dominance and parametrize the Lee--Yang contours that separate them. We conclude the section by resolving
the degenerate case of grand-canonical equilibrium states that feature an exceptional triple critical point.

The exact formula for the finite-time charge-current MGF $G_{c}(\lambda_{c}|t)$ in the hardcore automaton, computed for the grand-canonical
equilibrium states, is already known from Ref.~\cite{Krajnik22}. The computation can be easily adapted to bipartitioned initial ensembles.
To keep the notation in line with Ref.~\cite{Krajnik22}, we present the bivariate MGF $G(\lambda_{c},\lambda_{p}|t)$
as a double sum by summing over contribution of freely propagating \emph{holes} (as opposed to particles).

Let $l\equiv \ol{n}_{-}$ and $r\equiv \ol{n}_{+}$ denote the number
of \emph{holes} crossing the origin in a time interval $t$ starting from the left and right partitions, respectively, and similarly $n_{\mp}=|\Lambda_{\mp}|$ for particles. Note that a particle crossing the origin requires a hole passing in the opposite direction, and thus we have $|\Lambda_{\pm}|=[|l-r|\pm (l-r)]/2$. By averaging over initial configurations, the exact finite-time joint MGF $G(\lambda_{c},\lambda_{p}|t)$
is given by a double sum of the form
\begin{equation}
G(\lambda_{c},\lambda_{p}|t) = \hspace{-0.8em}
\sum_{\ol{n}_{-},\ol{n}_{+}=0}^{t}
\prod_{\epsilon \in \{\pm\}} \binom{t}{\ol{n}_{\epsilon}} \rho^{t}_{\epsilon}[\nu_{\epsilon}(\lambda_{p})]^{\ol{n}_{\epsilon}}
\mu^{|\Lambda_{\epsilon}|}_{\epsilon},
\label{eqn:G_double_sum}
\end{equation}
where $\mu_{\pm}(\lambda_{c})$ are the `dressed' counting fields (given by Eq.~\eqref{eqn:dressed_counting_fields}).
Under charge conjugation, ${\rm C}:b_{\pm}\mapsto -b_{\pm}$, the joint MGF transforms as
\begin{equation}
{\rm C}[G(\lambda_{c},\lambda_{p}|t)]=G(-\lambda_{c},\lambda_{p}|t).
\end{equation}
Similarly, space reflection ${\rm R}:x\mapsto -x$ interchanges the state parameters,
${\rm R} : b_{\pm} \mapsto b_{\mp}$ and ${\rm R} : \rho_{\pm} \mapsto \rho_{\mp}$, and hence the MGF satisfies
\begin{equation}
{\rm R}[G(\lambda_{c},\lambda_{p}|t)]=G(-\lambda_{c},-\lambda_{p}|t).
\end{equation}

The univariate MGF $G_{c}(\lambda_{c}|t)$ is simply obtained by marginalization, namely $G_{c}(\lambda_{c}|t)=G(\lambda_{c},0|t)$.
An alternative way to derive $G_{c}(\lambda_{c}|t)$ is to initially compute the PDF associated with transferred particles, which is easily accomplished by identifying the cumulative particle current ${\rm J}_{p} = r-l$, yielding
\begin{equation}
\mathcal{P}_{p}({\rm J}_{p}|t) = \sum_{\ol{n}_{-},\ol{n}_{+}=0}^{t}\delta_{{\rm J}_{p},\ol{n}_{+}-\ol{n}_{-}}
\prod_{\epsilon \in \{\pm\}}\binom{t}{\ol{n}_{\epsilon}}\nu^{\ol{n}_{\epsilon}}_{\epsilon}\rho^{t}_{\epsilon}.
\label{eqn:double_sum_Pp}
\end{equation}
The associated MGF $G_{p}(\lambda_{p}|t)$ is given by the Laplace transform,
$G_{p}(\lambda_{p}|t) = \mathcal{L}\left[\mathcal{P}_{p}({\rm J}_{p}|t) \right](-\lambda_{p})$, yielding
\begin{equation}
G_{p}(\lambda_{p}|t) = \sum_{\ol{n}_{-},\ol{n}_{+}=0}^{t}
\prod_{\epsilon \in \{\pm\}}\binom{t}{\ol{n}_{\epsilon}}[\nu_{\epsilon}(\lambda_{p})]^{\ol{n}_{\epsilon}}\rho^{t}_{\epsilon}.
\label{eqn:Gp}
\end{equation}
Applying the dressing operator $\mathfrak{D}_{G}$ (see Eq.~\eqref{eqn:Dg}), amounts to using the substitution rule \eqref{eqn:replacement_rule} in Eq.~\eqref{eqn:Gp}, which recovers Eq.~\eqref{eqn:G_double_sum}.

By applying the particle-hole transformation to Eq.~\eqref{eqn:double_sum_Pp}, we obtain the following equivalent double-sum representation of
$\mathcal{P}_{p}({\rm J}_{p}|t)$ involving summation over the transferred particles instead of vacancies
(cf. Eq.~\eqref{eqn:uncrestricted_dressing})
\begin{equation}
\mathcal{P}_{p}({\rm J}_{p}|t) = \sum_{n_{-},n_{+}=0}^{t}\delta_{{\rm J}_{p},n_{-}-n_{+}}
\prod_{\epsilon \in \{\pm\}}\binom{t}{n_{\epsilon}}
\ol{\rho}^{t}_{\epsilon}\nu^{-n_{\epsilon}}_{\epsilon}.
\end{equation}
The obtained result is very transparent and intuitive: computing $\mathcal{P}_{p}({\rm J}_{p}|t)$ for non-interacting
ballistically propagating particles entails summing over all the contributions from $n_{-}$ right movers from the left partition and subtracting $n_{+}$ left-movers from the right partition, weighted with appropriate statistical factors.

\subsection{Localization}

Asymptotic growth of MGF $G(\lambda|t)$ can be computed using Laplace's method of localization as we now describe. We begin by first listing
the general formulae, which we subsequently apply to our specific case.
Let $\mathscr{D}$ be a $d$-dimensional domain parameterized by coordinates ${\bf x}=(\xx_{1},\xx_{2},\ldots,\xx_{d})$.
We consider two dummy functions ${\rm f}({\bf x})$ and ${\rm g}({\bf x})$ and an integral of the type
$\mathfrak{G}_{{\rm g}}[{\rm f}] \equiv \int_{\mathscr{D}}\dd {\bf x}\,{\rm g}({\bf x})e^{t\,{\rm f}({\bf x})}$.
For large $t$, the integral localizes around the point 
${\bf x}_{0}$, $\nabla {\rm f}|_{{\bf x}_{0}}=0$, corresponding to the global maximum of 
${\rm f}$ in the bulk (i.e. interior) of the integration domain $\mathscr{D}$.
We have the following asymptotic approximation
\begin{equation}
\mathfrak{G}_{{\rm g}}[{\rm f}] \asymp \left(\frac{2\pi}{t}\right)^{d/2}
\frac{{\rm g}({\bf x}_{0}) e^{t {\rm f}(\bf{x}_0)}}{\sqrt{{\rm det}\mathscr{H}[{\rm f}]({\bf x}_{0})}},
\end{equation}
with $\mathscr{H}[{\rm f}]$ being the Hessian of function ${\rm f}$ evaluated at point ${\bf x}_{0}$.
When $\mathscr{D}$ is a compact domain, it might also happen that maximum is attained at the boundary $\partial \mathscr{D}$.
In this case, assuming a non-vanishing derivative in the normal direction to the boundary with unit normal ${\bf n}_{\perp}$, we use the following formula for ${\bf x}_{0}\in \partial \mathscr{D}$,
\begin{equation}
\mathfrak{G}_{g}[{\rm f}] \asymp \frac{1}{2\pi}\left(\frac{2\pi}{t}\right)^{\frac{d+1}{2}} \!\!
\frac{{\rm g}({\bf x}_{0})e^{t\,{\rm f}({\bf x}_{0})}}{|\nabla_{{\bf n}_{\perp}}{\rm f}({\bf x_{0}})|\sqrt{{\rm det}\mathscr{H}_{\parallel}[{\rm f}]({\bf x}_{0})}},
\end{equation}
with $\mathscr{H}_{\parallel}[{\rm f}]({\bf x_{0}})$ denoting the Hessian submatrix evaluated in the remaining (non-normal) coordinates (i.e. subspace orthogonal to ${\bf n}_{\perp}$) at ${\bf x_{0}}$.

Returning to our working example, we now infer the behavior of $G_{c,p}(\lambda_{c},\lambda_{p}|t)$ at large times.
We introduce the rescaled continuum coordinates
\begin{equation}
\frac{l}{t} \to \xx_{1},\qquad \frac{r}{t} \to \xx_{2},
\end{equation}
and subsequently convert the double sum in Eq.~\eqref{eqn:G_double_sum} into a two-dimensional ($d=2$) integral over
the square domain $\mathscr{D}_{\square}\equiv [0,1]^{2}$ and use Stirling's approximation $n! \asymp \sqrt{2\pi n}(n/e)^{n}$ applied
to the binomials in Eq.~\eqref{eqn:G_double_sum}, in the regime where all $t$, $l$ and their difference $t-l$ become large
\begin{equation}
\binom{t}{l} \asymp \sqrt{\frac{t}{2\pi l(t-l)}}\left(\frac{l}{t}\right)^{-l}\left(1-\frac{l}{t}\right)^{l-t}.
\end{equation}
In this manner we deduce the following exact asymptotic expression for the joint particle-charge MGF
\begin{equation}
G_{c,p}(\lambda_{c},\lambda_{p}|t) \asymp \frac{t}{2\pi} \int_{\mathscr{D}_{\square}} \dd {\mathbf x}
\frac{e^{t\,(f_{c}(\xx_{1},\xx_{2})+f_{p}(\xx_{1},\xx_{2}))}}{\prod_{i=1}^{2}\sqrt{\xx_{i}(1-\xx_{i})}},
\label{eqn:localized_MGF}
\end{equation}
with exponents
\begin{align}
e^{f_{c}(\xx_{1},\xx_{2})} &= \frac{\nu^{\xx}_{-}\nu^{\yy}_{+}
\prod_{\epsilon \in \{\pm\}}\mu^{(|\xx_{1}-\xx_{2}|+\epsilon(\xx_{1}-\xx_{2}))/2}_{\epsilon}}
{(\rho_{+}\rho_{-})^{-1}\prod_{i=1}^{2}\xx^{\xx_{i}}_{i}(1-\xx_{i})^{1-\xx_{i}}},\\
e^{f_{p}(\xx_{1},\xx_{2})} &= e^{\lambda_{p}(\xx_{2}-\xx_{1})}.
\end{align}

\medskip

\paragraph*{Bulk maxima.}
When $f({\bf x})$ attains its maximum in the bulk of $\mathscr{D}_{\square}$, we find two maxima corresponding to two critical points located at ${\bf x}_{\pm}\equiv (\xx_{1,\pm},\xx_{2,\pm})$, with coordinates
\begin{align}
\xx_{1,\pm} &= \frac{\nu_{-}(\lambda_{p})}{\nu_{-}(\lambda_{p}) + [\mu_{\pm}(\lambda_{c})]^{\mp 1}},\\
\xx_{2,\pm} &= \frac{\nu_{+}(\lambda_{p})}{\nu_{+}(\lambda_{p}) +[\mu_{\pm}(\lambda_{c})]^{\pm 1}}.
\end{align}
Critical points ${\bf x}_{\pm}$ are however not always simultaneously present.
In fact, depending on the ratio of densities $\kappa(\lambda_{p})=\sqrt{\nu_{+}(\lambda_{p})/\nu_{-}(\lambda_{p})}$,
extremal point ${\bf x}_{+}$ appears only below the diagonal of domain $\mathscr{D}_{\square}$,
namely for $\xx_{1,+}\geq \xx_{2,+}$, equivalent to inequality $\mu_{+}(\lambda_{c})\geq \kappa(\lambda_{p})$);
analogously ${\bf x}_{-}$ appears above the diagonal for $\xx_{1,-}\leq \xx_{2,-}$ or, equivalently,
$\mu_{-}(\lambda_{c})\geq \kappa^{-1}(\lambda_{p})$).
The two separatrices are determined by the conditions $\mu_{\pm}(\lambda_{\pm})=[\kappa(\lambda_{p})]^{\pm 1}$.
For example, for $\lambda_{p}=0$ this yields two solutions $\lambda^{(\epsilon)}_{\pm}$ (with $\epsilon \in \{\pm\}$) given by Eq.~\eqref{eqn:degenerate_minima}. At critical points ${\bf x}_{\pm}$, the exponent in Eq.~\eqref{eqn:localized_MGF} evaluates to
\begin{equation}
f^{(\pm)}_{c,p}(\lambda_{c},\lambda_{p}) \equiv [f_{c}(\lambda_{c})+f_{p}(\lambda_{p})]|_{{\bf x}_{\pm}},
\end{equation}
given by the expressions in Eq.~\eqref{eqn:f_pm}. Moreover, the Hessian precisely cancels out the square-root factor
in Eq.~\eqref{eqn:localized_MGF}.
Writing $\boldsymbol{\lambda}\equiv (\lambda_{c},\lambda_{p})$, the joint MGF therefore assumes the following simple asymptotic form
\begin{equation}
G_{c,p}(\boldsymbol{\lambda}|t) \asymp
\begin{cases} 
\hfill e^{t \, f^{(\pm)}_{c,p}(\boldsymbol{\lambda})}\hfill  & \boldsymbol{\lambda} \in \mathscr{E}_{\pm} \\
\hfill \sum_{\epsilon \in \{\pm\} }e^{t f^{(\epsilon)}_{c,p}(\boldsymbol{\lambda})} \hfill & \boldsymbol{\lambda} \in \mathscr{E}_{+,-}
\end{cases},
\end{equation}
where $\mathscr{E}_{\pm}$ and $\mathscr{E}_{+,-}$ are given by Eq.~\eqref{eqn:E_pm} and Eq.~\eqref{eqn:E_coexist}, respectively.
The above asymptotic formula does not apply for $\lambda_{c} \in \mathscr{I}_{0}$.

\medskip

\paragraph*{Localization along the diagonal of $\mathscr{D}_{\square}$.}

When both conditions $\mu_{+}(\lambda_{c})< \kappa(\lambda_{p})$ and $\mu_{-}(\lambda_{c})<\kappa^{-1}(\lambda_{p})$ are simultaneously satisfied, the interior of $\mathscr{D}_{\square}$ is devoid of any critical points.
In this case, function $f$ in the exponent in Eq.~\eqref{eqn:localized_MGF} attains a maximum on the diagonal
at ${\bf x}_{0}\equiv (\xx_{0},\xx_{0})$ with parallel and normal unit directions
${\bf n}_{\parallel}=\tfrac{1}{\sqrt{2}}(1,1)$ and ${\bf n}_{\perp}=\pm\tfrac{1}{\sqrt{2}}(1,-1)$
(for $\xx_{1}\gtrless \xx_{2}$), respectively. Solving for $\nabla_{{\bf n}_{\parallel}}f({\bf x})|_{{\bf x}_{0}}=0$, we find
$\xx_{0}=\sqrt{\nu_{-}\nu_{+}}/(1+\sqrt{\nu_{-}\nu_{+}})$, with a non-zero second derivative
$\nabla^{2}_{{\bf n}_{\parallel}}f({\bf x})|_{{\bf x}_{0}}=-(2+\sqrt{\nu_{-}\nu_{+}}+1/\sqrt{\nu_{-}\nu_{+}})$
and $\nabla_{{\bf n}_{\perp}}f({\bf x})|_{{\bf x}_{0}} = -\sqrt{2}\log{(\kappa^{\pm 1}(\lambda_{p})/\mu_{\pm}(\lambda_{c}))}$.
For $\boldsymbol{\lambda} \in \mathbb{R}\times \mathbb{R}$, the MGF $G_{c,p}(\boldsymbol{\lambda}|t)$ restricted to
intervals $\mathscr{I}_{k\in \{+,0,-\}}$ (see Eqs.~\eqref{eqn:selection_rules}) takes the form
\begin{equation}
G_{c,p}(\boldsymbol{\lambda} \in \mathscr{I}_{k}|t) \asymp W_{k}(\boldsymbol{\lambda})e^{t\,F_{k}(\boldsymbol{\lambda})},
\end{equation}
with bulk branches $F^{(\pm)}_{c,p}(\boldsymbol{\lambda})$ (see Eq.~\eqref{eqn:f_pm}) and flat branch
$F^{(0)}$ (see Eq.~\eqref{eqn:flat_branch}), respectively. The `weighting functions' read explicitly
\begin{align}
W_{\pm} &= 1, \\
W_{0}(\boldsymbol{\lambda}) &= \frac{\prod_{\epsilon \in \{\pm\}}[\log{(\kappa^{\epsilon}(\lambda_{p})\mu^{-1}_{\epsilon}(\lambda_{c}))}]^{-1}}
{[4\pi t(2+\sqrt{\nu_{-}\nu_{+}}+1/\sqrt{\nu_{-}\nu_{+}})]^{1/2}}.
\end{align}

\subsection{Scaled cumulants}

In this section, we compute the scaled cumulant $s_{n}\equiv s^{(c)}_{n}$ for the general case of unequal densities $\rho_{-}\neq \rho_{+}$. Invoking Bryc's regularity condition (which follows from the Lee--Yang analysis carried out in Sec.~\ref{sec:LeeYang} below),
the dominant branch \eqref{eqn:f_pm} is always faithful and consequently all cumulants $c_{n}(t)$ grow linearly with time.
In effect, scaled cumulants $s_{n}$ are precisely the Taylor series coefficients of the charge-current SCGF $F(\lambda)$.
The first two scaled cumulants are of particularly simple form
\begin{align}
s_{1} &= (\rho_{-} - \rho_{+})b_{\pm},\\
s_{2} &= \mp(\rho_{-}-\rho_{+}) - b^{2}_{\pm}(\rho^{2}_{+}+\rho^{2}_{-}-2\rho_{\mp}),
\end{align}
where the signature depends on which of the two bulk branches $F_{\pm}(\lambda)$ dominates at $\lambda=0$
(recall that $F_{0}$ can never be dominant at $\lambda=0$).

The degenerate case of equal densities, $\rho_{\pm}=\rho$, is rather exceptional. To begin with, the first cumulant $c_{1}(t)$ no longer grows linearly with time.
As discussed in Sec.~\ref{sec:setting}, the lack of Bryc regularity implies that the SCGF $F(\lambda)$ is no longer a faithful generating function. In Appendix \ref{sec:equal_densities}, we infer how cumulants $c_{n}(t)$ grow at large times with aid of localization and Fa\`{a} di Bruno's formula, establishing that $c_{n}\sim t^{n/2}$. To give a flavor, the first two cumulants read
\begin{align}
c_{1}(t) &\asymp (b_{-}-b_{+})\frac{\chi_{p}(\rho)}{\pi}\, t^{1/2},\\
c_{2}(t) &\asymp \chi_{p}(\rho)\big(b^{2}_{-}+b^{2}_{+}-\frac{1}{\pi}(b_{-}-b_{+})^{2}\big)\,t.
\end{align}
While in equilibrium, that is for $b_{-}=b_{+}$, all odd cumulants vanish identically, even cumulant still behave anomalously.

\subsection{Typical fluctuations}

We have already established that $G(\lambda|t)$ grows asymptotically with exponent $\alpha=1$. This means that typical values of the integrated charge current through the origin in a time interval $t$ are of the order ${\rm J}_{c}(t)\sim \sqrt{t}$.
By taking full advantage of the exact expression for the finite-time charge-current MGF $G(\lambda|t)$,
we are in a position to deduce the associated PDF, denoted subsequently by $\mathcal{P}^{\rm typ}(j)$
(here we have once again dropped the subscript 'c' by identifying $\lambda_{c}\to \lambda$ and $j_{c}\to j$).
In the following, we treat the generic case with unequal densities $\rho_{-}\neq \rho_{+}$.
The special case of equal densities is worked out in Appendix \ref{sec:equal_densities}.

Introducing the shifted cumulative charge current $\hat{\rm J}_{c}(t)\equiv {\rm J}_{c}(t)-c_{1}(t)$
by subtracting the first cumulant $c_{1}(t)$, we first define the dynamically rescaled time-dependent PDF
\begin{equation}
\mathcal{P}_{1/2}(\hat{\mathcal{J}}_{c}|t) \equiv t^{1/2}\mathcal{P}\big(\hat{\mathcal{J}}_{c}=t^{-1/2}\hat{\rm J}_{c}(t)|t\big),
\end{equation}
such that
\begin{equation}
e^{-\lambda\,c_{1}(t)}G(\lambda|t) = \int \dd \hat{\mathcal{J}}_{c}
\mathcal{P}_{1/2}(\hat{\mathcal{J}}_{c}|t) e^{\lambda\,t^{1/2}\hat{\mathcal{J}}_{c}}.
\end{equation}
This yields, in the limit of large times, the stationary PDF
\begin{equation}
\mathcal{P}^{\rm typ}(j)=\lim_{t\to \infty} \mathcal{P}_{1/2}(j=\mathcal{J}_{c}|t).
\end{equation}
By introducing a dynamically rescaled counting field $\eta$ via
\begin{equation}
\eta \equiv t^{1/2}\lambda,
\end{equation}
and using $\lim_{t\to \infty}e^{-t^{1/2}\eta \, c_{1}(t)}G\left(t^{-1/2}\eta|t\right) = e^{s_{2}\eta^{2}}$,
we can express $\mathcal{P}^{\rm typ}(j)$ via the inverse Laplace transform (acting on a function of variable $\eta$)
\begin{equation}
\mathcal{P}^{\rm typ}(j) = \mathfrak{L}^{-1}\left[e^{s_{2}\eta^{2}}\right](j),
\end{equation}
yielding a \emph{Gaussian} PDF
\begin{equation}
\mathcal{P}^{\rm typ}(j)=\frac{1}{\sqrt{2\pi \sigma^{2}_{\rm typ}}} \exp{\left[-\frac{j^{2}}{2\sigma^{2}_{\rm typ}}\right]},\quad
\sigma^{2}_{\rm typ}=s_{2}.
\end{equation}
Here $s_{2}=F^{\prime \prime}(0)$ is the second scaled cumulant $s_{2}$ of the (faithful) SCGF $F(\lambda)$.
Importantly, the obtained result implicitly depends on the dominant branch of $F_{\pm}(\lambda)$ at $\lambda=0$.
We have thus recovered the CLT behavior, as indeed guaranteed by Bryc regularity.

\begin{figure}[htb]
\includegraphics[width=\columnwidth]{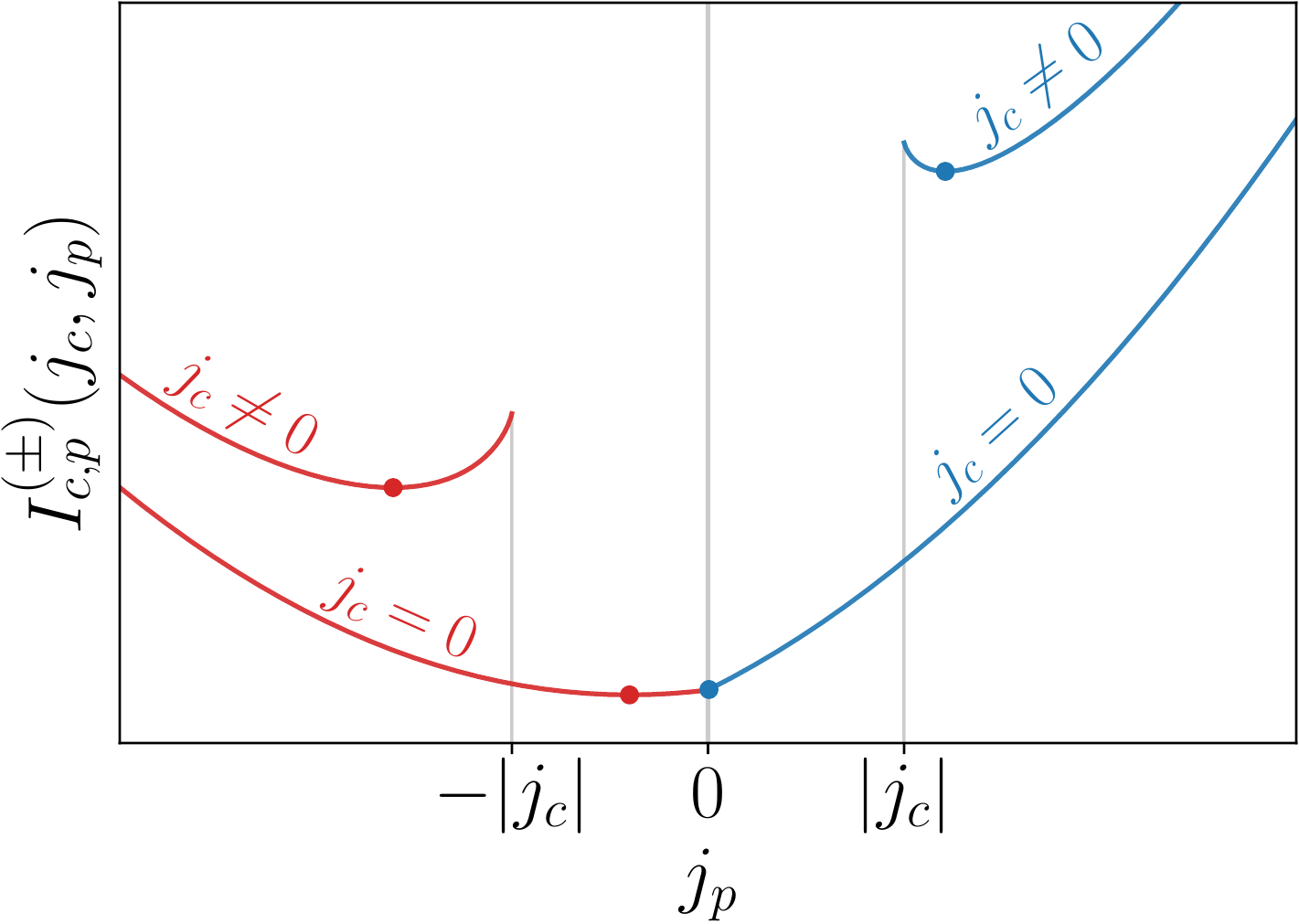}
\caption{Two branches of the joint particle-charge LDF $I_{c,p}(j_{c},j_{p})$. For $j_{c}\neq 0$, the two branches $I^{(\pm)}_{c,p}(j_{c},j_{p})$ are supported on disjoint domains $\mathscr{J}_{\pm}$ separated by a gap, attaining their minima $j^{\pm}_{p}(j_{c})$ in the interior of $\mathscr{J}_{\pm}$. For $j_{c}=0$, the gap between the two domains closes while preserving convexity of $I_{c,p}(j_{c},j_{p})$. This time, one of the branches (in Figure $\mathscr{J}_{-}$, shown in blue) attains the minimum at $j_{c}=0$ and since $j^{\pm}_{p}(j_{c})$ is not differentiable at $j_{c}=0$ the rate function $I_{c}(j_{c})$  \emph{can} acquire a corner point at the origin.}
\label{fig:minimization}
\end{figure}

\section{Dressing formalism}
\label{sec:dressing_technical}

\subsection{Particle rate function}

Here we carry out a formal analysis of the solutions to the optimization problem given below by Eq.~\eqref{eqn:convex_optimizations}.
Since now the particle and charge current both play a role and appear in the formulae simultaneously, we shall reinstate the subscript
indices to avoid ambiguities.  As we explain in turn, the computation of $I_{c}(j_{c})$ can be formulated as a nested, \emph{two-stage} optimization procedure. In the following, the LD rate function of the particle transfer $I_{p}(j_{p})$ provides an input to the outlined optimization.

Our starting point will be the following joint rate function
\begin{equation}
I_{c,p}(j_{c},j_{p}) = I_{c|p}(j_{c},j_{p})+I_{p}(j_{p}),
\end{equation}
allowing us to express $I_{c}(j_{c})$ through marginalization,
\begin{equation}
I_{c}(j_{c}) = {\rm inf}_{j_{p}}\{I_{c,p}(j_{c},j_{p})\},
\end{equation}
namely by minimizing the joint rate function over the domain $\mathscr{J}$ of time-integrated rescaled particle current $j_{p}$, with $j_{c}$ kept fixed.

By exploiting the fact that $|j_{c}|\leq |j_{p}|$, we split $\mathscr{J}$ into two \emph{disjoint} intervals $\mathscr{J}_{\pm}$ such that
$\mathscr{J}=\mathscr{J}_{-}\cup \mathscr{J}_{+}$. Assuming strict causality, such that $j_{p} \in [j^{\rm min}_{p},j^{\rm max}_{p}]$, we define 
\begin{equation}
\mathscr{J}_{+}(j_{c})\equiv -\big[j^{\rm min}_{p},|j_{c}|\big],\quad
\mathscr{J}_{-}(j_{c})\equiv \big[|j_{c}|,j^{\rm max}_{p}\big],
\end{equation}
and correspondingly split the bivariate function $I_{c,p}(j_{c},j_{p})$ into two separate branches
\begin{equation}
I^{(\pm)}_{c,p}(j_{c},j_{p}) \equiv I^{(\pm)}_{c|p}(j_{c},j_{p}) + I_{p}(j_{p}).
\end{equation}
The \emph{split} conditional rate functions
\begin{equation}
I^{(\pm)}_{c|p}(j_{c},j_{p}) \equiv I_{c|p}(j_{p}|j_{c})\Big|_{j_{p} \in \mathscr{J}_{\pm}},
\end{equation}
have the following explicit form
\begin{equation}
I^{(\pm)}_{c|p}(j_{c},j_{p}) = \sum_{\epsilon \in \{\pm\}} \frac{|j_{p}|}{2}(1+\epsilon \, \xi)\log{\left[\frac{1+\epsilon\,\xi}{1\mp \epsilon\,b_{\pm}}\right]}.
\end{equation}
The first stage of optimization involves minimizing the problems on the separate non-overlapping subdomains $\mathscr{J}_{\pm}$,
\begin{equation}
I^{(\pm)}_{c}(j_{c}) \equiv {\rm inf}_{j_{p} \in \mathscr{J}_{\pm}}\big\{I^{(\pm)}_{c,p}(j_{c},j_{p})\big\},
\label{eqn:convex_optimizations}
\end{equation} 
resulting in two branches of the rate function $I_c^{(\pm)}(j_c)$.
In the second stage of optimization, the physical rate function $I_{c}(j_{c})$ is determined by picking the minimum of the two branches $I^{(\pm)}_{c}$ for every given value of $j_{c}$ ,
\begin{equation}
I_{c}(j_{c}) = {\rm min}\{I^{(\pm)}_{c}(j_{c})\}.
\label{eqn:Ic_global_optimization}
\end{equation}
We shall not attempt to give a fully general classification of solutions. Instead, we proceed with certain mild simplifying assumptions
on the input rate function $I_{p}(j_{p})$. We require that $I_{p}(j_{p})$ is
a strictly convex and twice differentiable everywhere on its domain,
ensuring $I^{\prime \prime}_{p}(j_{p})\equiv \partial^{2}_{j_{p}}I_{p}(j_{p})\geq 0$.
Generic rate functions are expected to fulfill this technical assumptions, which appreciably simplifies the ongoing analysis.
If the particle dynamics itself sustains a dynamical phase transition,
there might of course be additional physical features that are not captured by our analysis. This is however beyond the scope of the present work.
Our objective is to understand how the single-file and inertness properties lead to anomalous fluctuations associated with the charge transfer.

We proceed by investigating the minimization problems given
by Eq.~\eqref{eqn:convex_optimizations}. Let $j^{\pm}_{p}$ denote the minima of $I^{(\pm)}_{c,p}(j_{c},j_{p})$.
The solutions to Eqs.~\eqref{eqn:convex_optimizations} are then given by $I^{(\pm)}_{c,p}(j_{c},j^\pm_{p})=I^{(\pm)}_{c}(j_{c})$. 
It is important to stress that minima $j^{\pm}_{p}$ are functions of $j_{c}$, i.e. $j_p^\pm = j_p^\pm(j_c)$.
Noting that $I^{(\pm)}_{c,p}(j_{c},j_{p})$ are differentiable in both arguments,
it follows that when $j^{\pm}_{p}(j_{c})$ are differentiable functions of variable $j_{c}$, then so are the functions $I^{(\pm)}_{c}(j_{c})$.

The second derivative with respect to the variable $j_{p}$ of the
conditional rate function $I_{c|p}(j_{c},j_{p})$, that is $I^{\prime \prime}_{c|p}(j_{c},j_{p})\equiv \partial^{2}_{j_{p}}I_{c|p}(j_{c},j_{p})$, reads explicitly (with $\xi = j_c/|j_p|$, see Eq.~\eqref{eqn:xi_def})
\begin{equation}
I^{\prime \prime}_{c|p}(j_{c},j_{p}) = \frac{1}{|j_{p}|}\frac{\xi^{2}}{1-\xi^{2}},
\label{eqn:2nd_derivative}
\end{equation}
and is \emph{non-negative} everywhere except at $j_{c}=0$ where it vanishes, $I^{\prime \prime}_{c|p}(0,j_{p})=0$. Using that $I_p(j_p)$ is convex, this further implies
that $I_{c,p}^{(\pm)}(j_c\neq0, j_p)$ is strictly convex as a function of $j_{p}$.
It then follows that the functions $I^{(\pm)}_{c,p}(j_{c},j_{p})$ possess unique minima at $j^{\pm}_{p}$, respectively. For this reason, we exclude for the moment the case $j_{c}=0$ and continue with the general case $j_{c}\neq 0$.

\onecolumngrid

\begin{figure}[htb]
	\includegraphics[width=0.9\textwidth]{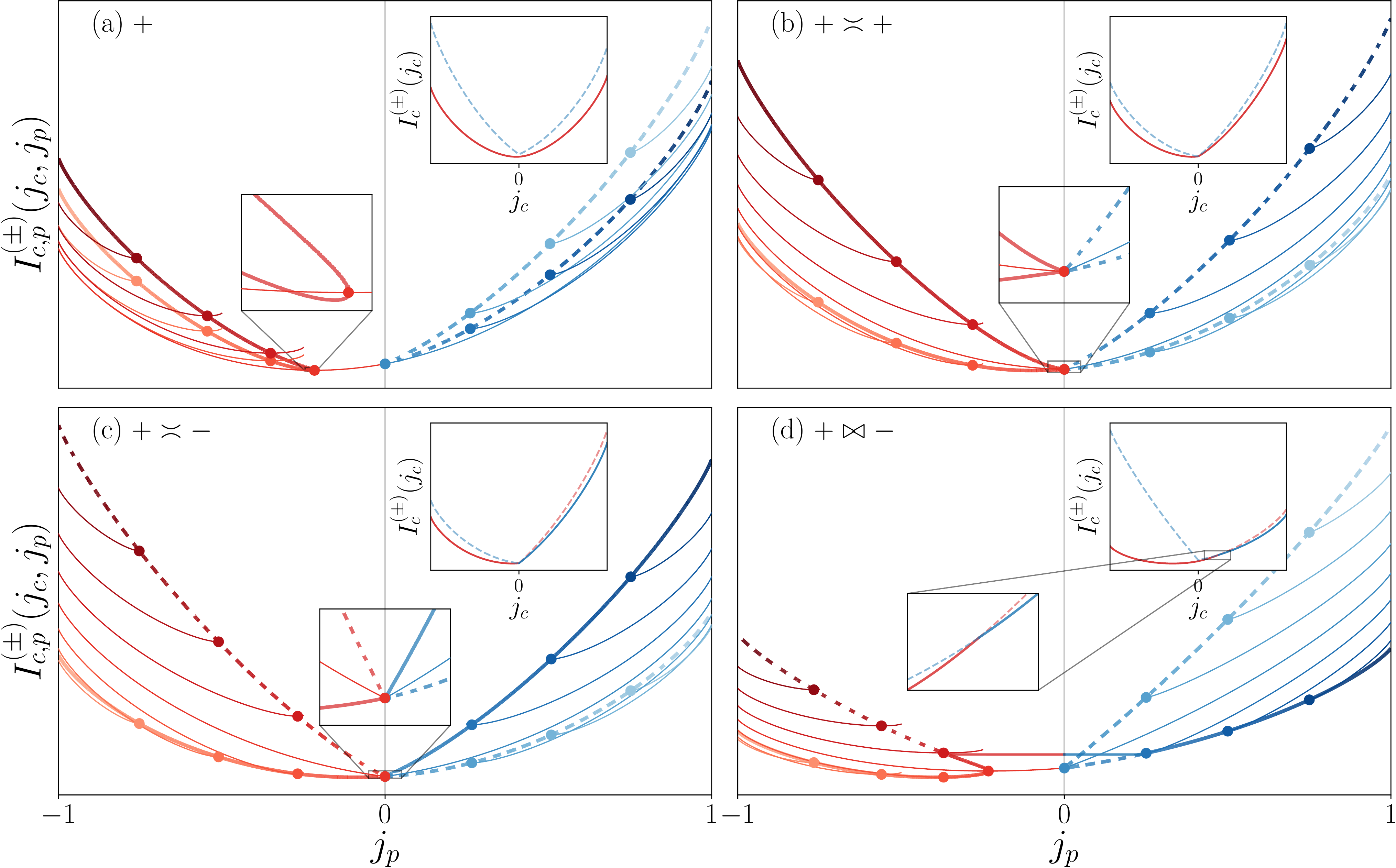}
	\caption{Two branches $I^{(\pm)}_{c,p}(j_{c},j_{p})$ (with $+$ and $-$ shown by red and blue curves, respectively) of the joint
		particle-charge LDF $I_{c,p}(j_{c},j_{p})$, depicted for different dynamical regimes of the hardcore automaton: (a) regular regime, (b) tunneling subregime to the same bulk branch, (c) tunneling subregime to the other bulk branch and (d) corner regime. Thin curves correspond to different values of $j_{c}$ with gray dots marking the location of minima. Thick curves indicate the flow of minima in each of the respective domains $\mathscr{J}_{\pm}$ while continuously varying $j_{c}$. 
		The global minima on $\mathscr{J}=\mathscr{J}_{+}\cup \mathscr{J}_{-}$ follow the solid thick lines, while the dashed thick lines trace the  subdominant minima. In regular regime (a), the global minimum always, i.e. for all $j_{c}$, belongs to one of the branches $I^{(\pm)}_{c,p}$, yielding a differentiable rate function $I_{c}(j_{c})$. Likewise, the global minimum belongs to the same branch in tunneling
		subregime $\llb + \asymp + \rrb$ shown in panel (b), except that at $j_{c}=0$ the minimum is found at $j_{p}=0$, resulting in a non-differentiable (corner) point in the rate function. Panel (c) shows tunneling to another bulk branch labeled by $\llb + \asymp - \rrb$, where at $j_{p}=0$ the global minimum shifts from $\llb + \rrb$ over to $\llb - \rrb$, also inducing a corner in the rate function. In the corner regime, shown in panel (d) for $\llb + \bowtie - \rrb$, the trajectories of the local minima $j^{\pm}_{p}(j_{c})$ do not meet at $j_{p}=0$. Instead global minimum jumps from one branch to another at the critical charge current $j_{\bowtie}$, producing
		a non-differentiable (corner) point in the physical rate function $I_{}(j_{c})$ at $j_{c}=j_{\bowtie}$. Insets show the two branches $I^{(\pm)}_{c}(j_{c})$ of the corresponding large-deviation rate function, with solid curves marking the physical rate function $I_{c}(j_{c})$.}
	\label{fig:minimization_examples}
\end{figure}

\twocolumngrid

Noting that for $|\xi| \to 1$ (i.e. when approaching the boundaries of domains $\mathscr{I}^\pm$ from the left/right respectively) the magnitudes of the first derivatives
\begin{equation}
I^{\prime(\pm)}_{c|p}(j_{c},j_{p}) = \pm \frac{1}{2}\log{\left(\frac{1-b^{2}_{\pm}}{1-\xi^{2}}\right)},
\label{eqn:1st_derivative}
\end{equation}
are unbounded and of different sign, we  conclude 
(using that $I_{c,p}^{(\pm)}$ is convex and $I^{\prime}_{p}(j_p)$ bounded) that the minima $j^{\pm}_{p}$ reside in the interior of domains $\mathscr{I}_{\pm}(j_{p})$ (see Fig.~\ref{fig:minimization}).
By invoking the implicit function theorem, it then follows from $I^{\prime \prime(\pm)}_{c,p}(j_{c}\neq 0,j_{p})>0$, that the minima $j^{\pm}_{p}(j_{c})$
are differentiable with respect to $j_{c}$, $j^{\pm}_{p}=j^{\pm}_{p}(j_{c})$ for $j_{c}\neq 0$. Consequently, both branches $I^{(\pm)}_{c}(j_{c}\neq 0)$ in Eq.~\eqref{eqn:convex_optimizations} are also differentiable functions.

Differentiability of $I_{c}(j_{c})$ is however no longer guaranteed at the origin $j_{c}=0$, representing a distinguished point. We note that at $j_{c}=0$  ($\xi=0$) we have a finite first derivative (see Eq.~\eqref{eqn:1st_derivative}) and a vanishing second derivative (see  Eq.~\eqref{eqn:2nd_derivative}). This means that the first derivatives of $I_{c,p}^{(\pm)}(0, j_{p})$ with respect to  $j_{p}$ can, depending on state parameters, be either negative or positive, respectively, and consequently the corresponding minima of $I_{c,p}^{(\pm)}(0, j_{p})$ \emph{can} be attained at the boundaries of the optimization domains $\mathscr{J}_{\pm}$, implying that $j^{\pm}_{p}=0$. Notice that even when $j_{c}=0$, the minima $j^{\pm}_{p}$ can still reside in the interior of $\mathscr{J}_{\pm}(0)$, in which case it follows immediately that $I^{(\pm)}_{c}(j_{c})$ are differentiable by essentially reiterating the argument for the $j_c \neq 0$ case. Otherwise, when the minimum of $I_{c,p}^{(\pm)}(0, j_p)$  is located precisely on the boundary $\partial \mathscr{J}_{\pm}$ of the minimization domains $\mathscr{J}_{\pm}$  (i.e. when $j^{\pm}_{p} = 0$), functions $j^{\pm}_{p}(j_{c})$ cease to be
differentiable at the origin, $j_{c}=0$, where they develop a (convex) corner.

Inserting it into Eqs.~\eqref{eqn:convex_optimizations}, we conclude that the branches $I^{\pm}_{c}(j_{c})$ are non-differentiable, also possessing a convex corner at $j_{c}=0$.

Having worked out the formal structure of $I^{\pm}_{c}(j_{c})$, we can now return to the outer optimization \eqref{eqn:Ic_global_optimization}. 
We first establish that at least one of the branches $I^{(\pm)}_{c}(j_{c})$ is not differentiable at $j_{c}=0$, which readily follows from the fact that $\lim_{j_{p}\to 0}I^{(+)}_{c|p}(0,j_{p})=\lim_{j_{p}\to 0}I^{(-)}_{c|p}(0,j_{p})$,
in turn implying that $I_{c,p}(0,j_{p})=I_{p}(j_{p})$ is continuous on its entire domain, i.e. for $j_{p} \in \mathscr{J}_{+}(0)\cup \mathscr{J}_{-}(0)=[-j^{\rm min}_{p},j^{\rm max}_{p}]$. Writing shortly $I^{\prime (\pm)}_{c,p}(0,j_{p})\equiv \partial_{j_{p}}I^{(\pm)}_{c,p}(0,j_{p})$ and taking into account that
\begin{equation}
\lim_{j_{p}\to 0^{\mp}}I^{\prime (\pm)}_{c,p}(0,j_{p})
=\lim_{j_{p}\to 0^{\mp}}I^{\prime}_{p}(j_{p}) \pm \tfrac{1}{2}\log{(1-b^{2}_{\pm})},
\end{equation}
in conjunction with differentiability of $I_{p}(j_{p})$ at the origin $j_{p}=0$, we infer the following inequality
\begin{equation}
\lim_{j_{p}\to 0^{+}}I^{\prime (-)}_{c,p}(0,j_{p})\geq \lim_{j_{p}\to 0^{-}}I^{\prime (+)}_{c,p}(0,j_{p}).
\end{equation}
Based on this, we conclude that $I_{c,p}(0,j_{p})$ is a convex function
on the entire domain of $j_{p}$, i.e. for $j_{p}\in [-j^{\rm min}_{p},j^{\rm max}_{p}]$, despite non-differentiability of the branches $I^{(\pm)}_{c}(j_{c})$ at $j_{c}=0$. Since a convex function has only one minimum, it follows that at least one of the minima of $I_{c,p}^{\pm}(0, j_p)$ occur at the boundary $j_p=0$, as illustrated in  Figure~\ref{fig:minimization}.

The outlined nested minimization problem, given by  Eq.~\eqref{eqn:convex_optimizations} and Eq.~\eqref{eqn:Ic_global_optimization}, thus gives rise to several qualitatively different types of rate functions $I_{c}(j_{c})$, There are four main cases which are exemplified in
Figure~\ref{fig:minimization_examples}.

\subsection{Moment generating function}
\label{sec:MGF_dressing}

\begin{figure}[htb]
\centering
\includegraphics[width=\columnwidth]{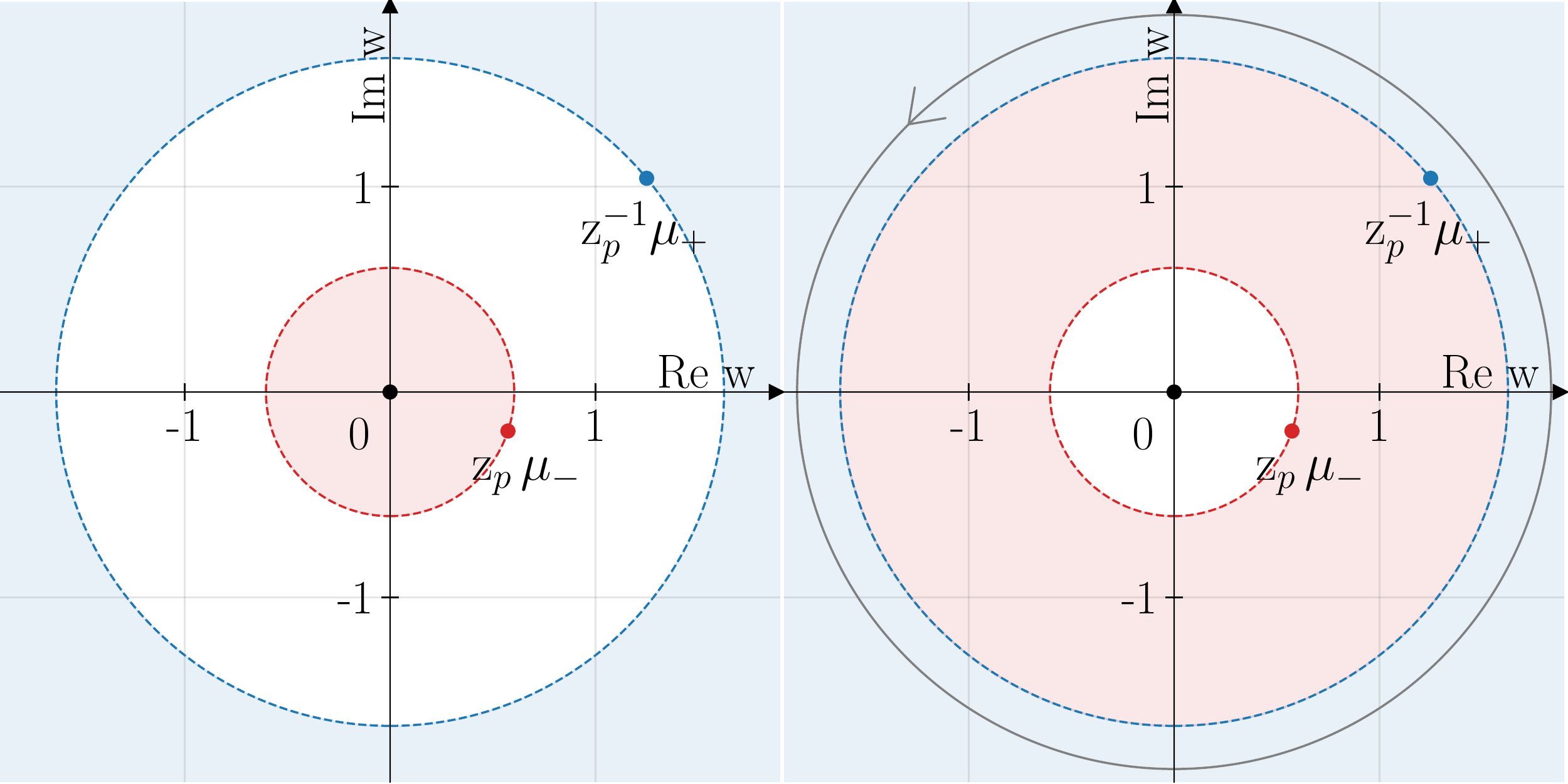}
\caption{Regions of convergence of integration kernels $K^{(-)}(\mathbf{z}|{\rm w})$ (blue) and
$K^{(+)}(\mathbf{z}|{\rm w}^{\pm1})$ (red) in complex ${\rm w}$-plane. Grey counter-clockwise circular contour represents
the integration path within the intersection of both regions of convergence.
The red and blue dots indicated the locations of poles of ${\rm w}^{-1}K^{(+)}(\mathbf{z}|{\rm w}^{-1})$ and
${\rm w}^{-1}K^{(-)}(\mathbf{z}|{\rm w})$, respectively, while ${\rm w}^{-1}K^{(0)}(\mathbf{z}|{\rm w})={\rm w}^{-1}$
is the simple pole at the origin (black dot).}
\label{fig:contour_integral}
\end{figure}

To evaluate $\mathfrak{D}_{G}$, defined in Eq.~\eqref{eqn:Dg}, the main step involves computing
the inverse bilateral Laplace transform
$\mathcal{P}({\rm J}_{p}|t)=\mathfrak{L}^{-1}[G_{p}({\rm w}^{-1}|t)]({\rm J}_{p})$.
This involves integrating along a counter-clockwise circular contour of radius $R$ within the region of convergence of
$G_p({\rm w}^{-1}|t)$,
\begin{equation}
\frac{1}{2\pi \ii}\oint_{|{\rm w}|=R} \frac{\dd {\rm w}}{\rm w} \,{\rm w}^{J_{p}}{G_{p}({\rm w}^{-1}|t)}.
\end{equation}
Using that PDF $\mathcal{P}({\rm J}_{p}|t)$ has finite support (implying that the region of convergence is $\mathbb{C}\setminus \{0,\infty\}$)
for any finite time, we can interchange the order of contour integration
and summations over ${\rm J}_{p}$ and ${\rm J}_{c}$ in $\mathfrak{D}_{G}$. Writing $\mathbf{z}\equiv (\zz_{c},\zz_{p})$,
we thus arrive at the following integral representation
\begin{equation}
G_{c,p}(\mathbf{z}|t)=\frac{1}{2\pi \ii}\oint_{|{\rm w}|=R}\frac{\dd {\rm w}}{{\rm w}}K(\mathbf{z}|{\rm w})G_{p}({\rm w}^{-1}|t),
\label{eqn:Gc_kernel}
\end{equation}
with the time-independent kernel
\begin{equation}
K(\mathbf{z}|{\rm w}) \equiv \sum_{{\rm J}_{p},{\rm J}_{c}}\zz^{{\rm J}_{c}}_{c}
\mathcal{P}_{c|p}({\rm J}_{c}|{\rm J}_{p})\zz^{{\rm J}_{p}}_{p}{\rm w}^{{\rm J}_{p}}.
\label{eqn:G_conditional}
\end{equation}
The trouble we now face with evaluating the contour integral in Eq.~\eqref{eqn:Gc_kernel} is that the kernel
$K(\mathbf{z}|{\rm w})$ has a vanishing region of convergence. This can be overcome by splitting the kernel 
into three separate pieces as follows. We introduce a pair of auxiliary projectors $\mathscr{P}_{\pm}$ such that, when
acting on a Laurent series in variable ${\rm w}$, they project out only the polynomial of (strictly) positive powers in ${\rm w}^{\pm 1}$.
Similarly, let $\mathscr{P}_{0}$ projects out only the ${\rm w}^{0}$ term, namely $K^{(0)}(\mathbf{z}|{\rm w})=1$.
In terms of the projectors we can perform the decomposition
\begin{equation}
K^{(k)}(\mathbf{z}|{\rm w}) \equiv \mathscr{P}_{k}[K(\mathbf{z}|{\rm w})],
\end{equation}
implying
\begin{equation}
G_{c,p}(\mathbf{z}|t)=\sum_{k\in\{+,0,-\}}G^{(k)}_{c,p}(\mathbf{z}|t).
\end{equation}
In particular, we have
\begin{equation}
G^{(0)}_{c,p}(\mathbf{z}|t) = \frac{1}{2\pi \ii} \oint_{|{\rm w}|=R} \frac{\dd {\rm w}}{\rm w } G_{p}({\rm w}^{-1}|t),
\label{G0_int}
\end{equation}
Crucially, positive and negative parts $K^{(\pm)}(\mathbf{z}|{\rm w})$, when viewed as functions of ${\rm w}$,
now enjoy finite regions of convergence around ${\rm w}=0$ and ${\rm w}=\infty$, respectively. To ensure that both regions of convergence have a non-vanishing overlap, we apply the variable transformation ${\rm w}\mapsto {\rm w}^{-1}$
to $K^{(+)}(\mathbf{z}|{\rm w})$ (as depicted in Fig.~\ref{fig:contour_integral}), obtaining the following representation
\begin{equation}
G^{(\pm)}_{c,p}(\mathbf{z}|t) = \oint_{|{\rm w}|=R} \!\!
\frac{\dd {\rm w}}{2\pi \ii{\rm w}}K^{(\pm)}(\mathbf{z}|{\rm w}^{\mp 1})G_{p}({\rm w}^{\pm 1}|t).
\label{eqn:Gpm_int}
\end{equation}

In terms of linear combinations ${\rm J}_{\pm}\equiv ({\rm J}_{p}\pm {\rm J}_{c})/2  \in \mathbb{Z}_{\geq 0}$, we then have
\begin{equation}
1+K^{(\pm)}(\mathbf{z}|{\rm w}^{\mp 1}) = \! \sum_{{\rm J}_{+},{\rm J}_{-}} \! \binom{{\rm J}_{p}}{{\rm J}_{+}}
\prod_{\epsilon \in \{\pm\}}\left[\frac{1\pm \epsilon \, b_{\mp}}{2\,{\rm w}}\frac{\zz^{\pm 1}_{p}}{\zz^{-\epsilon}_{c}}\right]^{{\rm J}_{\epsilon}},
\end{equation}
and by evaluating the sum arrive at a compact expression
\begin{equation}
K^{(\pm)}(\mathbf{z}|{\rm w}^{\mp 1}) = \frac{1}{1-\zz^{\pm 1}_{p}\mu_{\mp}(\zz_{c})/{\rm w}} - 1,
\end{equation}
with
\begin{equation}
|\zz^{\pm 1}_{p}\mu_{\mp}(\zz_{c})| < |{\rm w}|,
\end{equation}
where $\mu_{\pm}(\zz_{c})=\tfrac{1}{2}(\zz_{c} + \zz^{-1}_{c})\mp b_{\pm}\tfrac{1}{2}(\zz_{c} - \zz^{-1}_{c})$.
Notice that functions ${\rm w}^{-1}K^{(\pm)}(\mathbf{z}|{\rm w}^{\pm 1})$ in the integrand in Eq.~\eqref{eqn:Gpm_int}
involve two simple poles located at ${\rm w}=0$ and ${\rm w}=\zz_p^{\pm 1}\mu_{\mp}({\rm z}_c)$.
Using further that $G_{p}({\rm w}^{-1})$ is a Laurent series, the contour integrals can be evaluated simply by collecting the residues.
Specifically, a circular integration contour within the common region of convergence,
\begin{equation}
|{\rm w}| = R > \max\{|\zz_{p}\mu_{-}(\zz_{c})|, |\zz^{-1}_{p}\mu_{+}(\zz_{c})| \},
\end{equation}
encircles the three poles at ${\rm w}= 0$ and at ${\rm w} = \zz_p^{\pm 1}\mu_{\mp}({\rm z}_c)$, see Fig.~\ref{fig:contour_integral}.
Computing the integrals \eqref{eqn:Gpm_int} therefore amounts to simply replacing the positive ($+$) and negative ($-$)
powers of $\zz^{\pm n}_{p}$ with $z^{\pm n}_{p}[\mu_\mp(\zz)]^{n}$. As the outcome,
applying the dressing operator $\mathfrak{D}_{G}$ is equivalent to the replacement rule \eqref{eqn:replacement_rule}.

\section{The Lee--Yang theory}
\label{sec:LeeYang}

We now exhibit the dynamical critical phenomena in the framework of the Lee--Yang theory of phase transitions. To perform this analysis, we first express the MGF in terms of the exponential counting field $\zz \equiv e^{\lambda}$,
yielding a Laurent series
\begin{equation}
G({\rm z}|t) = \sum_{n=-t}^{t}g_{n}(t){\rm z}^{n}
= \mathscr{Z}_{0}(t)\,{\rm z}^{-t}\sum_{j=1}^{2t}\big({\rm z}-{\rm z}_{j}(t)\big).
\label{eqn:GLY}
\end{equation}
The MGF $G(z|t)$ involves $2t$ dynamical (i.e. time-dependent) zeros $\{\zz_{j}(t)\}$ called Lee--Yang zeros. Due to reality of ${\rm J}(t)$, we have that every every zero $\zz_{j}$ is paired with its complex-conjugate counterpart $\ol{\zz}_{j}$.
In equilibrium, the detailed balance condition forces the zeros to combine into pairs $\zz_{j}$ and $1/\zz_{j}$, implying a reflection symmetry $G(\zz|t)=G(\zz^{-1}|t)$.
To facilitate the asymptotic analysis of $F(\zz|t)=t^{-1}\log G(\zz|t)$, it proves useful to separate the contributions of
zeros, denoted hereafter by $F_{\rm LY}(\zz|t)$, from the `regular potential' $V(\zz|t)=-\log{(\zz)}+t^{-1}\log \mathscr{Z}_{0}(t)$
attributed to the non-polynomial prefactor in Eq.~\eqref{eqn:GLY}, i.e. perform the splitting $F(\zz|t)=V(\zz|t)+F_{\rm LY}(\zz|t)$.
The Lee--Yang zeros enter into the `singular part'
\begin{equation}
F_{\rm LY}(\zz|t)=\sum_{j=1}^{2t}\log(\zz-\zz_{j}(t)).
\end{equation}
in the form of logarithmic singularities.

The total number of zeros grows extensively with time $t$.
We are interested in asymptotic growth of $G(\zz|t)$ at late times. In this limit,
one commonly finds that the full fraction of zeros gradually approach each other before eventually forming condensates. One can expect (neglecting the possibility of accumulation points) the zeros to distribute along certain 
one-dimensional segments in the complex $\zz$-plane, or to (densely) concentrate inside certain two-dimensional domains. In the hardcore automaton, the latter scenario can be ruled out based on the asymptotic form of $G(z|t)$ given by Eq.~\eqref{eqn:G_branches}.
Notice that in the limit of large time any finite number (i.e.  of vanishing fraction, subextensive in $t$) of zeros may be harmlessly disregarded.

\subsection{Contour representation}

We now return to our main example -- the classical hardcore automaton. The task at hand is to explicitly determine the Lee--Yang contours. In turns out that the full fraction of zeros condense along a single contour or a union of contours which we denote hereafter by $\mathcal{C}_{\zz}$. 

Introducing the (line) density $\rho(\zz)$ of Lee--Yang zeros,
\begin{equation}
\rho(\zz) = \lim_{t\to \infty}\frac{1}{t}\sum_{j=1}^{2t}\delta(\zz-\zz_{j}(t)),
\label{eqn:line_density}
\end{equation}
distributed along a contour $\mathcal{C}_{\zz}$ (with normalization $\int_{\mathcal{C}_{\zz}} \dd \zz \rho(\zz) = 2$), the limiting SCGF $F_{\rm LY}(\zz)=\lim_{t \to \infty}F_{\rm LY}(\zz|t)$ can be cast as a \emph{contour} integral
\begin{equation}
F_{\rm LY}(\zz) = \int_{\mathcal{C}_{\zz}}\dd {\rm w}\,\mathscr{G}(\zz,{\rm w})\rho_{\zz}({\rm w}),
\end{equation}
with respect to $\rho({\rm w})$, where $\mathscr{G}(\zz,{\rm w})\equiv \log(\zz-{\rm w})$ is the Green's function. In terms of the arc-length parameter $s$,
the total number of zeros in an infinitesimal line element $\dd s$ at position $s$
on the contour $\mathcal{C}_{\zz}$ is $t\,\rho_{\zz}(s)\dd s \sim \mathcal{O}(t)$.

Exploiting the formal analogy to the problem of 2D electrostatics,
we further split $F_{\rm LY}(\zz)$ into two real-valued potentials, $F_{\rm LY}(\zz)=\Phi(\zz)+\ii \Psi(\zz)$.
Interpreting the density $\rho_{\zz}(\zz)$ as a source, the real component $\Phi(\zz)$ fulfils the Poisson equation $\tfrac{1}{2\pi}\nabla \Phi(\zz)=\rho_{\zz}(\zz)$. Potential $\Phi(\zz)$ is \emph{continuous} across $\mathcal{C}_{\zz}$. In contrast, $\Psi(\zz)$ experiences a jump,
\begin{equation}
\delta \Psi(\zz(s)) \equiv \Psi\big(\zz(s)+\ii 0\big)-\Psi\big(\zz(s)-\ii 0\big),
\end{equation}
proportional to the local density of zeros,
\begin{equation}
\rho_{\zz}(\zz(s)) = \frac{1}{2\pi}\frac{\dd}{\dd s}\delta\psi(\zz(s)), \quad \zz(s) \in \mathcal{C}_{\zz}.
\end{equation}
By convention, infinitesimal shifts $\pm \ii 0$ in the definition of $\delta \Psi$ pertain to values below (i.e. to the left) and above (to the right) of $\mathcal{C}_{\zz}$ upon tracing the contour in the positive direction of increasing $s$.

\subsection{Phase boundaries and the Stokes phenomenon}
\label{sec:Stokes}

The function $G(\zz|t)$, now understood as a complex function of complex counting field $\zz$, can exhibit different asymptotic behavior depending on the state parameters. As briefly discussed in Sec. \ref{sec:setting} and further expanded upon in Sec. \ref{sec:hardcore}, this can be traced back to coexistence of multiple competing extremal points (maxima) in the bulk of the integration domain $\mathscr{D}_{\square}$
of $G(\zz|t)$, attributed to individual (meta)stable branches $F_{k}(\zz)$.
To determine their `strength', one has to compare the respective `Morse potentials'
\begin{equation}
\phi_{k}(\zz) = {\rm Re}\, F_{k}(\zz),
\label{eqn:Morse}
\end{equation}
obtained by analytic continuation of $F_{k}(\zz \in \mathbb{R})$ to the complex $\zz$-plane. Individual maxima of $G(\zz|t)$ undergo a non-trivial motion upon varying the counting variable $\zz\in \mathbb{C}$,
and one can anticipate different competing maxima to dominate in different regions in the $\zz$-plane. Such behavior is generally known as the \emph{Stokes phenomenon}. Alternatively, for certain values of $\zz$ the dominant contribution to MGF $G(\zz|t)$ might not be due to localization around a bulk maximum but rather around a maximum at the boundary $\partial\mathscr{D}_{\square}$.

The exchange of dominance associated with dynamical phases $F_{k}(\zz)$ and $F_{\ell}(\zz)$ occurs along certain one-dimensional contours,
denoted by $\mathcal{C}^{(\zz)}_{k,\ell}$, separating apart their respective regions of dominance $\mathscr{R}_{k}$ and $\mathscr{R}_{\ell}$. To determine $\mathcal{C}^{(\zz)}_{k,\ell}$, we first identify the `extended anti-Stokes lines'
\begin{equation}
\mathcal{A}^{(\zz)}_{k,\ell} = \{\zz\in \mathbb{C}|\,\phi_{k}(\zz)-\phi_{\ell}(\zz)=0\}.
\label{eqn:antiStokes}
\end{equation}
Here `extended' signifies that $\mathcal{A}^{(\zz)}_{k,\ell}$ are not necessary the true anti-Stokes lines since we do not additionally account for the selection criteria, namely whether the compared $k$th and $\ell$th phases even coexist.
The latter can be inferred by analytic continuation of the selection rules, see Eq.~\eqref{eqn:selection_rules}. The Lee--Yang contour $\mathcal{C}$ is then given by a union of anti-Stokes lines $\mathcal{C}^{(\zz)}_{k,\ell}$,
\begin{equation}
\mathcal{C}_{\zz}=\bigcup_{k,\ell\in \{+,0,-\}}\mathcal{C}^{(\zz)}_{k,\ell}.
\end{equation} 
We reemphasize that $\mathcal{C}^{(\zz)}_{k,\ell}$ are only a part of the extended anti-Stokes lines $\mathcal{A}^{(\zz)}_{k,\ell}$, whereas the remaining `ghost' parts of $\mathcal{A}^{(\zz)}_{k,\ell}$ (depicted in figures by dashed lines, as e.g. shown in Figure \ref{fig:LY_mixed}) are of no physical relevance. We nonetheless choose to display them for convenience, given that full $\mathcal{A}^{(\zz)}_{k,\ell}$ in fact represent algebraic curves of finite genus.

\subsection{Equilibrium and exceptional triple point}
\label{sec:triple}

It turns out that several salient features of competing branches $F_{k}(z)$ neatly manifest themselves already in equilibrium states. However, equilibrium states are, as we corroborate in this section, rather singular. Here we perform the Lee--Yang analysis by putting $b_{\pm}\to b$ and $\rho_{\pm}\to \rho$, and writing shortly
$\chi \equiv \chi_{p}(\rho)=\rho(1-\rho)$.

To facilitate the computations, we find it most convenient to perform a change of variables $\zz \mapsto \upsilon(\zz)$
via a fractional linear transformation,
\begin{equation}
\upsilon(\zz) = \frac{\zz-1}{\zz+1}, 
\end{equation}
bijectively transforming $\zz$-plane to $\upsilon$-plane in such a way that the real $\zz$-axis gets mapped to the open interval $(-1,1)$
while the origin $\zz=0$ stays put, that is $\upsilon(0)=0$.
The preimages $\zz=\zz(\upsilon)$ of the edge points $\upsilon = \pm 1$ are $\zz(-1)=0$ and $\zz(1)=\infty$ (corresponding to $\lambda \to \pm \infty$). 
By the detailed balance condition, the set of zeros is invariant under reflection $\upsilon \mapsto -\upsilon$.

In the complex $\upsilon$-plane, the Lee--Yang contour by $\mathcal{C}_{\upsilon}$ is a union of two closed contours
$\mathcal{C}^{(\upsilon)}_{\pm,0}$, that is $\mathcal{C}_{\upsilon}=\mathcal{C}^{(\upsilon)}_{-,0}\cup \mathcal{C}^{(\upsilon)}_{+,0}$.
It is sufficient to focus on the single physical contour, say
$\mathcal{C}^{(\upsilon)}_{-,0}$ in the right-half of the $\upsilon$-plane
with ${\rm Re}(\upsilon)>0$. The other half of $\mathcal{C}_{\upsilon}$ is then simply obtained
by reflecting $\mathcal{C}^{(\upsilon)}_{-,0}$ across the imaginary axis, yielding $\mathcal{C}^{(\upsilon)}_{+,0}$.
We recall that in equilibrium the constant branch $F_{0}$ can never appear for $\upsilon \in \wp$.
In fact, the anti-Stokes contour $\mathcal{C}^{(\upsilon)}_{+,-}$ is trivial.

\medskip

\paragraph*{Finite bias.}

\begin{figure}[htb]
\includegraphics[width=\columnwidth]{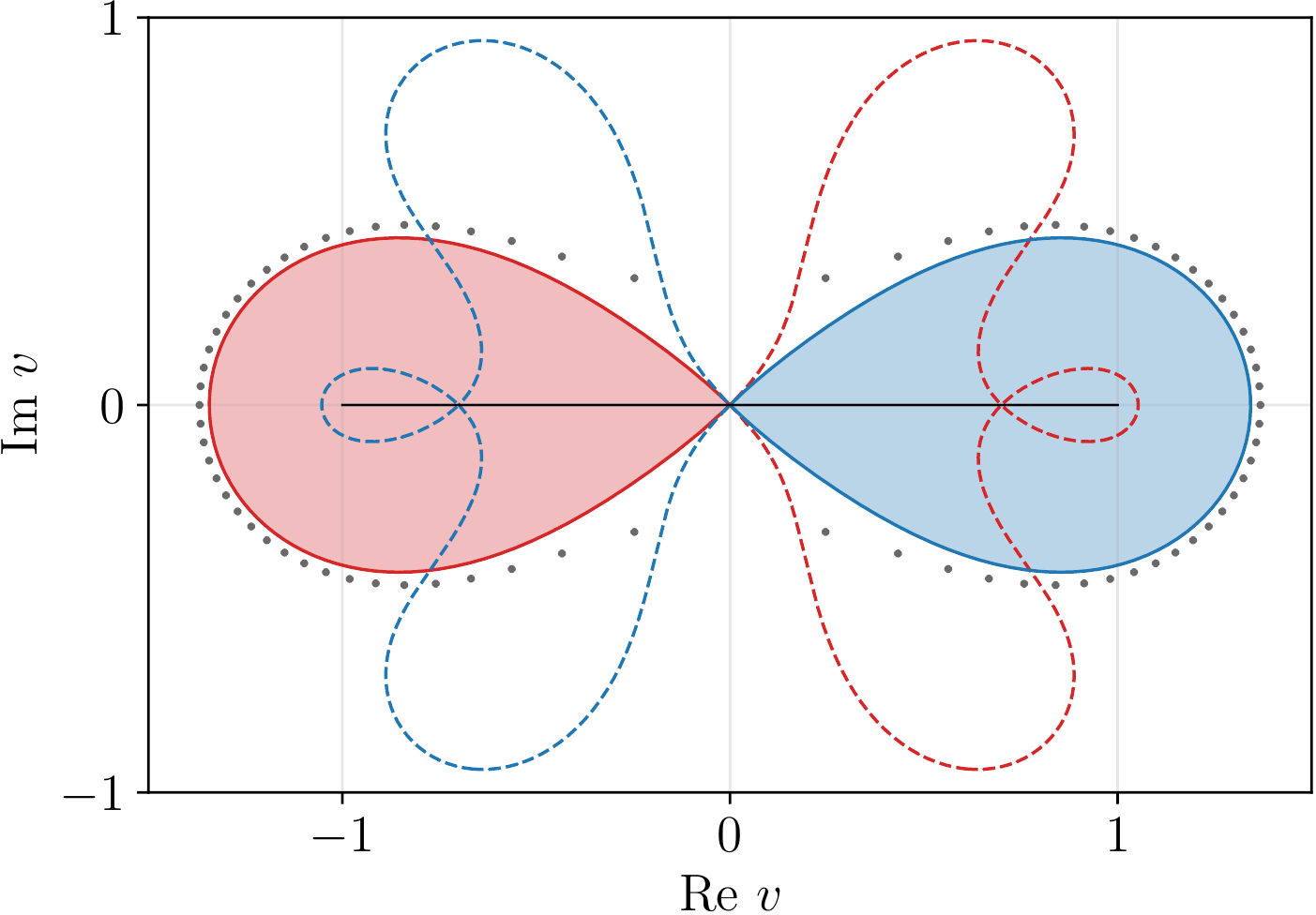}
\caption{Complex phase diagram in $\upsilon$-plane (shown for equilibrium at finite bias, with parameters $\rho_{\pm}=0.5$ and $b_{\pm}=0.7$), featuring
two bulk degenerate dynamical phases, $F_{+}(\upsilon)$ (red) and $F_{-}(\upsilon)$ (blue), and constant phase $F_{0}$ (white). 
The respective phase boundaries $\mathcal{C}^{(\upsilon)[b]}_{\pm,0}$ are shown by solid curves, while the ghost part of the extended anti-Strokes lines are marked by dashed curves. Gray dots show the Lee--Yang zeros for $t=34$.}
\label{fig:LY_equilibrium_bias}
\end{figure}

We initially consider a general equilibrium case with a non-vanishing bias in the range $0<b<1$.
The two bulk branches $F_{\pm}(\upsilon)=\log f_{\pm}(\upsilon)$ read explicitly
\begin{equation}
f_{\pm}(\upsilon) = 1 + \frac{4\chi \upsilon^{2}(b \mp \upsilon)^{2}}{(1-\upsilon^{2})(1 \mp 2b\upsilon+\upsilon^{2})}.
\end{equation}
The corresponding extended anti-Stokes lines $\mathcal{A}^{(\upsilon)[b]}_{\pm,0}$
are determined by the algebraic equations
\begin{equation}
|f_{\pm}(\upsilon)|=1.
\end{equation}
Writing $\upsilon \equiv \upsilon_{1}+\ii \upsilon_{2}$, the above condition specifies an
algebraic plane curve $P^{[b]}(\upsilon_{1},\upsilon_{2})=0$ of degree ${\deg}P^{[b]}=8$, with two real double (nodal) points
located on the real axis at $\upsilon=0$ and $\upsilon = -b$.
The latter however lie on the ghost part the Lee--Yang contour, see Figure \ref{fig:LY_equilibrium_bias}.
The condensate of zeros is entirely supported on two mirror-symmetric
contours $\mathcal{C}^{[b]}_{\upsilon,\pm}$ with $\upsilon \gtrless 0$,
representing closed contours in the shape of a droplet that emanate out from the double point at the origin at an angle $\pi/4$.

Having $F_{k}(\upsilon)$ available in a closed analytic form, one should in principle be able to find the expressions for density of zeros $\rho^{[b]}_{\upsilon}(s)$. We have not been able to derive a closed-form expression for the line density,
mainly because we lack a useful local parametrization of the physical contours $\mathcal{C}^{[b]}_{\upsilon,\pm}$. Nonetheless, the density near the critical point origin at $\upsilon_{0}=0$ can be readily obtained by employing a local series expansion $\upsilon_{2}=\upsilon_{1}+[2(b^{2}-1)/b]\upsilon^{2}_{1}+\mathcal{O}(\upsilon_1^3)$, allowing to subsequently compute $\nabla \psi\cdot \hat{t}|_{\upsilon=\upsilon_{1}}$ (with unit tangent $\hat{t}=(1,1)/\sqrt{2}$ at $\upsilon_{0}=0$) to the leading order
in the arc-length parameter $s$, yielding
\begin{equation}
\rho^{[b]}_{\upsilon}(s)=\frac{4\chi(\rho) b^{2}}{\pi}s+\mathcal{O}(s^2).
\end{equation}
We have thereby established \emph{linear} dependence near the critical point $\upsilon_{0}$ at small $s$, namely $\rho^{[b]}_{\upsilon}(s) \simeq s$, and the impact angle of $\pi/4$, indicating that the phase transition across the
anti-Stokes lines $\mathcal{A}^{(\upsilon)[b]}_{\pm,0}$ (and hence the physical phase boundary $\mathcal{C}_{\upsilon}$) is of the \emph{second order}.

\medskip

\begin{figure}[htb]
\includegraphics[width=\columnwidth]{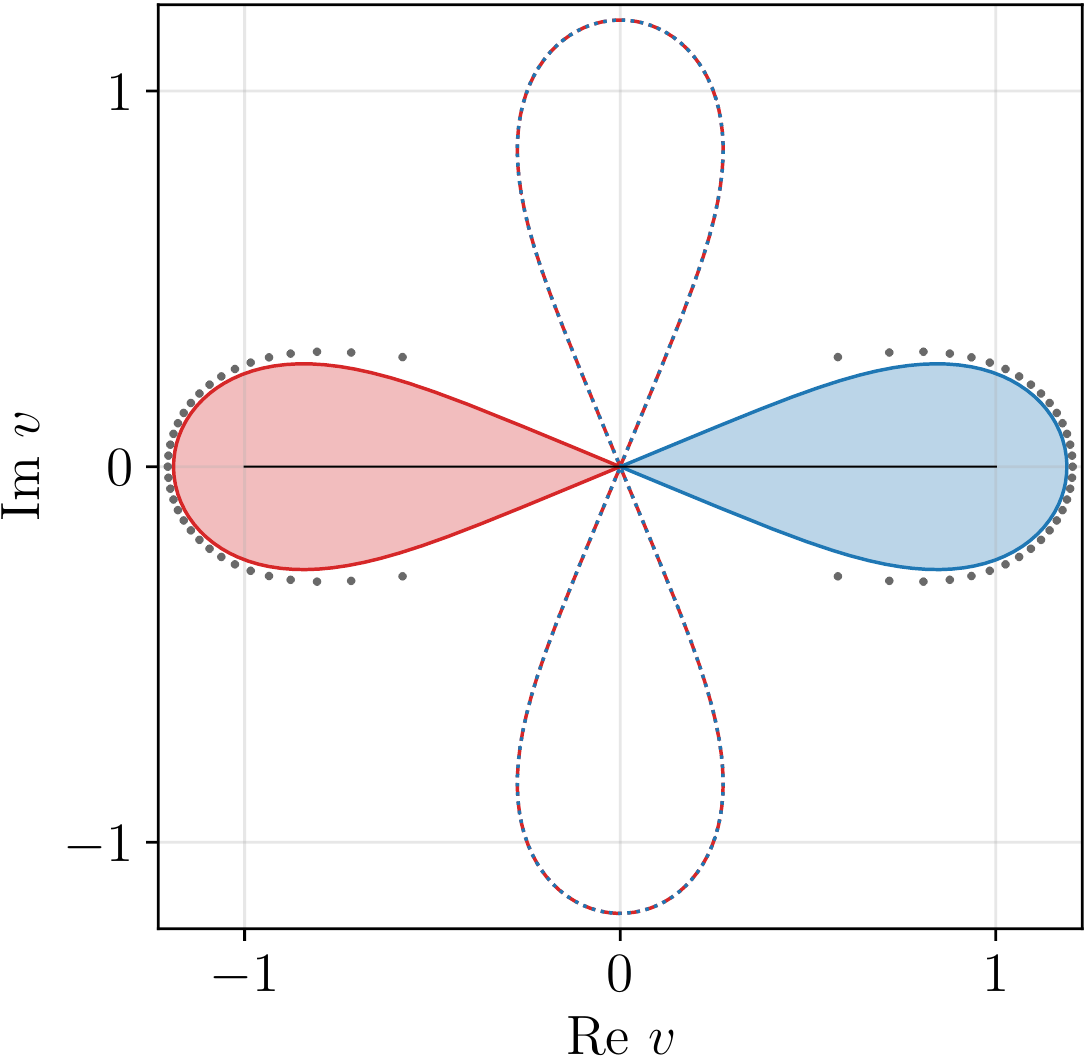}
\caption{Complex phase diagram in $\upsilon$-plane (shown for equilibrium with particle density $\rho=0.5$ an no bias, $b=0$), featuring two bulk degenerate dynamical phases, $F_{+}(\upsilon)$ (red) and $F_{-}(\upsilon)$ (blue), and constant phase $F_{0}$ (white). The respective phase boundaries $\mathcal{C}^{(\upsilon)[b]}_{\pm,0}$ are shown by solid curve. The ghost parts of the extended anti-Strokes lines, marked by dashed curves, overlap with each other. Gray dots show the Lee--Yang zeros for $t=34$.}
\label{fig:LY_propeller}
\end{figure}

\paragraph*{Zero bias.}

Upon taking the limit of zero bias, $b\to 0$, the MGF $G(\upsilon|t)$ acquires (next to retaining mirror symmetry $\upsilon \to -\upsilon$) an extra symmetry under charge conjugation $b\to -b$. As we now explain, this additional symmetry has important consequences. Firstly, the two phases $F_{\pm}(\upsilon)$ now become exactly degenerate, implying that at late times we have the following asymptotic form
\begin{equation}
G(\upsilon|t)\asymp e^{t\,F_{+}(\upsilon)}+e^{t\,F_{-}(\upsilon)}=2[f_{\pm}(\upsilon)]^{t},
\end{equation}
with
\begin{equation}
f(\upsilon) = \frac{(1-4\chi)\upsilon^{4}-1}{\upsilon^{4}-1}.
\end{equation}
Note also that the ghost parts of the anti-Stokes lines $\mathcal{A}^{(\upsilon)[b]}_{\pm,0}$ glue together into a single contour,
(as depicted in Figure \ref{fig:LY_propeller}) while the nodal points on the ghost contours approach the origin. Exactly for $b=0$, we end up with a curve in the shape of a four-blade `boat propeller', see Figure \ref{fig:LY_propeller}.
This symmetry reflects the fact that $G(\upsilon|t)$ is a polynomial of $\upsilon^{2}$ of degree $t$, allowing for a reduction by performing another variable substitution $w=\upsilon^{2}$.
This means that the algebraic curves $\mathcal{A}^{[b]}_{\pm,0}$ in the $\upsilon$-plane turn into quartic curves in the $w$-plane.
By accordingly parametrizing $w=w_{1}+\ii w_{2}$, the anti-Stokes line (once again determined by condition
$|f^{[0]}_{\pm}|=1$ with $f^{[0]}_{\pm}(w)=1+4\chi w^{2}/(1-w^{2})$ is equivalent to the algebraic curve
\begin{equation}
P^{[0]}(w_{1},w_{2}) \equiv (1-2\chi)(w^{2}_{2}+w^{2}_{1})^{2}+w^{2}_{2}-w^{2}_{1}=0.
\end{equation}
In practice, it proves convenient to use polar coordinates, namely writing
$w={\rm r} e^{\ii \vartheta}$, yielding the following compact parametrization of $P^{[0]}(w)=0$,
\begin{equation}
{\rm r}^{2}(\vartheta) = \frac{\cos{(2\vartheta)}}{1-2\chi}.
\end{equation} 
The reader can now recognize the famous lemniscate of Bernoulli -- a rational quartic algebraic curve of genus zero. It is important to stress that the physical Lee--Yang contour $\mathcal{C}^{[0]}_{w}$ is not the whole lemniscate,
but only the part lying in the right half-plane ${\rm Re}(w)\geq 0$, i.e. along the interval of angles $\vartheta \in [-\pi/4,\pi/4]$. Moreover, contour $\mathcal{C}^{[0]}_{w}$ encloses a compact region in the complex $w$-plane attributed to the region of dominance of (analytically continued) degenerate
bulk phases $F_{\pm}(w)$. As before, the exterior of $\mathcal{C}^{[0]}_{w}$ is the region of dominance of the flat branch.

This time, in the absence of bias, the line density $\rho_{w}(s)$, parametrized in terms of arc-length $s$, can actually be computed explicitly.
In the interior of $\mathcal{C}^{[0]}$, the $\psi$-potential
$\psi({\rm r},\vartheta)\equiv {\rm arctan}(\Upsilon({\rm r},\vartheta))$ can be parametrized explicitly
\begin{equation}
\Upsilon({\rm r},\vartheta) = \frac{4\chi {\rm r}^{2}\cos{(4\vartheta)}}{{\rm r}^{4}(1-4\chi)-2{\rm r}^{2}(1-2\chi)\cos{(2\vartheta)}+1},
\end{equation}
From the jump discontinuity at the phase boundary $\mathcal{A}^{[0]}_{w}$ we deduce the line density as a function
of polar angle $\vartheta$,
\begin{equation}
\rho_{w}(\vartheta) = \frac{1}{2\pi}\frac{16\chi(1-\chi)}{1-4\chi+8\chi^{2}+(4\chi-1)\cos{(4\vartheta)}},
\end{equation}
which now (by virtue of symmetry reduction) is normalized as $\int^{\pi/4}_{-\pi/4}\dd \vartheta \rho_{w}(\vartheta)=1$. In the arc-length parametrization, $\rho_{w}(s) = (\dd s/\dd \vartheta)^{-1}\rho_{w}(\vartheta)$, the line density takes the form
\begin{equation}
\rho_{w}(s) = \frac{4\sqrt{2}\chi}{\pi}\sqrt{\frac{\tfrac{\pi}{4}-\theta(s)}{1-2\chi}},
\end{equation}
where {
$\vartheta(s) = -{\rm am}\big((1-2\chi)^{-1/2}\,s-{\rm F}(\tfrac{\pi}{4},2),2\big)$ has been obtained by inverting $s(\vartheta)=(1-2\chi)^{-1/2}({\rm F}(\tfrac{\pi}{2},2)-{\rm F}(\vartheta,2))$
(with ${\rm F}(\vartheta,k)$ denoting the elliptic integral of the first kind
and ${\rm am}(u,k)$ the corresponding inverse function -- the Jacobi amplitude).}
By finally expanding $\rho_{w}(s)$ in a Taylor series, we once more find the \emph{linear} density of zeros near the origin,
\begin{equation}
\rho_{w}(s)=\frac{4\chi}{\pi}s+\mathcal{O}(s^{5}).
\end{equation}

\begin{figure}[htb]
\includegraphics[width=\columnwidth]{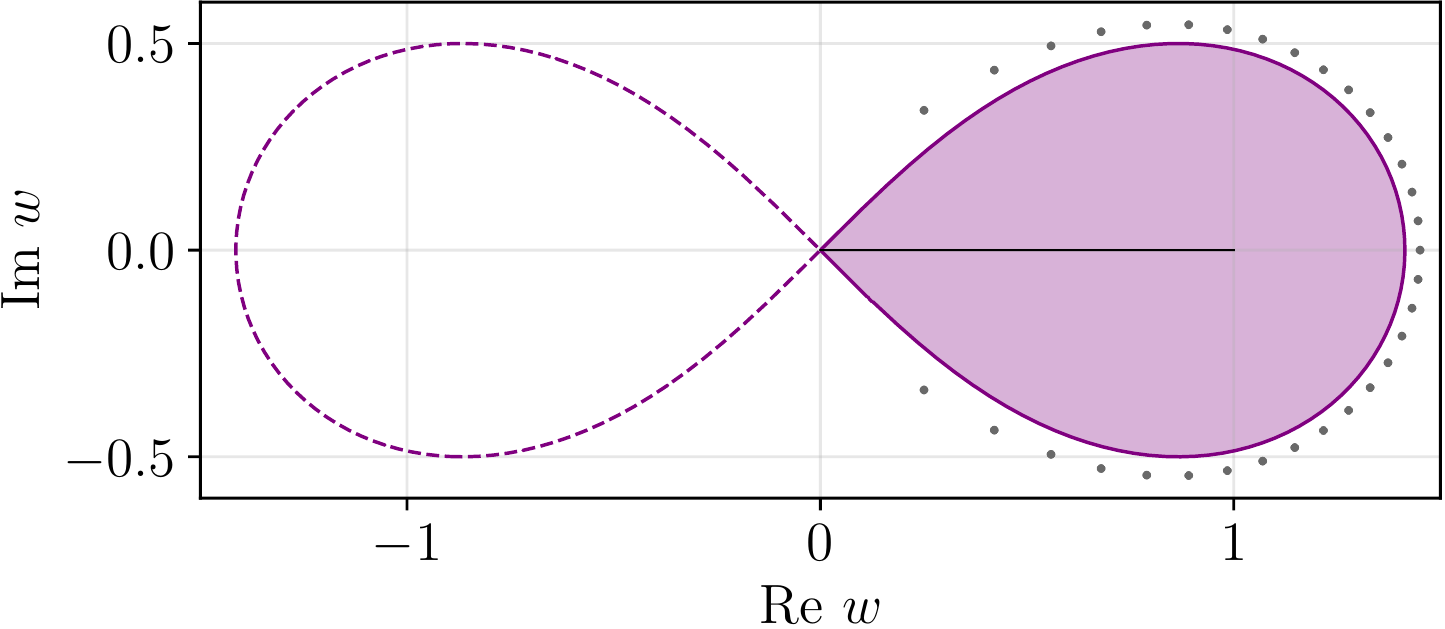}
\caption{Reduced complex phase diagram in $w$-plane (shown for equilibrium at zero bias with density $\rho=1/2$) with a single merged dynamical phase (purple) invariant under charge conjugation, with the background constant phase $F_{0}$ (white).  The extended anti-Stokes line is the lemniscate of Bernoulli, representing a quartic curve of genus zero.  The physical phase boundary, shown by solid curve, is the right half. gray dots show the Lee--Yang zeros for $t=34$.}
\label{fig:lemniscate}
\end{figure}

The main upshot of the above analysis is that \emph{in equilibrium}, dynamical phase transitions across the phase boundaries, $\mathcal{C}^{[b]}_{\upsilon,\pm}$ and $\mathcal{C}^{[b]}_{\upsilon,\pm}$, are always of the \emph{second} order, irrespective of bias. According to Ehrenfest classification, a transition is characterized as order $\mathfrak{n}\geq 1$ if the density $\rho(s)$ (in $\upsilon$-plane or, equivalently, $\zz$-plane) diminishes as
\begin{equation}
\rho(s) \sim s^{\mathfrak{n}-1},
\end{equation}
in the proximity of a critical point. A first-order transition is associated with non-zero constant density across the critical point $\upsilon_{\rm c}$ at $s=0$, while a second-order transition is characterized by a linearly decreasing
density as small $s$. By expanding the $F(\upsilon)$ phase around {$\upsilon_{\rm c}$} on both sides of the contour $\mathcal{C}^{[b]}_{\upsilon,\pm}$, namely
\begin{equation}
F_{k}(\upsilon) = \log f_{k}(\upsilon) = \sum_{n=0}^{\infty}\frac{f^{(k)}_{n}}{n!}(\upsilon-\upsilon_{\rm c}),
\end{equation}
the density of zeros $\rho(s)$ can be (to the leading order in $s$) read off from the jump discontinuity
\begin{equation}
\delta F_{k,\ell}(\upsilon) = \sum_{n\geq 0}\frac{\delta f^{(k,\ell)}_{n}}{n!}(\upsilon-\upsilon_{\rm c}).
\end{equation}
Using that $f^{(0)}_{2}=0$ and $\delta f^{(\pm,0)}_{2}=f^{(\pm)}_{2}-f^{(0)}_{2}$, we correctly retrieve
\begin{equation}
\rho(s) = \frac{\delta f^{(\pm,0)}_{2}}{\pi} s + \mathcal{O}(s^{2}).	
\end{equation}

Employing polar coordinates, $\upsilon(r,\varphi) = r\,e^{\ii \varphi}$, the \emph{impact angle}
can be determined by tracing the contours,
\begin{equation}
\varphi_{\rm imp} = \lim_{r\to 0}\varphi(r),\qquad \upsilon \in \mathcal{C}^{[b]}_{\pm,0},
\label{eqn:impact_angle}
\end{equation}
By requiring the $\phi(\upsilon)$ is continuous across the phase boundaries,
the ensuing condition $\sum_{n=\mathfrak{n}}^{\infty}r^{n}\cos{(n\varphi)}=0$ reduces, upon taking the limit $r\to 0$, to {$\cos{(\mathfrak{n}\varphi(r))}=0$},
yielding a finite set of $\mathfrak{n}$ admissible impact angles of the
form $\varphi^{(k)}_{\rm imp}=[(2k+1)/\mathfrak{n}]\tfrac{\pi}{2}$ ($k\in \mathbb{N}$).
For $\mathfrak{n}=2$ specifically, the allowed angles are $\varphi_{\rm imp} = \pm \pi/4$.

\medskip

\paragraph*{Exceptional triple point.}
Let us take a closer look at the emergent dynamical critical point $\upsilon_{t}$ located at the origin of the complex $\upsilon$-plane and examine its formal properties. This critical point is distinguished by the fact that is lies on two distinct Lee--Yang contours. We note that in equilibrium at finite density and any bias, the SCGF $F(\upsilon)$ in the physical interval $\upsilon \in \wp$ features a discontinuity at $\upsilon_{t}=0$.

For finite bias, $b>0$, SCGF $F(\upsilon)$ has a discontinuous \emph{third} derivative at $\upsilon_{\rm t}$. In this view, $\upsilon_{\rm t}$ is criticality of \emph{third order}, as already advocated in our previous work \cite{Krajnik22}. There is no controversy in this statement provided we stick with the conventional definition of a $\mathfrak{n}$-th order phase transition based on a discontinuity in the $\mathfrak{n}$-th derivative in the (dynamical) free-energy density (with all derivative or lower order being continuous). One should nevertheless be careful with reversing the logic. Contrary to a widely held belief,
\begin{quote}
\emph{an analytic real-valued (thermodynamic or dynamical) free energy $F(\upsilon)$ does not preclude emergence of criticality nor do the Lee--Yang zeros approaching and colliding with the physical axis necessarily render the free energy singular.}
\end{quote} 
Indeed, $F(\upsilon)$ in equilibrium at zero bias provides an explicit counterexample.

The outlined Lee--Yang analysis helps resolving this subtle aspect.
Let us first recall that in equilibrium, $F(\upsilon)$ for $\upsilon \in \wp$ 
involves only two bulk branches $F_{\pm}$ that are degenerate in magnitude, while $F_{0}$ branch is entirely invisible. By contrast, any neighbourhood within a finite radius encompasses \emph{three} distinct phases, indicating
that $\upsilon_{\rm t}$ is indeed a \emph{triple} point. It is not an ordinary triple point though, but rather an \emph{exceptional} one; there are only two distinct phase boundaries (contours) emanating out of $\upsilon_{t}$.

We wish to emphasize that only upon analytically continuing $F(\upsilon)$ into complex $\upsilon$-plane we have enough information at hand for an unambiguous classification of a (dynamical) criticality. Specifically, since there are three phases involved, there are likewise three types of phase transitions: the pair of transitions
$F_{\pm}\leftrightarrow F_{0}$ across the contours $\mathcal{C}^{[b]}_{\upsilon,\pm}$
(expelling the phase $F_{0}$ away from the physical interval), closing into the critical point $\upsilon_{t}$ at the origin at an angle of $\pm \pi/4$, and the `direct transition' $F_{+}\leftrightarrow F_{-}$ between the degenerate bulk phases mediated by $\upsilon_{t}$. In the former case, we find the difference of dynamical free energies,
\begin{equation}
\delta F_{\pm,0}(\upsilon) = \sum_{n=0}^{\infty}\frac{\delta f^{(\pm,0)}_{n}}{n!}\upsilon^{n},\qquad \delta f^{(\pm,0)}_{n<2}=0,
\end{equation}
is discontinuous at the second order, whereas in the case of $\delta F_{+,-}(\upsilon)$
we instead find, for finite bias $b>0$, a third-order discontinuity.
By contrast, for $b=0$ we find $\delta F_{+,-}\equiv 0$ by virtue of the charge-conjugation symmetry.

To briefly summarize: in equilibrium states, $\upsilon_{\rm t}=0$ is (dynamical) critical triple point lying at the intersection of two
distinct (extended) anti-Stokes lines (separating $F_{\pm}$ phases encoding large deviations of the positive and negative charge transfer from the flat branch $F_{0}$) where two different (albeit degenerate) dynamical phases touch at a point.
Upon crossing from $F_{-}$ to $F_{+}$ through a singular (nodal) point one passes across two phase boundaries at the same time, see Fig.~\ref{fig:LY_propeller} and Fig.~\ref{fig:lemniscate}. We are unaware of a thermodynamic counterpart of such singular behavior.

\section{Fluctuations at equal densities}
\label{sec:equal_densities}

\paragraph*{Asymptotic moments.}
To infer the asymptotic behavior of $c_{n}(t)$ at equal density ($\rho_{-}=\rho_{+}$),
we first compute the asymptotic of moments of the MGF $G(\lambda|t)$,
\begin{equation}
G^{(n)}\equiv \frac{\dd^{n}G(\lambda|t)}{\dd \lambda^{n}}\Big|_{\lambda=0},
\end{equation}
with the double-sum representation
\begin{equation}
G^{(n)}(t) = \rho^{2t}\sum_{l=0}^{t}\sum_{r=0}^{t}\binom{t}{l}\binom{t}{r}\nu^{l+r}\gamma b_{\gamma}\,d^{|l-r|}_{n}(b),
\label{eqn:G_moments}
\end{equation}
where $\gamma = {\rm sgn}(l-r)$ and $d^{k}_{n}(b)=[\dd^{n}\mu^{k}_{b}(\lambda)/\dd \lambda^{n}]_{\lambda=0}$ with 
\begin{equation}
\mu_{b}(\lambda)\equiv \cosh{(\lambda)}-b\sinh{(\lambda)}.
\end{equation}
We separate the computation into several steps. To begin with, we define a sequence of auxiliary double sums
\begin{equation}
{\rm S}_{n} = \rho^{2t}\sum_{l=0}^{t}\sum^{t}_{r=0}\binom{t}{l}\binom{t}{r} \nu^{l+r}|l-r|^{n}.
\label{eqn:S_functions}
\end{equation}
Rescaling the coordinates as $l/t \to \xx_{1}$, $r/t \to \xx_{2}$ and approximating binomials using the Stirling's approximation, the double
sum can be cast as an integral over the square domain $\mathscr{D}_{\square}$
\begin{equation}
{\rm S}_{n} \asymp \frac{t^{n+1}}{2\pi}\int_{\mathscr{D}_{\square}}\dd \xx_{1}\,\dd \xx_{2}\,h_{n}(\xx_{1},\xx_{2})e^{t\,f(\xx_{1},\xx_{2})},
\end{equation} 
with $h_{n}(\xx_{1},\xx_{2}) = |\xx_{1}-\xx_{2}|^{n}/\sqrt{\xx_{1}(1-\xx_{1})\xx_{2}(1-\xx_{2})}$ and
\begin{equation}
f(\xx_{1},\xx_{2}) = \log{\left[\frac{\rho^{2}\nu^{\xx_{1}+\xx_{2}}}{\prod_{i=1}^{2}\xx_{i}^{\xx_{i}}(1-\xx_{i})^{1-\xx_{i}}}\right]}.
\end{equation}
The critical point ${\bf x}_{\star}=(\xx_{1,\star},\xx_{2,\star})$, determined from the solution to $\nabla f({\bf x}_{\star})=0$,
sits at $\xx_{1,\star}=\xx_{2,\star}=1-\rho$, while ${\rm det}H[f]({\bf x}_{\star})=[\rho(1-\rho)]^{-2}$.
Here we run in an obstacle, as functions $h_{n}(\xx_{1},\xx_{2})$ in the integrand evaluate to zero, i.e. $h_{n}({\bf x}_{\star})=0$.
This can be overcome by developing $f(\xx_{1},\xx_{2})$ as a Taylor series up to quadratic order about the critical points
\begin{equation}
f(\xx_{1},\xx_{2}) \approx \frac{[\xx_{1}-(1-\rho)]^{2}+[\xx_{2}-(1-\rho)]^{2}}{2\rho(1-\rho)},
\end{equation}
and subsequently performing integration in the rotated frame, with coordinates $\uu \equiv \xx_{1}+\xx_{2}$, $\vv \equiv \xx_{1}-\xx_{2}$, yielding
\begin{equation}
{\rm S}_{n} \asymp \frac{t^{n+1}}{\pi}\int \dd \uu \int_{\vv>0} \dd \vv \mathscr{S}(\uu,\vv),
\end{equation}
with
\begin{equation}
\mathscr{S} \equiv \frac{\vv^{n}
e^{-t[(\uu-\uu_{\star})^{2}+\vv^{2}]/4\chi_{p}}}{\sqrt{(\uu^{2}-\vv^{2})(1-\tfrac{1}{2}(\uu+\vv))(1-\tfrac{1}{2}(\uu-\vv))}}.
\end{equation}
Since both maxima lie symmetrically about $\vv=0$, it sufficient to keep only the $\vv>0$ region. Using that the expression under the
square root localizes to the value $\chi^{-1}_{p}$ and, anticipating localization, extending the range of integration, we have
\begin{equation}
{\rm S}_{n}(t) \asymp \frac{t^{n+1}}{2\pi\chi_{p}}\int^{\infty}_{-\infty}\dd \uu\,e^{-\frac{t(\uu-\uu_{\star})}{4\chi_{p}}}
\int^{\infty}_{0}\dd \vv\,\vv^{n}\,e^{-\frac{t\vv^{2}}{4\chi_{p}}}.
\end{equation}
After computing the Gaussian integrals we arrive at a compact expression
\begin{equation}
{\rm S}_{n}(t) \asymp {\rm s}_{n}(\chi_{p}\,t)^{n/2},\qquad {\rm s}_{n}\equiv \frac{2^{n-1}}{\sqrt{\pi}}\Gamma\left(\frac{n+1}{2}\right).
\end{equation}

The asymptotics of cumulants $c_{n}(t)$ can be deduced from that of the moments $G^{(n)}$.
Writing shortly $g^{(p)}(x)\equiv \dd^{p}g/\dd x^{p}$, and introducing the incomplete Bell polynomials $B_{n,m}=B_{n,m}(\{x_{j}\}_{j=1}^{n-m+1})$
with the generating series
\begin{equation}
\exp{\left[z\sum_{k=1}^{\infty}x_{k}\frac{y^{k}}{k!}\right]}
= 1 + \sum_{n=1}^{\infty}\frac{y^{n}}{n!}\sum_{m=1}^{n}z^{m}B_{n,m},
\end{equation}
the asymptotic behavior of higher moments $G^{(n)}(t)$ can be expressed
in terms of $S_{m}(t)$ with aid of the Fa\'{a} di Bruno's formula
\begin{equation}
\frac{\dd^{n}{\rm f}({\rm g}(x))}{\dd x^{n}} = \sum_{m=1}^{n}{\rm f}^{(m)}\big({\rm g}(x)\big)B_{n,m},
\label{eqn:Faa_di_Bruno}
\end{equation} 
where $B_{n,m}\equiv B_{n,m}\big(\{{\rm g}^{(p)}\}_{p=1}^{n-m+1}\big)$.
Coefficients $d^{k}_{n}$ are polynomials in variable $k$ of degree $n/2$,
\begin{equation}
d^{k}_{n}=\sum_{m=1}^{n}\frac{k}{(k-m)!}B_{n,m}(-b,1,-b,1,\ldots).
\end{equation}
Replacing monomials $k^{m}$ by ${\rm S}_{m}$ in Eq.~\eqref{eqn:G_moments} we obtain $G^{(n)}(\{S_{j}\})$.
With another application of Eq.~\eqref{eqn:Faa_di_Bruno}, we finally obtain
\begin{equation}
c_{n}(t) \asymp \sum_{m=1}^{n} (-1)^{m-1}(m-1)!\,B_{n,m}\big(\{G^{(j)}(t)\}\big).
\end{equation}

\paragraph*{Generating function for asymptotic cumulants.}

Introducing dynamically rescaled counting fields,
\begin{equation}
\omega_{\pm}(\lambda) \equiv b_{\pm}(\chi_{p}t)^{1/2}\lambda,
\end{equation}
the asymptotic moments
\begin{equation}
G^{(n)}(t) \asymp {\rm s}_{n}\big(b^{n}_{-} + (-b_{+})^{n}\big)(\chi_{p}\,t)^{n/2},
\end{equation} 
can be resumed into the following generating series
\begin{equation}
\Omega\big(\{\omega_{\pm}(\lambda)\}\big)
= \log\left[\frac{E_{1/2}(\omega_{-})+E_{1/2}(-\omega_{+})}{2}\right],
\label{eqn:generating_series_equal}
\end{equation}
where $E_{1/2}(\pm x) = e^{x^{2}}[1 \pm {\rm erf}(x)]$.
Asymptotics of the cumulants is retrieved as
\begin{equation}
	c_{n}(t)\asymp \frac{\dd^{n}}{\dd \lambda^{n}}\Omega(\{\eta_{\pm}(\lambda)\})|_{\lambda=0}.
\end{equation}

\paragraph*{Typical fluctuations.}
Setting the densities $\rho_{\pm}$ to be equal (while leaving $b_{\pm}$ arbitrary),
the PDF $\mathcal{P}^{(\rho)}_{\rm typ}(j)$ of typical charge-current fluctuations can be recovered from Eq.~\eqref{eqn:generating_series_equal}.
By dynamically rescaling the generating function $\Omega\big(\{\omega_{\pm}(\lambda)\}\big)$ with $\lambda\to t^{-1/2}\eta$, we
explicitly compute
\begin{equation}
\mathcal{G}^{(\rho)}(\eta) \equiv \lim_{t\to \infty}\left[e^{-c_{1}(t)t^{1/2}\eta}G(\lambda(\eta)|t)\right].
\end{equation}
In terms of $\omega_{\pm}=\sqrt{\chi_{p}}b_{\pm}\eta$, we find
\begin{equation}
\log \mathcal{G}^{(\rho)}(\eta) = \frac{\omega_{+}-\omega_{-}}{\sqrt{\pi}}\Omega({\omega_{\pm}}).
\end{equation}
The PDF $\mathcal{P}^{(\rho)}_{\rm typ}(j_{c})$ is given implicitly by the inverse Laplace transform
$\mathfrak{L}^{-1}[\mathcal{G}^{(\rho)}(\eta)]$ of the MGF, reading compactly
\begin{equation}
\mathcal{G}^{(\rho)}(\eta) =  e^{\pi^{-1/2}(\omega_{+}-\omega_{-})} \,\frac{1}{2}\sum_{\epsilon \in \{\pm\}}E_{1/2}(-\epsilon\,\omega_{\epsilon}).
\end{equation}
Notice that the extra multiplicative factor relative to the generating series, cf. Eq.~\eqref{eqn:generating_series_equal},
comes from the first cumulant $c_{1}(t)$ (i.e. average time-integrated current) which has been subtracted.
The results for the biased equilibrium ensembles is recovered in the limit $b_{\pm}\to b$, yielding a Gaussian MGF
$G^{(\rho)}(\eta)=\exp{[b^{2}\chi_{p}\eta^{2}]}$ and hence the PDF is a normal distribution with variance $\sigma=2\chi_p b^2$.

\bibliography{FluctuationSymmetry}

\begin{thebibliography}{114}%
\makeatletter
\providecommand \@ifxundefined [1]{%
 \@ifx{#1\undefined}
}%
\providecommand \@ifnum [1]{%
 \ifnum #1\expandafter \@firstoftwo
 \else \expandafter \@secondoftwo
 \fi
}%
\providecommand \@ifx [1]{%
 \ifx #1\expandafter \@firstoftwo
 \else \expandafter \@secondoftwo
 \fi
}%
\providecommand \natexlab [1]{#1}%
\providecommand \enquote  [1]{``#1''}%
\providecommand \bibnamefont  [1]{#1}%
\providecommand \bibfnamefont [1]{#1}%
\providecommand \citenamefont [1]{#1}%
\providecommand \href@noop [0]{\@secondoftwo}%
\providecommand \href [0]{\begingroup \@sanitize@url \@href}%
\providecommand \@href[1]{\@@startlink{#1}\@@href}%
\providecommand \@@href[1]{\endgroup#1\@@endlink}%
\providecommand \@sanitize@url [0]{\catcode `\\12\catcode `\$12\catcode
  `\&12\catcode `\#12\catcode `\^12\catcode `\_12\catcode `\%12\relax}%
\providecommand \@@startlink[1]{}%
\providecommand \@@endlink[0]{}%
\providecommand \url  [0]{\begingroup\@sanitize@url \@url }%
\providecommand \@url [1]{\endgroup\@href {#1}{\urlprefix }}%
\providecommand \urlprefix  [0]{URL }%
\providecommand \Eprint [0]{\href }%
\providecommand \doibase [0]{https://doi.org/}%
\providecommand \selectlanguage [0]{\@gobble}%
\providecommand \bibinfo  [0]{\@secondoftwo}%
\providecommand \bibfield  [0]{\@secondoftwo}%
\providecommand \translation [1]{[#1]}%
\providecommand \BibitemOpen [0]{}%
\providecommand \bibitemStop [0]{}%
\providecommand \bibitemNoStop [0]{.\EOS\space}%
\providecommand \EOS [0]{\spacefactor3000\relax}%
\providecommand \BibitemShut  [1]{\csname bibitem#1\endcsname}%
\let\auto@bib@innerbib\@empty
\bibitem [{\citenamefont {Kardar}\ \emph {et~al.}(1986)\citenamefont {Kardar},
  \citenamefont {Parisi},\ and\ \citenamefont {Zhang}}]{KPZ}%
  \BibitemOpen
  \bibfield  {author} {\bibinfo {author} {\bibfnamefont {M.}~\bibnamefont
  {Kardar}}, \bibinfo {author} {\bibfnamefont {G.}~\bibnamefont {Parisi}},\
  and\ \bibinfo {author} {\bibfnamefont {Y.-C.}\ \bibnamefont {Zhang}},\
  }\bibfield  {title} {\bibinfo {title} {{Dynamic Scaling of Growing
  Interfaces}},\ }\href {https://doi.org/10.1103/PhysRevLett.56.889} {\bibfield
   {journal} {\bibinfo  {journal} {Phys. Rev. Lett.}\ }\textbf {\bibinfo
  {volume} {56}},\ \bibinfo {pages} {889} (\bibinfo {year} {1986})}\BibitemShut
  {NoStop}%
\bibitem [{\citenamefont {Pr{\"a}hofer}\ and\ \citenamefont
  {Spohn}(2004)}]{Prahofer}%
  \BibitemOpen
  \bibfield  {author} {\bibinfo {author} {\bibfnamefont {M.}~\bibnamefont
  {Pr{\"a}hofer}}\ and\ \bibinfo {author} {\bibfnamefont {H.}~\bibnamefont
  {Spohn}},\ }\bibfield  {title} {\bibinfo {title} {Exact scaling functions for
  one-dimensional stationary kpz growth},\ }\href
  {https://link.springer.com/article/10.1023/B:JOSS.0000019810.21828.fc}
  {\bibfield  {journal} {\bibinfo  {journal} {Journal of statistical physics}\
  }\textbf {\bibinfo {volume} {115}},\ \bibinfo {pages} {255} (\bibinfo {year}
  {2004})}\BibitemShut {NoStop}%
\bibitem [{\citenamefont {Mendl}\ and\ \citenamefont
  {Spohn}(2015)}]{MendlSpohn15}%
  \BibitemOpen
  \bibfield  {author} {\bibinfo {author} {\bibfnamefont {C.~B.}\ \bibnamefont
  {Mendl}}\ and\ \bibinfo {author} {\bibfnamefont {H.}~\bibnamefont {Spohn}},\
  }\bibfield  {title} {\bibinfo {title} {Current fluctuations for anharmonic
  chains in thermal equilibrium},\ }\href@noop {} {\bibfield  {journal}
  {\bibinfo  {journal} {Journal of Statistical Mechanics: Theory and
  Experiment}\ }\textbf {\bibinfo {volume} {2015}},\ \bibinfo {pages} {P03007}
  (\bibinfo {year} {2015})}\BibitemShut {NoStop}%
\bibitem [{\citenamefont {{G. M. Sch\"{u}tz}}(2018)}]{Schutz18}%
  \BibitemOpen
  \bibfield  {author} {\bibinfo {author} {\bibnamefont {{G. M. Sch\"{u}tz}}},\
  }\bibfield  {title} {\bibinfo {title} {{On the Fibonacci Universality Classes
  in Nonlinear Fluctuating Hydrodynamics}},\ }in\ \href
  {https://doi.org/10.1007/978-3-319-99689-9_2} {\emph {\bibinfo {booktitle}
  {{Springer Proceedings in Mathematics \& Statistics}}}}\ (\bibinfo
  {publisher} {Springer International Publishing},\ \bibinfo {year} {2018})\
  pp.\ \bibinfo {pages} {149--167}\BibitemShut {NoStop}%
\bibitem [{\citenamefont {Popkov}\ \emph {et~al.}(2015)\citenamefont {Popkov},
  \citenamefont {Schadschneider}, \citenamefont {Schmidt},\ and\ \citenamefont
  {Schütz}}]{Popkov15}%
  \BibitemOpen
  \bibfield  {author} {\bibinfo {author} {\bibfnamefont {V.}~\bibnamefont
  {Popkov}}, \bibinfo {author} {\bibfnamefont {A.}~\bibnamefont
  {Schadschneider}}, \bibinfo {author} {\bibfnamefont {J.}~\bibnamefont
  {Schmidt}},\ and\ \bibinfo {author} {\bibfnamefont {G.~M.}\ \bibnamefont
  {Schütz}},\ }\bibfield  {title} {\bibinfo {title} {{Fibonacci family of
  dynamical universality classes}},\ }\href
  {https://doi.org/10.1073/pnas.1512261112} {\bibfield  {journal} {\bibinfo
  {journal} {Proceedings of the National Academy of Sciences}\ }\textbf
  {\bibinfo {volume} {112}},\ \bibinfo {pages} {12645} (\bibinfo {year}
  {2015})}\BibitemShut {NoStop}%
\bibitem [{\citenamefont {Singh}\ \emph {et~al.}(2021)\citenamefont {Singh},
  \citenamefont {Ware}, \citenamefont {Vasseur},\ and\ \citenamefont
  {Friedman}}]{Singh21}%
  \BibitemOpen
  \bibfield  {author} {\bibinfo {author} {\bibfnamefont {H.}~\bibnamefont
  {Singh}}, \bibinfo {author} {\bibfnamefont {B.~A.}\ \bibnamefont {Ware}},
  \bibinfo {author} {\bibfnamefont {R.}~\bibnamefont {Vasseur}},\ and\ \bibinfo
  {author} {\bibfnamefont {A.~J.}\ \bibnamefont {Friedman}},\ }\bibfield
  {title} {\bibinfo {title} {{Subdiffusion and Many-Body Quantum Chaos with
  Kinetic Constraints}},\ }\bibfield  {journal} {\bibinfo  {journal} {Physical
  Review Letters}\ }\textbf {\bibinfo {volume} {127}},\ \href
  {https://doi.org/10.1103/physrevlett.127.230602}
  {10.1103/physrevlett.127.230602} (\bibinfo {year} {2021})\BibitemShut
  {NoStop}%
\bibitem [{\citenamefont {Sala}\ \emph {et~al.}(2020)\citenamefont {Sala},
  \citenamefont {Rakovszky}, \citenamefont {Verresen}, \citenamefont {Knap},\
  and\ \citenamefont {Pollmann}}]{Sala20}%
  \BibitemOpen
  \bibfield  {author} {\bibinfo {author} {\bibfnamefont {P.}~\bibnamefont
  {Sala}}, \bibinfo {author} {\bibfnamefont {T.}~\bibnamefont {Rakovszky}},
  \bibinfo {author} {\bibfnamefont {R.}~\bibnamefont {Verresen}}, \bibinfo
  {author} {\bibfnamefont {M.}~\bibnamefont {Knap}},\ and\ \bibinfo {author}
  {\bibfnamefont {F.}~\bibnamefont {Pollmann}},\ }\bibfield  {title} {\bibinfo
  {title} {{Ergodicity Breaking Arising from Hilbert Space Fragmentation in
  Dipole-Conserving Hamiltonians}},\ }\bibfield  {journal} {\bibinfo  {journal}
  {Physical Review X}\ }\textbf {\bibinfo {volume} {10}},\ \href
  {https://doi.org/10.1103/physrevx.10.011047} {10.1103/physrevx.10.011047}
  (\bibinfo {year} {2020})\BibitemShut {NoStop}%
\bibitem [{\citenamefont {Morningstar}\ \emph {et~al.}(2020)\citenamefont
  {Morningstar}, \citenamefont {Khemani},\ and\ \citenamefont
  {Huse}}]{Morningstar20}%
  \BibitemOpen
  \bibfield  {author} {\bibinfo {author} {\bibfnamefont {A.}~\bibnamefont
  {Morningstar}}, \bibinfo {author} {\bibfnamefont {V.}~\bibnamefont
  {Khemani}},\ and\ \bibinfo {author} {\bibfnamefont {D.~A.}\ \bibnamefont
  {Huse}},\ }\bibfield  {title} {\bibinfo {title} {{Kinetically constrained
  freezing transition in a dipole-conserving system}},\ }\bibfield  {journal}
  {\bibinfo  {journal} {Physical Review B}\ }\textbf {\bibinfo {volume}
  {101}},\ \href {https://doi.org/10.1103/physrevb.101.214205}
  {10.1103/physrevb.101.214205} (\bibinfo {year} {2020})\BibitemShut {NoStop}%
\bibitem [{\citenamefont {Zhang}(2020)}]{Zhang20}%
  \BibitemOpen
  \bibfield  {author} {\bibinfo {author} {\bibfnamefont {P.}~\bibnamefont
  {Zhang}},\ }\bibfield  {title} {\bibinfo {title} {{Subdiffusion in strongly
  tilted lattice systems}},\ }\bibfield  {journal} {\bibinfo  {journal}
  {Physical Review Research}\ }\textbf {\bibinfo {volume} {2}},\ \href
  {https://doi.org/10.1103/physrevresearch.2.033129}
  {10.1103/physrevresearch.2.033129} (\bibinfo {year} {2020})\BibitemShut
  {NoStop}%
\bibitem [{\citenamefont {Iaconis}\ \emph {et~al.}(2021)\citenamefont
  {Iaconis}, \citenamefont {Lucas},\ and\ \citenamefont
  {Nandkishore}}]{Iaconis21}%
  \BibitemOpen
  \bibfield  {author} {\bibinfo {author} {\bibfnamefont {J.}~\bibnamefont
  {Iaconis}}, \bibinfo {author} {\bibfnamefont {A.}~\bibnamefont {Lucas}},\
  and\ \bibinfo {author} {\bibfnamefont {R.}~\bibnamefont {Nandkishore}},\
  }\bibfield  {title} {\bibinfo {title} {{Multipole conservation laws and
  subdiffusion in any dimension}},\ }\bibfield  {journal} {\bibinfo  {journal}
  {Physical Review E}\ }\textbf {\bibinfo {volume} {103}},\ \href
  {https://doi.org/10.1103/physreve.103.022142} {10.1103/physreve.103.022142}
  (\bibinfo {year} {2021})\BibitemShut {NoStop}%
\bibitem [{\citenamefont {Gromov}\ \emph {et~al.}(2020)\citenamefont {Gromov},
  \citenamefont {Lucas},\ and\ \citenamefont {Nandkishore}}]{Gromov20}%
  \BibitemOpen
  \bibfield  {author} {\bibinfo {author} {\bibfnamefont {A.}~\bibnamefont
  {Gromov}}, \bibinfo {author} {\bibfnamefont {A.}~\bibnamefont {Lucas}},\ and\
  \bibinfo {author} {\bibfnamefont {R.~M.}\ \bibnamefont {Nandkishore}},\
  }\bibfield  {title} {\bibinfo {title} {Fracton hydrodynamics},\ }\href
  {https://doi.org/10.1103/PhysRevResearch.2.033124} {\bibfield  {journal}
  {\bibinfo  {journal} {Phys. Rev. Research}\ }\textbf {\bibinfo {volume}
  {2}},\ \bibinfo {pages} {033124} (\bibinfo {year} {2020})}\BibitemShut
  {NoStop}%
\bibitem [{\citenamefont {Grosvenor}\ \emph {et~al.}(2021)\citenamefont
  {Grosvenor}, \citenamefont {Hoyos}, \citenamefont {Pe\~na Benitez},\ and\
  \citenamefont {Sur\'owka}}]{Grosvenor21}%
  \BibitemOpen
  \bibfield  {author} {\bibinfo {author} {\bibfnamefont {K.~T.}\ \bibnamefont
  {Grosvenor}}, \bibinfo {author} {\bibfnamefont {C.}~\bibnamefont {Hoyos}},
  \bibinfo {author} {\bibfnamefont {F.}~\bibnamefont {Pe\~na Benitez}},\ and\
  \bibinfo {author} {\bibfnamefont {P.}~\bibnamefont {Sur\'owka}},\ }\bibfield
  {title} {\bibinfo {title} {Hydrodynamics of ideal fracton fluids},\ }\href
  {https://doi.org/10.1103/PhysRevResearch.3.043186} {\bibfield  {journal}
  {\bibinfo  {journal} {Phys. Rev. Research}\ }\textbf {\bibinfo {volume}
  {3}},\ \bibinfo {pages} {043186} (\bibinfo {year} {2021})}\BibitemShut
  {NoStop}%
\bibitem [{\citenamefont {Bertini}\ \emph {et~al.}(2021)\citenamefont
  {Bertini}, \citenamefont {Heidrich-Meisner}, \citenamefont {Karrasch},
  \citenamefont {Prosen}, \citenamefont {Steinigeweg},\ and\ \citenamefont
  {{\v{Z}}nidari{\v{c}}}}]{transport_review}%
  \BibitemOpen
  \bibfield  {author} {\bibinfo {author} {\bibfnamefont {B.}~\bibnamefont
  {Bertini}}, \bibinfo {author} {\bibfnamefont {F.}~\bibnamefont
  {Heidrich-Meisner}}, \bibinfo {author} {\bibfnamefont {C.}~\bibnamefont
  {Karrasch}}, \bibinfo {author} {\bibfnamefont {T.}~\bibnamefont {Prosen}},
  \bibinfo {author} {\bibfnamefont {R.}~\bibnamefont {Steinigeweg}},\ and\
  \bibinfo {author} {\bibfnamefont {M.}~\bibnamefont {{\v{Z}}nidari{\v{c}}}},\
  }\bibfield  {title} {\bibinfo {title} {Finite-temperature transport in
  one-dimensional quantum lattice models},\ }\bibfield  {journal} {\bibinfo
  {journal} {Reviews of Modern Physics}\ }\textbf {\bibinfo {volume} {93}},\
  \href {https://doi.org/10.1103/revmodphys.93.025003}
  {10.1103/revmodphys.93.025003} (\bibinfo {year} {2021})\BibitemShut {NoStop}%
\bibitem [{\citenamefont {Doyon}\ and\ \citenamefont {Spohn}(2017)}]{DS17}%
  \BibitemOpen
  \bibfield  {author} {\bibinfo {author} {\bibfnamefont {B.}~\bibnamefont
  {Doyon}}\ and\ \bibinfo {author} {\bibfnamefont {H.}~\bibnamefont {Spohn}},\
  }\bibfield  {title} {\bibinfo {title} {{Drude Weight for the Lieb-Liniger
  Bose Gas}},\ }\href {https://doi.org/10.21468/SciPostPhys.3.6.039} {\bibfield
   {journal} {\bibinfo  {journal} {SciPost Phys.}\ }\textbf {\bibinfo {volume}
  {3}},\ \bibinfo {pages} {039} (\bibinfo {year} {2017})}\BibitemShut {NoStop}%
\bibitem [{\citenamefont {Ilievski}\ and\ \citenamefont {{De
  Nardis}}(2017{\natexlab{a}})}]{IN17}%
  \BibitemOpen
  \bibfield  {author} {\bibinfo {author} {\bibfnamefont {E.}~\bibnamefont
  {Ilievski}}\ and\ \bibinfo {author} {\bibfnamefont {J.}~\bibnamefont {{De
  Nardis}}},\ }\bibfield  {title} {\bibinfo {title} {Ballistic transport in the
  one-dimensional hubbard model: The hydrodynamic approach},\ }\href
  {https://doi.org/10.1103/PhysRevB.96.081118} {\bibfield  {journal} {\bibinfo
  {journal} {Phys. Rev. B}\ }\textbf {\bibinfo {volume} {96}},\ \bibinfo
  {pages} {081118} (\bibinfo {year} {2017}{\natexlab{a}})}\BibitemShut
  {NoStop}%
\bibitem [{\citenamefont {Castro-Alvaredo}\ \emph {et~al.}(2016)\citenamefont
  {Castro-Alvaredo}, \citenamefont {Doyon},\ and\ \citenamefont
  {Yoshimura}}]{CDY16}%
  \BibitemOpen
  \bibfield  {author} {\bibinfo {author} {\bibfnamefont {O.~A.}\ \bibnamefont
  {Castro-Alvaredo}}, \bibinfo {author} {\bibfnamefont {B.}~\bibnamefont
  {Doyon}},\ and\ \bibinfo {author} {\bibfnamefont {T.}~\bibnamefont
  {Yoshimura}},\ }\bibfield  {title} {\bibinfo {title} {{Emergent Hydrodynamics
  in Integrable Quantum Systems Out of Equilibrium}},\ }\href
  {https://doi.org/10.1103/PhysRevX.6.041065} {\bibfield  {journal} {\bibinfo
  {journal} {Phys. Rev. X}\ }\textbf {\bibinfo {volume} {6}},\ \bibinfo {pages}
  {041065} (\bibinfo {year} {2016})}\BibitemShut {NoStop}%
\bibitem [{\citenamefont {Bertini}\ \emph {et~al.}(2016)\citenamefont
  {Bertini}, \citenamefont {Collura}, \citenamefont {{De Nardis}},\ and\
  \citenamefont {Fagotti}}]{BCDF16}%
  \BibitemOpen
  \bibfield  {author} {\bibinfo {author} {\bibfnamefont {B.}~\bibnamefont
  {Bertini}}, \bibinfo {author} {\bibfnamefont {M.}~\bibnamefont {Collura}},
  \bibinfo {author} {\bibfnamefont {J.}~\bibnamefont {{De Nardis}}},\ and\
  \bibinfo {author} {\bibfnamefont {M.}~\bibnamefont {Fagotti}},\ }\bibfield
  {title} {\bibinfo {title} {{Transport in Out-of-Equilibrium $XXZ$ Chains:
  Exact Profiles of Charges and Currents}},\ }\href
  {https://doi.org/10.1103/PhysRevLett.117.207201} {\bibfield  {journal}
  {\bibinfo  {journal} {Phys. Rev. Lett.}\ }\textbf {\bibinfo {volume} {117}},\
  \bibinfo {pages} {207201} (\bibinfo {year} {2016})}\BibitemShut {NoStop}%
\bibitem [{\citenamefont {Ilievski}\ and\ \citenamefont {{De
  Nardis}}(2017{\natexlab{b}})}]{IN_Drude}%
  \BibitemOpen
  \bibfield  {author} {\bibinfo {author} {\bibfnamefont {E.}~\bibnamefont
  {Ilievski}}\ and\ \bibinfo {author} {\bibfnamefont {J.}~\bibnamefont {{De
  Nardis}}},\ }\bibfield  {title} {\bibinfo {title} {{Microscopic Origin of
  Ideal Conductivity in Integrable Quantum Models}},\ }\bibfield  {journal}
  {\bibinfo  {journal} {Physical Review Letters}\ }\textbf {\bibinfo {volume}
  {119}},\ \href {https://doi.org/10.1103/physrevlett.119.020602}
  {10.1103/physrevlett.119.020602} (\bibinfo {year}
  {2017}{\natexlab{b}})\BibitemShut {NoStop}%
\bibitem [{\citenamefont {Ilievski}\ \emph {et~al.}(2018)\citenamefont
  {Ilievski}, \citenamefont {{De Nardis}}, \citenamefont {Medenjak},\ and\
  \citenamefont {Prosen}}]{Ilievski18}%
  \BibitemOpen
  \bibfield  {author} {\bibinfo {author} {\bibfnamefont {E.}~\bibnamefont
  {Ilievski}}, \bibinfo {author} {\bibfnamefont {J.}~\bibnamefont {{De
  Nardis}}}, \bibinfo {author} {\bibfnamefont {M.}~\bibnamefont {Medenjak}},\
  and\ \bibinfo {author} {\bibfnamefont {T.}~\bibnamefont {Prosen}},\
  }\bibfield  {title} {\bibinfo {title} {Superdiffusion in one-dimensional
  quantum lattice models},\ }\bibfield  {journal} {\bibinfo  {journal}
  {Physical Review Letters}\ }\textbf {\bibinfo {volume} {121}},\ \href
  {https://doi.org/10.1103/physrevlett.121.230602}
  {10.1103/physrevlett.121.230602} (\bibinfo {year} {2018})\BibitemShut
  {NoStop}%
\bibitem [{\citenamefont {Ljubotina}\ \emph {et~al.}(2019)\citenamefont
  {Ljubotina}, \citenamefont {{\v Z}nidari{\v c}},\ and\ \citenamefont
  {Prosen}}]{Ljubotina19}%
  \BibitemOpen
  \bibfield  {author} {\bibinfo {author} {\bibfnamefont {M.}~\bibnamefont
  {Ljubotina}}, \bibinfo {author} {\bibfnamefont {M.}~\bibnamefont {{\v
  Z}nidari{\v c}}},\ and\ \bibinfo {author} {\bibfnamefont {T.}~\bibnamefont
  {Prosen}},\ }\bibfield  {title} {\bibinfo {title} {{Kardar-Parisi-Zhang
  Physics in the Quantum Heisenberg Magnet}},\ }\bibfield  {journal} {\bibinfo
  {journal} {Physical Review Letters}\ }\textbf {\bibinfo {volume} {122}},\
  \href {https://doi.org/10.1103/physrevlett.122.210602}
  {10.1103/physrevlett.122.210602} (\bibinfo {year} {2019})\BibitemShut
  {NoStop}%
\bibitem [{\citenamefont {De~Nardis}\ \emph {et~al.}(2019)\citenamefont
  {De~Nardis}, \citenamefont {Medenjak}, \citenamefont {Karrasch},\ and\
  \citenamefont {Ilievski}}]{PhysRevLett.123.186601}%
  \BibitemOpen
  \bibfield  {author} {\bibinfo {author} {\bibfnamefont {J.}~\bibnamefont
  {De~Nardis}}, \bibinfo {author} {\bibfnamefont {M.}~\bibnamefont {Medenjak}},
  \bibinfo {author} {\bibfnamefont {C.}~\bibnamefont {Karrasch}},\ and\
  \bibinfo {author} {\bibfnamefont {E.}~\bibnamefont {Ilievski}},\ }\bibfield
  {title} {\bibinfo {title} {{Anomalous Spin Diffusion in One-Dimensional
  Antiferromagnets}},\ }\href {https://doi.org/10.1103/PhysRevLett.123.186601}
  {\bibfield  {journal} {\bibinfo  {journal} {Phys. Rev. Lett.}\ }\textbf
  {\bibinfo {volume} {123}},\ \bibinfo {pages} {186601} (\bibinfo {year}
  {2019})}\BibitemShut {NoStop}%
\bibitem [{\citenamefont {Ilievski}\ \emph {et~al.}(2021)\citenamefont
  {Ilievski}, \citenamefont {Nardis}, \citenamefont {Gopalakrishnan},
  \citenamefont {Vasseur},\ and\ \citenamefont {Ware}}]{superuniversality}%
  \BibitemOpen
  \bibfield  {author} {\bibinfo {author} {\bibfnamefont {E.}~\bibnamefont
  {Ilievski}}, \bibinfo {author} {\bibfnamefont {J.~D.}\ \bibnamefont
  {Nardis}}, \bibinfo {author} {\bibfnamefont {S.}~\bibnamefont
  {Gopalakrishnan}}, \bibinfo {author} {\bibfnamefont {R.}~\bibnamefont
  {Vasseur}},\ and\ \bibinfo {author} {\bibfnamefont {B.}~\bibnamefont
  {Ware}},\ }\bibfield  {title} {\bibinfo {title} {{Superuniversality of
  Superdiffusion}},\ }\bibfield  {journal} {\bibinfo  {journal} {Physical
  Review X}\ }\textbf {\bibinfo {volume} {11}},\ \href
  {https://doi.org/10.1103/physrevx.11.031023} {10.1103/physrevx.11.031023}
  (\bibinfo {year} {2021})\BibitemShut {NoStop}%
\bibitem [{\citenamefont {Bulchandani}\ \emph {et~al.}(2021)\citenamefont
  {Bulchandani}, \citenamefont {Gopalakrishnan},\ and\ \citenamefont
  {Ilievski}}]{superdiffusion_review}%
  \BibitemOpen
  \bibfield  {author} {\bibinfo {author} {\bibfnamefont {V.~B.}\ \bibnamefont
  {Bulchandani}}, \bibinfo {author} {\bibfnamefont {S.}~\bibnamefont
  {Gopalakrishnan}},\ and\ \bibinfo {author} {\bibfnamefont {E.}~\bibnamefont
  {Ilievski}},\ }\bibfield  {title} {\bibinfo {title} {Superdiffusion in spin
  chains},\ }\href {https://doi.org/10.1088/1742-5468/ac12c7} {\bibfield
  {journal} {\bibinfo  {journal} {Journal of Statistical Mechanics: Theory and
  Experiment}\ }\textbf {\bibinfo {volume} {2021}},\ \bibinfo {pages} {084001}
  (\bibinfo {year} {2021})}\BibitemShut {NoStop}%
\bibitem [{\citenamefont {Bertini}\ \emph {et~al.}(2001)\citenamefont
  {Bertini}, \citenamefont {Sole}, \citenamefont {Gabrielli}, \citenamefont
  {Jona-Lasinio},\ and\ \citenamefont {Landim}}]{Bertini01}%
  \BibitemOpen
  \bibfield  {author} {\bibinfo {author} {\bibfnamefont {L.}~\bibnamefont
  {Bertini}}, \bibinfo {author} {\bibfnamefont {A.~D.}\ \bibnamefont {Sole}},
  \bibinfo {author} {\bibfnamefont {D.}~\bibnamefont {Gabrielli}}, \bibinfo
  {author} {\bibfnamefont {G.}~\bibnamefont {Jona-Lasinio}},\ and\ \bibinfo
  {author} {\bibfnamefont {C.}~\bibnamefont {Landim}},\ }\bibfield  {title}
  {\bibinfo {title} {{Fluctuations in Stationary Nonequilibrium States of
  Irreversible Processes}},\ }\bibfield  {journal} {\bibinfo  {journal}
  {Physical Review Letters}\ }\textbf {\bibinfo {volume} {87}},\ \href
  {https://doi.org/10.1103/physrevlett.87.040601}
  {10.1103/physrevlett.87.040601} (\bibinfo {year} {2001})\BibitemShut
  {NoStop}%
\bibitem [{\citenamefont {Bertini}\ \emph {et~al.}(2002)\citenamefont
  {Bertini}, \citenamefont {Sole}, \citenamefont {Gabrielli}, \citenamefont
  {Jona-Lasinio},\ and\ \citenamefont {Landim}}]{Bertini02}%
  \BibitemOpen
  \bibfield  {author} {\bibinfo {author} {\bibfnamefont {L.}~\bibnamefont
  {Bertini}}, \bibinfo {author} {\bibfnamefont {A.~D.}\ \bibnamefont {Sole}},
  \bibinfo {author} {\bibfnamefont {D.}~\bibnamefont {Gabrielli}}, \bibinfo
  {author} {\bibfnamefont {G.}~\bibnamefont {Jona-Lasinio}},\ and\ \bibinfo
  {author} {\bibfnamefont {C.}~\bibnamefont {Landim}},\ }\bibfield  {title}
  {\bibinfo {title} {Macroscopic fluctuation theory for stationary
  non-equilibrium states},\ }\href {https://doi.org/10.1023/a:1014525911391}
  {\bibfield  {journal} {\bibinfo  {journal} {Journal of Statistical Physics}\
  }\textbf {\bibinfo {volume} {107}},\ \bibinfo {pages} {635} (\bibinfo {year}
  {2002})}\BibitemShut {NoStop}%
\bibitem [{\citenamefont {{Jinho Baik and Eric M. Rains}}(2000)}]{BaikRains}%
  \BibitemOpen
  \bibfield  {author} {\bibinfo {author} {\bibnamefont {{Jinho Baik and Eric M.
  Rains}}},\ }\bibfield  {title} {\bibinfo {title} {{Limiting distributions for
  a polynuclear growth model with external sources}},\ }\href
  {https://doi.org/10.1023/a:1018615306992} {\bibfield  {journal} {\bibinfo
  {journal} {Journal of Statistical Physics}\ }\textbf {\bibinfo {volume}
  {100}},\ \bibinfo {pages} {523} (\bibinfo {year} {2000})}\BibitemShut
  {NoStop}%
\bibitem [{\citenamefont {Takeuchi}(2018)}]{Takeuchi18}%
  \BibitemOpen
  \bibfield  {author} {\bibinfo {author} {\bibfnamefont {K.~A.}\ \bibnamefont
  {Takeuchi}},\ }\bibfield  {title} {\bibinfo {title} {{An appetizer to modern
  developments on the Kardar--Parisi--Zhang universality class}},\ }\href
  {https://doi.org/10.1016/j.physa.2018.03.009} {\bibfield  {journal} {\bibinfo
   {journal} {Physica A: Statistical Mechanics and its Applications}\ }\textbf
  {\bibinfo {volume} {504}},\ \bibinfo {pages} {77} (\bibinfo {year}
  {2018})}\BibitemShut {NoStop}%
\bibitem [{\citenamefont {de~Gier}\ and\ \citenamefont
  {Essler}(2005)}]{deGier05}%
  \BibitemOpen
  \bibfield  {author} {\bibinfo {author} {\bibfnamefont {J.}~\bibnamefont
  {de~Gier}}\ and\ \bibinfo {author} {\bibfnamefont {F.~H.~L.}\ \bibnamefont
  {Essler}},\ }\bibfield  {title} {\bibinfo {title} {{Bethe Ansatz Solution of
  the Asymmetric Exclusion Process with Open Boundaries}},\ }\bibfield
  {journal} {\bibinfo  {journal} {Physical Review Letters}\ }\textbf {\bibinfo
  {volume} {95}},\ \href {https://doi.org/10.1103/physrevlett.95.240601}
  {10.1103/physrevlett.95.240601} (\bibinfo {year} {2005})\BibitemShut
  {NoStop}%
\bibitem [{\citenamefont {Krajnik}\ \emph {et~al.}(2020)\citenamefont
  {Krajnik}, \citenamefont {Ilievski},\ and\ \citenamefont
  {Prosen}}]{MatrixModels}%
  \BibitemOpen
  \bibfield  {author} {\bibinfo {author} {\bibfnamefont {{\v Z}.}~\bibnamefont
  {Krajnik}}, \bibinfo {author} {\bibfnamefont {E.}~\bibnamefont {Ilievski}},\
  and\ \bibinfo {author} {\bibfnamefont {T.}~\bibnamefont {Prosen}},\
  }\bibfield  {title} {\bibinfo {title} {{Integrable Matrix Models in Discrete
  Space-Time}},\ }\href {https://doi.org/10.21468/SciPostPhys.9.3.038}
  {\bibfield  {journal} {\bibinfo  {journal} {SciPost Phys.}\ }\textbf
  {\bibinfo {volume} {9}},\ \bibinfo {pages} {38} (\bibinfo {year}
  {2020})}\BibitemShut {NoStop}%
\bibitem [{\citenamefont {Krajnik}\ \emph
  {et~al.}(2022{\natexlab{a}})\citenamefont {Krajnik}, \citenamefont
  {Ilievski},\ and\ \citenamefont {Prosen}}]{KIP22}%
  \BibitemOpen
  \bibfield  {author} {\bibinfo {author} {\bibfnamefont {{\v{Z}}.}~\bibnamefont
  {Krajnik}}, \bibinfo {author} {\bibfnamefont {E.}~\bibnamefont {Ilievski}},\
  and\ \bibinfo {author} {\bibfnamefont {T.}~\bibnamefont {Prosen}},\
  }\bibfield  {title} {\bibinfo {title} {{Absence of Normal Fluctuations in an
  Integrable Magnet}},\ }\href {https://doi.org/10.1103/PhysRevLett.128.090604}
  {\bibfield  {journal} {\bibinfo  {journal} {Phys. Rev. Lett.}\ }\textbf
  {\bibinfo {volume} {128}},\ \bibinfo {pages} {090604} (\bibinfo {year}
  {2022}{\natexlab{a}})}\BibitemShut {NoStop}%
\bibitem [{\citenamefont {Bryc}(1993)}]{Bryc93}%
  \BibitemOpen
  \bibfield  {author} {\bibinfo {author} {\bibfnamefont {W.}~\bibnamefont
  {Bryc}},\ }\bibfield  {title} {\bibinfo {title} {{A remark on the connection
  between the large deviation principle and the central limit theorem}},\
  }\href {https://doi.org/10.1016/0167-7152(93)90012-8} {\bibfield  {journal}
  {\bibinfo  {journal} {Statistics {\&} Probability Letters}\ }\textbf
  {\bibinfo {volume} {18}},\ \bibinfo {pages} {253} (\bibinfo {year}
  {1993})}\BibitemShut {NoStop}%
\bibitem [{\citenamefont {Jak{\v{s}}i{\'{c}}}\ \emph
  {et~al.}(2012)\citenamefont {Jak{\v{s}}i{\'{c}}}, \citenamefont {Ogata},
  \citenamefont {Pautrat},\ and\ \citenamefont {Pillet}}]{Jaksic12}%
  \BibitemOpen
  \bibfield  {author} {\bibinfo {author} {\bibfnamefont {V.}~\bibnamefont
  {Jak{\v{s}}i{\'{c}}}}, \bibinfo {author} {\bibfnamefont {Y.}~\bibnamefont
  {Ogata}}, \bibinfo {author} {\bibfnamefont {Y.}~\bibnamefont {Pautrat}},\
  and\ \bibinfo {author} {\bibfnamefont {C.-A.}\ \bibnamefont {Pillet}},\
  }\bibfield  {title} {\bibinfo {title} {{Entropic fluctuations in quantum
  statistical mechanics{\textemdash}an introduction}},\ }in\ \href
  {https://doi.org/10.1093/acprof:oso/9780199652495.003.0004} {\emph {\bibinfo
  {booktitle} {Quantum Theory from Small to Large Scales}}}\ (\bibinfo
  {publisher} {Oxford University Press},\ \bibinfo {year} {2012})\ pp.\
  \bibinfo {pages} {213--410}\BibitemShut {NoStop}%
\bibitem [{\citenamefont {Feldmeier}\ \emph {et~al.}(2022)\citenamefont
  {Feldmeier}, \citenamefont {Witczak-Krempa},\ and\ \citenamefont
  {Knap}}]{Feldmeier22}%
  \BibitemOpen
  \bibfield  {author} {\bibinfo {author} {\bibfnamefont {J.}~\bibnamefont
  {Feldmeier}}, \bibinfo {author} {\bibfnamefont {W.}~\bibnamefont
  {Witczak-Krempa}},\ and\ \bibinfo {author} {\bibfnamefont {M.}~\bibnamefont
  {Knap}},\ }\bibfield  {title} {\bibinfo {title} {Emergent tracer dynamics in
  constrained quantum systems},\ }\href
  {https://doi.org/10.1103/PhysRevB.106.094303} {\bibfield  {journal} {\bibinfo
   {journal} {Phys. Rev. B}\ }\textbf {\bibinfo {volume} {106}},\ \bibinfo
  {pages} {094303} (\bibinfo {year} {2022})}\BibitemShut {NoStop}%
\bibitem [{\citenamefont {Krajnik}\ \emph
  {et~al.}(2022{\natexlab{b}})\citenamefont {Krajnik}, \citenamefont {Schmidt},
  \citenamefont {Pasquier}, \citenamefont {Ilievski},\ and\ \citenamefont
  {Prosen}}]{Krajnik22}%
  \BibitemOpen
  \bibfield  {author} {\bibinfo {author} {\bibfnamefont {{\v{Z}}.}~\bibnamefont
  {Krajnik}}, \bibinfo {author} {\bibfnamefont {J.}~\bibnamefont {Schmidt}},
  \bibinfo {author} {\bibfnamefont {V.}~\bibnamefont {Pasquier}}, \bibinfo
  {author} {\bibfnamefont {E.}~\bibnamefont {Ilievski}},\ and\ \bibinfo
  {author} {\bibfnamefont {T.}~\bibnamefont {Prosen}},\ }\bibfield  {title}
  {\bibinfo {title} {Exact anomalous current fluctuations in a deterministic
  interacting model},\ }\href {https://doi.org/10.1103/PhysRevLett.128.160601}
  {\bibfield  {journal} {\bibinfo  {journal} {Phys. Rev. Lett.}\ }\textbf
  {\bibinfo {volume} {128}},\ \bibinfo {pages} {160601} (\bibinfo {year}
  {2022}{\natexlab{b}})}\BibitemShut {NoStop}%
\bibitem [{\citenamefont {Bertini}\ \emph {et~al.}(2015)\citenamefont
  {Bertini}, \citenamefont {Sole}, \citenamefont {Gabrielli}, \citenamefont
  {Jona-Lasinio},\ and\ \citenamefont {Landim}}]{MFT}%
  \BibitemOpen
  \bibfield  {author} {\bibinfo {author} {\bibfnamefont {L.}~\bibnamefont
  {Bertini}}, \bibinfo {author} {\bibfnamefont {A.~D.}\ \bibnamefont {Sole}},
  \bibinfo {author} {\bibfnamefont {D.}~\bibnamefont {Gabrielli}}, \bibinfo
  {author} {\bibfnamefont {G.}~\bibnamefont {Jona-Lasinio}},\ and\ \bibinfo
  {author} {\bibfnamefont {C.}~\bibnamefont {Landim}},\ }\bibfield  {title}
  {\bibinfo {title} {Macroscopic fluctuation theory},\ }\href
  {https://doi.org/10.1103/revmodphys.87.593} {\bibfield  {journal} {\bibinfo
  {journal} {Reviews of Modern Physics}\ }\textbf {\bibinfo {volume} {87}},\
  \bibinfo {pages} {593} (\bibinfo {year} {2015})}\BibitemShut {NoStop}%
\bibitem [{\citenamefont {Doyon}\ and\ \citenamefont
  {Myers}(2019)}]{DoyonMyers20}%
  \BibitemOpen
  \bibfield  {author} {\bibinfo {author} {\bibfnamefont {B.}~\bibnamefont
  {Doyon}}\ and\ \bibinfo {author} {\bibfnamefont {J.}~\bibnamefont {Myers}},\
  }\bibfield  {title} {\bibinfo {title} {{Fluctuations in Ballistic Transport
  from Euler Hydrodynamics}},\ }\href
  {https://doi.org/10.1007/s00023-019-00860-w} {\bibfield  {journal} {\bibinfo
  {journal} {Annales Henri Poincar{\'{e}}}\ }\textbf {\bibinfo {volume} {21}},\
  \bibinfo {pages} {255} (\bibinfo {year} {2019})}\BibitemShut {NoStop}%
\bibitem [{\citenamefont {Myers}\ \emph {et~al.}(2020)\citenamefont {Myers},
  \citenamefont {Bhaseen}, \citenamefont {Harris},\ and\ \citenamefont
  {Doyon}}]{MBHD20}%
  \BibitemOpen
  \bibfield  {author} {\bibinfo {author} {\bibfnamefont {J.}~\bibnamefont
  {Myers}}, \bibinfo {author} {\bibfnamefont {J.}~\bibnamefont {Bhaseen}},
  \bibinfo {author} {\bibfnamefont {R.~J.}\ \bibnamefont {Harris}},\ and\
  \bibinfo {author} {\bibfnamefont {B.}~\bibnamefont {Doyon}},\ }\bibfield
  {title} {\bibinfo {title} {Transport fluctuations in integrable models out of
  equilibrium},\ }\bibfield  {journal} {\bibinfo  {journal} {{SciPost}
  Physics}\ }\textbf {\bibinfo {volume} {8}},\ \href
  {https://doi.org/10.21468/scipostphys.8.1.007} {10.21468/scipostphys.8.1.007}
  (\bibinfo {year} {2020})\BibitemShut {NoStop}%
\bibitem [{\citenamefont {Perfetto}\ and\ \citenamefont
  {Doyon}(2021)}]{PerfettoDoyon21}%
  \BibitemOpen
  \bibfield  {author} {\bibinfo {author} {\bibfnamefont {G.}~\bibnamefont
  {Perfetto}}\ and\ \bibinfo {author} {\bibfnamefont {B.}~\bibnamefont
  {Doyon}},\ }\bibfield  {title} {\bibinfo {title} {Euler-scale dynamical
  fluctuations in non-equilibrium interacting integrable systems},\ }\bibfield
  {journal} {\bibinfo  {journal} {{SciPost} Physics}\ }\textbf {\bibinfo
  {volume} {10}},\ \href {https://doi.org/10.21468/scipostphys.10.5.116}
  {10.21468/scipostphys.10.5.116} (\bibinfo {year} {2021})\BibitemShut
  {NoStop}%
\bibitem [{\citenamefont {Doyon}\ \emph {et~al.}(2023)\citenamefont {Doyon},
  \citenamefont {Perfetto}, \citenamefont {Sasamoto},\ and\ \citenamefont
  {Yoshimura}}]{BMFT}%
  \BibitemOpen
  \bibfield  {author} {\bibinfo {author} {\bibfnamefont {B.}~\bibnamefont
  {Doyon}}, \bibinfo {author} {\bibfnamefont {G.}~\bibnamefont {Perfetto}},
  \bibinfo {author} {\bibfnamefont {T.}~\bibnamefont {Sasamoto}},\ and\
  \bibinfo {author} {\bibfnamefont {T.}~\bibnamefont {Yoshimura}},\ }\bibfield
  {title} {\bibinfo {title} {{Ballistic macroscopic fluctuation theory}},\
  }\href {https://doi.org/10.21468/SciPostPhys.15.4.136} {\bibfield  {journal}
  {\bibinfo  {journal} {SciPost Phys.}\ }\textbf {\bibinfo {volume} {15}},\
  \bibinfo {pages} {136} (\bibinfo {year} {2023})}\BibitemShut {NoStop}%
\bibitem [{\citenamefont {Gopalakrishnan}\ \emph {et~al.}(2024)\citenamefont
  {Gopalakrishnan}, \citenamefont {Morningstar}, \citenamefont {Vasseur},\ and\
  \citenamefont {Khemani}}]{Sarang_FCS}%
  \BibitemOpen
  \bibfield  {author} {\bibinfo {author} {\bibfnamefont {S.}~\bibnamefont
  {Gopalakrishnan}}, \bibinfo {author} {\bibfnamefont {A.}~\bibnamefont
  {Morningstar}}, \bibinfo {author} {\bibfnamefont {R.}~\bibnamefont
  {Vasseur}},\ and\ \bibinfo {author} {\bibfnamefont {V.}~\bibnamefont
  {Khemani}},\ }\bibfield  {title} {\bibinfo {title} {Distinct universality
  classes of diffusive transport from full counting statistics},\ }\href
  {https://doi.org/10.1103/PhysRevB.109.024417} {\bibfield  {journal} {\bibinfo
   {journal} {Phys. Rev. B}\ }\textbf {\bibinfo {volume} {109}},\ \bibinfo
  {pages} {024417} (\bibinfo {year} {2024})}\BibitemShut {NoStop}%
\bibitem [{\citenamefont {Krajnik}\ \emph {et~al.}(2024)\citenamefont
  {Krajnik}, \citenamefont {Schmidt}, \citenamefont {Ilievski},\ and\
  \citenamefont {Prosen}}]{KSIP23}%
  \BibitemOpen
  \bibfield  {author} {\bibinfo {author} {\bibfnamefont {{\v{Z}}.}~\bibnamefont
  {Krajnik}}, \bibinfo {author} {\bibfnamefont {J.}~\bibnamefont {Schmidt}},
  \bibinfo {author} {\bibfnamefont {E.}~\bibnamefont {Ilievski}},\ and\
  \bibinfo {author} {\bibfnamefont {T.}~\bibnamefont {Prosen}},\ }\bibfield
  {title} {\bibinfo {title} {Dynamical criticality of magnetization transfer in
  integrable spin chains},\ }\href
  {https://doi.org/10.1103/PhysRevLett.132.017101} {\bibfield  {journal}
  {\bibinfo  {journal} {Phys. Rev. Lett.}\ }\textbf {\bibinfo {volume} {132}},\
  \bibinfo {pages} {017101} (\bibinfo {year} {2024})}\BibitemShut {NoStop}%
\bibitem [{\citenamefont {Kormos}\ \emph {et~al.}(2022)\citenamefont {Kormos},
  \citenamefont {V\"or\"os},\ and\ \citenamefont {Zar\'and}}]{Kormos22}%
  \BibitemOpen
  \bibfield  {author} {\bibinfo {author} {\bibfnamefont {M.}~\bibnamefont
  {Kormos}}, \bibinfo {author} {\bibfnamefont {D.}~\bibnamefont {V\"or\"os}},\
  and\ \bibinfo {author} {\bibfnamefont {G.}~\bibnamefont {Zar\'and}},\
  }\bibfield  {title} {\bibinfo {title} {{Finite-temperature dynamics in gapped
  one-dimensional models in the sine-Gordon family}},\ }\href
  {https://doi.org/10.1103/PhysRevB.106.205151} {\bibfield  {journal} {\bibinfo
   {journal} {Phys. Rev. B}\ }\textbf {\bibinfo {volume} {106}},\ \bibinfo
  {pages} {205151} (\bibinfo {year} {2022})}\BibitemShut {NoStop}%
\bibitem [{\citenamefont {Altshuler}\ \emph {et~al.}(2006)\citenamefont
  {Altshuler}, \citenamefont {Konik},\ and\ \citenamefont
  {Tsvelik}}]{Altshuler06}%
  \BibitemOpen
  \bibfield  {author} {\bibinfo {author} {\bibfnamefont {B.}~\bibnamefont
  {Altshuler}}, \bibinfo {author} {\bibfnamefont {R.}~\bibnamefont {Konik}},\
  and\ \bibinfo {author} {\bibfnamefont {A.}~\bibnamefont {Tsvelik}},\
  }\bibfield  {title} {\bibinfo {title} {{Low temperature correlation functions
  in integrable models: Derivation of the large distance and time asymptotics
  from the form factor expansion}},\ }\href
  {https://doi.org/10.1016/j.nuclphysb.2006.01.022} {\bibfield  {journal}
  {\bibinfo  {journal} {Nuclear Physics B}\ }\textbf {\bibinfo {volume}
  {739}},\ \bibinfo {pages} {311} (\bibinfo {year} {2006})}\BibitemShut
  {NoStop}%
\bibitem [{\citenamefont {Krapivsky}\ \emph {et~al.}(2014)\citenamefont
  {Krapivsky}, \citenamefont {Mallick},\ and\ \citenamefont
  {Sadhu}}]{Krapivsky14}%
  \BibitemOpen
  \bibfield  {author} {\bibinfo {author} {\bibfnamefont {P.}~\bibnamefont
  {Krapivsky}}, \bibinfo {author} {\bibfnamefont {K.}~\bibnamefont {Mallick}},\
  and\ \bibinfo {author} {\bibfnamefont {T.}~\bibnamefont {Sadhu}},\ }\bibfield
   {title} {\bibinfo {title} {{Large Deviations in Single-File Diffusion}},\
  }\bibfield  {journal} {\bibinfo  {journal} {Physical Review Letters}\
  }\textbf {\bibinfo {volume} {113}},\ \href
  {https://doi.org/10.1103/physrevlett.113.078101}
  {10.1103/physrevlett.113.078101} (\bibinfo {year} {2014})\BibitemShut
  {NoStop}%
\bibitem [{\citenamefont {Krapivsky}\ \emph
  {et~al.}(2015{\natexlab{a}})\citenamefont {Krapivsky}, \citenamefont
  {Mallick},\ and\ \citenamefont {Sadhu}}]{Krapivsky15}%
  \BibitemOpen
  \bibfield  {author} {\bibinfo {author} {\bibfnamefont {P.~L.}\ \bibnamefont
  {Krapivsky}}, \bibinfo {author} {\bibfnamefont {K.}~\bibnamefont {Mallick}},\
  and\ \bibinfo {author} {\bibfnamefont {T.}~\bibnamefont {Sadhu}},\ }\bibfield
   {title} {\bibinfo {title} {Dynamical properties of single-file diffusion},\
  }\href {https://doi.org/10.1088/1742-5468/2015/09/p09007} {\bibfield
  {journal} {\bibinfo  {journal} {Journal of Statistical Mechanics: Theory and
  Experiment}\ }\textbf {\bibinfo {volume} {2015}},\ \bibinfo {pages} {P09007}
  (\bibinfo {year} {2015}{\natexlab{a}})}\BibitemShut {NoStop}%
\bibitem [{\citenamefont {Krapivsky}\ \emph
  {et~al.}(2015{\natexlab{b}})\citenamefont {Krapivsky}, \citenamefont
  {Mallick},\ and\ \citenamefont {Sadhu}}]{Krapivsky_tagged}%
  \BibitemOpen
  \bibfield  {author} {\bibinfo {author} {\bibfnamefont {P.~L.}\ \bibnamefont
  {Krapivsky}}, \bibinfo {author} {\bibfnamefont {K.}~\bibnamefont {Mallick}},\
  and\ \bibinfo {author} {\bibfnamefont {T.}~\bibnamefont {Sadhu}},\ }\bibfield
   {title} {\bibinfo {title} {{Tagged Particle in Single-File Diffusion}},\
  }\href {https://doi.org/10.1007/s10955-015-1291-0} {\bibfield  {journal}
  {\bibinfo  {journal} {Journal of Statistical Physics}\ }\textbf {\bibinfo
  {volume} {160}},\ \bibinfo {pages} {885} (\bibinfo {year}
  {2015}{\natexlab{b}})}\BibitemShut {NoStop}%
\bibitem [{\citenamefont {Imamura}\ \emph {et~al.}(2017)\citenamefont
  {Imamura}, \citenamefont {Mallick},\ and\ \citenamefont
  {Sasamoto}}]{Imamura17}%
  \BibitemOpen
  \bibfield  {author} {\bibinfo {author} {\bibfnamefont {T.}~\bibnamefont
  {Imamura}}, \bibinfo {author} {\bibfnamefont {K.}~\bibnamefont {Mallick}},\
  and\ \bibinfo {author} {\bibfnamefont {T.}~\bibnamefont {Sasamoto}},\
  }\bibfield  {title} {\bibinfo {title} {{Large Deviations of a Tracer in the
  Symmetric Exclusion Process}},\ }\bibfield  {journal} {\bibinfo  {journal}
  {Physical Review Letters}\ }\textbf {\bibinfo {volume} {118}},\ \href
  {https://doi.org/10.1103/physrevlett.118.160601}
  {10.1103/physrevlett.118.160601} (\bibinfo {year} {2017})\BibitemShut
  {NoStop}%
\bibitem [{\citenamefont {Khemani}\ \emph {et~al.}(2020)\citenamefont
  {Khemani}, \citenamefont {Hermele},\ and\ \citenamefont
  {Nandkishore}}]{Khemani20}%
  \BibitemOpen
  \bibfield  {author} {\bibinfo {author} {\bibfnamefont {V.}~\bibnamefont
  {Khemani}}, \bibinfo {author} {\bibfnamefont {M.}~\bibnamefont {Hermele}},\
  and\ \bibinfo {author} {\bibfnamefont {R.}~\bibnamefont {Nandkishore}},\
  }\bibfield  {title} {\bibinfo {title} {{Localization from Hilbert space
  shattering: From theory to physical realizations}},\ }\bibfield  {journal}
  {\bibinfo  {journal} {Physical Review B}\ }\textbf {\bibinfo {volume}
  {101}},\ \href {https://doi.org/10.1103/physrevb.101.174204}
  {10.1103/physrevb.101.174204} (\bibinfo {year} {2020})\BibitemShut {NoStop}%
\bibitem [{\citenamefont {Mallick}(2015)}]{Mallick15}%
  \BibitemOpen
  \bibfield  {author} {\bibinfo {author} {\bibfnamefont {K.}~\bibnamefont
  {Mallick}},\ }\bibfield  {title} {\bibinfo {title} {{The exclusion process: A
  paradigm for non-equilibrium behaviour}},\ }\href
  {https://doi.org/10.1016/j.physa.2014.07.046} {\bibfield  {journal} {\bibinfo
   {journal} {Physica A: Statistical Mechanics and its Applications}\ }\textbf
  {\bibinfo {volume} {418}},\ \bibinfo {pages} {17} (\bibinfo {year}
  {2015})}\BibitemShut {NoStop}%
\bibitem [{\citenamefont {Medenjak}\ \emph {et~al.}(2017)\citenamefont
  {Medenjak}, \citenamefont {Klobas},\ and\ \citenamefont
  {Prosen}}]{Medenjak17}%
  \BibitemOpen
  \bibfield  {author} {\bibinfo {author} {\bibfnamefont {M.}~\bibnamefont
  {Medenjak}}, \bibinfo {author} {\bibfnamefont {K.}~\bibnamefont {Klobas}},\
  and\ \bibinfo {author} {\bibfnamefont {T.}~\bibnamefont {Prosen}},\
  }\bibfield  {title} {\bibinfo {title} {Diffusion in deterministic interacting
  lattice systems},\ }\href {https://doi.org/10.1103/PhysRevLett.119.110603}
  {\bibfield  {journal} {\bibinfo  {journal} {Phys. Rev. Lett.}\ }\textbf
  {\bibinfo {volume} {119}},\ \bibinfo {pages} {110603} (\bibinfo {year}
  {2017})}\BibitemShut {NoStop}%
\bibitem [{\citenamefont {Klobas}\ \emph {et~al.}(2018)\citenamefont {Klobas},
  \citenamefont {Medenjak},\ and\ \citenamefont {Prosen}}]{Klobas2018}%
  \BibitemOpen
  \bibfield  {author} {\bibinfo {author} {\bibfnamefont {K.}~\bibnamefont
  {Klobas}}, \bibinfo {author} {\bibfnamefont {M.}~\bibnamefont {Medenjak}},\
  and\ \bibinfo {author} {\bibfnamefont {T.}~\bibnamefont {Prosen}},\
  }\bibfield  {title} {\bibinfo {title} {Exactly solvable deterministic lattice
  model of crossover between ballistic and diffusive transport},\ }\href
  {https://doi.org/10.1088/1742-5468/aae853} {\bibfield  {journal} {\bibinfo
  {journal} {Journal of Statistical Mechanics: Theory and Experiment}\ }\textbf
  {\bibinfo {volume} {2018}},\ \bibinfo {pages} {123202} (\bibinfo {year}
  {2018})}\BibitemShut {NoStop}%
\bibitem [{\citenamefont {Medenjak}\ \emph {et~al.}(2019)\citenamefont
  {Medenjak}, \citenamefont {Popkov}, \citenamefont {Prosen}, \citenamefont
  {Ragoucy},\ and\ \citenamefont {Vanicat}}]{Medenjak19}%
  \BibitemOpen
  \bibfield  {author} {\bibinfo {author} {\bibfnamefont {M.}~\bibnamefont
  {Medenjak}}, \bibinfo {author} {\bibfnamefont {V.}~\bibnamefont {Popkov}},
  \bibinfo {author} {\bibfnamefont {T.}~\bibnamefont {Prosen}}, \bibinfo
  {author} {\bibfnamefont {E.}~\bibnamefont {Ragoucy}},\ and\ \bibinfo {author}
  {\bibfnamefont {M.}~\bibnamefont {Vanicat}},\ }\bibfield  {title} {\bibinfo
  {title} {Two-species hardcore reversible cellular automaton: matrix ansatz
  for dynamics and nonequilibrium stationary state},\ }\bibfield  {journal}
  {\bibinfo  {journal} {{SciPost} Physics}\ }\textbf {\bibinfo {volume} {6}},\
  \href {https://doi.org/10.21468/scipostphys.6.6.074}
  {10.21468/scipostphys.6.6.074} (\bibinfo {year} {2019})\BibitemShut {NoStop}%
\bibitem [{\citenamefont {Medenjak}(2022)}]{Medenjak_OTOC}%
  \BibitemOpen
  \bibfield  {author} {\bibinfo {author} {\bibfnamefont {M.}~\bibnamefont
  {Medenjak}},\ }\bibfield  {title} {\bibinfo {title} {Operator spreading in
  quantum hardcore gases},\ }\href {https://doi.org/10.1088/1751-8121/ac8fc4}
  {\bibfield  {journal} {\bibinfo  {journal} {Journal of Physics A:
  Mathematical and Theoretical}\ }\textbf {\bibinfo {volume} {55}},\ \bibinfo
  {pages} {404002} (\bibinfo {year} {2022})}\BibitemShut {NoStop}%
\bibitem [{\citenamefont {Gombor}\ and\ \citenamefont {Pozsgay}(2022)}]{GP22}%
  \BibitemOpen
  \bibfield  {author} {\bibinfo {author} {\bibfnamefont {T.}~\bibnamefont
  {Gombor}}\ and\ \bibinfo {author} {\bibfnamefont {B.}~\bibnamefont
  {Pozsgay}},\ }\bibfield  {title} {\bibinfo {title} {{Superintegrable cellular
  automata and dual unitary gates from Yang-Baxter maps}},\ }\bibfield
  {journal} {\bibinfo  {journal} {{SciPost} Physics}\ }\textbf {\bibinfo
  {volume} {12}},\ \href {https://doi.org/10.21468/scipostphys.12.3.102}
  {10.21468/scipostphys.12.3.102} (\bibinfo {year} {2022})\BibitemShut
  {NoStop}%
\bibitem [{\citenamefont {Arndt}\ \emph {et~al.}(1998)\citenamefont {Arndt},
  \citenamefont {Heinzel},\ and\ \citenamefont {Rittenberg}}]{AHR98}%
  \BibitemOpen
  \bibfield  {author} {\bibinfo {author} {\bibfnamefont {P.~F.}\ \bibnamefont
  {Arndt}}, \bibinfo {author} {\bibfnamefont {T.}~\bibnamefont {Heinzel}},\
  and\ \bibinfo {author} {\bibfnamefont {V.}~\bibnamefont {Rittenberg}},\
  }\bibfield  {title} {\bibinfo {title} {{Spontaneous breaking of translational
  invariance in one-dimensional stationary states on a ring}},\ }\href
  {https://doi.org/10.1088/0305-4470/31/2/001} {\bibfield  {journal} {\bibinfo
  {journal} {Journal of Physics A: Mathematical and General}\ }\textbf
  {\bibinfo {volume} {31}},\ \bibinfo {pages} {L45} (\bibinfo {year}
  {1998})}\BibitemShut {NoStop}%
\bibitem [{\citenamefont {Arratia}(1983)}]{Arratia83}%
  \BibitemOpen
  \bibfield  {author} {\bibinfo {author} {\bibfnamefont {R.}~\bibnamefont
  {Arratia}},\ }\bibfield  {title} {\bibinfo {title} {{The Motion of a Tagged
  Particle in the Simple Symmetric Exclusion System on $Z$}},\ }\href
  {https://doi.org/10.1214/aop/1176993602} {\bibfield  {journal} {\bibinfo
  {journal} {The Annals of Probability}\ }\textbf {\bibinfo {volume} {11}},\
  \bibinfo {pages} {362 } (\bibinfo {year} {1983})}\BibitemShut {NoStop}%
\bibitem [{\citenamefont {Helfand}(1960)}]{Helfand60}%
  \BibitemOpen
  \bibfield  {author} {\bibinfo {author} {\bibfnamefont {E.}~\bibnamefont
  {Helfand}},\ }\bibfield  {title} {\bibinfo {title} {{Transport Coefficients
  from Dissipation in a Canonical Ensemble}},\ }\href
  {https://doi.org/10.1103/physrev.119.1} {\bibfield  {journal} {\bibinfo
  {journal} {Physical Review}\ }\textbf {\bibinfo {volume} {119}},\ \bibinfo
  {pages} {1} (\bibinfo {year} {1960})}\BibitemShut {NoStop}%
\bibitem [{\citenamefont {Ellis}(2006)}]{Ellis_book}%
  \BibitemOpen
  \bibfield  {author} {\bibinfo {author} {\bibfnamefont {R.~S.}\ \bibnamefont
  {Ellis}},\ }\href {https://doi.org/10.1007/3-540-29060-5} {\emph {\bibinfo
  {title} {Entropy, Large Deviations, and Statistical Mechanics}}}\ (\bibinfo
  {publisher} {Springer Berlin Heidelberg},\ \bibinfo {year}
  {2006})\BibitemShut {NoStop}%
\bibitem [{\citenamefont {Touchette}(2009)}]{Touchette_LDT}%
  \BibitemOpen
  \bibfield  {author} {\bibinfo {author} {\bibfnamefont {H.}~\bibnamefont
  {Touchette}},\ }\bibfield  {title} {\bibinfo {title} {The large deviation
  approach to statistical mechanics},\ }\href
  {https://doi.org/10.1016/j.physrep.2009.05.002} {\bibfield  {journal}
  {\bibinfo  {journal} {Physics Reports}\ }\textbf {\bibinfo {volume} {478}},\
  \bibinfo {pages} {1} (\bibinfo {year} {2009})}\BibitemShut {NoStop}%
\bibitem [{\citenamefont {Esposito}\ \emph {et~al.}(2009)\citenamefont
  {Esposito}, \citenamefont {Harbola},\ and\ \citenamefont
  {Mukamel}}]{Esposito_review}%
  \BibitemOpen
  \bibfield  {author} {\bibinfo {author} {\bibfnamefont {M.}~\bibnamefont
  {Esposito}}, \bibinfo {author} {\bibfnamefont {U.}~\bibnamefont {Harbola}},\
  and\ \bibinfo {author} {\bibfnamefont {S.}~\bibnamefont {Mukamel}},\
  }\bibfield  {title} {\bibinfo {title} {Nonequilibrium fluctuations,
  fluctuation theorems, and counting statistics in quantum systems},\ }\href
  {https://doi.org/10.1103/revmodphys.81.1665} {\bibfield  {journal} {\bibinfo
  {journal} {Reviews of Modern Physics}\ }\textbf {\bibinfo {volume} {81}},\
  \bibinfo {pages} {1665} (\bibinfo {year} {2009})}\BibitemShut {NoStop}%
\bibitem [{\citenamefont {Derrida}\ and\ \citenamefont
  {Gerschenfeld}(2009)}]{Derrida09}%
  \BibitemOpen
  \bibfield  {author} {\bibinfo {author} {\bibfnamefont {B.}~\bibnamefont
  {Derrida}}\ and\ \bibinfo {author} {\bibfnamefont {A.}~\bibnamefont
  {Gerschenfeld}},\ }\bibfield  {title} {\bibinfo {title} {{Current
  Fluctuations in One Dimensional Diffusive Systems with a Step Initial Density
  Profile}},\ }\href {https://doi.org/10.1007/s10955-009-9830-1} {\bibfield
  {journal} {\bibinfo  {journal} {Journal of Statistical Physics}\ }\textbf
  {\bibinfo {volume} {137}},\ \bibinfo {pages} {978} (\bibinfo {year}
  {2009})}\BibitemShut {NoStop}%
\bibitem [{\citenamefont {Evans}\ \emph {et~al.}(1993)\citenamefont {Evans},
  \citenamefont {Cohen},\ and\ \citenamefont {Morriss}}]{ECM93}%
  \BibitemOpen
  \bibfield  {author} {\bibinfo {author} {\bibfnamefont {D.~J.}\ \bibnamefont
  {Evans}}, \bibinfo {author} {\bibfnamefont {E.~G.~D.}\ \bibnamefont
  {Cohen}},\ and\ \bibinfo {author} {\bibfnamefont {G.~P.}\ \bibnamefont
  {Morriss}},\ }\bibfield  {title} {\bibinfo {title} {{Probability of Second
  Law Violations in Shearing Steady States}},\ }\href
  {https://doi.org/10.1103/physrevlett.71.3616} {\bibfield  {journal} {\bibinfo
   {journal} {Physical Review Letters}\ }\textbf {\bibinfo {volume} {71}},\
  \bibinfo {pages} {3616} (\bibinfo {year} {1993})}\BibitemShut {NoStop}%
\bibitem [{\citenamefont {Gallavotti}\ and\ \citenamefont
  {Cohen}(1995)}]{GC95}%
  \BibitemOpen
  \bibfield  {author} {\bibinfo {author} {\bibfnamefont {G.}~\bibnamefont
  {Gallavotti}}\ and\ \bibinfo {author} {\bibfnamefont {E.~G.~D.}\ \bibnamefont
  {Cohen}},\ }\bibfield  {title} {\bibinfo {title} {{Dynamical Ensembles in
  Nonequilibrium Statistical Mechanics}},\ }\href
  {https://doi.org/10.1103/physrevlett.74.2694} {\bibfield  {journal} {\bibinfo
   {journal} {Physical Review Letters}\ }\textbf {\bibinfo {volume} {74}},\
  \bibinfo {pages} {2694} (\bibinfo {year} {1995})}\BibitemShut {NoStop}%
\bibitem [{\citenamefont {Kurchan}(1998)}]{Kurchan98}%
  \BibitemOpen
  \bibfield  {author} {\bibinfo {author} {\bibfnamefont {J.}~\bibnamefont
  {Kurchan}},\ }\bibfield  {title} {\bibinfo {title} {{Fluctuation theorem for
  stochastic dynamics}},\ }\href {https://doi.org/10.1088/0305-4470/31/16/003}
  {\bibfield  {journal} {\bibinfo  {journal} {Journal of Physics A:
  Mathematical and General}\ }\textbf {\bibinfo {volume} {31}},\ \bibinfo
  {pages} {3719} (\bibinfo {year} {1998})}\BibitemShut {NoStop}%
\bibitem [{\citenamefont {Lebowitz}\ and\ \citenamefont {Spohn}(1999)}]{LS99}%
  \BibitemOpen
  \bibfield  {author} {\bibinfo {author} {\bibfnamefont {J.~L.}\ \bibnamefont
  {Lebowitz}}\ and\ \bibinfo {author} {\bibfnamefont {H.}~\bibnamefont
  {Spohn}},\ }\bibfield  {title} {\bibinfo {title} {{A Gallavotti–Cohen-type
  symmetry in the large deviation functional for stochastic dynamics}},\ }\href
  {https://doi.org/10.1023/a:1004589714161} {\bibfield  {journal} {\bibinfo
  {journal} {Journal of Statistical Physics}\ }\textbf {\bibinfo {volume}
  {95}},\ \bibinfo {pages} {333} (\bibinfo {year} {1999})}\BibitemShut
  {NoStop}%
\bibitem [{\citenamefont {Hurtado}\ \emph {et~al.}(2011)\citenamefont
  {Hurtado}, \citenamefont {P{\'{e}}rez-Espigares}, \citenamefont {del Pozo},\
  and\ \citenamefont {Garrido}}]{Hurtado11}%
  \BibitemOpen
  \bibfield  {author} {\bibinfo {author} {\bibfnamefont {P.~I.}\ \bibnamefont
  {Hurtado}}, \bibinfo {author} {\bibfnamefont {C.}~\bibnamefont
  {P{\'{e}}rez-Espigares}}, \bibinfo {author} {\bibfnamefont {J.~J.}\
  \bibnamefont {del Pozo}},\ and\ \bibinfo {author} {\bibfnamefont {P.~L.}\
  \bibnamefont {Garrido}},\ }\bibfield  {title} {\bibinfo {title} {{Symmetries
  in fluctuations far from equilibrium}},\ }\href
  {https://doi.org/10.1073/pnas.1013209108} {\bibfield  {journal} {\bibinfo
  {journal} {Proceedings of the National Academy of Sciences}\ }\textbf
  {\bibinfo {volume} {108}},\ \bibinfo {pages} {7704} (\bibinfo {year}
  {2011})}\BibitemShut {NoStop}%
\bibitem [{\citenamefont {Maes}(1999)}]{Maes99}%
  \BibitemOpen
  \bibfield  {author} {\bibinfo {author} {\bibfnamefont {C.}~\bibnamefont
  {Maes}},\ }\bibfield  {title} {\bibinfo {title} {{The fluctuation theorem as
  a Gibbs property}},\ }\href {https://doi.org/10.1023/a:1004541830999}
  {\bibfield  {journal} {\bibinfo  {journal} {Journal of Statistical Physics}\
  }\textbf {\bibinfo {volume} {95}},\ \bibinfo {pages} {367} (\bibinfo {year}
  {1999})}\BibitemShut {NoStop}%
\bibitem [{\citenamefont {Andrieux}\ and\ \citenamefont
  {Gaspard}(2006)}]{Andrieux06}%
  \BibitemOpen
  \bibfield  {author} {\bibinfo {author} {\bibfnamefont {D.}~\bibnamefont
  {Andrieux}}\ and\ \bibinfo {author} {\bibfnamefont {P.}~\bibnamefont
  {Gaspard}},\ }\bibfield  {title} {\bibinfo {title} {Fluctuation theorem for
  transport in mesoscopic systems},\ }\href
  {https://doi.org/10.1088/1742-5468/2006/01/p01011} {\bibfield  {journal}
  {\bibinfo  {journal} {Journal of Statistical Mechanics: Theory and
  Experiment}\ }\textbf {\bibinfo {volume} {2006}},\ \bibinfo {pages} {P01011}
  (\bibinfo {year} {2006})}\BibitemShut {NoStop}%
\bibitem [{\citenamefont {Andrieux}\ and\ \citenamefont
  {Gaspard}(2008)}]{Andrieux08}%
  \BibitemOpen
  \bibfield  {author} {\bibinfo {author} {\bibfnamefont {D.}~\bibnamefont
  {Andrieux}}\ and\ \bibinfo {author} {\bibfnamefont {P.}~\bibnamefont
  {Gaspard}},\ }\bibfield  {title} {\bibinfo {title} {{The fluctuation theorem
  for currents in semi-Markov processes}},\ }\href
  {https://doi.org/10.1088/1742-5468/2008/11/p11007} {\bibfield  {journal}
  {\bibinfo  {journal} {Journal of Statistical Mechanics: Theory and
  Experiment}\ }\textbf {\bibinfo {volume} {2008}},\ \bibinfo {pages} {P11007}
  (\bibinfo {year} {2008})}\BibitemShut {NoStop}%
\bibitem [{\citenamefont {Gaspard}(2013)}]{Gaspard13}%
  \BibitemOpen
  \bibfield  {author} {\bibinfo {author} {\bibfnamefont {P.}~\bibnamefont
  {Gaspard}},\ }\bibfield  {title} {\bibinfo {title} {{Multivariate fluctuation
  relations for currents}},\ }\href
  {https://doi.org/10.1088/1367-2630/15/11/115014} {\bibfield  {journal}
  {\bibinfo  {journal} {New Journal of Physics}\ }\textbf {\bibinfo {volume}
  {15}},\ \bibinfo {pages} {115014} (\bibinfo {year} {2013})}\BibitemShut
  {NoStop}%
\bibitem [{\citenamefont {Bodineau}\ and\ \citenamefont
  {Derrida}(2004)}]{BodineauDerrida04}%
  \BibitemOpen
  \bibfield  {author} {\bibinfo {author} {\bibfnamefont {T.}~\bibnamefont
  {Bodineau}}\ and\ \bibinfo {author} {\bibfnamefont {B.}~\bibnamefont
  {Derrida}},\ }\bibfield  {title} {\bibinfo {title} {{Current Fluctuations in
  Nonequilibrium Diffusive Systems: An Additivity Principle}},\ }\bibfield
  {journal} {\bibinfo  {journal} {Physical Review Letters}\ }\textbf {\bibinfo
  {volume} {92}},\ \href {https://doi.org/10.1103/physrevlett.92.180601}
  {10.1103/physrevlett.92.180601} (\bibinfo {year} {2004})\BibitemShut
  {NoStop}%
\bibitem [{\citenamefont {Derrida}(2007)}]{Derrida07}%
  \BibitemOpen
  \bibfield  {author} {\bibinfo {author} {\bibfnamefont {B.}~\bibnamefont
  {Derrida}},\ }\bibfield  {title} {\bibinfo {title} {Non-equilibrium steady
  states: fluctuations and large deviations of the density and of the
  current},\ }\href {https://doi.org/10.1088/1742-5468/2007/07/p07023}
  {\bibfield  {journal} {\bibinfo  {journal} {Journal of Statistical Mechanics:
  Theory and Experiment}\ }\textbf {\bibinfo {volume} {2007}},\ \bibinfo
  {pages} {P07023} (\bibinfo {year} {2007})}\BibitemShut {NoStop}%
\bibitem [{\citenamefont {Cuetara}\ \emph {et~al.}(2013)\citenamefont
  {Cuetara}, \citenamefont {Esposito}, \citenamefont {Schaller},\ and\
  \citenamefont {Gaspard}}]{BulnesCuetara13}%
  \BibitemOpen
  \bibfield  {author} {\bibinfo {author} {\bibfnamefont {G.~B.}\ \bibnamefont
  {Cuetara}}, \bibinfo {author} {\bibfnamefont {M.}~\bibnamefont {Esposito}},
  \bibinfo {author} {\bibfnamefont {G.}~\bibnamefont {Schaller}},\ and\
  \bibinfo {author} {\bibfnamefont {P.}~\bibnamefont {Gaspard}},\ }\bibfield
  {title} {\bibinfo {title} {Effective fluctuation theorems for electron
  transport in a double quantum dot coupled to a quantum point contact},\
  }\bibfield  {journal} {\bibinfo  {journal} {Physical Review B}\ }\textbf
  {\bibinfo {volume} {88}},\ \href {https://doi.org/10.1103/physrevb.88.115134}
  {10.1103/physrevb.88.115134} (\bibinfo {year} {2013})\BibitemShut {NoStop}%
\bibitem [{\citenamefont {Gaspard}\ and\ \citenamefont
  {Gerritsma}(2007)}]{GG07}%
  \BibitemOpen
  \bibfield  {author} {\bibinfo {author} {\bibfnamefont {P.}~\bibnamefont
  {Gaspard}}\ and\ \bibinfo {author} {\bibfnamefont {E.}~\bibnamefont
  {Gerritsma}},\ }\bibfield  {title} {\bibinfo {title} {{The stochastic
  chemomechanics of the ATPase molecular motor}},\ }\href
  {https://doi.org/10.1016/j.jtbi.2007.03.034} {\bibfield  {journal} {\bibinfo
  {journal} {Journal of Theoretical Biology}\ }\textbf {\bibinfo {volume}
  {247}},\ \bibinfo {pages} {672} (\bibinfo {year} {2007})}\BibitemShut
  {NoStop}%
\bibitem [{\citenamefont {{De Nardis}}\ \emph {et~al.}(2018)\citenamefont {{De
  Nardis}}, \citenamefont {Bernard},\ and\ \citenamefont {Doyon}}]{DeNardis18}%
  \BibitemOpen
  \bibfield  {author} {\bibinfo {author} {\bibfnamefont {J.}~\bibnamefont {{De
  Nardis}}}, \bibinfo {author} {\bibfnamefont {D.}~\bibnamefont {Bernard}},\
  and\ \bibinfo {author} {\bibfnamefont {B.}~\bibnamefont {Doyon}},\ }\bibfield
   {title} {\bibinfo {title} {Hydrodynamic diffusion in integrable systems},\
  }\bibfield  {journal} {\bibinfo  {journal} {Physical Review Letters}\
  }\textbf {\bibinfo {volume} {121}},\ \href
  {https://doi.org/10.1103/physrevlett.121.160603}
  {10.1103/physrevlett.121.160603} (\bibinfo {year} {2018})\BibitemShut
  {NoStop}%
\bibitem [{\citenamefont {Nardis}\ \emph {et~al.}(2022)\citenamefont {Nardis},
  \citenamefont {Doyon}, \citenamefont {Medenjak},\ and\ \citenamefont
  {Panfil}}]{GHD_review}%
  \BibitemOpen
  \bibfield  {author} {\bibinfo {author} {\bibfnamefont {J.~D.}\ \bibnamefont
  {Nardis}}, \bibinfo {author} {\bibfnamefont {B.}~\bibnamefont {Doyon}},
  \bibinfo {author} {\bibfnamefont {M.}~\bibnamefont {Medenjak}},\ and\
  \bibinfo {author} {\bibfnamefont {M.}~\bibnamefont {Panfil}},\ }\bibfield
  {title} {\bibinfo {title} {Correlation functions and transport coefficients
  in generalised hydrodynamics},\ }\href
  {https://doi.org/10.1088/1742-5468/ac3658} {\bibfield  {journal} {\bibinfo
  {journal} {Journal of Statistical Mechanics: Theory and Experiment}\ }\textbf
  {\bibinfo {volume} {2022}},\ \bibinfo {pages} {014002} (\bibinfo {year}
  {2022})}\BibitemShut {NoStop}%
\bibitem [{\citenamefont {Biskup}\ \emph {et~al.}(2000)\citenamefont {Biskup},
  \citenamefont {Borgs}, \citenamefont {Chayes}, \citenamefont {Kleinwaks},\
  and\ \citenamefont {Koteck{\'{y}}}}]{Biskup00}%
  \BibitemOpen
  \bibfield  {author} {\bibinfo {author} {\bibfnamefont {M.}~\bibnamefont
  {Biskup}}, \bibinfo {author} {\bibfnamefont {C.}~\bibnamefont {Borgs}},
  \bibinfo {author} {\bibfnamefont {J.~T.}\ \bibnamefont {Chayes}}, \bibinfo
  {author} {\bibfnamefont {L.~J.}\ \bibnamefont {Kleinwaks}},\ and\ \bibinfo
  {author} {\bibfnamefont {R.}~\bibnamefont {Koteck{\'{y}}}},\ }\bibfield
  {title} {\bibinfo {title} {{General Theory of Lee-Yang Zeros in Models with
  First-Order Phase Transitions}},\ }\href
  {https://doi.org/10.1103/physrevlett.84.4794} {\bibfield  {journal} {\bibinfo
   {journal} {Physical Review Letters}\ }\textbf {\bibinfo {volume} {84}},\
  \bibinfo {pages} {4794} (\bibinfo {year} {2000})}\BibitemShut {NoStop}%
\bibitem [{\citenamefont {Costeniuc}\ \emph {et~al.}(2005)\citenamefont
  {Costeniuc}, \citenamefont {Ellis},\ and\ \citenamefont
  {Touchette}}]{Costeniuc05}%
  \BibitemOpen
  \bibfield  {author} {\bibinfo {author} {\bibfnamefont {M.}~\bibnamefont
  {Costeniuc}}, \bibinfo {author} {\bibfnamefont {R.~S.}\ \bibnamefont
  {Ellis}},\ and\ \bibinfo {author} {\bibfnamefont {H.}~\bibnamefont
  {Touchette}},\ }\bibfield  {title} {\bibinfo {title} {{Complete analysis of
  phase transitions and ensemble equivalence for the
  Curie{\textendash}Weiss{\textendash}Potts model}},\ }\href
  {https://doi.org/10.1063/1.1904507} {\bibfield  {journal} {\bibinfo
  {journal} {Journal of Mathematical Physics}\ }\textbf {\bibinfo {volume}
  {46}},\ \bibinfo {pages} {063301} (\bibinfo {year} {2005})}\BibitemShut
  {NoStop}%
\bibitem [{\citenamefont {Costeniuc}\ \emph {et~al.}(2006)\citenamefont
  {Costeniuc}, \citenamefont {Ellis}, \citenamefont {Touchette},\ and\
  \citenamefont {Turkington}}]{Costeniuc06}%
  \BibitemOpen
  \bibfield  {author} {\bibinfo {author} {\bibfnamefont {M.}~\bibnamefont
  {Costeniuc}}, \bibinfo {author} {\bibfnamefont {R.~S.}\ \bibnamefont
  {Ellis}}, \bibinfo {author} {\bibfnamefont {H.}~\bibnamefont {Touchette}},\
  and\ \bibinfo {author} {\bibfnamefont {B.}~\bibnamefont {Turkington}},\
  }\bibfield  {title} {\bibinfo {title} {{Generalized canonical ensembles and
  ensemble equivalence}},\ }\bibfield  {journal} {\bibinfo  {journal} {Physical
  Review E}\ }\textbf {\bibinfo {volume} {73}},\ \href
  {https://doi.org/10.1103/physreve.73.026105} {10.1103/physreve.73.026105}
  (\bibinfo {year} {2006})\BibitemShut {NoStop}%
\bibitem [{\citenamefont {Baek}\ \emph {et~al.}(2017)\citenamefont {Baek},
  \citenamefont {Kafri},\ and\ \citenamefont {Lecomte}}]{Baek17}%
  \BibitemOpen
  \bibfield  {author} {\bibinfo {author} {\bibfnamefont {Y.}~\bibnamefont
  {Baek}}, \bibinfo {author} {\bibfnamefont {Y.}~\bibnamefont {Kafri}},\ and\
  \bibinfo {author} {\bibfnamefont {V.}~\bibnamefont {Lecomte}},\ }\bibfield
  {title} {\bibinfo {title} {{Dynamical Symmetry Breaking and Phase Transitions
  in Driven Diffusive Systems}},\ }\bibfield  {journal} {\bibinfo  {journal}
  {Physical Review Letters}\ }\textbf {\bibinfo {volume} {118}},\ \href
  {https://doi.org/10.1103/physrevlett.118.030604}
  {10.1103/physrevlett.118.030604} (\bibinfo {year} {2017})\BibitemShut
  {NoStop}%
\bibitem [{\citenamefont {Baek}\ \emph {et~al.}(2018)\citenamefont {Baek},
  \citenamefont {Kafri},\ and\ \citenamefont {Lecomte}}]{Baek18}%
  \BibitemOpen
  \bibfield  {author} {\bibinfo {author} {\bibfnamefont {Y.}~\bibnamefont
  {Baek}}, \bibinfo {author} {\bibfnamefont {Y.}~\bibnamefont {Kafri}},\ and\
  \bibinfo {author} {\bibfnamefont {V.}~\bibnamefont {Lecomte}},\ }\bibfield
  {title} {\bibinfo {title} {{Dynamical phase transitions in the current
  distribution of driven diffusive channels}},\ }\href
  {https://doi.org/10.1088/1751-8121/aaa8f9} {\bibfield  {journal} {\bibinfo
  {journal} {Journal of Physics A: Mathematical and Theoretical}\ }\textbf
  {\bibinfo {volume} {51}},\ \bibinfo {pages} {105001} (\bibinfo {year}
  {2018})}\BibitemShut {NoStop}%
\bibitem [{\citenamefont {Chaganty}(1997)}]{Chaganty97}%
  \BibitemOpen
  \bibfield  {author} {\bibinfo {author} {\bibfnamefont {N.~R.}\ \bibnamefont
  {Chaganty}},\ }\bibfield  {title} {\bibinfo {title} {{Large Deviations for
  Joint Distributions and Statistical Applications}},\ }\href
  {http://www.jstor.org/stable/25051147} {\bibfield  {journal} {\bibinfo
  {journal} {Sankhyā: The Indian Journal of Statistics, Series A (1961-2002)}\
  }\textbf {\bibinfo {volume} {59}},\ \bibinfo {pages} {147} (\bibinfo {year}
  {1997})}\BibitemShut {NoStop}%
\bibitem [{\citenamefont {Haubold}\ \emph {et~al.}(2011)\citenamefont
  {Haubold}, \citenamefont {Mathai},\ and\ \citenamefont {Saxena}}]{Haubold11}%
  \BibitemOpen
  \bibfield  {author} {\bibinfo {author} {\bibfnamefont {H.~J.}\ \bibnamefont
  {Haubold}}, \bibinfo {author} {\bibfnamefont {A.~M.}\ \bibnamefont
  {Mathai}},\ and\ \bibinfo {author} {\bibfnamefont {R.~K.}\ \bibnamefont
  {Saxena}},\ }\bibfield  {title} {\bibinfo {title} {{Mittag-Leffler Functions
  and Their Applications}},\ }\href {https://doi.org/10.1155/2011/298628}
  {\bibfield  {journal} {\bibinfo  {journal} {Journal of Applied Mathematics}\
  }\textbf {\bibinfo {volume} {2011}},\ \bibinfo {pages} {1} (\bibinfo {year}
  {2011})}\BibitemShut {NoStop}%
\bibitem [{\citenamefont {Mainardi}(1997)}]{Mainardi_book}%
  \BibitemOpen
  \bibfield  {author} {\bibinfo {author} {\bibfnamefont {F.}~\bibnamefont
  {Mainardi}},\ }\bibfield  {title} {\bibinfo {title} {{Fractional Calculus}},\
  }in\ \href {https://doi.org/10.1007/978-3-7091-2664-6_7} {\emph {\bibinfo
  {booktitle} {Fractals and Fractional Calculus in Continuum Mechanics}}}\
  (\bibinfo  {publisher} {Springer Vienna},\ \bibinfo {year} {1997})\ pp.\
  \bibinfo {pages} {291--348}\BibitemShut {NoStop}%
\bibitem [{\citenamefont {Gorenflo}\ \emph {et~al.}(2020)\citenamefont
  {Gorenflo}, \citenamefont {Kilbas}, \citenamefont {Mainardi},\ and\
  \citenamefont {Rogosin}}]{Gorenflo_book}%
  \BibitemOpen
  \bibfield  {author} {\bibinfo {author} {\bibfnamefont {R.}~\bibnamefont
  {Gorenflo}}, \bibinfo {author} {\bibfnamefont {A.~A.}\ \bibnamefont
  {Kilbas}}, \bibinfo {author} {\bibfnamefont {F.}~\bibnamefont {Mainardi}},\
  and\ \bibinfo {author} {\bibfnamefont {S.}~\bibnamefont {Rogosin}},\ }\href
  {https://doi.org/10.1007/978-3-662-61550-8} {\emph {\bibinfo {title}
  {{Mittag-Leffler Functions, Related Topics and Applications}}}}\ (\bibinfo
  {publisher} {Springer Berlin Heidelberg},\ \bibinfo {year}
  {2020})\BibitemShut {NoStop}%
\bibitem [{\citenamefont {Mainardi}\ and\ \citenamefont
  {Consiglio}(2020)}]{Mainardi20}%
  \BibitemOpen
  \bibfield  {author} {\bibinfo {author} {\bibfnamefont {F.}~\bibnamefont
  {Mainardi}}\ and\ \bibinfo {author} {\bibfnamefont {A.}~\bibnamefont
  {Consiglio}},\ }\bibfield  {title} {\bibinfo {title} {{The Wright Functions
  of the Second Kind in Mathematical Physics}},\ }\href
  {https://doi.org/10.3390/math8060884} {\bibfield  {journal} {\bibinfo
  {journal} {Mathematics}\ }\textbf {\bibinfo {volume} {8}},\ \bibinfo {pages}
  {884} (\bibinfo {year} {2020})}\BibitemShut {NoStop}%
\bibitem [{\citenamefont {Barkai}\ \emph {et~al.}(2000)\citenamefont {Barkai},
  \citenamefont {Metzler},\ and\ \citenamefont {Klafter}}]{PhysRevE.61.132}%
  \BibitemOpen
  \bibfield  {author} {\bibinfo {author} {\bibfnamefont {E.}~\bibnamefont
  {Barkai}}, \bibinfo {author} {\bibfnamefont {R.}~\bibnamefont {Metzler}},\
  and\ \bibinfo {author} {\bibfnamefont {J.}~\bibnamefont {Klafter}},\
  }\bibfield  {title} {\bibinfo {title} {From continuous time random walks to
  the fractional fokker-planck equation},\ }\href
  {https://doi.org/10.1103/PhysRevE.61.132} {\bibfield  {journal} {\bibinfo
  {journal} {Phys. Rev. E}\ }\textbf {\bibinfo {volume} {61}},\ \bibinfo
  {pages} {132} (\bibinfo {year} {2000})}\BibitemShut {NoStop}%
\bibitem [{\citenamefont {Barkai}(2002)}]{Barkai2002}%
  \BibitemOpen
  \bibfield  {author} {\bibinfo {author} {\bibfnamefont {E.}~\bibnamefont
  {Barkai}},\ }\bibfield  {title} {\bibinfo {title} {Ctrw pathways to the
  fractional diffusion equation},\ }\href
  {https://doi.org/https://doi.org/10.1016/S0301-0104(02)00533-5} {\bibfield
  {journal} {\bibinfo  {journal} {Chemical Physics}\ }\textbf {\bibinfo
  {volume} {284}},\ \bibinfo {pages} {13} (\bibinfo {year} {2002})}\BibitemShut
  {NoStop}%
\bibitem [{\citenamefont {Touchette}\ \emph {et~al.}(2010)\citenamefont
  {Touchette}, \citenamefont {Harris},\ and\ \citenamefont {Tailleur}}]{THT10}%
  \BibitemOpen
  \bibfield  {author} {\bibinfo {author} {\bibfnamefont {H.}~\bibnamefont
  {Touchette}}, \bibinfo {author} {\bibfnamefont {R.~J.}\ \bibnamefont
  {Harris}},\ and\ \bibinfo {author} {\bibfnamefont {J.}~\bibnamefont
  {Tailleur}},\ }\bibfield  {title} {\bibinfo {title} {First-order phase
  transitions from poles in asymptotic representations of partition
  functions},\ }\bibfield  {journal} {\bibinfo  {journal} {Physical Review E}\
  }\textbf {\bibinfo {volume} {81}},\ \href
  {https://doi.org/10.1103/physreve.81.030101} {10.1103/physreve.81.030101}
  (\bibinfo {year} {2010})\BibitemShut {NoStop}%
\bibitem [{\citenamefont {Dinwoodie}\ and\ \citenamefont
  {Zabell}(1992)}]{Dinwoodie92}%
  \BibitemOpen
  \bibfield  {author} {\bibinfo {author} {\bibfnamefont {I.~H.}\ \bibnamefont
  {Dinwoodie}}\ and\ \bibinfo {author} {\bibfnamefont {S.~L.}\ \bibnamefont
  {Zabell}},\ }\bibfield  {title} {\bibinfo {title} {{Large Deviations for
  Exchangeable Random Vectors}},\ }\href
  {https://doi.org/10.1214/aop/1176989683} {\bibfield  {journal} {\bibinfo
  {journal} {The Annals of Probability}\ }\textbf {\bibinfo {volume} {20}},\
  \bibinfo {pages} {1147 } (\bibinfo {year} {1992})}\BibitemShut {NoStop}%
\bibitem [{\citenamefont {Lesovik}\ and\ \citenamefont {Levitov}(1994)}]{LL94}%
  \BibitemOpen
  \bibfield  {author} {\bibinfo {author} {\bibfnamefont {G.~B.}\ \bibnamefont
  {Lesovik}}\ and\ \bibinfo {author} {\bibfnamefont {L.~S.}\ \bibnamefont
  {Levitov}},\ }\bibfield  {title} {\bibinfo {title} {{Noise in an ac biased
  junction: Nonstationary Aharonov-Bohm effect}},\ }\href
  {https://doi.org/10.1103/physrevlett.72.538} {\bibfield  {journal} {\bibinfo
  {journal} {Physical Review Letters}\ }\textbf {\bibinfo {volume} {72}},\
  \bibinfo {pages} {538} (\bibinfo {year} {1994})}\BibitemShut {NoStop}%
\bibitem [{\citenamefont {Levitov}\ \emph {et~al.}(1996)\citenamefont
  {Levitov}, \citenamefont {Lee},\ and\ \citenamefont {Lesovik}}]{LLL96}%
  \BibitemOpen
  \bibfield  {author} {\bibinfo {author} {\bibfnamefont {L.~S.}\ \bibnamefont
  {Levitov}}, \bibinfo {author} {\bibfnamefont {H.}~\bibnamefont {Lee}},\ and\
  \bibinfo {author} {\bibfnamefont {G.~B.}\ \bibnamefont {Lesovik}},\
  }\bibfield  {title} {\bibinfo {title} {Electron counting statistics and
  coherent states of electric current},\ }\href
  {https://doi.org/10.1063/1.531672} {\bibfield  {journal} {\bibinfo  {journal}
  {Journal of Mathematical Physics}\ }\textbf {\bibinfo {volume} {37}},\
  \bibinfo {pages} {4845} (\bibinfo {year} {1996})}\BibitemShut {NoStop}%
\bibitem [{\citenamefont {Bernard}\ and\ \citenamefont {Doyon}(2013)}]{BD13}%
  \BibitemOpen
  \bibfield  {author} {\bibinfo {author} {\bibfnamefont {D.}~\bibnamefont
  {Bernard}}\ and\ \bibinfo {author} {\bibfnamefont {B.}~\bibnamefont
  {Doyon}},\ }\bibfield  {title} {\bibinfo {title} {Time-reversal symmetry and
  fluctuation relations in non-equilibrium quantum steady states},\ }\href
  {https://doi.org/10.1088/1751-8113/46/37/372001} {\bibfield  {journal}
  {\bibinfo  {journal} {Journal of Physics A: Mathematical and Theoretical}\
  }\textbf {\bibinfo {volume} {46}},\ \bibinfo {pages} {372001} (\bibinfo
  {year} {2013})}\BibitemShut {NoStop}%
\bibitem [{\citenamefont {Bernard}\ and\ \citenamefont {Doyon}(2014)}]{BD15}%
  \BibitemOpen
  \bibfield  {author} {\bibinfo {author} {\bibfnamefont {D.}~\bibnamefont
  {Bernard}}\ and\ \bibinfo {author} {\bibfnamefont {B.}~\bibnamefont
  {Doyon}},\ }\bibfield  {title} {\bibinfo {title} {{Non-Equilibrium Steady
  States in Conformal Field Theory}},\ }\href
  {https://doi.org/10.1007/s00023-014-0314-8} {\bibfield  {journal} {\bibinfo
  {journal} {Annales Henri Poincar{\'{e}}}\ }\textbf {\bibinfo {volume} {16}},\
  \bibinfo {pages} {113} (\bibinfo {year} {2014})}\BibitemShut {NoStop}%
\bibitem [{\citenamefont {Yang}\ and\ \citenamefont {Lee}(1952)}]{LeeYang52}%
  \BibitemOpen
  \bibfield  {author} {\bibinfo {author} {\bibfnamefont {C.~N.}\ \bibnamefont
  {Yang}}\ and\ \bibinfo {author} {\bibfnamefont {T.~D.}\ \bibnamefont {Lee}},\
  }\bibfield  {title} {\bibinfo {title} {{Statistical Theory of Equations of
  State and Phase Transitions. I. Theory of Condensation}},\ }\href
  {https://doi.org/10.1103/physrev.87.404} {\bibfield  {journal} {\bibinfo
  {journal} {Physical Review}\ }\textbf {\bibinfo {volume} {87}},\ \bibinfo
  {pages} {404} (\bibinfo {year} {1952})}\BibitemShut {NoStop}%
\bibitem [{\citenamefont {Suzuki}(1967)}]{Suzuki67}%
  \BibitemOpen
  \bibfield  {author} {\bibinfo {author} {\bibfnamefont {M.}~\bibnamefont
  {Suzuki}},\ }\bibfield  {title} {\bibinfo {title} {{A Theory of the Second
  Order Phase Transitions in Spin Systems. {II}}},\ }\href
  {https://doi.org/10.1143/ptp.38.1225} {\bibfield  {journal} {\bibinfo
  {journal} {Progress of Theoretical Physics}\ }\textbf {\bibinfo {volume}
  {38}},\ \bibinfo {pages} {1225} (\bibinfo {year} {1967})}\BibitemShut
  {NoStop}%
\bibitem [{\citenamefont {Janke}\ and\ \citenamefont
  {Kenna}(2002)}]{JankeKenna02}%
  \BibitemOpen
  \bibfield  {author} {\bibinfo {author} {\bibfnamefont {W.}~\bibnamefont
  {Janke}}\ and\ \bibinfo {author} {\bibfnamefont {R.}~\bibnamefont {Kenna}},\
  }\bibfield  {title} {\bibinfo {title} {{Phase transition strengths from the
  density of partition function zeroes}},\ }\href
  {https://doi.org/10.1016/s0920-5632(01)01881-3} {\bibfield  {journal}
  {\bibinfo  {journal} {Nuclear Physics B - Proceedings Supplements}\ }\textbf
  {\bibinfo {volume} {106-107}},\ \bibinfo {pages} {905} (\bibinfo {year}
  {2002})}\BibitemShut {NoStop}%
\bibitem [{\citenamefont {Bena}\ \emph {et~al.}(2005)\citenamefont {Bena},
  \citenamefont {Droz},\ and\ \citenamefont {Lipowski}}]{Bena05}%
  \BibitemOpen
  \bibfield  {author} {\bibinfo {author} {\bibfnamefont {I.}~\bibnamefont
  {Bena}}, \bibinfo {author} {\bibfnamefont {M.}~\bibnamefont {Droz}},\ and\
  \bibinfo {author} {\bibfnamefont {A.}~\bibnamefont {Lipowski}},\ }\bibfield
  {title} {\bibinfo {title} {{Statistical Mechanics of Equilibrium and
  Nonequilibrium Phase Transitions: the Yang{\textendash}Lee Formalism}},\
  }\href {https://doi.org/10.1142/s0217979205032759} {\bibfield  {journal}
  {\bibinfo  {journal} {International Journal of Modern Physics B}\ }\textbf
  {\bibinfo {volume} {19}},\ \bibinfo {pages} {4269} (\bibinfo {year}
  {2005})}\BibitemShut {NoStop}%
\bibitem [{\citenamefont {Blythe}\ and\ \citenamefont
  {Evans}(2003)}]{Blythe03}%
  \BibitemOpen
  \bibfield  {author} {\bibinfo {author} {\bibfnamefont {R.}~\bibnamefont
  {Blythe}}\ and\ \bibinfo {author} {\bibfnamefont {M.}~\bibnamefont {Evans}},\
  }\bibfield  {title} {\bibinfo {title} {{The Lee-Yang theory of equilibrium
  and nonequilibrium phase transitions}},\ }\href
  {https://doi.org/10.1590/s0103-97332003000300008} {\bibfield  {journal}
  {\bibinfo  {journal} {Brazilian Journal of Physics}\ }\textbf {\bibinfo
  {volume} {33}},\ \bibinfo {pages} {464} (\bibinfo {year} {2003})}\BibitemShut
  {NoStop}%
\bibitem [{\citenamefont {Itzykson}\ \emph {et~al.}(1983)\citenamefont
  {Itzykson}, \citenamefont {Pearson},\ and\ \citenamefont
  {Zuber}}]{Itzykson83}%
  \BibitemOpen
  \bibfield  {author} {\bibinfo {author} {\bibfnamefont {C.}~\bibnamefont
  {Itzykson}}, \bibinfo {author} {\bibfnamefont {R.}~\bibnamefont {Pearson}},\
  and\ \bibinfo {author} {\bibfnamefont {J.}~\bibnamefont {Zuber}},\ }\bibfield
   {title} {\bibinfo {title} {{Distribution of zeros in Ising and gauge
  models}},\ }\href {https://doi.org/10.1016/0550-3213(83)90499-6} {\bibfield
  {journal} {\bibinfo  {journal} {Nuclear Physics B}\ }\textbf {\bibinfo
  {volume} {220}},\ \bibinfo {pages} {415} (\bibinfo {year}
  {1983})}\BibitemShut {NoStop}%
\bibitem [{\citenamefont {Pisani}\ and\ \citenamefont
  {Smith}(1993)}]{PisaniSmith93}%
  \BibitemOpen
  \bibfield  {author} {\bibinfo {author} {\bibfnamefont {C.}~\bibnamefont
  {Pisani}}\ and\ \bibinfo {author} {\bibfnamefont {E.~R.}\ \bibnamefont
  {Smith}},\ }\bibfield  {title} {\bibinfo {title} {{Lee-Yang zeros and stokes
  phenomenon in a model with a wetting transition}},\ }\href
  {https://doi.org/10.1007/bf01048040} {\bibfield  {journal} {\bibinfo
  {journal} {Journal of Statistical Physics}\ }\textbf {\bibinfo {volume}
  {72}},\ \bibinfo {pages} {51} (\bibinfo {year} {1993})}\BibitemShut {NoStop}%
\bibitem [{\citenamefont {Rakovszky}\ \emph {et~al.}(2020)\citenamefont
  {Rakovszky}, \citenamefont {Sala}, \citenamefont {Verresen}, \citenamefont
  {Knap},\ and\ \citenamefont {Pollmann}}]{Rakovszky20}%
  \BibitemOpen
  \bibfield  {author} {\bibinfo {author} {\bibfnamefont {T.}~\bibnamefont
  {Rakovszky}}, \bibinfo {author} {\bibfnamefont {P.}~\bibnamefont {Sala}},
  \bibinfo {author} {\bibfnamefont {R.}~\bibnamefont {Verresen}}, \bibinfo
  {author} {\bibfnamefont {M.}~\bibnamefont {Knap}},\ and\ \bibinfo {author}
  {\bibfnamefont {F.}~\bibnamefont {Pollmann}},\ }\bibfield  {title} {\bibinfo
  {title} {{Statistical localization: From strong fragmentation to strong edge
  modes}},\ }\bibfield  {journal} {\bibinfo  {journal} {Physical Review B}\
  }\textbf {\bibinfo {volume} {101}},\ \href
  {https://doi.org/10.1103/physrevb.101.125126} {10.1103/physrevb.101.125126}
  (\bibinfo {year} {2020})\BibitemShut {NoStop}%
\bibitem [{\citenamefont {Zadnik}\ and\ \citenamefont
  {Fagotti}(2021)}]{Zadnik_foldedI}%
  \BibitemOpen
  \bibfield  {author} {\bibinfo {author} {\bibfnamefont {L.}~\bibnamefont
  {Zadnik}}\ and\ \bibinfo {author} {\bibfnamefont {M.}~\bibnamefont
  {Fagotti}},\ }\bibfield  {title} {\bibinfo {title} {{The Folded Spin-1/2
  {XXZ} Model: I. Diagonalisation, Jamming, and Ground State Properties}},\
  }\bibfield  {journal} {\bibinfo  {journal} {{SciPost} Physics Core}\ }\textbf
  {\bibinfo {volume} {4}},\ \href
  {https://doi.org/10.21468/scipostphyscore.4.2.010}
  {10.21468/scipostphyscore.4.2.010} (\bibinfo {year} {2021})\BibitemShut
  {NoStop}%
\bibitem [{\citenamefont {Zadnik}\ \emph {et~al.}(2021)\citenamefont {Zadnik},
  \citenamefont {Bidzhiev},\ and\ \citenamefont {Fagotti}}]{Zadnik_foldedII}%
  \BibitemOpen
  \bibfield  {author} {\bibinfo {author} {\bibfnamefont {L.}~\bibnamefont
  {Zadnik}}, \bibinfo {author} {\bibfnamefont {K.}~\bibnamefont {Bidzhiev}},\
  and\ \bibinfo {author} {\bibfnamefont {M.}~\bibnamefont {Fagotti}},\
  }\bibfield  {title} {\bibinfo {title} {{The folded spin-1/2 {XXZ} model:
  {II}. Thermodynamics and hydrodynamics with a minimal set of charges}},\
  }\bibfield  {journal} {\bibinfo  {journal} {{SciPost} Physics}\ }\textbf
  {\bibinfo {volume} {10}},\ \href
  {https://doi.org/10.21468/scipostphys.10.5.099}
  {10.21468/scipostphys.10.5.099} (\bibinfo {year} {2021})\BibitemShut
  {NoStop}%
\bibitem [{\citenamefont {Pozsgay}(2021)}]{Pozsgay_folded}%
  \BibitemOpen
  \bibfield  {author} {\bibinfo {author} {\bibfnamefont {B.}~\bibnamefont
  {Pozsgay}},\ }\bibfield  {title} {\bibinfo {title} {{A
  Yang{\textendash}Baxter integrable cellular automaton with a four site update
  rule}},\ }\href {https://doi.org/10.1088/1751-8121/ac1dbf} {\bibfield
  {journal} {\bibinfo  {journal} {Journal of Physics A: Mathematical and
  Theoretical}\ }\textbf {\bibinfo {volume} {54}},\ \bibinfo {pages} {384001}
  (\bibinfo {year} {2021})}\BibitemShut {NoStop}%
\bibitem [{\citenamefont {Pozsgay}\ \emph {et~al.}(2021)\citenamefont
  {Pozsgay}, \citenamefont {Gombor}, \citenamefont {Hutsalyuk}, \citenamefont
  {Jiang}, \citenamefont {Pristy{\'{a}}k},\ and\ \citenamefont
  {Vernier}}]{Pozsgay21}%
  \BibitemOpen
  \bibfield  {author} {\bibinfo {author} {\bibfnamefont {B.}~\bibnamefont
  {Pozsgay}}, \bibinfo {author} {\bibfnamefont {T.}~\bibnamefont {Gombor}},
  \bibinfo {author} {\bibfnamefont {A.}~\bibnamefont {Hutsalyuk}}, \bibinfo
  {author} {\bibfnamefont {Y.}~\bibnamefont {Jiang}}, \bibinfo {author}
  {\bibfnamefont {L.}~\bibnamefont {Pristy{\'{a}}k}},\ and\ \bibinfo {author}
  {\bibfnamefont {E.}~\bibnamefont {Vernier}},\ }\bibfield  {title} {\bibinfo
  {title} {{Integrable spin chain with Hilbert space fragmentation and solvable
  real-time dynamics}},\ }\bibfield  {journal} {\bibinfo  {journal} {Physical
  Review E}\ }\textbf {\bibinfo {volume} {104}},\ \href
  {https://doi.org/10.1103/physreve.104.044106} {10.1103/physreve.104.044106}
  (\bibinfo {year} {2021})\BibitemShut {NoStop}%
\bibitem [{\citenamefont {Gamayun}\ \emph {et~al.}(2023)\citenamefont
  {Gamayun}, \citenamefont {Hutsalyuk}, \citenamefont {Pozsgay},\ and\
  \citenamefont {Zvonarev}}]{Gamayun23}%
  \BibitemOpen
  \bibfield  {author} {\bibinfo {author} {\bibfnamefont {O.}~\bibnamefont
  {Gamayun}}, \bibinfo {author} {\bibfnamefont {A.}~\bibnamefont {Hutsalyuk}},
  \bibinfo {author} {\bibfnamefont {B.}~\bibnamefont {Pozsgay}},\ and\ \bibinfo
  {author} {\bibfnamefont {M.~B.}\ \bibnamefont {Zvonarev}},\ }\bibfield
  {title} {\bibinfo {title} {{Finite temperature spin diffusion in the Hubbard
  model in the strong coupling limit}},\ }\href
  {https://doi.org/10.21468/SciPostPhys.15.2.073} {\bibfield  {journal}
  {\bibinfo  {journal} {SciPost Phys.}\ }\textbf {\bibinfo {volume} {15}},\
  \bibinfo {pages} {073} (\bibinfo {year} {2023})}\BibitemShut {NoStop}%
\bibitem [{\citenamefont {Schmidt}\ and\ \citenamefont
  {Schadschneider}(2022)}]{KPZSchSch2022}%
  \BibitemOpen
  \bibfield  {author} {\bibinfo {author} {\bibfnamefont {J.}~\bibnamefont
  {Schmidt}}\ and\ \bibinfo {author} {\bibfnamefont {A.}~\bibnamefont
  {Schadschneider}},\ }\bibfield  {title} {\bibinfo {title} {{Mirror symmetry
  breakdown in the Kardar–Parisi–Zhang universality class}},\ }\href
  {https://arxiv.org/abs/2205.06062} {\bibfield  {journal} {\bibinfo  {journal}
  {arXiv preprint arXiv:2205.06062}\ } (\bibinfo {year} {2022})}\BibitemShut
  {NoStop}%
\bibitem [{\citenamefont {Flindt}\ and\ \citenamefont
  {Garrahan}(2013)}]{Flindt13}%
  \BibitemOpen
  \bibfield  {author} {\bibinfo {author} {\bibfnamefont {C.}~\bibnamefont
  {Flindt}}\ and\ \bibinfo {author} {\bibfnamefont {J.~P.}\ \bibnamefont
  {Garrahan}},\ }\bibfield  {title} {\bibinfo {title} {{Trajectory Phase
  Transitions, Lee-Yang Zeros, and High-Order Cumulants in Full Counting
  Statistics}},\ }\href {https://doi.org/10.1103/PhysRevLett.110.050601}
  {\bibfield  {journal} {\bibinfo  {journal} {Phys. Rev. Lett.}\ }\textbf
  {\bibinfo {volume} {110}},\ \bibinfo {pages} {050601} (\bibinfo {year}
  {2013})}\BibitemShut {NoStop}%
\bibitem [{\citenamefont {Peng}\ \emph {et~al.}(2015)\citenamefont {Peng},
  \citenamefont {Zhou}, \citenamefont {Wei}, \citenamefont {Cui}, \citenamefont
  {Du},\ and\ \citenamefont {Liu}}]{Peng15}%
  \BibitemOpen
  \bibfield  {author} {\bibinfo {author} {\bibfnamefont {X.}~\bibnamefont
  {Peng}}, \bibinfo {author} {\bibfnamefont {H.}~\bibnamefont {Zhou}}, \bibinfo
  {author} {\bibfnamefont {B.-B.}\ \bibnamefont {Wei}}, \bibinfo {author}
  {\bibfnamefont {J.}~\bibnamefont {Cui}}, \bibinfo {author} {\bibfnamefont
  {J.}~\bibnamefont {Du}},\ and\ \bibinfo {author} {\bibfnamefont {R.-B.}\
  \bibnamefont {Liu}},\ }\bibfield  {title} {\bibinfo {title} {{Experimental
  Observation of Lee-Yang Zeros}},\ }\bibfield  {journal} {\bibinfo  {journal}
  {Physical Review Letters}\ }\textbf {\bibinfo {volume} {114}},\ \href
  {https://doi.org/10.1103/physrevlett.114.010601}
  {10.1103/physrevlett.114.010601} (\bibinfo {year} {2015})\BibitemShut
  {NoStop}%
\bibitem [{\citenamefont {Brandner}\ \emph {et~al.}(2017)\citenamefont
  {Brandner}, \citenamefont {Maisi}, \citenamefont {Pekola}, \citenamefont
  {Garrahan},\ and\ \citenamefont {Flindt}}]{Brandner17}%
  \BibitemOpen
  \bibfield  {author} {\bibinfo {author} {\bibfnamefont {K.}~\bibnamefont
  {Brandner}}, \bibinfo {author} {\bibfnamefont {V.~F.}\ \bibnamefont {Maisi}},
  \bibinfo {author} {\bibfnamefont {J.~P.}\ \bibnamefont {Pekola}}, \bibinfo
  {author} {\bibfnamefont {J.~P.}\ \bibnamefont {Garrahan}},\ and\ \bibinfo
  {author} {\bibfnamefont {C.}~\bibnamefont {Flindt}},\ }\bibfield  {title}
  {\bibinfo {title} {{Experimental Determination of Dynamical Lee-Yang
  Zeros}},\ }\bibfield  {journal} {\bibinfo  {journal} {Physical Review
  Letters}\ }\textbf {\bibinfo {volume} {118}},\ \href
  {https://doi.org/10.1103/physrevlett.118.180601}
  {10.1103/physrevlett.118.180601} (\bibinfo {year} {2017})\BibitemShut
  {NoStop}%
\bibitem [{\citenamefont {Deger}\ \emph {et~al.}(2018)\citenamefont {Deger},
  \citenamefont {Brandner},\ and\ \citenamefont {Flindt}}]{Deger18}%
  \BibitemOpen
  \bibfield  {author} {\bibinfo {author} {\bibfnamefont {A.}~\bibnamefont
  {Deger}}, \bibinfo {author} {\bibfnamefont {K.}~\bibnamefont {Brandner}},\
  and\ \bibinfo {author} {\bibfnamefont {C.}~\bibnamefont {Flindt}},\
  }\bibfield  {title} {\bibinfo {title} {{Lee-Yang zeros and large-deviation
  statistics of a molecular zipper}},\ }\bibfield  {journal} {\bibinfo
  {journal} {Physical Review E}\ }\textbf {\bibinfo {volume} {97}},\ \href
  {https://doi.org/10.1103/physreve.97.012115} {10.1103/physreve.97.012115}
  (\bibinfo {year} {2018})\BibitemShut {NoStop}%
\bibitem [{\citenamefont {Deger}\ and\ \citenamefont {Flindt}(2019)}]{DF19}%
  \BibitemOpen
  \bibfield  {author} {\bibinfo {author} {\bibfnamefont {A.}~\bibnamefont
  {Deger}}\ and\ \bibinfo {author} {\bibfnamefont {C.}~\bibnamefont {Flindt}},\
  }\bibfield  {title} {\bibinfo {title} {{Determination of universal critical
  exponents using Lee-Yang theory}},\ }\bibfield  {journal} {\bibinfo
  {journal} {Physical Review Research}\ }\textbf {\bibinfo {volume} {1}},\
  \href {https://doi.org/10.1103/physrevresearch.1.023004}
  {10.1103/physrevresearch.1.023004} (\bibinfo {year} {2019})\BibitemShut
  {NoStop}%
\bibitem [{\citenamefont {Kist}\ \emph {et~al.}(2021)\citenamefont {Kist},
  \citenamefont {Lado},\ and\ \citenamefont
  {Flindt}}]{PhysRevResearch.3.033206}%
  \BibitemOpen
  \bibfield  {author} {\bibinfo {author} {\bibfnamefont {T.}~\bibnamefont
  {Kist}}, \bibinfo {author} {\bibfnamefont {J.~L.}\ \bibnamefont {Lado}},\
  and\ \bibinfo {author} {\bibfnamefont {C.}~\bibnamefont {Flindt}},\
  }\bibfield  {title} {\bibinfo {title} {Lee-yang theory of criticality in
  interacting quantum many-body systems},\ }\href
  {https://doi.org/10.1103/PhysRevResearch.3.033206} {\bibfield  {journal}
  {\bibinfo  {journal} {Phys. Rev. Research}\ }\textbf {\bibinfo {volume}
  {3}},\ \bibinfo {pages} {033206} (\bibinfo {year} {2021})}\BibitemShut
  {NoStop}%
\end{thebibliography}%

\end{document}